\documentclass[a4paper,11pt]{article}
\pdfoutput=1 

\usepackage{jheppub} 

\usepackage[T1]{fontenc} 

\usepackage{verbatim}
\usepackage{graphicx}

\usepackage{sectsty}
\allsectionsfont{\boldmath}

\newcommand{\eq}[1]{Eq.~(\ref{#1})}
\newcommand{\bib}[1]{Ref.~\cite{#1}}
\newcommand{\refs}[1]{Refs.~\cite{#1}}
\newcommand{\bibs}[1]{\cite{#1}}
\newcommand{\fig}[1]{Fig.~\ref{#1}}
\newcommand{\tab}[1]{Table~\ref{#1}}

\newcommand{\sect}[1]{Section~\ref{#1}}

\newcommand{\appen}[1]{Appendix~\ref{#1}}

\newcommand{\bea}{\begin{eqnarray}}
\newcommand{\eea}{\end{eqnarray}}

\newcommand{\eps}{\epsilon}

\newcommand{\crn}{\nonumber \\}

\newcommand{\fr}{\frac}

\newcommand{\gev}{{\unskip\,\text{GeV}}}
\newcommand{\tev}{{\unskip\,\text{TeV}}}

\newcommand{\fb}{{\unskip\,\text{fb}}}

\title{Fiducial polarization observables in hadronic WZ production: A
  next-to-leading order QCD+EW study}

\author[a,b]{Julien BAGLIO,}
\author[b,c,1]{LE Duc Ninh,\note{Corresponding author.}}

\affiliation[a]{Institut f\"{u}r Theoretische Physik, Eberhard Karls Universit\"{a}t
T\"{u}bingen,\\ Auf der Morgenstelle 14, D-72076 T\"{u}bingen, Germany}
\affiliation[b]{Institute For Interdisciplinary Research in Science and Education,\\  
ICISE, 590000 Quy Nhon, Vietnam}
\affiliation[c]{Humboldt-Universit\"at zu Berlin, Institut f\"ur Physik,\\
Newtonstra{\ss}e 15, D-12489 Berlin, Germany}

\emailAdd{julien.baglio@uni-tuebingen.de}
\emailAdd{ldninh@ifirse.icise.vn}

\preprint{IFIRSE-TH-2018-2}      

\abstract{We present a study at next-to-leading-order (NLO) of the
  process $pp \to W^\pm Z \to \ell \nu_l \ell'^+ \ell'^-$, where
  $\ell,\ell' =e, \mu$, at the Large Hadron Collider. We include the
  full NLO QCD corrections and the NLO electroweak (EW) corrections in
  the double-pole approximation. We define eight fiducial polarization coefficients 
  directly constructed from the polar-azimuthal angular distribution of the decay leptons. 
  These coefficients depend strongly 
  on the kinematical cuts on the transverse momentum or rapidity of the individual leptons.  
  Similarly, fiducial polarization fractions are also defined and they can be directly 
  related to the fiducial coefficients. We perform a detailed analysis of the
  NLO QCD+EW fiducial polarization observables including theoretical
  uncertainties stemming from the scale variation and parton
  distribution function uncertainties, using the fiducial phase space
  defined by the ATLAS and CMS experiments. We provide results in
  the helicity coordinate system and in the Collins-Soper coordinate
  system, at a center-of-mass energy of 13 TeV. The EW corrections are found to be 
  important in two of the angular coefficients related to the $Z$ boson, irrespective of 
  the kinematical cuts or the coordinate system. Meanwhile, 
  those EW corrections are very small for the $W^\pm$ bosons.}

\begin{document}
\maketitle
\flushbottom

\section{Introduction}
\label{sect:intro}

Since the Large Hadron Collider (LHC) at CERN has started to operate,
the production of electroweak gauge bosons has been extensively
studied both by theorists and experimentalists. With the accumulation
of data we can reach high precision measurements, thus probing new
physics effects in non-trivial observables such as in the polarization
of the gauge bosons. $W$
bosons only interact with left-handed quarks, while $Z$ bosons
interact with both left- and right-handed quarks, but with different
coupling strengths. This means that $W$ and $Z$ bosons produced at
hadron colliders are in principle polarized and that the angular
distributions of the final-state leptons display an asymmetry that
reflects the polarization of the underlying gauge bosons. 

The polarization of gauge bosons produced in hadron collider processes 
has been studied in the literature. At the LHC, $W$ bosons are produced 
abundantly in top quark decays or in association with jets, where the later 
channel is characterized by high transverse momentum $W$ bosons. 
The polarization of $W$ boson in the top quark decay has been measured 
by ATLAS \cite{Aad:2012ky} and CMS \cite{Khachatryan:2016fky}. 
For $W+\text{jets}$, polarization measurements have also been
performed by CMS \cite{Chatrchyan:2011ig} and ATLAS
\cite{ATLAS:2012au}. Similar studies for $Z$ boson polarization in
$Z+\text{jets}$ channel have been presented by CMS
\cite{Khachatryan:2015paa} and ATLAS \cite{Aad:2016izn}. Recent
theoretical studies for $W+\text{jets}$ have been presented in
Refs.~\cite{Bern:2011ie,Stirling:2012zt}. 

The study in \bib{Bern:2011ie} uses the helicity coordinate system in
which the angular observables for the $W$ boson are defined, namely
the $W$ boson rest frame where the $z$ direction is defined as the
direction of the $W$ boson in the laboratory frame. Another popular
coordinate system has been previously introduced in
\bib{Collins:1977iv}, called the Collins-Soper coordinate system, in
which the $z$ direction is defined as the bisection of the flight
direction of the two incoming protons in the $W$ boson rest frame. It
is noted that both ATLAS and CMS use the helicity coordinate system
for $W+\text{jets}$ and Collins-Soper coordinate system for
$Z+\text{jets}$. Following \bib{Collins:1977iv} there has been a
number of phenomenological studies of the spin-density matrix of the
$W$ boson~\cite{Lam:1980uc,Mirkes:1992hu,Aguilar-Saavedra:2015yza} as
well as of the $Z$
boson~\cite{Mirkes:1994eb,Aguilar-Saavedra:2017zkn}, that relate to
the corresponding angular coefficients. One-loop QCD effects have
also been studied in Refs.~\cite{Hagiwara:1984hi,Hagiwara:2006qe} and
up to next-to-next-to-leading order (NNLO) in QCD in Drell-Yan $Z$
production~\cite{Karlberg:2014qua}.

The production of $W^\pm Z$ at a hadron collider has been extensively
studied in the literature. For on-shell (OS) production,
next-to-leading order (NLO) QCD corrections have been calculated in
Refs.~\cite{Ohnemus:1991gb,Frixione:1992pj}. The full NLO electroweak (EW)
corrections including quark-photon induced correction, which is now
recognized to be important, were first calculated in
\bib{Baglio:2013toa}. The virtual and real photon emission corrections
have been also been calculated in \bib{Bierweiler:2013dja}, almost at
the same time. NNLO QCD corrections for both on-shell and off-shell
cases have been presented in \refs{Grazzini:2016swo,Grazzini:2017ckn} and very recently
full NLO EW corrections including off-shell effects for $3\ell\nu$ final
state have been calculated in \bib{Biedermann:2017oae}, which confirms
the importance of the quark-photon induced correction. We note that
full NLO QCD calculations including full off-shell and
spin-correlation effects for leptonic final states have been
implemented in computer programs such as {\tt MCFM}
\cite{Campbell:1999ah} and {\tt VBFNLO} \cite{Arnold:2008rz}. Recent
measurements of the cross section at $13\tev$ have been performed by
ATLAS \cite{Aaboud:2016yus} and CMS
\cite{Khachatryan:2016tgp}. Results for kinematical distributions at
$8\tev$ have also been presented by ATLAS \cite{Aad:2016ett} and CMS
\cite{Khachatryan:2016poo}.

The study of gauge boson polarization effects in $W^\pm Z$ production
together with other processes also started quite a while ago with
leading-order (LO) predictions in the
eighties~\cite{Bilchak:1984gv,Willenbrock:1987xz}. A more modern study
of polarization of gauge bosons produced at the LHC via various
channels including $WZ$ has been performed in
\bib{Stirling:2012zt}. To the best of our knowledge, no detailed study
of NLO QCD and EW corrections on polarization
observables in $WZ$ production at a hadron collider has been
performed.

Compared to $V+\text{jets}$ (with $V=W,Z$) production, the cross
sections for diboson channels are much smaller, therefore polarization
effects are much more difficult to be measured. However, very recently
ATLAS has presented a study of angular observables in $W^\pm Z$
production at the 13 TeV LHC \cite{ATLAS:2018ogj,Aaboud:2019gxl}. This indicates that
it is now possible to perform detailed studies and comparisons with
measurements for polarization observables in diboson production at the
LHC.

In the experiments, the polarization observables are measured using polar-azimuthal 
angular distribution of a decay charged lepton. In the first step, this distribution 
is measured in the fiducial phase space using cuts on the transverse momentum and rapidity 
of the decay lepton. The off-shell, interference and radiation effects are here included.  
Experimentalists then fit this distribution using a template 
fitting method to find the polarization fractions, see e.g. \bib{ATLAS:2012au}. 
The helicity templates are calculated using Monte-Carlo generators. 
For processes where on-shell effects are dominant 
(e.g. Drell-Yan or diboson production), 
we expect that the measurements are not so far away from the on-shell approximated values.  
In this context, it is important to note that the choice of the coordinate system is 
important as the results depend on it. 

From the theory side, the 
polar-azimuthal angular distribution of the decay lepton can 
also be calculated with the same fiducial cuts and with those 
off-shell, interference and radiation effects included. 
To compare to the measurements, we then have to do the same template fitting method. 
This is not easy to do in practice and we do not know of any theoretical papers 
doing this step. The simplest thing for theorists to do is to use the on-shell approximation 
or using the angular distribution of the decay lepton with an inclusive phase-space cut 
(i.e. without restriction on the individual decay lepton phase space) 
as done e.g. in \refs{Bern:2011ie,Stirling:2012zt}. However, we expect that 
this can only provide a rough comparison to the measurements. 

We discuss in this paper a set of {\em fiducial} polarization 
observables \footnote{These observables are also discussed in \bib{Stirling:2012zt}, 
where they are called projection results. See also \bib{Mirkes:1994eb}.} which are defined using the same 
polar-azimuthal angular distribution of the decay lepton with arbitrary fiducial cuts, parameterized also 
by eight coefficients. These coefficients are not the usual polarization angular coefficients, 
and hence are called fiducial angular coefficients in this paper. From these coefficients, three 
fiducial fractions can be easily calculated. 
In the limit of an inclusive phase-space cut, e.g. $66 < m_{\ell'^+ \ell'^-} < 116\gev$, 
the two notions of fiducial angular coefficients and inclusive angular coefficients 
coincide. The differences between them are thus due to the kinematical cuts 
on the individual decay leptons. We will see therefore some similarities between them. 
We will also show that the fiducial longitudinal polarization fraction
calculated in the helicity coordinate system decreases at large
$p_{T,V}$, as the inclusive polarization fraction does according to the
equivalence theorem. 

The goal of this paper is to provide NLO QCD+EW predictions for the fiducial 
polarization observables in the process $p p \to W^\pm_{} Z \to \ell
\nu_l \ell'^+ \ell'^-$ channel at the 13 TeV LHC, where $\ell,\ell'
=e, \mu$. The NLO QCD corrections will be calculated using the program
{\tt VBFNLO}~\cite{Arnold:2008rz,Arnold:2011wj,Baglio:2014uba}
including full off-shell effects, while the EW corrections will be
calculated using a double-pole approximation (DPA). Spin-correlation
effects are fully taken into account in the EW corrections, but the
off-shell effects are missing. We will build our approximation on a
minimal extension of the OS $2\to 2$ calculation presented in
\bib{Baglio:2013toa}. In order to judge how good our approximation is,
we will also compare the results of our DPA with the full results
presented in \bib{Biedermann:2017oae}. We will provide results for the
fiducial cuts defined by ATLAS \cite{Aaboud:2016yus} and CMS
\cite{Khachatryan:2016tgp} at $13\tev$, in both the helicity and
Collins-Soper coordinate systems. Theoretical errors including both
parton distribution function (PDF) and scale uncertainties are
calculated. As a by-product, we present also results for fiducial
cross sections and standard kinematical distributions at NLO QCD+EW
with full theoretical uncertainties.

An advantage of our DPA calculation, compared to the full calculation,
is that EW corrections to the production and to the decay of a gauge
boson (either $W$ or $Z$) are completely separated, because off-shell
effects are neglected. 
The photon radiation off the decay lepton effects on polarization observables are
interesting because it helps us to know whether the results obtained 
using an on-shell gauge boson production approximation are 
good estimates of the measurements. 
The effects of NLO EW corrections to the decay mode and to the production
mode will be separately presented in this work. Effects from the
quark-photon induced contribution, which is sensitive to the photon
distribution function, will be separated as well.

The paper is organized as follows. In \sect{sect:cal} we present the
calculational details, in particular discussing the DPA calculation
and defining the fiducial polarization observables. In \sect{sect:numset} we
provide our numerical setup and how the theoretical uncertainties are
calculated. Since the paper is very long with a lot of numerical
results, we present predictions for the $W^+Z$ channel with ATLAS and
CMS fiducial cuts in the main sections. Similar results for the $W^-Z$
channel are provided in the appendices. Our predictions for the
fiducial cross sections and differential distributions are presented
in \sect{sect:numset:ewcorrections}. Results for the fiducial polarization
observables are provided in \sect{sect:numres} and conclusions are
given in \sect{sect:conclusion}. \appen{appen:DPA_details} provides
the details of our NLO EW calculation in the DPA. Kinematical
distributions for the $W^-Z$ channels are given in
\appen{appen:kin_dist_Wm}, and numerical results for the fiducial polarization
observables for the $W^-Z$ channel in
\appen{appen:numres_Wm}. Finally,
\appen{appen:off_shell_NLOEW_effects} contains the results of the fiducial
polarization observables with various EW correction effects
separated. Off-shell effects at LO can also be seen there by comparing
the full LO results to the DPA LO ones.

\section{Calculational details}
\label{sect:cal}

We consider the process
\bea
p + p \to \ell_1(k_1) + \ell_2(k_2) + \ell_3
(k_3) + \ell_4(k_4) + X,
\label{eq:proc1}
\eea
where the final-state leptons can be either $e^+\nu_e\mu^+\mu^-$ or 
$e^-\bar{\nu}_e\mu^+\mu^-$.  

\begin{figure}[t!]
\includegraphics[width=\textwidth]{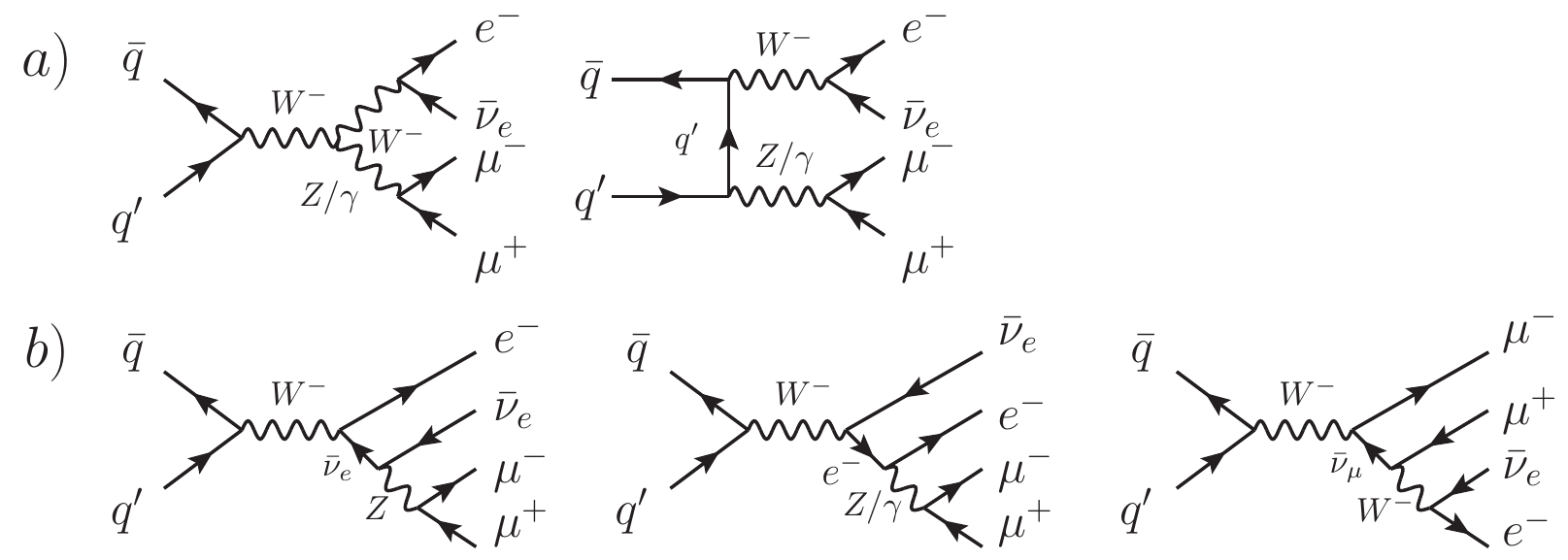}
\caption{LO Feynman diagrams for the partonic process $\bar{q}q' \to
  e^- \bar{\nu}_e\, \mu^+\mu^-$. The diagrams for the opposite-charge
  process $\bar{q}q' \to e^+ \nu_e\, \mu^+\mu^-$ are similar but with
  reversed fermion flows in the final state. a) Doubly-resonant $W^-
  Z/W^-\gamma$ diagrams; b) Diagrams with singly-resonant electroweak
  gauge bosons.
  \label{diags:LO}}
\end{figure}

At LO and NLO in QCD, we will consider the full contributions: The
double-pole contributions with intermediate-state $V_1 V_2 = W^\pm Z$
as well as the off-shell contributions with singly-resonant
electroweak-gauge-boson states. There are no contributions from the
third-generation quarks in the initial state. The main production
mechanism at proton-proton colliders proceeds via quark-antiquark
annihilations as shown in \fig{diags:LO}. The representative Feynman
diagrams for the double-pole contributions are displayed in
\fig{diags:LO}a) while the singly-resonant contributions are displayed
in \fig{diags:LO}b). These  contributions have been known for
decades~\cite{Ohnemus:1991gb,Frixione:1992pj}.

The LO hadronic differential cross section is calculated by a
convolution between the partonic quark-antiquark annihilation
differential cross section $d\hat{\sigma}_{\rm LO}^{\bar{q}q'}$ and
the PDFs of the first- and second-generation quarks in the proton. The
PDFs, denoted $\bar{q}$ and $q'$ below, are functions of the momentum
fraction $x$ carried by the quark in the corresponding proton, and of
the factorization scale $\mu_F$ which defines the scale at which this
convolution is performed. The LO hadronic cross section reads
\begin{align}
d\sigma_{\rm LO} = \int d x_1 d x_2 \Big[ \bar{q}(x_1,\mu_F)
  q'(x_2,\mu_F) d\hat{\sigma}_{\rm LO}^{\bar{q}q'} + (1\leftrightarrow
  2)\Big].
\end{align}
In the following we will present the NLO QCD corrections and the tools
that have been used for the calculation of the LO and NLO differential
cross sections, and then we will focus specifically on the calculation
of the EW corrections in the double-pole-approximation framework.

\subsection{NLO QCD corrections}
\label{sect:cal:qcd}
The NLO QCD corrections can be divided into the virtual corrections
containing one gluon loop, and the real corrections in which one extra
parton (quark, antiquark, or gluon) is included in the final state. As
the final state we consider is purely leptonic, the virtual gluon can
only run between the initial-state quark/antiquark pair.

The NLO QCD corrections have been calculated for on-shell production
for the first time in \refs{Ohnemus:1991gb,Frixione:1992pj}, and then
extended in
\refs{Ohnemus:1994ff,Dixon:1998py,Dixon:1999di,Campbell:1999ah,Campbell:2011bn}
to include full off-shell effects and spin correlations. The NNLO QCD
corrections have been calculated in \bib{Grazzini:2016swo} and have
been found to be of the order of an $8\%$ to $11\%$ increase of the
cross section, depending on the collider energy. We limit our analysis
in this paper at NLO, hence we do not include the NNLO QCD
contributions in the final analysis. As the perturbative development
is truncated at a fixed order, the cross section depends on the two
unphysical scales $\mu_R$ and $\mu_F$, the former being the scale
entering the loop functions and at which the strong coupling constant
$\alpha_s$ is calculated, the latter being the scale at which the PDFs
are evaluated and occurring in the real corrections. Our central scale
choice is the natural scale of the process, $\displaystyle \mu_R =
\mu_F = \mu_0 \equiv \left(M_W + M_Z\right)/2$. The pattern of the
NNLO corrections also motivate the value chosen for the central scale,
as they are moderate and positive for this value of $\mu_0$.

We use the computer program {\tt VBFNLO
  2.7.1}~\bibs{Arnold:2008rz,Arnold:2011wj,Baglio:2014uba} to
calculate both the LO and NLO cross sections and kinematical
distributions. The implementation of the QCD corrections in this
program is based on the Catani-Seymour dipole subtraction
algorithm~\bibs{Catani:1996vz} to combine the virtual and the real contributions. We will use the Hessian NNLO PDF set {\tt
  LUXqed17\char`_plus\char`_PDF4LHC15\char`_nnlo\char`_30}~\bibs{Manohar:2016nzj,Manohar:2017eqh}
based on the Monte-Carlo fit {\tt
  PDF4LHC15}~\bibs{Butterworth:2015oua,Dulat:2015mca,Harland-Lang:2014zoa,Ball:2014uwa,Gao:2013bia,Carrazza:2015aoa,Watt:2012tq},
using the QED evolution of the splitting functions described in
\bib{deFlorian:2015ujt}. We use the library {\tt LHAPDF
  6}~\bibs{Buckley:2014ana} and $\alpha_s\left(M_Z^2\right) = 0.118$
as given by {\tt
  LUXqed17\char`_plus\char`_PDF4LHC15\char`_nnlo\char`_30}. It is
noted that the same PDF set is also used for LO and NLO EW
results. The LO calculation has also been cross-checked against an
independent calculation which is also part of the NLO EW contribution
discussed in the next sub-section.

\subsection{NLO EW corrections in the double-pole approximation}
\label{sect:cal:ew}

In the framework of the double-pole approximation, the amplitude is 
built using an on-shell gauge boson approximation. This is important to
guarantee that the final result is gauge invariant. Therefore,
\eq{eq:proc1} is now approximated as follows
\bea
p + p \to V_1(q_1) + V_2(q_2) \to \ell_1(k_1) + \ell_2(k_2) + \ell_3
(k_3) + \ell_4(k_4) + X,
\label{eq:proc1_VV}
\eea
where the intermediate gauge bosons ($V_1=W^\pm$, $V_2=Z$) are massive
and the momenta satisfy the following relations:
\bea
q_1 = k_1 + k_2, \quad q_2 = k_3 + k_4, \quad k_i^2 = 0,\; i=1,4,
\label{eq:mom1}
\eea
At the partonic level we have
\bea
\bar{q}(p_1) + q'(p_2) \to V_1(q_1) + V_2(q_2) \to \ell_1(k_1) + \ell_2(k_2)
+ \ell_3 (k_3) + \ell_4(k_4).
\label{eq:proc2}
\eea
Since $V_j$ (with $j=1,2$ denoting the two gauge bosons) are on-shell,
we have to map the momenta ($k_i$, $q_{1,2}$) to an OS momentum basis
($\hat{k}_i$, $\hat{q}_{1,2}$) that has the following properties
\bea
\hat{q}_1^2 &=& M_{V_1}^2, \quad \hat{q}_2^2 = M_{V_2}^2,\crn
\hat{q}_1 &=& \hat{k}_1 + \hat{k}_2, \quad \hat{q}_2 = \hat{k}_3 +
\hat{k}_4, \quad \hat{k}_i^2 = 0.
\label{eq:mom2}
\eea
This mapping is not unique. However, it has been pointed out in
\bib{Denner:2000bj} that different mapping choices lead to differences
of the order of $\alpha\Gamma_V/(\pi M_V)$. Details of the OS mappings
used in this paper are provided in
\appen{appen:DPA_details}. Comparisons of different mappings are
presented in \tab{tab:compare_OSmap_LO} for LO cross sections and in
\tab{tab:compare_decayV_NLOEW} for NLO EW corrections.

At LO, the amplitude in the DPA is defined as (see
e.g. \bib{Denner:2000bj})
\bea
\mathcal{A}_\text{LO,DPA}^{\bar{q}q'\to V_1V_2\to 4l} = \fr{1}{Q_1Q_2}
\sum_{\lambda_1,\lambda_2}
\mathcal{A}_\text{LO}^{\bar{q}q'\to V_1V_2}\mathcal{A}_\text{LO}^{V_1\to
    \ell_1\ell_2}\mathcal{A}_\text{LO}^{V_2\to \ell_3\ell_4}
,\label{eq:LO_DPA}
\eea
where 
\bea
Q_j = q_j^2 - M_{V_j}^2 + iM_{V_j}\Gamma_{V_j}.
\label{eq:Qi_def}
\eea
We note that all helicity amplitudes in the numerator are calculated
with the DPA kinematics denoted by a hat. The polarization vectors in
the production and decay amplitudes are physical by definition. They
satisfy the following condition 
\bea
\sum_{\lambda_j=1}^{3}\eps^\mu(\hat{q}_j,\lambda_j)\eps^{*\nu}(\hat{q}_j,\lambda_j) 
= -g^{\mu\nu} + \fr{\hat{q}_j^\mu \hat{q}_j^\nu}{M_{V_j}^2}.
\eea 
It is important that the same definition is used for the polarization
vectors in the production and decay amplitudes. In this way, all spin
correlations are properly taken into account. It is obvious that the
definition in \eq{eq:LO_DPA} is gauge invariant because all the
amplitudes on the right-hand side are individually gauge
invariant. The helicity amplitudes for the production part
$\mathcal{A}_\text{LO}^{\bar{q}q'\to V_1V_2}$ have been calculated in
our previous work \cite{Baglio:2013toa} and can be taken over. For the
decay amplitudes, we use the program {\tt FormCalc}
\cite{Hahn:1998yk,Hahn:2000kx} to generate them.

For integrated cross section, we have to take care of the two
resonances of the intermediate gauge bosons, i.e. the denominator in
\eq{eq:LO_DPA}. Even though it is integrable due to finite value of
the widths, the phase-space integration can be more efficiently done
using an appropriate mapping to smooth out the Breit-Wigner
distributions. This step is also done in the {\tt VBFNLO} program.

From the above definition, it becomes clear that the DPA is limited by
the following factors. Not all Feynman diagrams are included, only the
ones that are enhanced by two resonant weak bosons are selected,
off-shell effects are missing, and the kinematics, which enter the
matrix elements, are not exact. In particular, the DPA is only valid
when the partonic center-of-mass energy is high enough, i.e.
\bea
\sqrt{\hat{s}} = \sqrt{(p_1+p_2)^2} > M_W + M_Z.
\label{eq:cond_DPA}
\eea

\begin{figure}[t!]
\includegraphics[width=\textwidth]{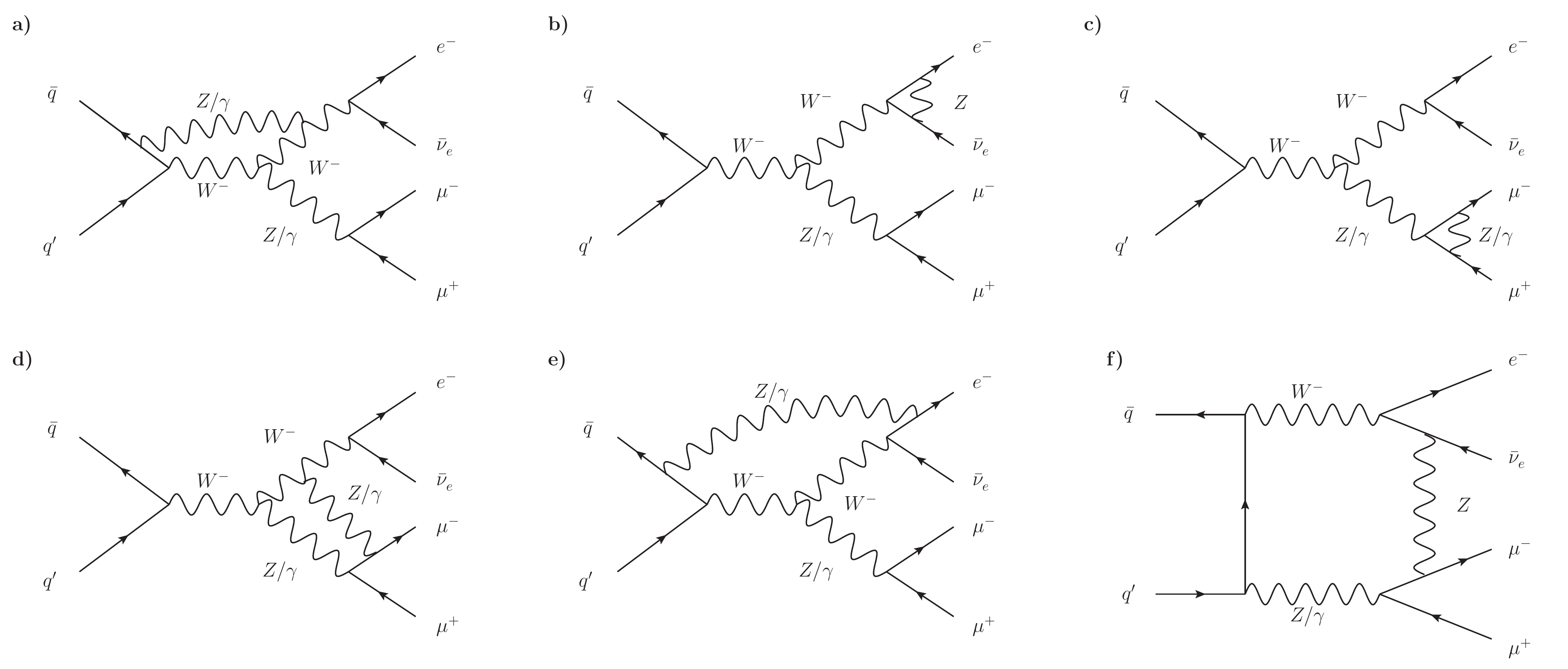}
\caption{Representative Feynman diagrams for the NLO EW virtual corrections for the partonic process $\bar{q}q' \to
  e^- \bar{\nu}_e\, \mu^+\mu^-$. Factorizable diagrams are in the top row, while 
non-factorizable ones in the bottom row. 
  \label{diags:NLOEW}}
\end{figure}
We use the same principles to build the NLO EW corrections in the
DPA. For this, we have to calculate the virtual and real
corrections. EW corrections to both production and decay parts are
separately included. However, non-factorizable corrections are neglected, since they are expected
to be very small \bibs{Beenakker:1997bp,Denner:1997ia,Denner:1998rh}. 
The non-factorizable contribution includes all Feynman diagrams that 
are not parts of the on-shell $WZ$ production group or of the on-shell $V$ decay groups. 
These diagrams are displayed in \fig{diags:NLOEW}, where a) belongs to the 
$WZ$ production group, b) the $W$ decay group, c) the $Z$ decay group, and d), e), f) 
the non-factorizable group. The corresponding photon-emission and quark-photon induced 
diagrams of the factorizable groups are fully taken into account. 

The master formulas for
the virtual, real-photon emission, and quark-photon induced amplitudes
are schematically written as follows,
\begin{align}
\delta\mathcal{A}_\text{virt,DPA}^{\bar{q}q'\to V_1V_2 \to 4l} &=\fr{1}{Q_1Q_2}
\sum_{\lambda_1,\lambda_2}
\left(\delta\mathcal{A}_\text{virt}^{\bar{q}q'\to
    V_1V_2}\mathcal{A}_\text{LO}^{V_1\to
    \ell_1\ell_2}\mathcal{A}_\text{LO}^{V_2\to
    \ell_3\ell_4}\right.\crn
&\left.+\mathcal{A}_\text{LO}^{\bar{q}q'\to
    V_1V_2}\delta\mathcal{A}_\text{virt}^{V_1\to
    \ell_1\ell_2}\mathcal{A}_\text{LO}^{V_2\to
    \ell_3\ell_4}
+\mathcal{A}_\text{LO}^{\bar{q}q'\to
    V_1V_2}\mathcal{A}_\text{LO}^{V_1\to
    \ell_1\ell_2}\delta\mathcal{A}_\text{virt}^{V_2\to
    \ell_3\ell_4}
\right),\label{eq:virt_DPA}\\
\delta\mathcal{A}_\text{rad,DPA}^{\bar{q}q'\to V_1V_2 \to 4l\gamma} &=
\sum_{\lambda_1,\lambda_2}
\left(\fr{\delta\mathcal{A}_\text{rad}^{\bar{q}q'\to
    V_1V_2\gamma}\mathcal{A}_\text{LO}^{V_1\to
    \ell_1\ell_2}\mathcal{A}_\text{LO}^{V_2\to \ell_3\ell_4}}{Q_1Q_2}\right.\crn
&\left.+\fr{\mathcal{A}_\text{LO}^{\bar{q}q'\to
    V_1V_2}\delta\mathcal{A}_\text{rad}^{V_1\to
    \ell_1\ell_2\gamma}\mathcal{A}_\text{LO}^{V_2\to \ell_3\ell_4}}{Q'_1Q_2}
+\fr{\mathcal{A}_\text{LO}^{\bar{q}q'\to
    V_1V_2}\mathcal{A}_\text{LO}^{V_1\to
    \ell_1\ell_2}\delta\mathcal{A}_\text{rad}^{V_2\to \ell_3\ell_4\gamma}}{Q_1Q'_2}\right),
\label{eq:rad_DPA}\\
\delta\mathcal{A}_\text{ind,DPA}^{q\gamma\to V_1V_2 \to 4l q'} &=
\sum_{\lambda_1,\lambda_2}
\fr{\delta\mathcal{A}_\text{ind}^{q\gamma\to
    V_1V_2q'}\mathcal{A}_\text{LO}^{V_1\to
    \ell_1\ell_2}\mathcal{A}_\text{LO}^{V_2\to \ell_3\ell_4}}{Q_1Q_2},
\label{eq:ind_DPA}
\end{align}
where the correction amplitudes
$\delta\mathcal{A}_\text{virt}^{\bar{q}q'\to V_1V_2}$,
$\delta\mathcal{A}_\text{rad}^{\bar{q}q'\to V_1V_2\gamma}$, and
$\delta\mathcal{A}_\text{ind}^{q\gamma\to V_1V_2q'}$ have been
calculated in the OS production calculation in \bib{Baglio:2013toa}
and are reused here. The missing pieces related to the corrections to
the decay amplitudes are generated again by {\tt FormCalc} and
combined together using the dipole-subtraction method
\cite{Catani:1996vz,Dittmaier:1999mb,Basso:2015gca}.
The new variables $Q'_1$ and $Q'_2$ are defined as in 
\eq{eq:Qi_def} but with the gauge-boson momenta being 
reconstructed from the $1 \to 3$ decays. 
For the cross-section contributions, we have
\begin{align}
\Delta\sigma_\text{virt,DPA}^{\bar{q}q'\to V_1V_2 \to 4l} &\propto
2\text{Re}\left[\delta\mathcal{A}_\text{virt,DPA}^{\bar{q}q'\to V_1V_2 \to 4l} \mathcal{A}_\text{LO,DPA}^{\bar{q}q'\to V_1V_2 \to 4l \star}\right]
\label{eq_sigma_virtEW},\\
\Delta\sigma_\text{rad,DPA}^{\bar{q}q'\to V_1V_2 \to 4l\gamma} &\propto
|\delta\mathcal{A}_\text{rad,DPA}^{\bar{q}q'\to V_1V_2 \to 4l\gamma}|^2\;\; \text{with interference terms neglected}
\label{eq_sigma_radEW},\\
\Delta\sigma_\text{ind,DPA}^{q\gamma\to V_1V_2 \to 4l q'} &\propto 
|\delta\mathcal{A}_\text{ind,DPA}^{q\gamma\to V_1V_2 \to 4l q'}|^2 
\label{eq_sigma_indEW}.
\end{align}
For later use, the following corrections coming from the radiative decays are defined. 
\begin{align}
\Delta\sigma^\text{virt}_{\text{d}V_1} &\propto 
2\text{Re}\left[\left(\sum_{\lambda_1,\lambda_2}
\fr{\mathcal{A}_\text{LO}^{\bar{q}q'\to
    V_1V_2}\delta\mathcal{A}_\text{virt}^{V_1\to
    \ell_1\ell_2}\mathcal{A}_\text{LO}^{V_2\to
    \ell_3\ell_4}}{Q_1Q_2}\right) \mathcal{A}_\text{LO,DPA}^{\bar{q}q'\to V_1V_2 \to 4l \star}\right]
\label{eq_sigma_V1decayEW_virt},\\
\Delta\sigma^\text{rad}_{\text{d}V_1} &\propto 
|\sum_{\lambda_1,\lambda_2}\fr{\mathcal{A}_\text{LO}^{\bar{q}q'\to
    V_1V_2}\delta\mathcal{A}_\text{rad}^{V_1\to
    \ell_1\ell_2\gamma}\mathcal{A}_\text{LO}^{V_2\to \ell_3\ell_4}}{Q'_1Q_2}|^2
\label{eq_sigma_V1decayEW_rad},\\
\Delta\sigma^\text{virt}_{\text{d}V_2} &\propto 
2\text{Re}\left[\left(\sum_{\lambda_1,\lambda_2}
\fr{\mathcal{A}_\text{LO}^{\bar{q}q'\to
    V_1V_2}\mathcal{A}_\text{LO}^{V_1\to
    \ell_1\ell_2}\delta\mathcal{A}_\text{virt}^{V_2\to
    \ell_3\ell_4}}{Q_1Q_2}\right) \mathcal{A}_\text{LO,DPA}^{\bar{q}q'\to V_1V_2 \to 4l \star}\right]
\label{eq_sigma_V2decayEW_virt},\\
\Delta\sigma^\text{rad}_{\text{d}V_2} &\propto 
|\sum_{\lambda_1,\lambda_2}\fr{\mathcal{A}_\text{LO}^{\bar{q}q'\to
    V_1V_2}\mathcal{A}_\text{LO}^{V_1\to
    \ell_1\ell_2}\delta\mathcal{A}_\text{rad}^{V_2\to \ell_3\ell_4\gamma}}{Q_1Q'_2}|^2
\label{eq_sigma_V2decayEW_rad}.
\end{align}
Further technical details on how the momenta and the amplitudes are calculated are provided in \appen{appen:DPA_details}.

From the above terms we define schematically some important EW
corrections to understand various effects as follows
\begin{align}
\delta_{\bar{q}q'} &= \left(\Delta\sigma_\text{virt,DPA}^{\bar{q}q'\to V_1V_2 \to 4l} + \Delta\sigma_\text{rad,DPA}^{\bar{q}q'\to V_1V_2 \to 4l\gamma}\right)/\sigma_\text{LO},\label{eq:qq_corrections}\\
\delta_{q\gamma} &= \Delta\sigma_\text{ind,DPA}^{q\gamma\to V_1V_2 \to 4l q'}/\sigma_\text{LO},\label{eq:qgamma_corrections}\\ 
\delta_\text{NLOEW} &= \delta_{\bar{q}q'} + \delta_{q\gamma},\label{eq:NLOEW_corrections}\\
\delta_{\text{d}V_1} &= (\Delta\sigma^\text{virt}_{\text{d}V_1} + \Delta\sigma^\text{rad}_{\text{d}V_1} )/\sigma_\text{LO},\label{eq:Vdecay_corrections}\\
\delta_{\text{p}V_1} &= \delta_\text{NLOEW} - \delta_{\text{d}V_1},
\label{eq:Vproduction_corrections}
\end{align}
and $\delta_{\text{d}V_2}$, $\delta_{\text{p}V_2}$ for the second
gauge boson are similarly defined. $\delta_{q\gamma}$ is interesting
because it is sensitive to the photon PDF and can be large. This
correction is also provided in \bib{Biedermann:2017oae}, so that a
numerical comparison will be later performed. $\delta_{\text{d}V_j}$
(total correction to $V_j$ decay) and $\delta_{\text{p}V_j}$ (total
correction to $V_j$ production) are interesting for polarization
observables of the $V_j$ boson as mentioned in the introduction. These
effects will be presented in
\tab{table:Xsection_Born_EW_cor_ATLAS_CMS} and in
\appen{appen:off_shell_NLOEW_effects}.

In the calculation of polarization observables, the LO results must be
always included. By default, our NLO EW results are the sum of the
full LO results and the EW corrections calculated in the DPA. 
In addition, if not explicitly mentioned, NLO means NLO QCD and EW. 

\subsection{Definition of fiducial polarization observables}
\label{sect:cal:polar_observables}

At LO in the DPA, the angular distribution of a final-state lepton 
created by an on-shell massive gauge boson is described as 
\begin{align}
\fr{d\sigma}{\sigma d\!\cos\theta d\phi} &= \fr{3}{16\pi}
\Big[ 
(1+\cos^2\theta) + A_0 \fr{1}{2}(1-3\cos^2\theta)
+ A_1 \sin(2\theta)\cos\phi  \crn
& + A_2 \fr{1}{2} \sin^2\theta \cos(2\phi)
+ A_3 \sin\theta\cos\phi + A_4 \cos\theta \crn
& + A_5 \sin^2\theta \sin(2\phi) 
+ A_6 \sin(2\theta) \sin\phi + A_7 \sin\theta \sin\phi
\Big],\label{eq:definition_Ai_tot}
\end{align}
where $A_{0-7}$ are dimensionless angular coefficients, $\theta$ and
$\phi$ are the lepton polar and azimuthal angles, respectively, in the
rest frame of the massive gauge boson in a particular coordinate
system that needs to be specified. In the case of the charged lepton
coming from the $W$ decay, we set the notation $\theta = \theta_3$, $\phi = \phi_3$. 
For the negatively charged lepton coming from the $Z$ decay, we set $\theta =
\theta_6$, $\phi = \phi_6$. We note that the rest frame of the $W$ bosons 
cannot be reconstructed in experiments as the longitudinal momentum 
of the neutrino is unknown. However, this can be calculated if the 
on-shell condition $(p_e + p_{\nu_e})^2 = M_W^2$ is imposed and then choosing 
the solution with smaller magnitude \cite{ATLAS:2018ogj}. This step is not done 
in the present paper but should be done if comparisons with measurements are performed. 

It is important to note that \eq{eq:definition_Ai_tot} is only correct if there is 
no restriction on the phase space of the individual leptons. 
We will use the term polarization observables to refer to 
the coefficients $A_i$ or the polarization fractions below defined. 

Polarizations of the gauge bosons can be described using a spin-density matrix. 
In the DPA and at LO, for the process at hand, this matrix reads
\begin{align}
\rho^{V_j}_{\lambda\lambda'} = C\sum_{s_q,s_l} 
\mathcal{A}_{\bar{q}q'\to l_il'_i V_j}^*(\lambda,s_q,s_l)\mathcal{A}_{\bar{q}q'\to l_il'_i V_j}(\lambda',s_q,s_l),
\label{eq:spin_matrix_DPA}
\end{align}
where $s_q$ and $s_l$ denote the set of quark and lepton helicities,
respectively and $C$ is a normalization factor determined from the
condition $\text{Tr}(\rho^{V_j})=1$. Since $\rho^{V_j}$ is
Hermitian, the spin-density matrix is parameterized by eight coefficients,
equivalent to the definition by \eq{eq:definition_Ai_tot}. It is noted
that the matrix $\rho^{V_j}$ is independent of the decay $V_j \to l_j
l'_j$. However, when we calculate the angular coefficients by using
\eq{eq:definition_Ai_tot} at NLO EW, they receive contributions from
EW corrections to the decays. These effects are therefore interesting
and deserve special attention.  

At LO in the EW coupling and for a single $V$ resonance, direct relations between the
angular coefficients and the elements of the spin-density matrix of
the gauge boson can be proved as shown in
\refs{Aguilar-Saavedra:2015yza,Ballestrero:2017bxn} for the $W$ boson and in
\bib{Aguilar-Saavedra:2017zkn} for the $Z$ boson. This is the reason
why the angular coefficients are also called spin or polarization
observables. We give here the explicit relations between the angular
coefficients and the spin-density matrix elements $\rho_{ij}$ with
$i,j=+,0,-$, that can be directly read off from the results of
\refs{Aguilar-Saavedra:2015yza,Aguilar-Saavedra:2017zkn},
\begin{align}
A_0 &= 2 \rho_{00},\; A_1 = \fr{1}{\sqrt{2}}(\rho_{+0}-\rho_{-0}+\rho_{0+}-\rho_{0-}),\crn
A_2 &= 2(\rho_{+-} + \rho_{-+}),\; A_3 = \sqrt{2}b(\rho_{+0} + \rho_{-0} + \rho_{0+} + \rho_{0-}),\crn
A_4 &= 2b(\rho_{++} - \rho_{--}),\; A_5 = \fr{1}{i}(\rho_{-+} - \rho_{+-}),\crn
A_6 &= -\fr{1}{i\sqrt{2}}(\rho_{+0} + \rho_{-0} - \rho_{0+} - \rho_{0-}),\; 
A_7 = \fr{\sqrt{2}b}{i}(\rho_{0+} - \rho_{0-} - \rho_{+0} + \rho_{-0}),
\label{eq:relations_Ai_spin_matrix}
\end{align}  
where $b=1$ for the $W^\pm$ bosons and $b=-c$ for the $Z$ boson, with
\bea
c = \fr{g_L^2 - g_R^2}{g_L^2 + g_R^2} =
\fr{1-4s^2_W}{1-4s^2_W+8s^4_W},\quad s^2_W = 1 - \fr{M_W^2}{M_Z^2}.
\label{eq:def_c}
\eea
Numerically, we have $c\approx 0.21$. Since $A_i$ are real, we see
that $A_5$, $A_6$, $A_7$ come from the imaginary part of the
spin-density matrix elements. 

We can also calculate the three coefficients $f_L$, $f_R$, and
$f_0$, called fiducial polarization fractions and defined as
(see e.g. \bibs{Bern:2011ie, Stirling:2012zt})  
\begin{align}
\fr{d\sigma}{\sigma d\!\cos\theta_3} &\equiv \fr{3}{8}
\Big[ 
(1\mp \cos\theta_3)^2 f^{W^{\pm}}_L + (1\pm \cos\theta_3)^2 f^{W^{\pm}}_R 
+ 2 \sin^2\theta_3 f^{W^{\pm}}_0 \Big],\crn
\fr{d\sigma}{\sigma d\!\cos\theta_6} &\equiv \fr{3}{8} 
\Big[ 
(1 + \cos^2\theta_6 + 2c\cos\theta_6) f^{Z}_L + (1 + \cos^2\theta_6 - 2c\cos\theta_6) f^{Z}_R
+ 2 \sin^2\theta_6 f^{Z}_0 \Big],
\label{eq:def_fLR0}
\end{align}
where the upper signs are for $W^+$ and the lower signs are for $W^-$,
$c$ defined in \eq{eq:def_c} ocurring because the $Z$ boson decays
into both left- and right-handed muons.

To see the relations between the polarization fractions and angular
coefficients, we perform the integration over $\phi \in [0,2\pi]$ of
\eq{eq:definition_Ai_tot}. We obtain
\bea
\fr{d\sigma}{\sigma d\!\cos\theta} = \fr{3}{8} 
\Big[ 
(1+\cos^2\theta) + A_0 \fr{1}{2}(1-3\cos^2\theta) + A_4 \cos\theta
\Big].
\label{eq:definition_Ai_cos}
\eea
Defining the expectation of observables $f(\theta)$ and $g(\theta,\phi)$ as
\begin{align}
\langle f(\theta) \rangle &= \int_{-1}^{1} d\!\cos\theta f(\theta) \fr{1}{\sigma}
\fr{d\sigma}{d\!\cos\theta},\label{eq:defs_expectation_2D}\\ 
\langle g(\theta,\phi) \rangle &= \int_{-1}^{1} d\!\cos\theta \int_{0}^{2\pi} d\phi
g(\theta,\phi)\fr{d\sigma}{\sigma d\!\cos\theta d\phi},
\label{eq:defs_expectation_3D} 
\end{align}
which can be calculated from $\cos\theta$ and $\cos\theta$-$\phi$ distributions, we have
\begin{align}
A_0 &= 4 - \langle 10\cos^2\theta\rangle, \; A_1 = \langle 5\sin 2\theta\cos\phi \rangle, 
\; A_2 = \langle 10\sin^2\theta \cos 2\phi \rangle, \; A_3 = \langle 4 \sin\theta\cos\phi \rangle,\crn 
\; A_4 &= \langle 4\cos\theta\rangle, \; A_5 = \langle 5\sin^2\theta\sin 2\phi \rangle, 
\; A_6 = \langle 5\sin 2\theta \sin\phi \rangle, \; A_7 = \langle 4\sin\theta \sin\phi \rangle,
\label{eq:cal_Ai}
\end{align}
which agree with \bib{Bern:2011ie}. We then obtain (see also \bib{Bern:2011ie} for the $W^\pm$ bosons and
\bib{Stirling:2012zt} for the $Z$ boson),
\begin{align}
f^{W^{\pm}}_L 
  &= -\fr{1}{2} \mp \langle\cos\theta_3\rangle +
    \fr{5}{2}\langle\cos^2\theta_3\rangle, \; 
f^{W^{\pm}}_R = -\fr{1}{2} \pm \langle\cos\theta_3\rangle +
    \fr{5}{2}\langle\cos^2\theta_3\rangle, \crn 
f^{W^{\pm}}_0 &= 2 - 5 \langle\cos^2\theta_3\rangle,\label{eq:cal_fLR0_W}\\
f^Z_L 
  &= -\fr{1}{2} + \fr{1}{c}\langle\cos\theta_6\rangle +
    \fr{5}{2}\langle\cos^2\theta_6\rangle, \;
f^Z_R 
  = -\fr{1}{2} - \fr{1}{c}\langle\cos\theta_6\rangle +
    \fr{5}{2}\langle\cos^2\theta_6\rangle, \crn
f^Z_0 
  &= 2 - 5 \langle\cos^2\theta_6\rangle,
\label{eq:cal_fLR0_Z}
\end{align}
which satisfy $f_L + f_R + f_0 = 1$. The relations between the
polarization fractions and angular coefficients read 
\begin{align}
f^{W^{\pm}}_L 
  &= \fr{1}{4}(2- A^{W^{\pm}}_0 \mp A^{W^{\pm}}_4),\;\; f^{W^{\pm}}_R
    = \fr{1}{4}(2- A^{W^{\pm}}_0 \pm A^{W^{\pm}}_4),\;\;
f^{W^{\pm}}_0 = \fr{1}{2}A^{W^{\pm}}_0,\crn
f^Z_L 
  &= \fr{1}{4}(2- A^Z_0 + \fr{1}{c}A^Z_4),\;\; f^Z_R = \fr{1}{4}(2-
    A^Z_0 - \fr{1}{c}A^Z_4),\;\;
f^Z_0 = \fr{1}{2}A^Z_0.\label{eq:relations_fL0R_Ai}
\end{align}

In the present work, we will go beyond the DPA and beyond LO. 
Realistic cuts on the individual lepton momenta as used by ATLAS 
or CMS are also required. The eight angular coefficients 
are therefore no longer enough to describe the angular distributions~\cite{Stirling:2012zt,Ballestrero:2017bxn}. 
However, the equations from (\ref{eq:defs_expectation_2D}) to (\ref{eq:relations_fL0R_Ai}) 
can still be used. The coefficients $A_i$ are now defined as the 
projections of realistic angular distributions calculated with full matrix elements 
at any order in perturbation theory and with arbitrary cuts on the individual leptons. 
This definition has been used and discussed in \bib{Stirling:2012zt}. 

To distinguish with the usual polarization observables used for the 
case of full lepton phase space such as in $W+\text{jets}$ production \cite{Bern:2011ie}, 
we will refer to those as inclusive polarization observables. When cuts on the individual 
lepton momenta are used, we call them fiducial polarization observables or fiducial angular 
coefficients. When moving from the full phase space to fiducial phase space, the cuts on $p_{T,\ell}$ and 
$\eta_{\ell}$ reduce event fraction at $|\cos\theta_{\ell,V}|\approx 1$. Therefore, the fiducial 
longitudinal fractions $f_{0}^V$ are larger than the corresponding inclusive fractions.

The effects of EW corrections on the gauge-boson decays for the fiducial 
polarization coefficients will be shown in \appen{appen:off_shell_NLOEW_effects}, where
effects from EW corrections to the production process $\bar{q}q'\to
l_il'_i V_j$  are also presented. In this appendix, one can also
compare the full LO to the DPA LO results to see the  off-shell
effects, which are not present in the DPA approximation.

It is important to note that $\sigma$ in \eq{eq:definition_Ai_tot} and
\eq{eq:def_fLR0} can be  replaced by a differential distribution such
as \cite{Stirling:2012zt}
\bea
\sigma \to \fr{d\sigma}{dp_{T,V}},\; \fr{d\sigma}{dp_{T,\ell}},\; \ldots . 
\eea
From the $\cos\theta$-$p_{T,V}$ distribution we can calculate
$d\sigma_{L,R,0}/dp_{T,V}$. In this paper, we will show fiducial polarization
results for $p_{T,W}$, $p_{T,Z}$, $\eta_Z$ (pseudo-rapidity), and
$y_Z$ (rapidity) distributions as the corresponding results for the inclusive polarization 
fractions exist, while it is not the case for the individual lepton momentum distributions. 
Results for $d\sigma/dp_{T,\ell}$ 
and for $d\sigma/dy_{\ell}$ in $W+\text{jets}$ production have been presented 
in \bib{Stirling:2012zt}. We however do not discuss them in this work. 

\begin{figure}[!t]
  \centering
  \begin{tabular}{cc}
    \includegraphics[width=0.45\textwidth]{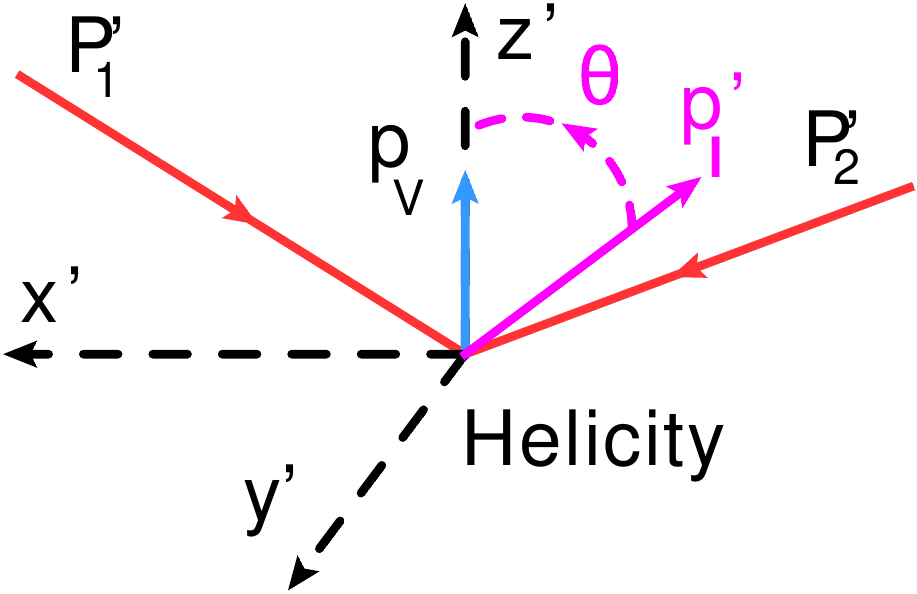}&
    \includegraphics[width=0.45\textwidth]{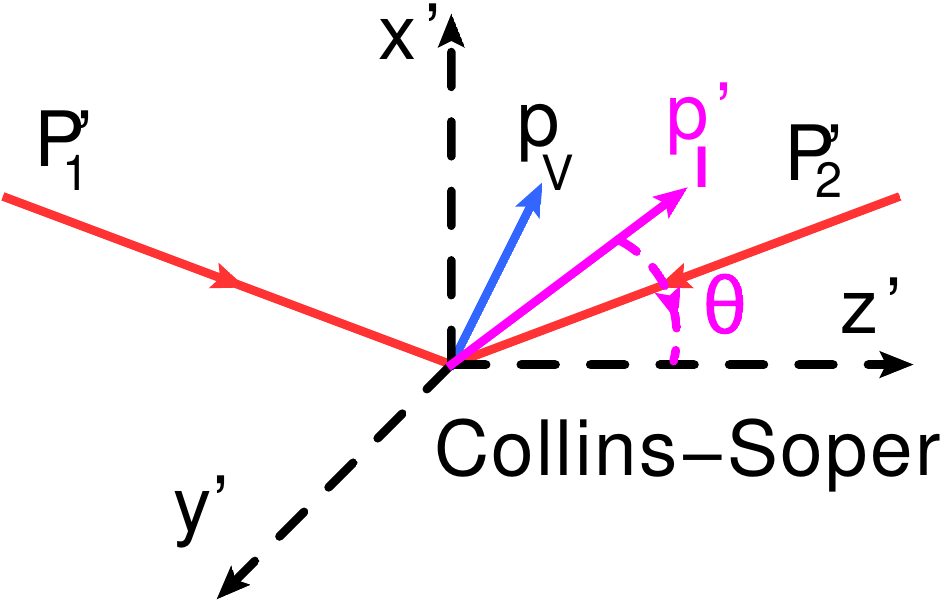}
  \end{tabular}
  \caption{Representation of the HE coordinate systems (left) and the
    CS coordinate systems (right), in the rest frame of the vector
    boson $V$ under consideration. The corresponding angle $\theta$ for
    the charged lepton is also defined.}
  \label{figs:frame}
\end{figure}

We now address the issue of choosing a coordinate system. In this work, 
we will consider and compare two coordinate systems:
\begin{itemize}
\item {\it Helicity (HE) coordinate system:}
  This coordinate system is defined in \bib{Bern:2011ie}, where the
  $z'$-axis (the prime is used to denote the gauge-boson rest frame)
  is along the momentum of the gauge boson in the laboratory frame
  ($p_V$). The exact definitions of $x'$ and $y'$ axes are given in
  \bib{Bern:2011ie} and a representation of the HE coordinate system
  is depicted in \fig{figs:frame} (left).
\item {\it Collins-Soper (CS) coordinate system:} This coordinate
  system was defined in \bib{Collins:1977iv}. We use in our paper the
  convention followed by \bib{Karlberg:2014qua,Aad:2016izn}. The
  $z'$-axis is defined as follows. Let $P_{1} = (E,0,0,+E)$ and $P_{2}
  = (E,0,0,-E)$ are the momenta of the two protons in the laboratory
  frame. Then $P'_{1}$ and $P'_{2}$ are the corresponding momenta in
  the gauge boson rest frame. The $z'$-axis is the bisector of
  $\vec{P}_{1}$ and $-\vec{P}_{2}$. Furthermore, the $z'$-axis points
  into the hemisphere of $p_V$. The $x'-z'$ plane is the plane of
  $P'_{1}$ and $P'_{2}$. The $x'$-axis is perpendicular to the
  $z'$-axis and points into the hemisphere of $-(\vec{P'}_{1} +
  \vec{P'}_{2})$. The coordinate system is right-handed, which defines
  the $y'$-axis. A representation of the CS coordinate system is
  depicted in \fig{figs:frame} (right).
\end{itemize}
\section{Numerical setup and theoretical uncertainties}
\label{sect:numset}
In this section we specify our input parameters, which are used to obtain numerical 
results presented in \sect{sect:numset:ewcorrections}, \sect{sect:numres}, and \appen{appen:off_shell_NLOEW_effects}. 
Our best results at NLO QCD+EW are calculated using the full NLO QCD matrix elements combined with the NLO EW corrections 
calculated using the DPA.
\subsection{Input parameters and definition of kinematical cuts}
\label{sect:numset:input}

The input parameters are
\begin{eqnarray}
G_{\mu} = 1.16637\times 10^{-5} \gev^{-2}, \,
M_W=80.385 \gev, \, 
M_Z = 91.1876 \gev, \crn 
\Gamma_W = 2.085\gev, \, \Gamma_Z = 2.4952\gev, \, 
M_t = 173 \gev, \, M_H=125\gev, 
\label{eq:param-setup}
\end{eqnarray}
essentially based on \bib{Agashe:2014kda} and are the same as the ones
used in \bib{Baglio:2013toa}. The masses of the leptons and the light
quarks, {\it i.e.} all but the top mass, are approximated as
zero. This is justified because our results are insensitive to those
small masses. The electromagnetic coupling is calculated as 
$\alpha_{G_\mu}=\sqrt{2}G_\mu M_W^2(1-M_W^2/M_Z^2)/\pi$. 

We will give results for the LHC running at a center-of-mass 
energy $\sqrt{s} = 13\,\tev$. We will consider ATLAS and CMS cuts for
their corresponding fiducial phase space, both for
$e^+\nu_e\,\mu^+\mu^-$ and $e^-\bar{\nu}_e\,\mu^+\mu^-$ final states.
We treat the extra parton occurring in the NLO QCD corrections
inclusively and we do not apply any jet cuts. We also consider the
possibility of lepton-photon recombination, where we redefine the
momentum of a given charged lepton $\ell$ as being $p'_\ell = p_\ell +
p_\gamma$ if $\Delta R(\ell,\gamma) \equiv
\sqrt{(\Delta\eta)^2+(\Delta\phi)^2}< 0.1$ for ATLAS and CMS
cuts. 
This recombination is done before applying the following kinematical cuts related to 
the charged leptons.  
For ATLAS we also define this transverse mass, 
\begin{align}
  m_{T,W} \equiv \sqrt{2p_{T,\nu}p_{T,e}[1-\cos\Delta\phi(e,\nu)]},
\end{align}
where $\Delta\phi(e,\nu)$ is the angle between the electron and the
neutrino in the transverse plane~\bibs{Aaboud:2016yus}. The sets of
cuts are identical for $e^+\nu_e\,\mu^+\mu^-$ and
$e^-\bar{\nu}_e\,\mu^+\mu^-$ final states. We use $\ell$ for either $e$
or $\mu$.

The cuts for ATLAS at $13\,\tev$ are given in
\bib{Aaboud:2016yus}. The CMS cuts for $13\,\tev$ are given in
\bib{Khachatryan:2016tgp}. The complete list of the cuts we have used for
our analysis is summarized in \tab{table:cutsfid}.
\begin{table}[ht!]
  \renewcommand{\arraystretch}{1.5}
  \begin{center}
    \small
    \begin{tabular}{|c|c|}\hline
      ATLAS fiducial
      & CMS fiducial\\
      \hline
      $\begin{matrix}
        p_{T,\mu} > 15\gev, \quad p_{T,e} > 20\gev, \quad 
        |\eta_\ell|<2.5,\\
        \Delta R\left(e,\mu^\pm\right) > 0.3, \quad \Delta
        R\left(\mu^+,\mu^-\right) > 0.2,\\
        \left|m_{\mu^+\mu^-} - M_Z\right| < 10\gev, \quad m_{T,W} >
        30\gev
      \end{matrix}$
      & $\begin{matrix}
        p^\text{leading}_{T,\mu} > 20\gev, \quad
        p^\text{sub-leading}_{T,\mu} > 10\gev,\\
        p_{T,e} > 20\gev, \quad |\eta_{\ell}|<2.5,\\
        \quad 60\gev < m_{\mu^+\mu^-} < 120\gev
      \end{matrix}$\\
      \hline
    \end{tabular}
    \caption{\small List of the cuts used in our fiducial-phase-space
      analysis at the 13 TeV LHC, depending on the experiment under
      consideration. They are extracted from
      Refs.~\cite{Aaboud:2016yus,Khachatryan:2016tgp}.}
    \label{table:cutsfid}
  \end{center}
\end{table}


\subsection{Theoretical uncertainties}
\label{sect:uncertainties}

We consider two sources of theoretical uncertainties in this work. The
first uncertainty we consider comes from the truncation of the
perturbative expansion at a given order. This truncation leads to a
dependence of the cross section on two unphysical scales, the
renormalization scale $\mu_R^{}$ and the factorization scale
$\mu_F^{}$. We evaluate the scale uncertainties by varying
independently these two scales as $n\mu_0/2$ with $n=1,2,4$ and $\mu_0
= (M_W + M_Z)/2$ being our central scale choice. We further use the
constraint $1/2 \leq\mu_R/\mu_F \leq 2$ to limit the number of scale
choices to seven at NLO QCD. It is noted that $\mu_R$ does not appear
at LO and hence there are only three possibilities for choosing
$\mu_F$.

The second uncertainty we consider is due to the uncertainties in the
determination of the PDFs. We use the 30 Hessian error sets provided
by {\tt
  LUXqed17\char`_plus\char`_PDF4LHC15\char`_nnlo} 
to calculate the PDF uncertainty as~\cite{Butterworth:2015oua}
\begin{align}
\Delta_{}^{\rm PDF}\sigma = \sqrt{\sum_{i=1}^{30}
  \left(\sigma_i^{}-\sigma_0^{}\right)_{}^2},
\label{eq:pdferror}
\end{align}
where $\sigma_{i}^{}$ stands for the calculation using the $i$th error
set and $\sigma_0^{}$ stands for the calculation using the central
set. Note that \eq{eq:pdferror} will also be used for the calculation
of the PDF uncertainty affecting the polarization observables and
their associated kinematical distributions presented in
\sect{sect:numres}.

In the numerical results, we will present also the NLO QCD+EW
predictions calculated with the central PDF set and with the central
scale $\mu_0$. The PDF and scale uncertainties are calculated using
the NLO QCD results. This is acceptable as long as the EW corrections
are small. In the ideal case, the EW corrections should also be
calculated for all members of the PDF set and for all scale choices,
and then the NLO QCD+EW errors would be computed from there. This is
more important for the polarization observables as the combination of
QCD and EW corrections is not a linear summation, see the discussion
in \sect{sect:numres:polar_observables_nloqcdew}. This step is however
not done in the present work as it requires a lot of computing power,
since the calculation of the EW corrections is much more complicated
than the QCD one. Besides, the information we would get when
performing such a full analysis is expected not to be substantially
different from our current analysis, given the size of the
uncertainties (see \sect{sect:numres} and \appen{appen:numres_Wm}).

In the following the PDF errors are indicated in round brackets,
statistical errors in square brackets, and the scale errors being
asymmetric are indicated using upper and lower superscripts, unless
otherwise stated.

It is noted that the calculation of the statistical error for
polarization observables defined by \eq{eq:defs_expectation_2D} and
\eq{eq:defs_expectation_3D} are nontrivial as the correlations between
different bins are unknown. As a simple exercise, one can try to
calculate the total cross section and its statistical error from a
two-dimensional distribution, say the LO $\cos\theta^\text{HE}_e$
distribution shown in \fig{fig:dist_cos_theta_HEL_CS_e_muon_Wp_cms},
and compare them with the known results in
\tab{table:Xsection_Born_EW_cor_ATLAS_CMS}. One will see that there is
agreement for the central value but not for the error. If we sum the
errors of all the bins linearly, it overshoots the true value (that is
the one given in the table) by a factor of $4$. If summation in
quadrature is used, then it undershoots by a factor of
$1/2$. Moreover, for three-dimensional distributions, the statistical
error for each bin is not known in many Monte-Carlo programs including
the {\tt VBFNLO} code. And it gets worse when higher-order corrections
are included, because the calculation of the statistical error for
every bin becomes more difficult and therefore unreliable.

Fortunately, in the framework of Monte-Carlo method, there is a simple
way to get an estimate for the statistical errors for any observables
independently of the complexity of the calculation, that is to use
different random-number seeds to get a list of central values. From
this list the mean value and an estimate for the statistical error are
obtained as follows 
\begin{align}
\bar{\sigma} = \fr{1}{N}\sum_{i=1}^{N} \sigma_i, \quad \Delta\sigma =
  \fr{1}{\sqrt{N-1}}\sqrt{\sum_{i=1}^{N} (\sigma_i - \bar{\sigma})^2},
\label{eq:seed_average}
\end{align}
where $N$ is the number of random seeds. Using this method for the
$\cos\theta^\text{HE}_e$ distribution above mentioned, we get the LO
cross section, for the $e^+\nu_e\mu^+\mu^-$ final state with the ATLAS
fiducial cuts, $19.345 \pm 0.003\fb$ with $N=10$, which is in good
agreement with the value of $19.344 \pm 0.002\fb$ obtained by using
the standard Monte-Carlo integration method with one random seed. This
method also helps to smooth out statistical fluctuations, thereby
giving nicer plots for distributions. All numerical results for
kinematical distributions and polarization observables presented in
this paper are obtained using this method with $N=50$ $(10)$ for the
NLO QCD (EW) results. For polarization observables, there are actually
two ways to do the seed average. One method is to do the seed average
for all relevant distributions to get the seed-combined distributions
first. Then from these combined distributions we proceed to calculate
polarization observables. The second method is to calculate the
polarization observables for every seed first, then combine these
observables using \eq{eq:seed_average}. We have checked that both
methods give the same central values, but the second way provides also
the statistical errors.

Finally, we remark that the statistical errors are very small compared
to the PDF and scale uncertainties. They will be therefore not
indicated, unless where necessary.

\section{Results for fiducial cross sections and kinematical
  distributions}
\label{sect:numset:ewcorrections}

We present in this section our results for the cross section in the
fiducial phase space, including scale and parton distribution function
(PDF) uncertainties, as well as a selection of kinematical
distributions. Our results are presented in such a way that direct
comparisons between the full calculation of \bib{Biedermann:2017oae}
and our DPA for the NLO EW corrections are possible. For this
comparison, it is noted that the input parameter scheme of
\bib{Biedermann:2017oae} is different from ours as
follows. \bib{Biedermann:2017oae} uses the complex mass scheme, which
introduces a shift in $s_W^2$ and other parameters due to the widths
of the $W$ and $Z$ bosons. They used a non-diagonal CKM matrix, taking
into account the effect of the Cabibbo mixing angle. Lastly, they used
the first version of {\tt
  LUXqed\char`_plus\char`_PDF4LHC15\char`_nnlo\char`_100} for PDF set,
while we use the latest version. The difference on the photon PDF is
very minor \cite{Manohar:2017eqh}, and since both versions use {\tt
  PDF4LHC15\char`_nnlo} the differences on the quark and gluon PDFs
should be also negligible. These effects on the fiducial cross
sections are quantified at the end of \sect{sect:fid_xsection}. The
total difference is small, of about $1\%$ for the ATLAS fiducial cross
sections at full LO. 

\subsection{Fiducial cross sections}
\label{sect:fid_xsection}
We start this subsection by presenting a comparison of our results for
the cross sections with the experimental measurements from ATLAS and
CMS. It is important to note that our predictions are not the
state-of-the-art as we are not including the NNLO QCD corrections that
would amount to $\simeq  +10\%$~\cite{Grazzini:2017ckn}. Nevertheless
we wish to do an NLO comparison to confirm that we do use the same
setup as ATLAS and CMS.

The latest ATLAS results, obtained with 36 fb$^{-1}_{}$ of data, allow
for a comparison channel by channel. In Table~4 of \bib{ATLAS:2018ogj}
we find the following results,
\begin{align}
\sigma_{e^+_{} \mu^+_{}\mu^-_{}}^{\rm ATLASfid} = 36.7 \pm
  2.5\,\,\mathrm{fb}, \quad \sigma_{e^-_{} \mu^+_{}\mu^-_{}}^{\rm
  ATLASfid} = 25.7 \pm 2.1\,\,\mathrm{fb},
\label{eq:atlasxs}
\end{align}
to be compared to our NLO QCD+EW results,
\begin{align}
\sigma_{e^+_{} \mu^+_{}\mu^-_{}}^{\rm th,ATLASfid} = 34.7 \pm 0.5\,
  ({\rm PDF}) +1.8/\!-1.5\,({\rm scale}) - 0.8\, ({\rm NLOEW})\,\,\mathrm{fb},\nonumber\\
\sigma_{e^-_{} \mu^+_{}\mu^-_{}}^{\rm th,ATLASfid} =24.1 \pm
  0.4\,({\rm PDF}) +1.3/\!-1.1\,({\rm scale}) - 0.6\, ({\rm NLOEW})\,\,\mathrm{fb}.
\label{eq:atlasxsth}
\end{align}
Comparing \eq{eq:atlasxs} and \eq{eq:atlasxsth} we get about $1\sigma$
agreement between ATLAS experimental results and our theoretical
predictions at NLO.

The latest CMS results we get in the literature have been obtained
with 2.3 fb$^{-1}_{}$ of data and collect together $W^+_{}$ and
$W^-_{}$ channels as well as all leptonic decay
modes~\cite{Khachatryan:2016tgp}, reading
\begin{align}
\sigma_{W^\pm_{}Z}^{\rm CMSfid} = 258\pm 30 \,\,\mathrm{fb}.
\label{eq:cmsxs}
\end{align}
Our NLO QCD+EW predictions for the leptonic channel $e^\pm_{} \nu_e^{}\,
\mu^+_{}\mu^-_{}$ read
\begin{align}
\sigma_{e^+_{} \mu^+_{}\mu^-_{}}^{\rm th,CMSfid} = 44.7 \pm 0.7\,({\rm
  PDF}) +2.4/\!-1.9\,({\rm scale}) - 0.5\, ({\rm NLOEW}) \,\,\mathrm{fb},\nonumber\\
\sigma_{e^-_{} \mu^+_{}\mu^-_{}}^{\rm th,CMSfid} =30.7 \pm 0.5\,({\rm
  PDF}) +1.7/\!-1.4\,({\rm scale}) - 0.3\, ({\rm NLOEW})\,\,\mathrm{fb}.
\label{eq:cmsxsth1}
\end{align}
We can combine the results in \eq{eq:cmsxsth1} by assuming the same
cross-section for the four different leptonic modes $e^\pm
\mu^+_{}\mu^-_{}$, $\mu^\pm e^+_{}e^-_{}$, $e^\pm e^+_{} e^-_{}$,
$\mu^\pm \mu^+_{}\mu^-_{}$, and adding our predictions for $W^+_{}$
and $W^-_{}$ channels. Adding the uncertainties in
quadrature, we finally obtain at NLO QCD+EW
\begin{align}
\sigma_{W^\pm_{}Z}^{\rm th,CMSfid} = 302\pm 2\,({\rm PDF})
  +6/\!-5\,({\rm scale}) - 3\, ({\rm NLOEW})\,\mathrm{fb}.
\label{eq:cmsxsth2}
\end{align}
We compare our prediction in \eq{eq:cmsxsth2} with the
experimental result in \eq{eq:cmsxs} and obtain a $1.3\sigma$
agreement at NLO.
\begin{table}[ht!]
 \renewcommand{\arraystretch}{1.3}
  \begin{center}
\setlength\tabcolsep{0.05cm}
\fontsize{11.0}{11.0}
\begin{tabular}{|c|c|c|c|c|c|c|c|c|}\hline
Cut  &  Process  & LO~[fb] & DPA LO~[fb] & $\delta_{\bar{q}q'}$(\%) & $\delta_{q\gamma}$(\%) & $\delta_{\text{d}W}$(\%) & $\delta_{\text{d}Z}$(\%) & $\delta_\text{NLOEW}$(\%) 
\\\hline
ATLAS fid.  &  $e^+\nu_e\mu^+\mu^-$ & $19.344[2]$ & $18.740[2]$ & $-6.10$ & $+1.76$ & $-1.20$ & $-3.55$ & $-4.34$ \\
\hline
ATLAS fid.  &  $e^-\bar{\nu}_e\mu^+\mu^-$ & $13.001[1]$ & $12.987[1]$ & $-6.11$ & $+1.86$ & $-1.15$ & $-3.55$ & $-4.25$ \\
\hline
CMS fid.  &  $e^+\nu_e\mu^+\mu^-$ & $24.6225[4]$ & $23.510[2]$ & $-3.63$ & $+1.77$ & $-1.09$ & $-1.19$ & $-1.86$ \\
\hline
CMS fid.  &  $e^-\bar{\nu}_e\mu^+\mu^-$ & $16.3205[2]$ & $16.157[1]$ & $-3.62$ & $+1.90$ & $-1.05$ & $-1.16$ & $-1.72$ \\
\hline
\end{tabular}
\caption{\small Born cross sections in fb obtained using the 
full and DPA matrix elements with the ATLAS and CMS fiducial cuts. The EW corrections in percentage calculated using DPA normalized 
to the LO are also shown.}
\label{table:Xsection_Born_EW_cor_ATLAS_CMS}
\end{center}
\end{table}

To shed light on the goodness of the DPA and to compare with the full
results of \bib{Biedermann:2017oae}, we present in
\tab{table:Xsection_Born_EW_cor_ATLAS_CMS} the LO, DPA LO, and the NLO
EW corrections calculated using DPA relative to the LO results with
the ATLAS and CMS fiducial cuts. The definitions of the corrections $\delta_{\bar{q}q'}$
and $\delta_{q\gamma}$ are the same as those in
\bib{Biedermann:2017oae} and are given in
\eq{eq:defs_EW_cor}.  
We observe that the DPA cross sections are
smaller than the full results, with the difference about $-3\,(-5)\%$
for the $W^+Z$ channel with the ATLAS (CMS) fiducial cuts. For the
$W^-Z$ case, the differences are much smaller. The corrections to the
decays of the $W$ and $Z$ bosons, defined in \sect{sect:cal:ew}, are
also separately shown. These are new compared to our previous results
for OS production \cite{Baglio:2013toa}.

Before commenting on the differences between our results and those of
\bib{Biedermann:2017oae}, it is important to know the effects of the
differences in the input parameter schemes as above mentioned. We have
checked at the full LO with the ATLAS fiducial cuts that setting the
Cabibbo angle to zero as in this work increases the cross section
about $0.7\,(0.9)\%$ compared to the case of
$\sin\theta_\text{c}=0.225$ as used in \bib{Biedermann:2017oae} for
the $W^+Z$ ($W^-Z$) channels. We have also implemented the
complex-mass scheme of \bib{Biedermann:2017oae}, i.e. taking into
account the shifts in $M_V$, $\Gamma_V$ (with $V=W,Z$),
$\alpha_{G_\mu}$, $s^2_W$ due to $\Gamma_V^\text{os}$ as in their
Section 3.1, and obtained that this effect is about $0.2\%$ for both
channels, with the complex-mass scheme cross sections being
smaller. The differences due to different versions of the PDF set are
completely negligible as expected. Overall, the differences in the
input parameters between \bib{Biedermann:2017oae} and this work are
about $1\%$ for both channels, which are essentially the sum of those
two effects.

Comparing to the results of \bib{Biedermann:2017oae} for the ATLAS
fiducial cuts, we see good agreement for $\delta_{q\gamma}$ and 
$\delta_{\bar{q}q'}$ individually. 
As seen in the following
\sect{sect:numset:kinematicdist}, similar agreement is also obtained for 
several kinematical distributions. 
Comparisons between the DPA with non-factorizable corrections included
and the full results for the case of $pp \to W^+ W^- \to \nu_\mu \mu^+ e^- \bar{\nu}_e + X$ 
and of $e^+ e^- \to W^+ W^- \to$ 4 fermions 
have been presented in \bib{Biedermann:2016guo} and \bib{Denner:2005es}, 
respectively. Good agreement has also been observed there.

\subsection{Kinematical distributions: $W^+Z$ channel}
\label{sect:numset:kinematicdist}

\begin{figure}[ht!]
  \centering
  \begin{tabular}{cc}
  \includegraphics[width=0.48\textwidth]{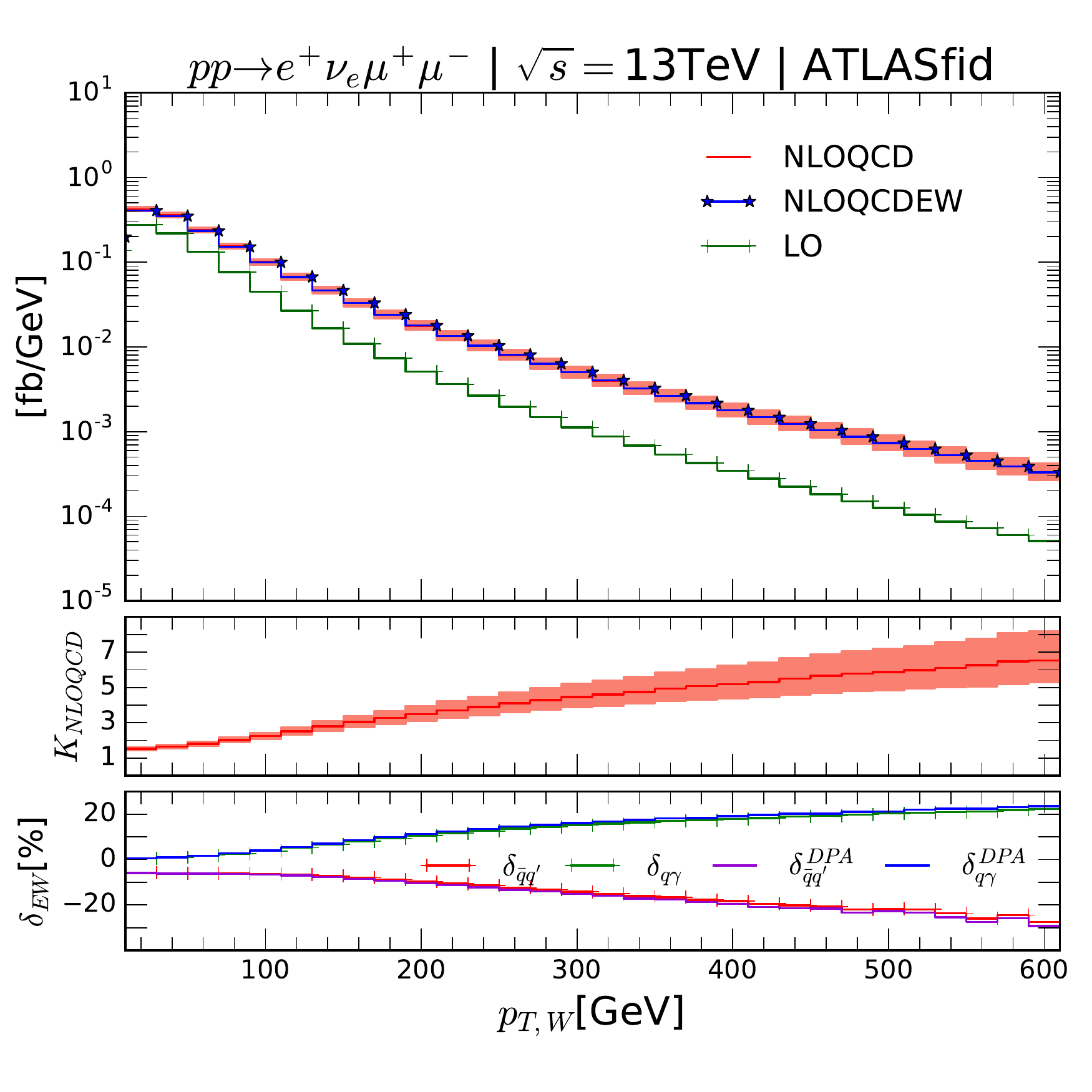}& 
  \includegraphics[width=0.48\textwidth]{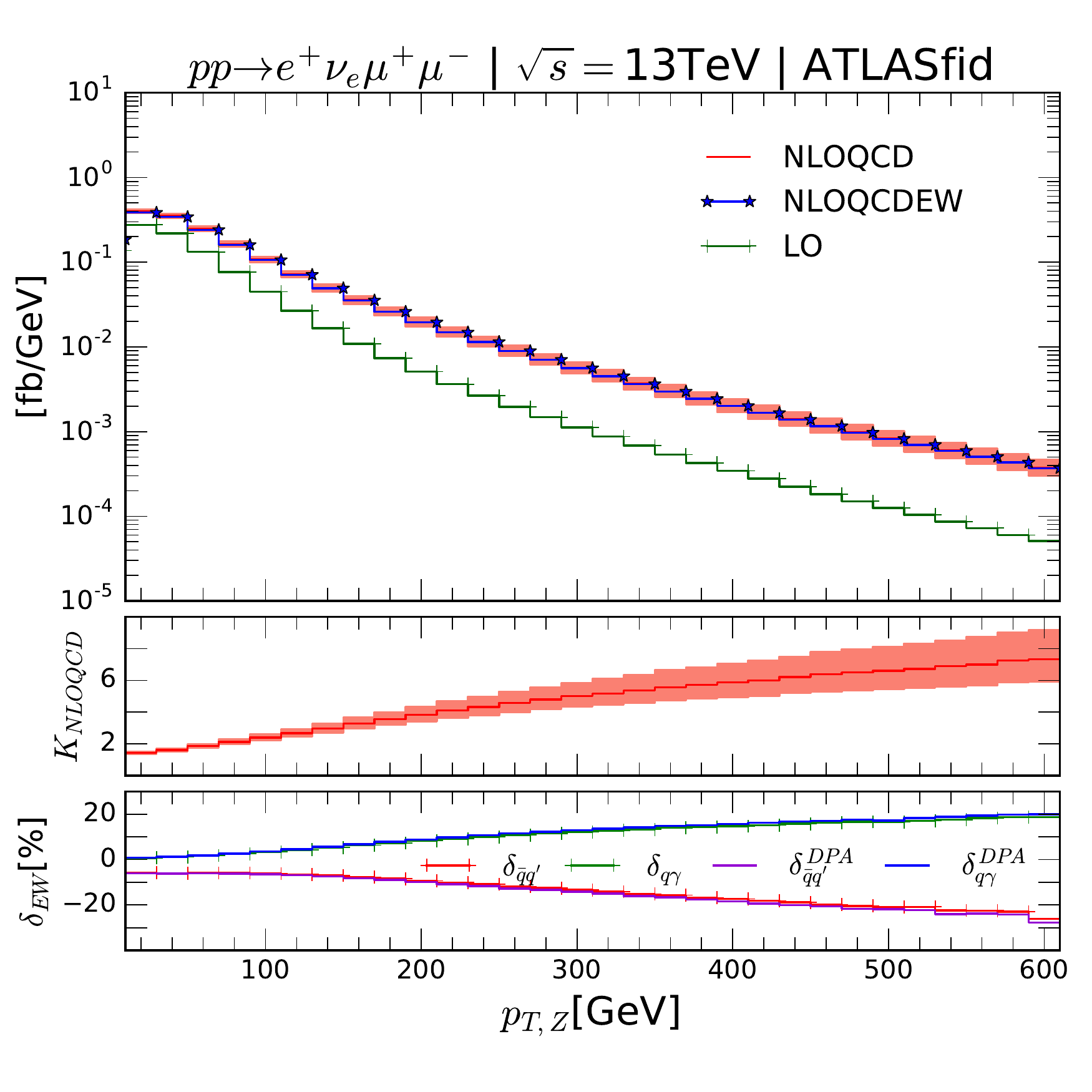}\\
  \includegraphics[width=0.48\textwidth]{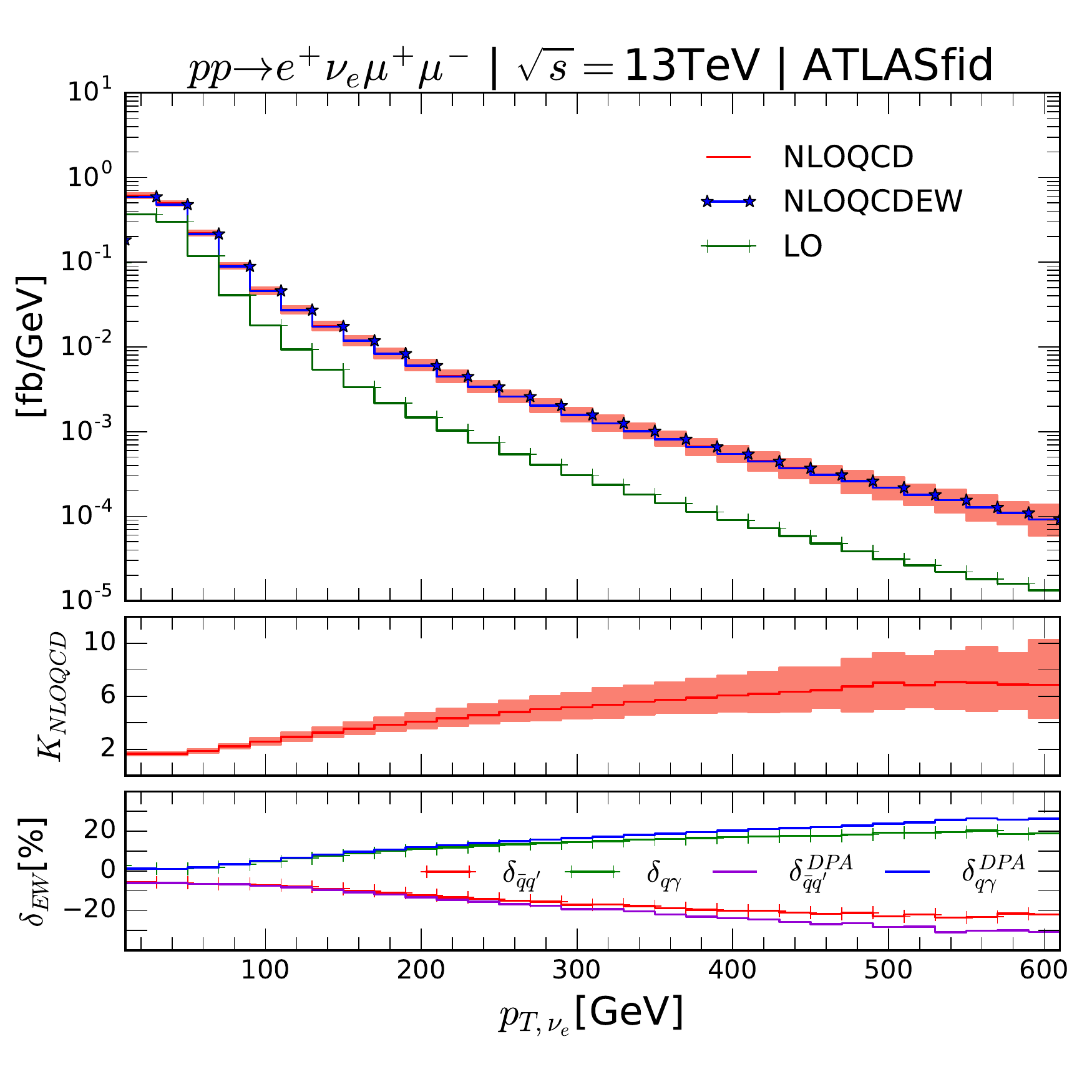}& 
  \includegraphics[width=0.48\textwidth]{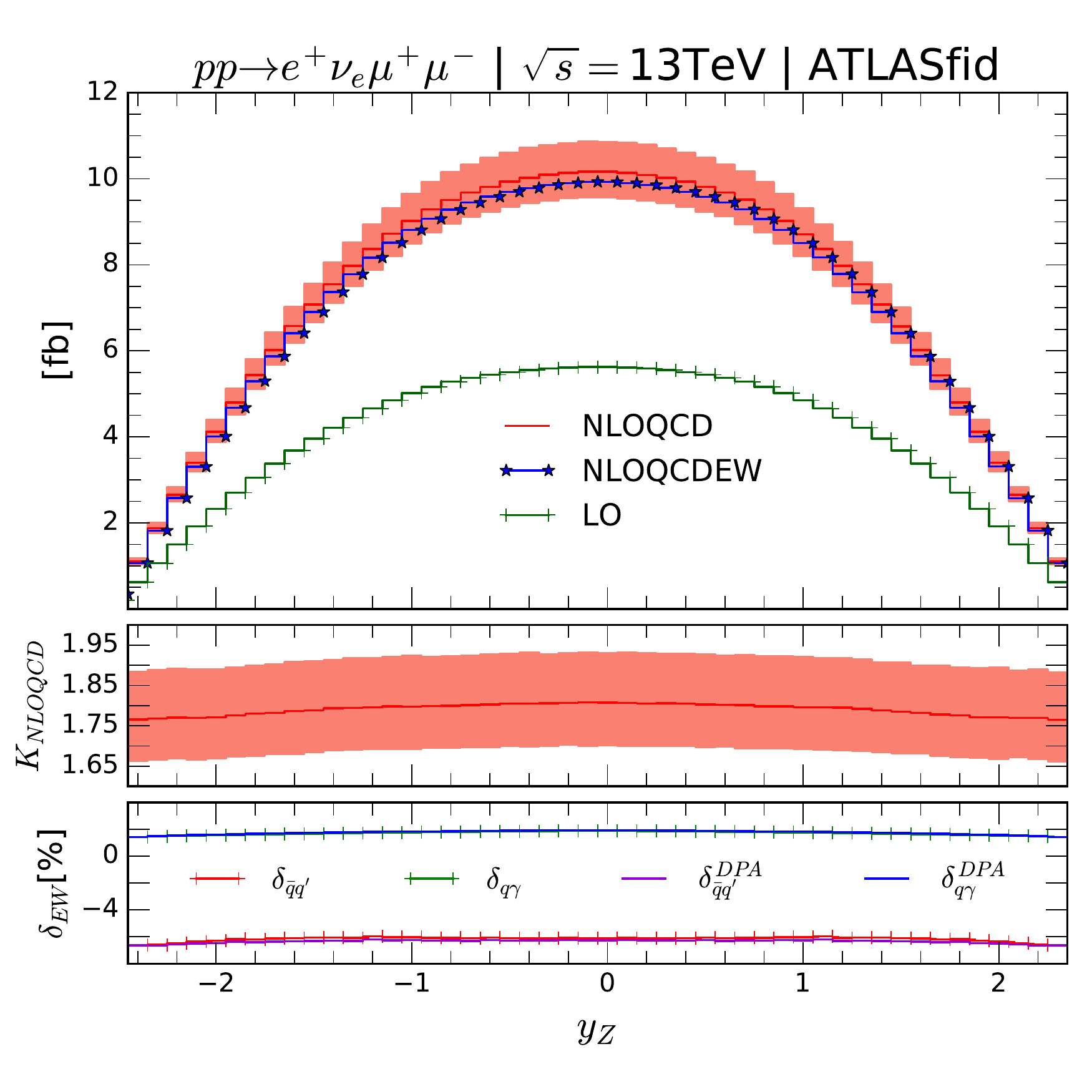}
  \end{tabular}
  \caption{Distributions of the transverse momentum of the $e^+\nu_e$ (top left) and 
    $\mu^+\mu^-$ (top right) systems and of the neutrino (bottom left)
    in the processes $p p\to e^+_{} \nu_e^{}\,
    \mu^+_{}\mu^-_{} +X$ at the 13 TeV LHC with the 
    ATLAS fiducial cuts. Rapidity distribution of the $\mu^+\mu^-$ system 
    is also displayed at the bottom right corner. 
    The upper panels show the absolute values of 
    the cross sections at LO (in green), NLO QCD (red), and NLO QCD+EW (blue). 
    The middle panels display the ratio of the NLO QCD cross sections to the
    corresponding LO ones. The bands indicate the total theoretical uncertainty calculated 
    as a linear sum of PDF and scale uncertainties at NLO QCD. The bottom panels show the 
    NLO EW corrections (see text) calculated using DPA relative to the full LO (marked with plus signs) and DPA LO cross sections.}
  \label{fig:dist_pT_W_Z_y_W_Z_Wp_atlas}
\end{figure}

\begin{figure}[ht!]
  \centering
  \begin{tabular}{cc}
  \includegraphics[width=0.48\textwidth]{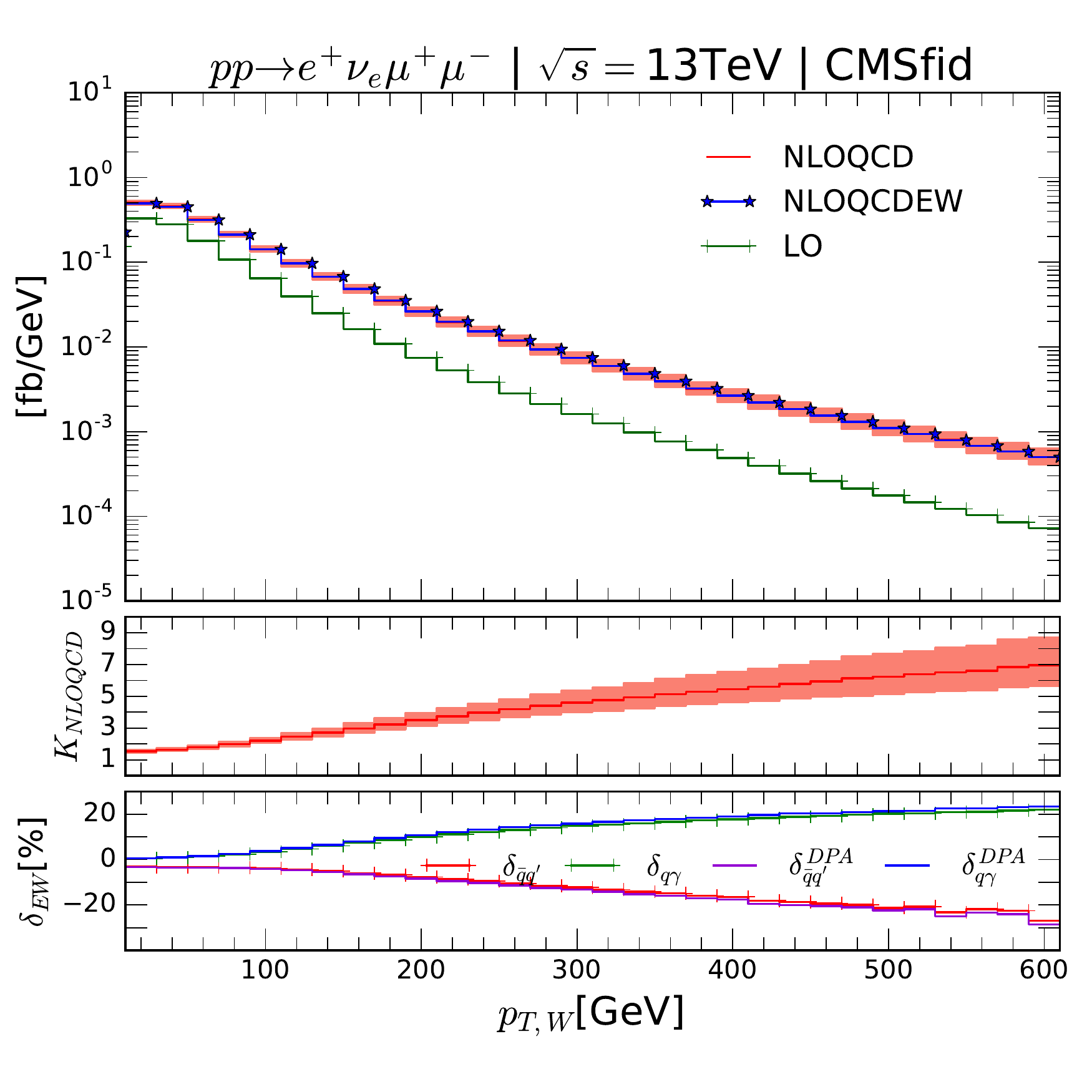}& 
  \includegraphics[width=0.48\textwidth]{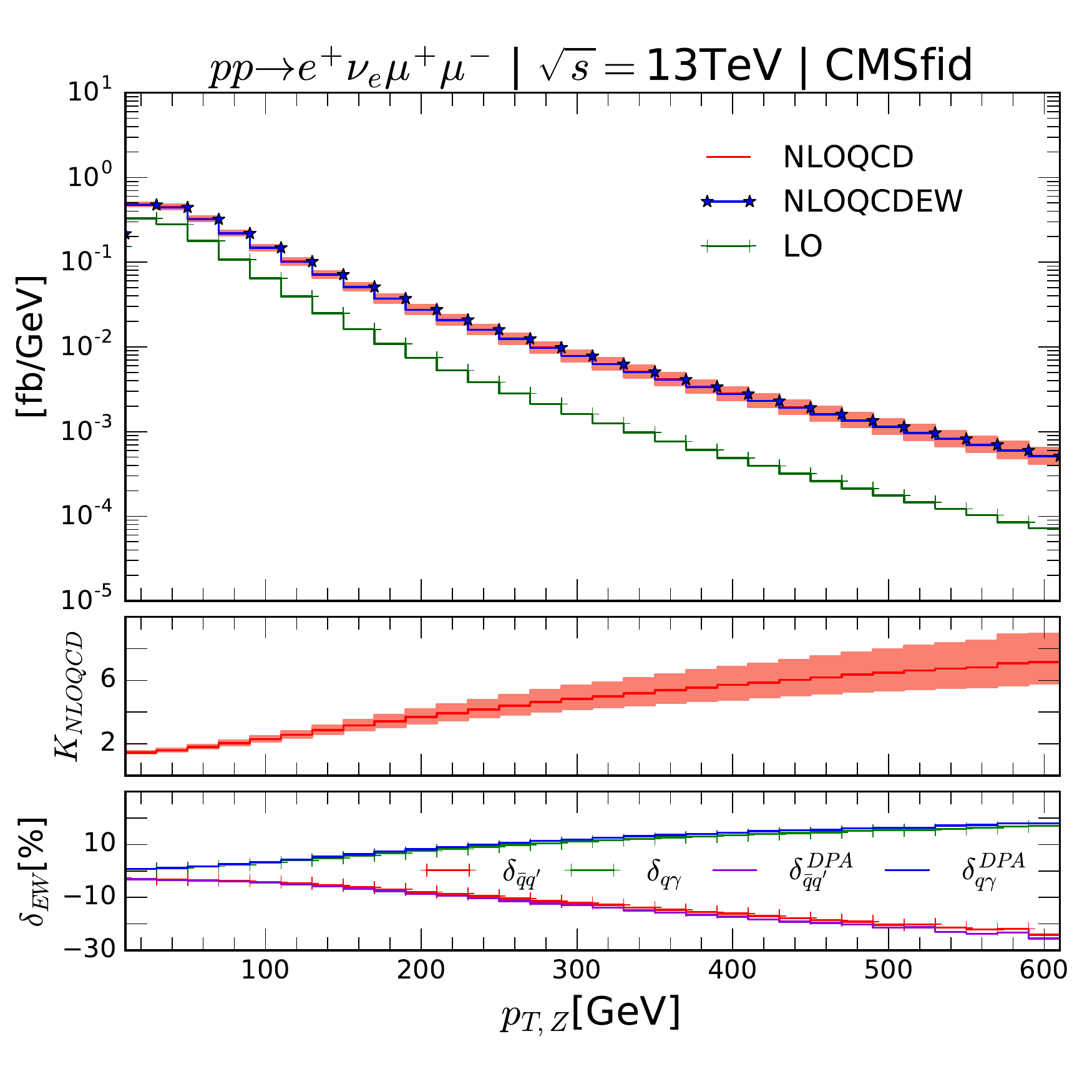}\\
  \includegraphics[width=0.48\textwidth]{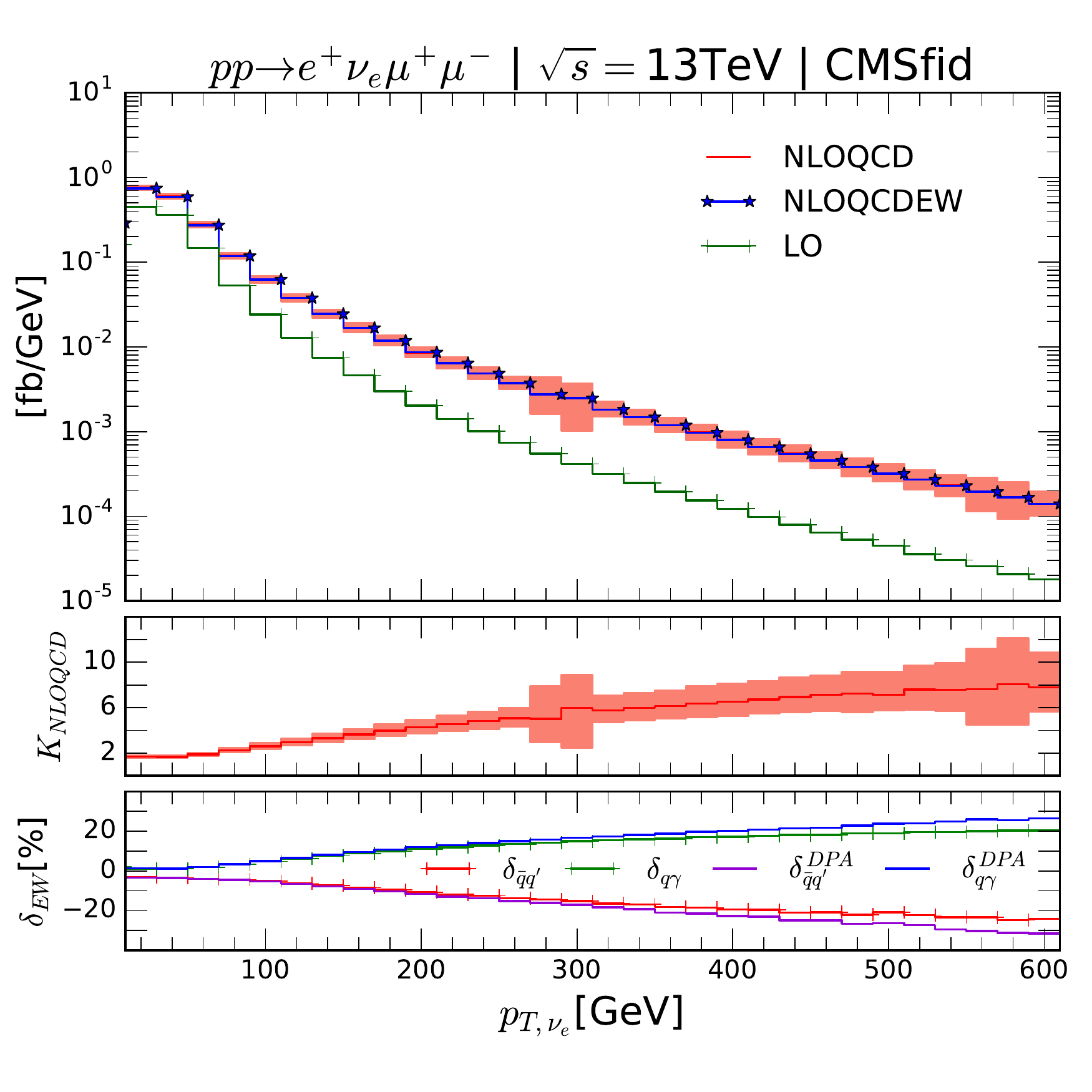}& 
  \includegraphics[width=0.48\textwidth]{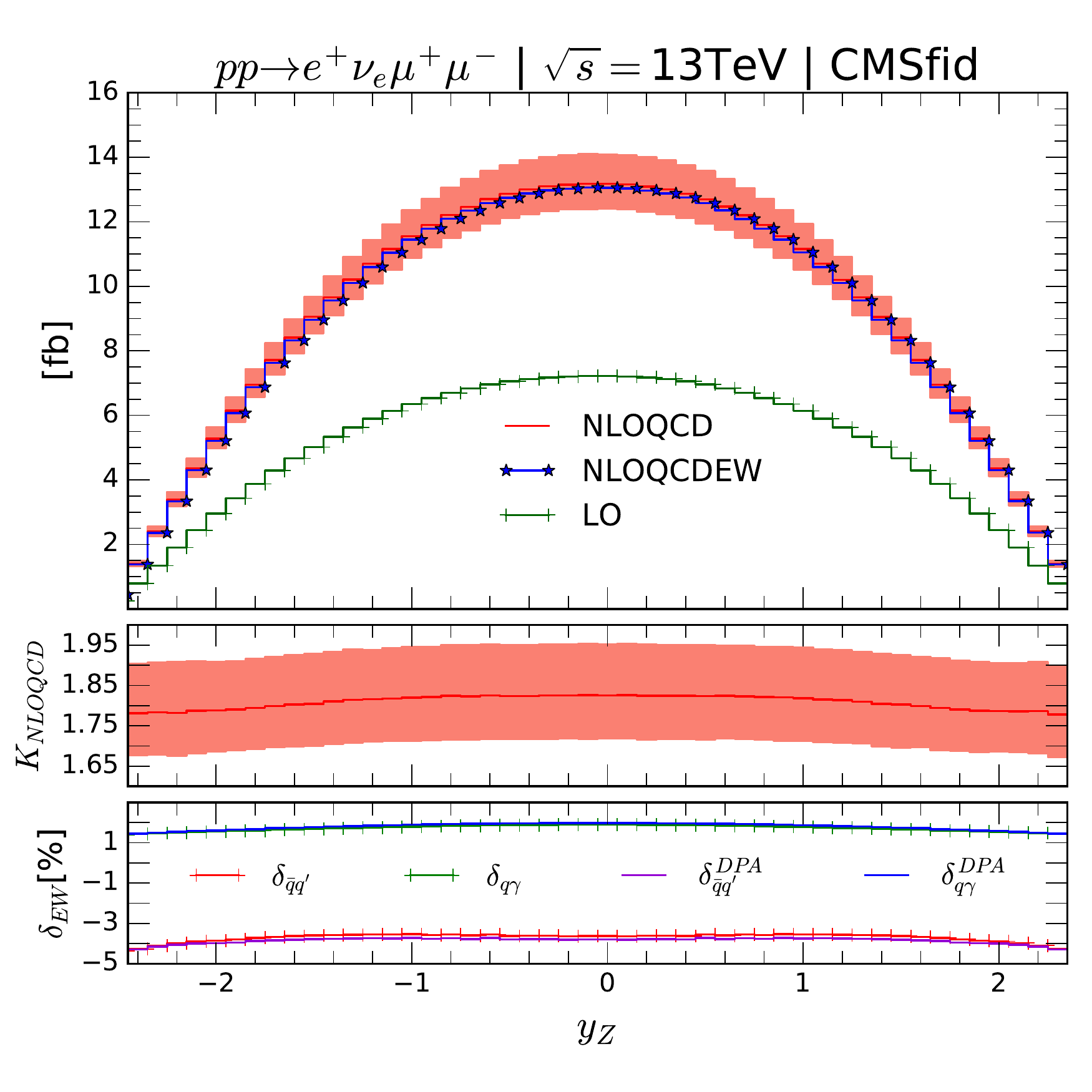}
  \end{tabular}
  \caption{Same as \fig{fig:dist_pT_W_Z_y_W_Z_Wp_atlas} but with the CMS fiducial cuts.}
  \label{fig:dist_pT_W_Z_y_W_Z_Wp_cms}
\end{figure}

In order to get more insight into the theoretical uncertainties
affecting the process we study a selection of differential
distributions including the scale and PDF uncertainties at NLO
QCD. We limit our discussion to the $W^+Z$ channel and
  present the corresponding plots for the $W^-Z$ channel in
  \appen{appen:kin_dist_Wm}. We first start with the transverse
momentum distributions of the $e^+\nu_e$ and $\mu^+_{}\mu^-_{}$
systems and we display in \fig{fig:dist_pT_W_Z_y_W_Z_Wp_atlas} the
predictions using the ATLAS fiducial cuts and in
\fig{fig:dist_pT_W_Z_y_W_Z_Wp_cms} the predictions using the CMS fiducial
cuts. The transverse momentum distribution of the neutrino and
rapidity distribution of the $\mu^+_{}\mu^-_{}$ system are also shown
in those figures (bottom row). In both cases we also display the total
theoretical uncertainty calculated as a linear sum of PDF and scale
uncertainties at NLO QCD, shown as bands around the central prediction
calculated at $\mu_F = \mu_R = (M_W+M_Z)/2$. We follow the
recommendations of \bib{Dittmaier:2011ti} to combine PDF and scale
uncertainties linearly. This procedure was also implemented in
\bib{Aad:2016ett} by ATLAS. Cross sections at NLO QCD+EW are also
displayed in blue, and  the LO predictions in green, in the top
panels.
\begin{figure}[ht!]
  \centering
  \begin{tabular}{cc}
  \includegraphics[width=0.48\textwidth]{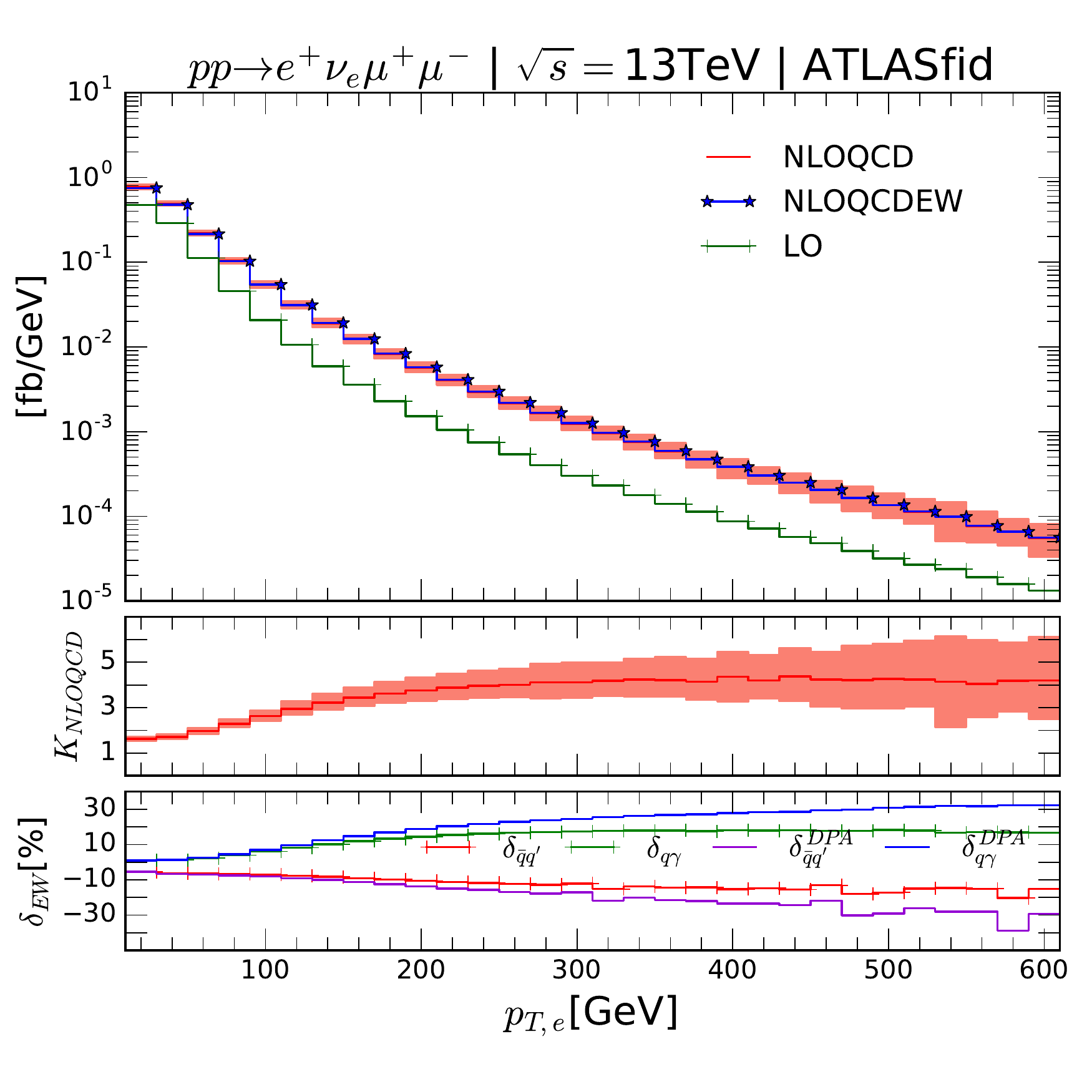}& 
  \includegraphics[width=0.48\textwidth]{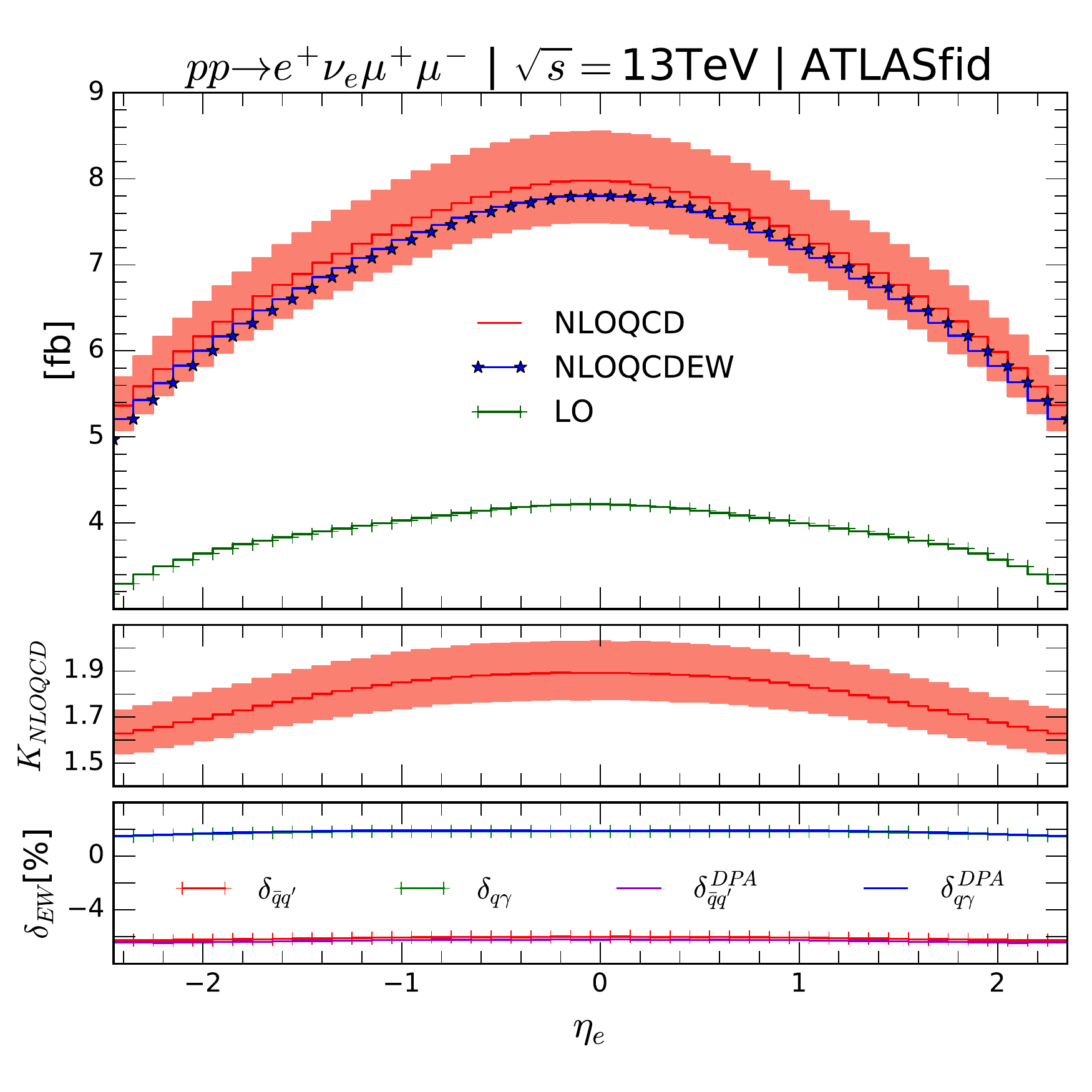}\\
  \includegraphics[width=0.48\textwidth]{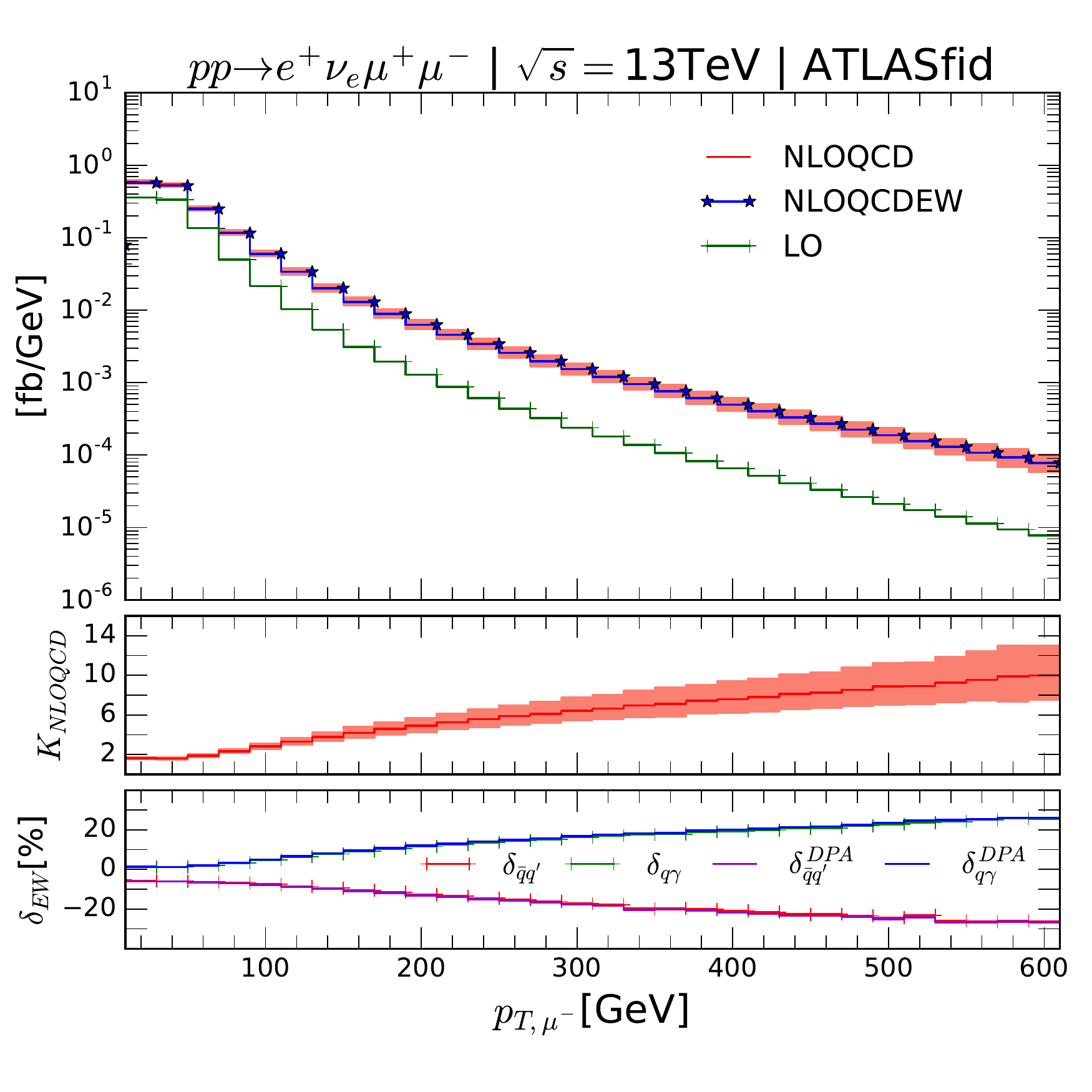}& 
  \includegraphics[width=0.48\textwidth]{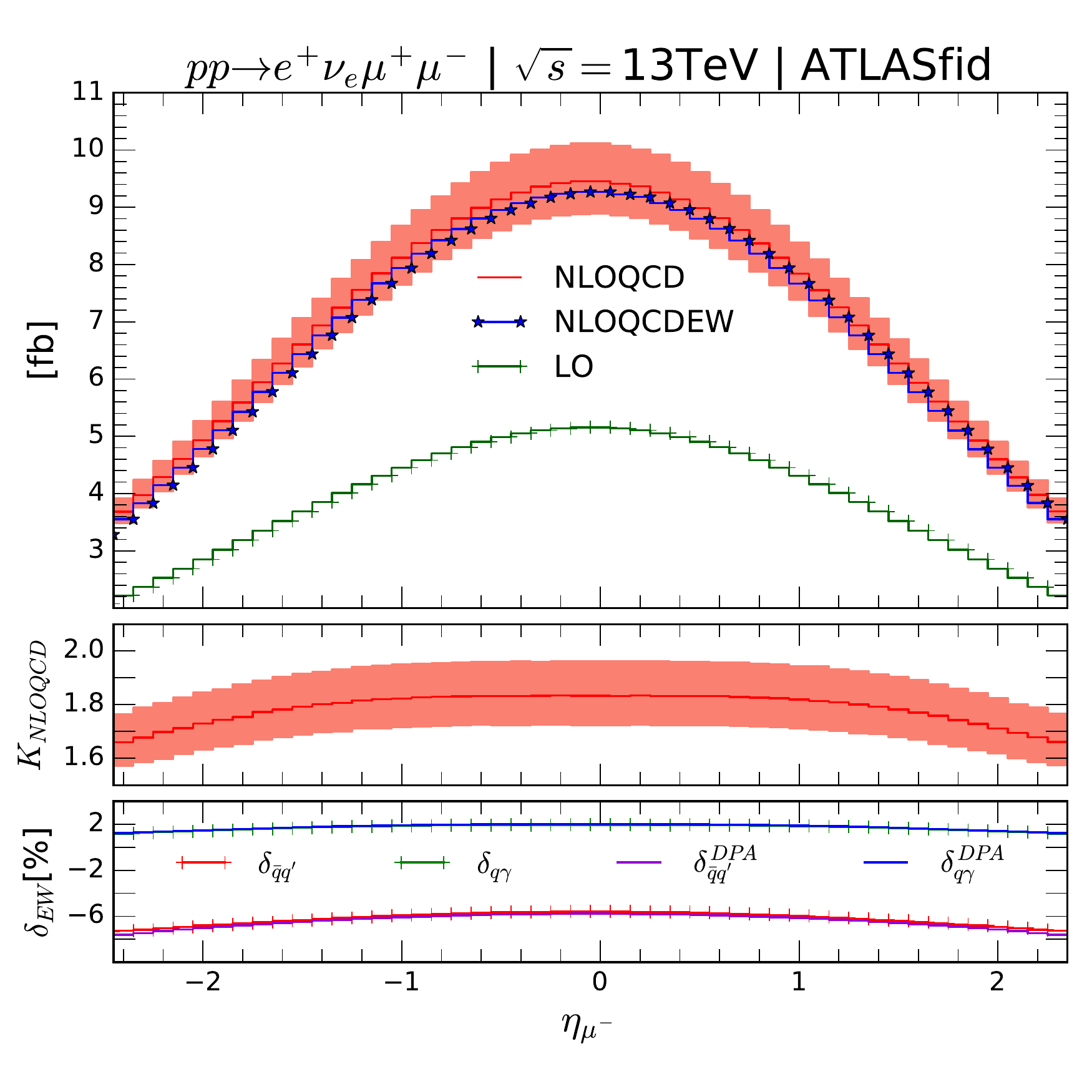}
  \end{tabular}
  \caption{Same as \fig{fig:dist_pT_W_Z_y_W_Z_Wp_atlas} but for the transverse momentum and pseudo-rapidity 
  distributions of the positron (top row) and the muon (bottom row).}
  \label{fig:dist_pT_e_mu_eta_e_mu_Wp_atlas}
\end{figure}
\begin{figure}[ht!]
  \centering
  \begin{tabular}{cc}
  \includegraphics[width=0.48\textwidth]{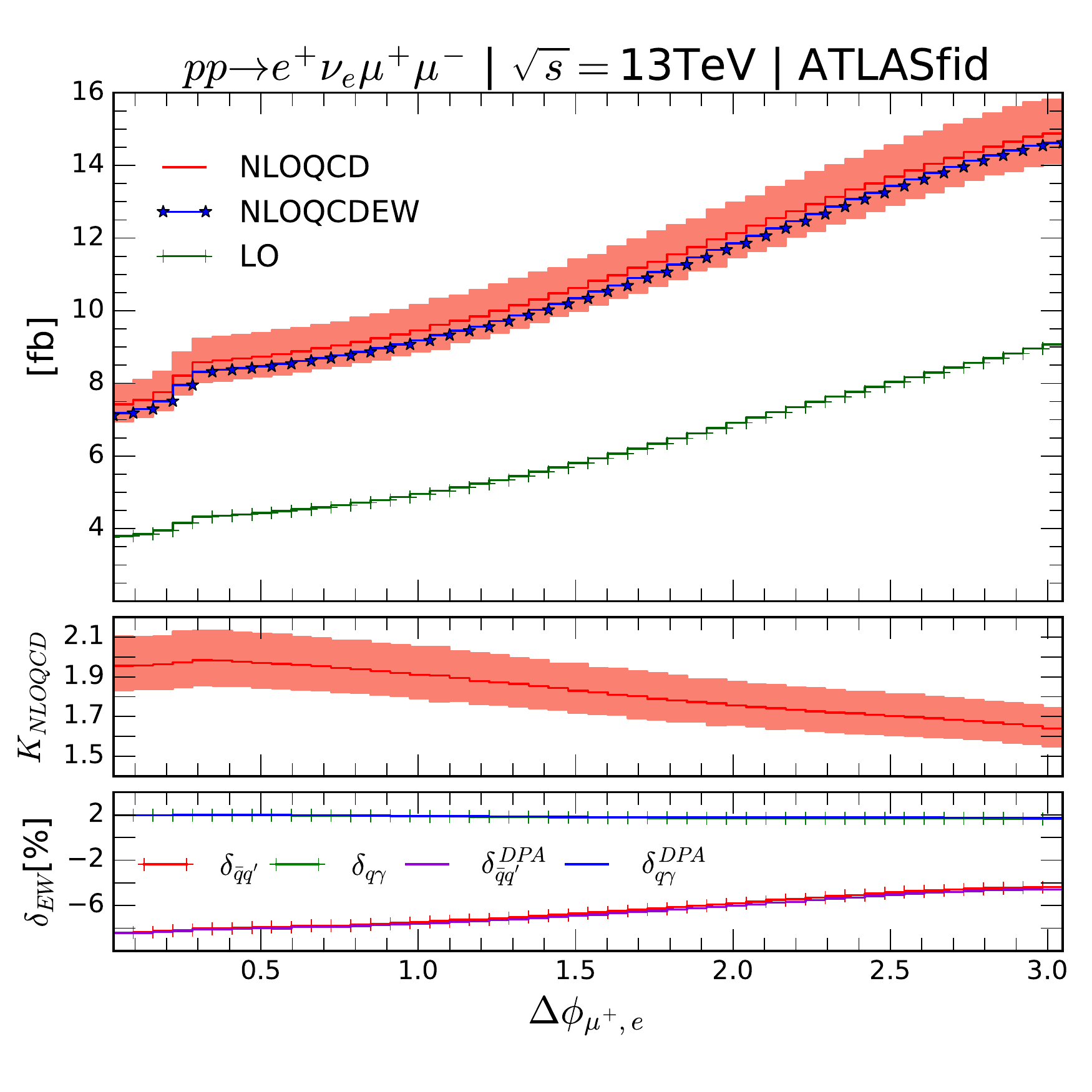}& 
  \includegraphics[width=0.48\textwidth]{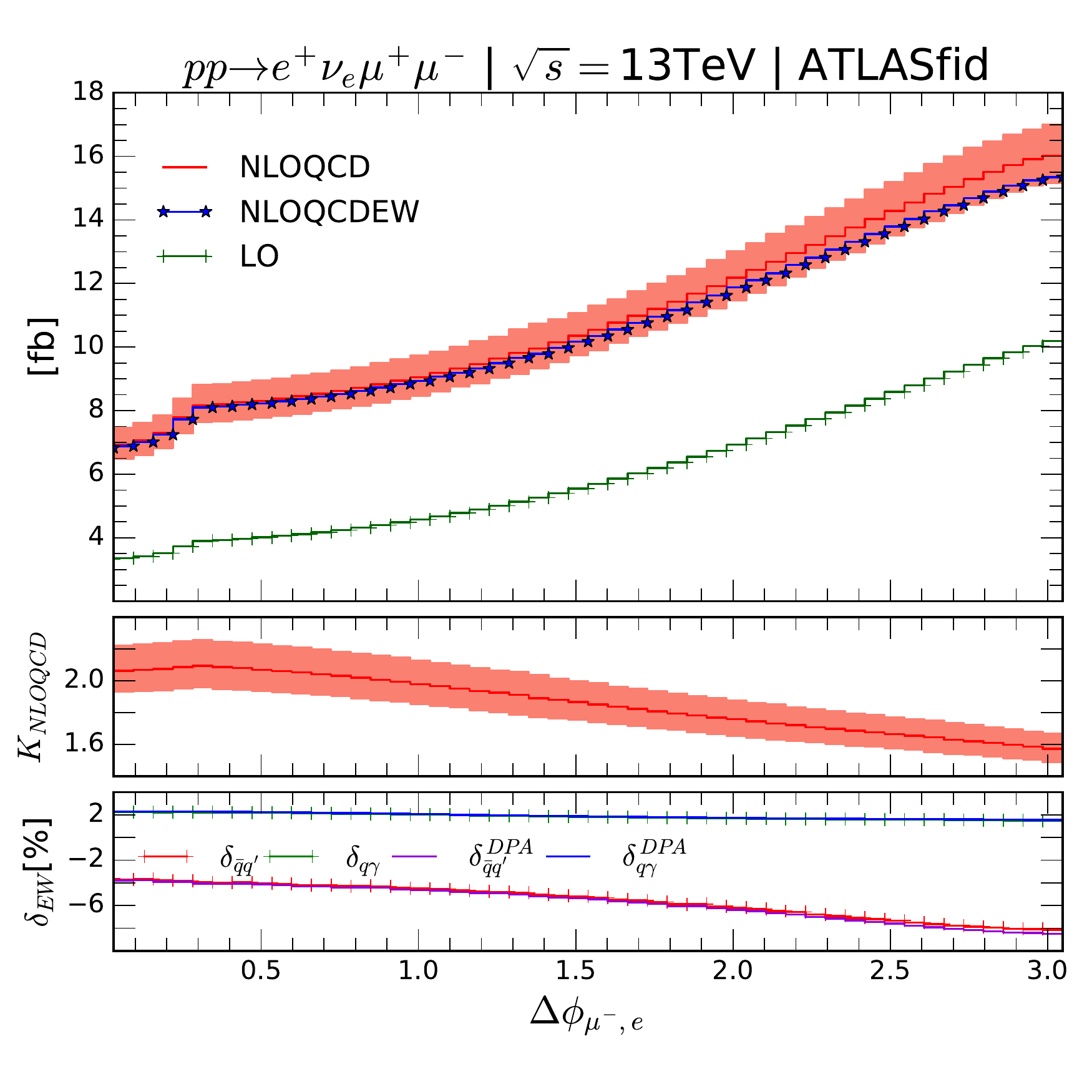}\\
  \includegraphics[width=0.48\textwidth]{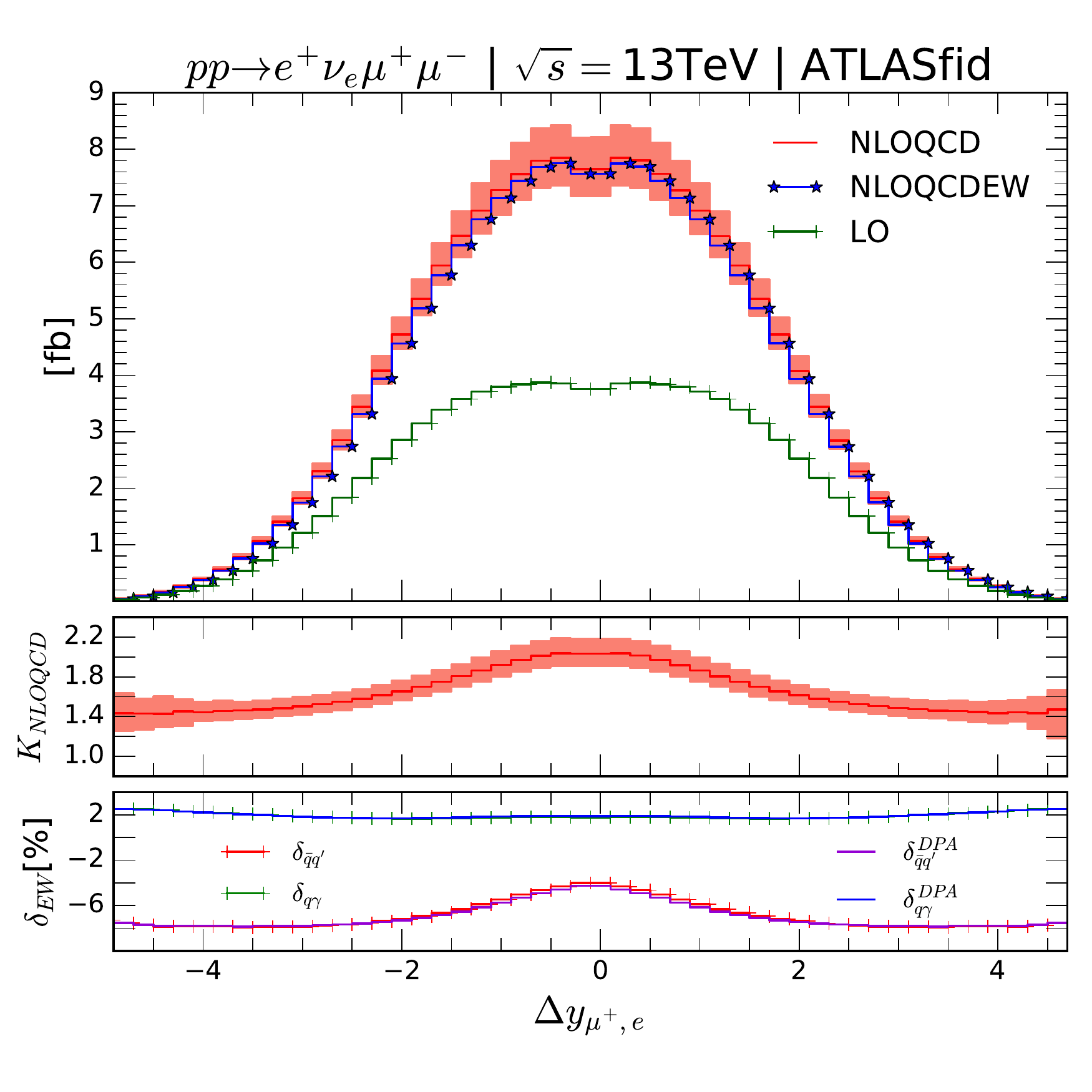}& 
  \includegraphics[width=0.48\textwidth]{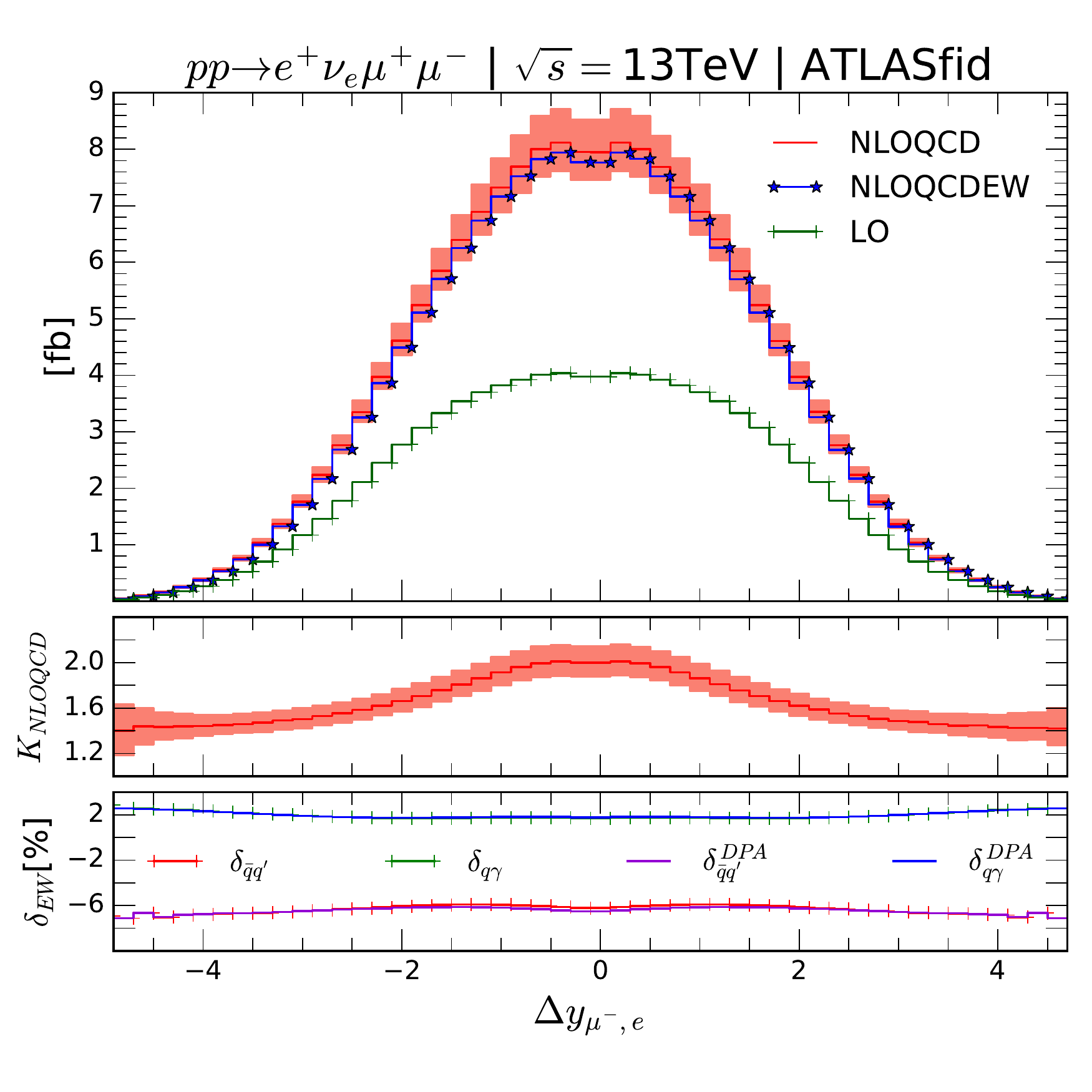}
  \end{tabular}
  \caption{Same as \fig{fig:dist_pT_W_Z_y_W_Z_Wp_atlas} but for the azimuthal-angle difference (top row) 
   and the rapidity difference (bottom row) between the $\mu^+$ and $e^+$ (left column) and between the 
   $\mu^-$ and $e^+$ (right column).}
  \label{fig:dist_Delta_phi_y_Wp_atlas}
\end{figure}
\begin{figure}[ht!]
  \centering
  \begin{tabular}{cc}
    \includegraphics[width=0.48\textwidth]{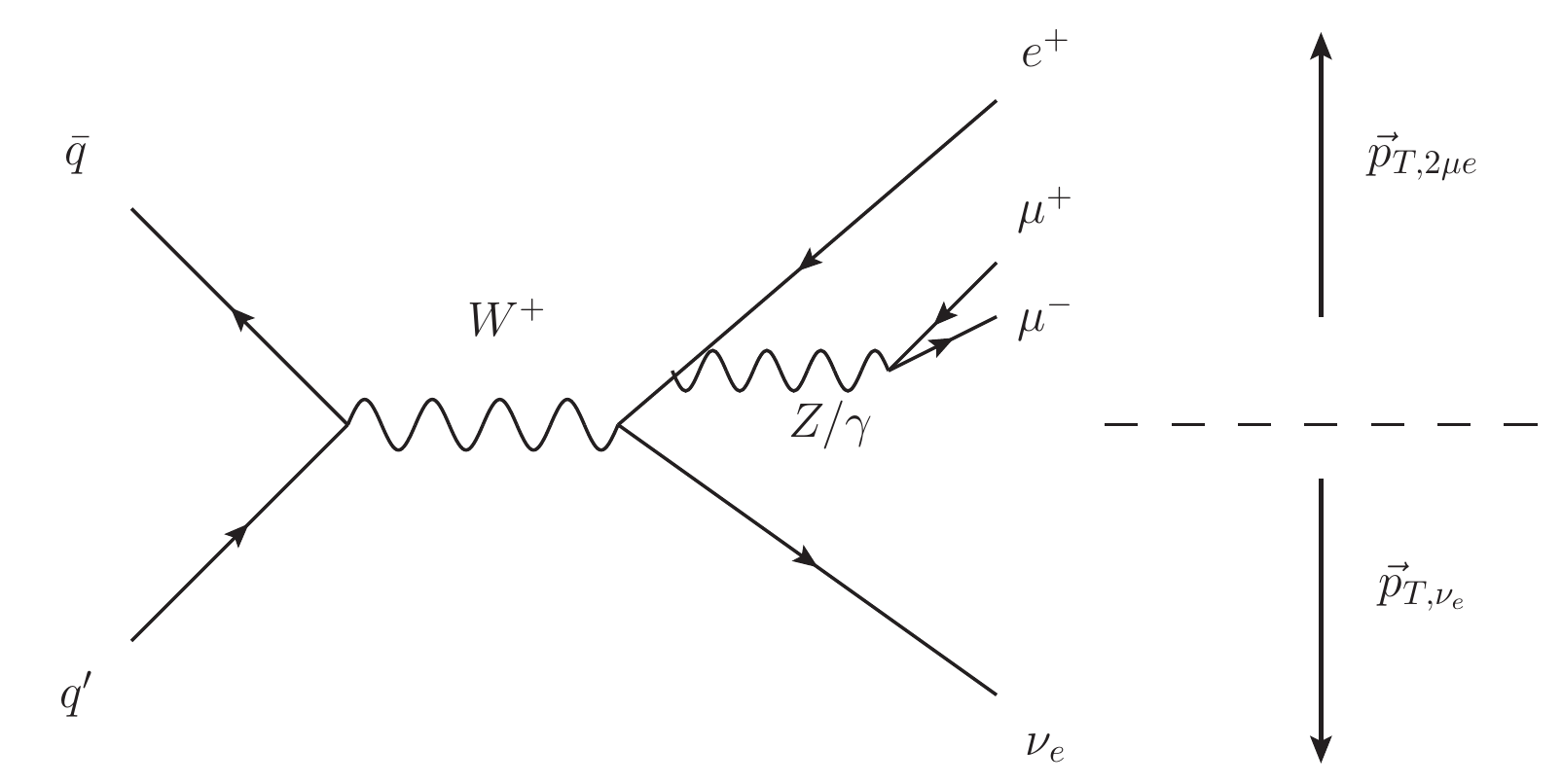}&
    \includegraphics[width=0.48\textwidth]{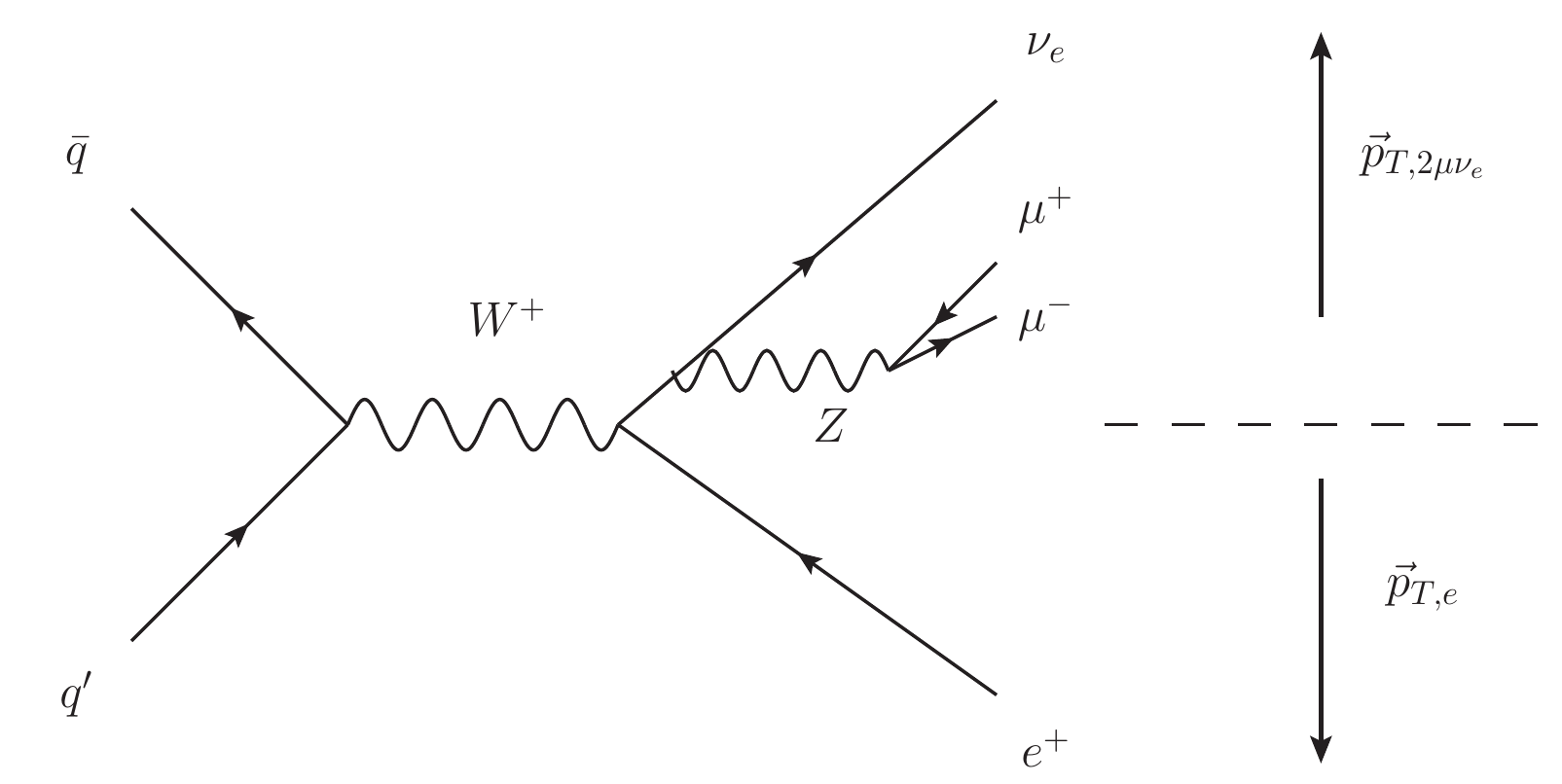}
  \end{tabular}
  \caption{Illustration of the diagrammatic structures non present in
    the DPA and that dominate the $p_{T,e}$ distribution (left) and
    the $p_{T,\nu}$ distribution (right), for high transverse momentum.}
  \label{fig:diags_pT_structure}
\end{figure}
\begin{figure}[ht!]
  \centering
  \begin{tabular}{cc}
  \includegraphics[width=0.48\textwidth]{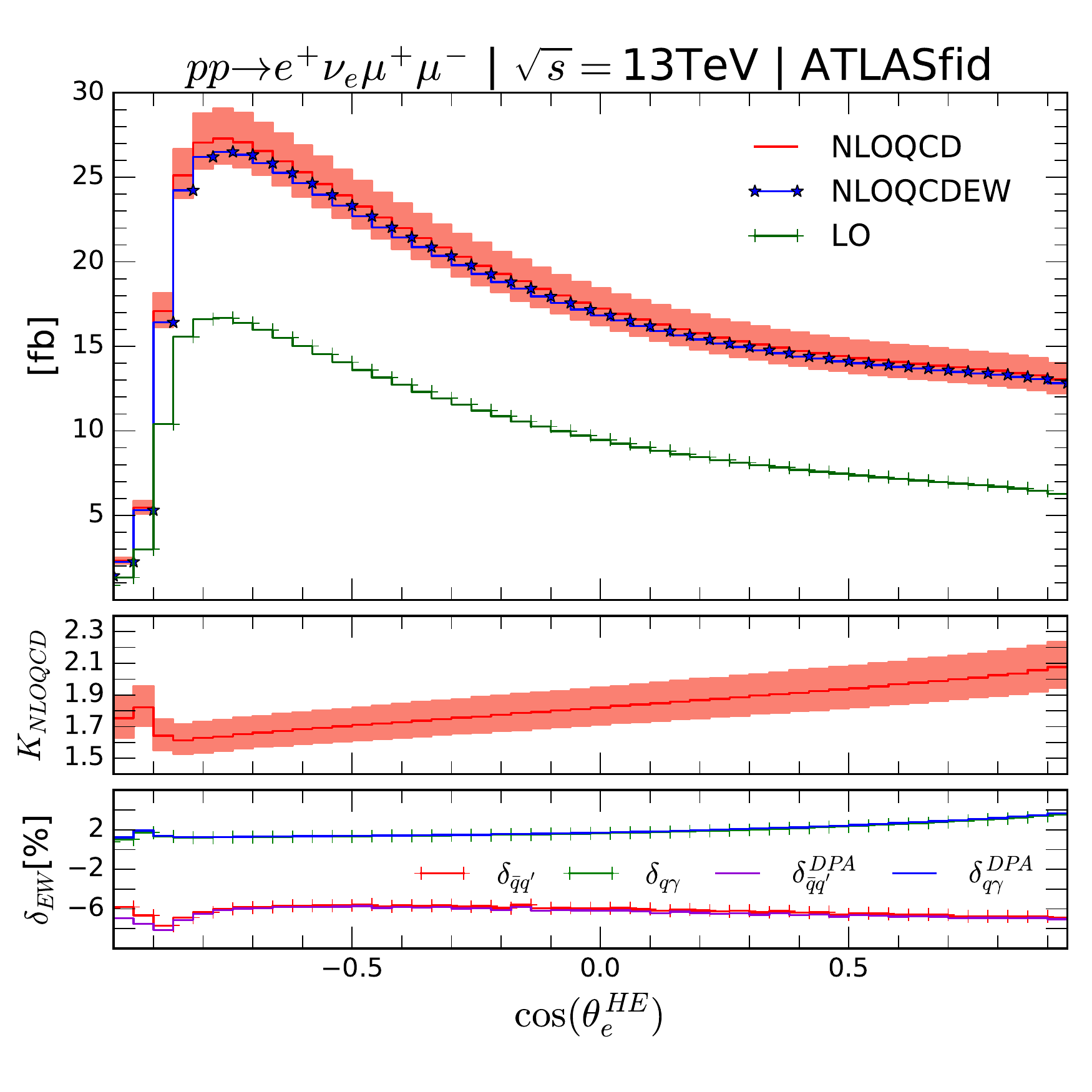}&
  \includegraphics[width=0.48\textwidth]{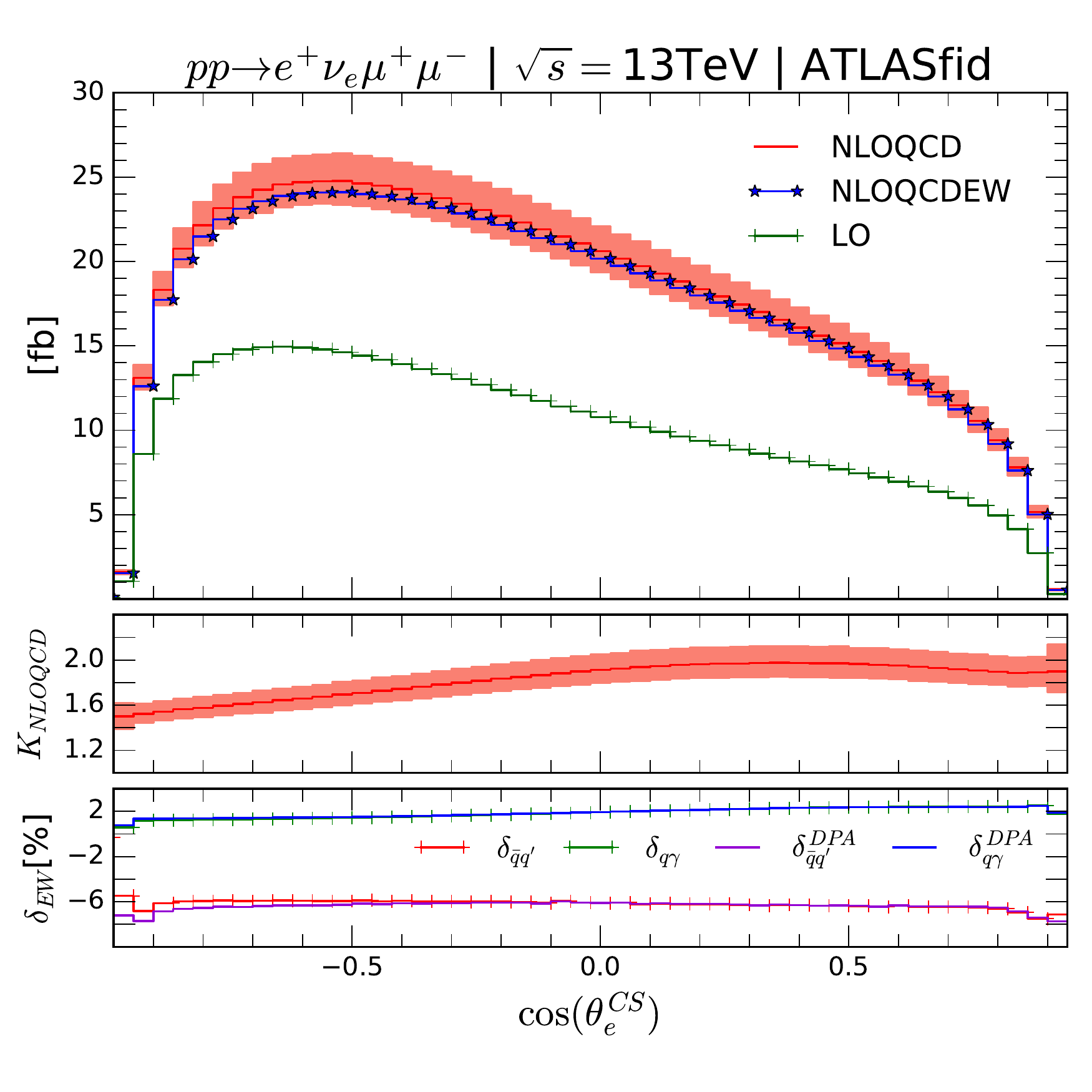}\\
  \includegraphics[width=0.48\textwidth]{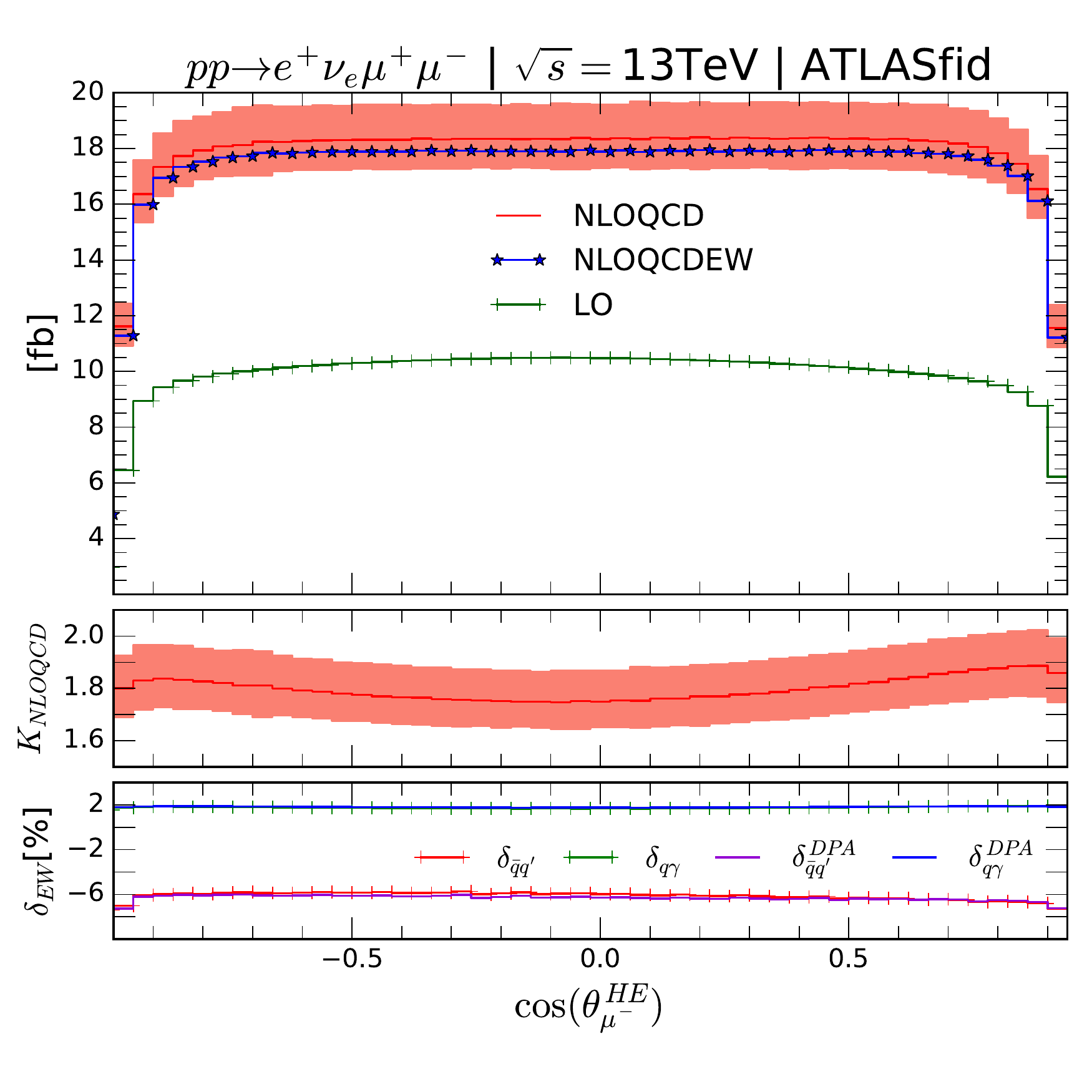}&
  \includegraphics[width=0.48\textwidth]{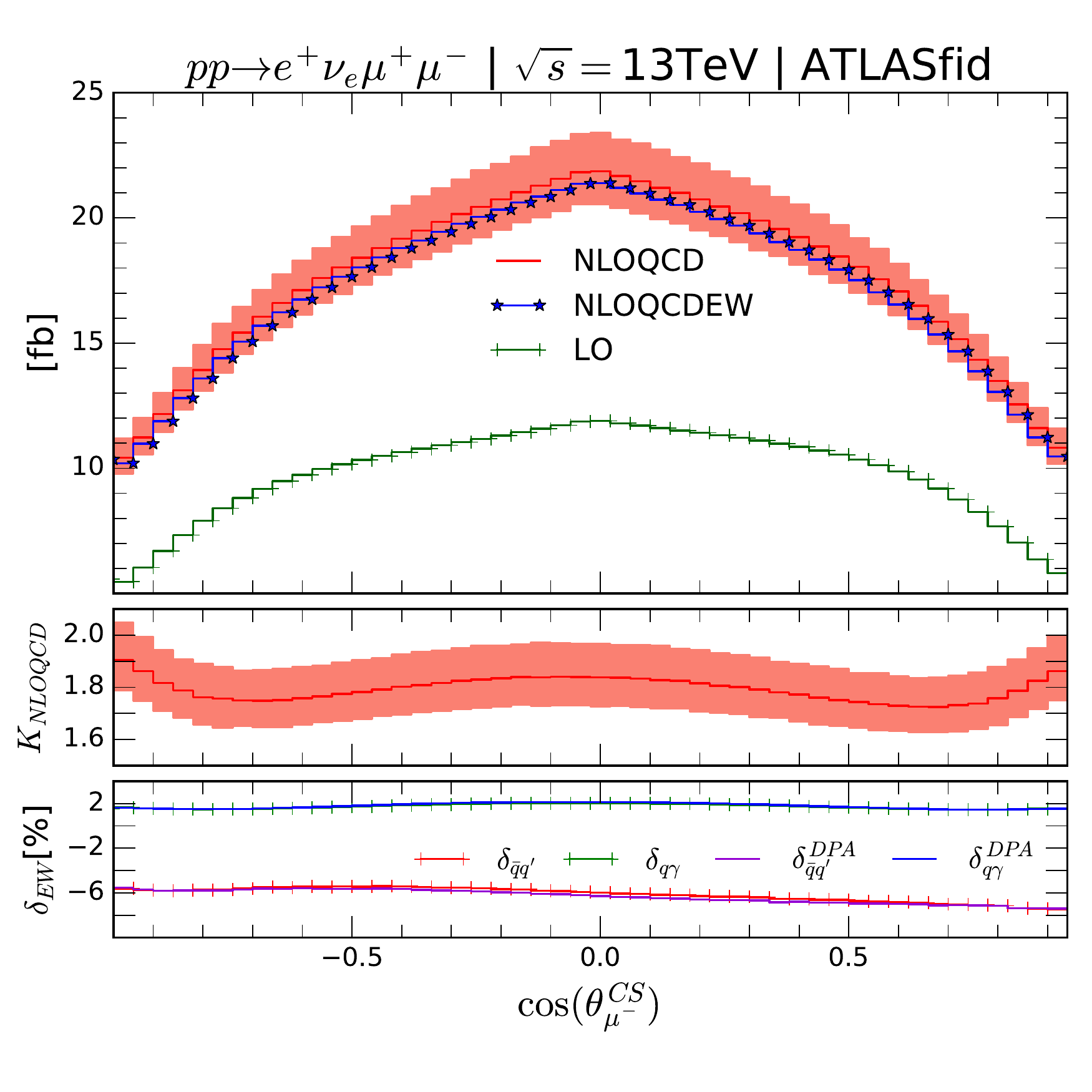}
  \end{tabular}
  \caption{Same as \fig{fig:dist_pT_W_Z_y_W_Z_Wp_atlas} but for the $\cos\theta$ distributions of the electron (top row) 
   calculated in the Helicity (left) and Collins-Soper (right) coordinate systems. The same distributions for 
   the muon are shown in the bottom row.}
   \label{fig:dist_cos_theta_HEL_CS_e_muon_Wp_atlas}
\end{figure}
\begin{figure}[ht!]
  \centering
  \begin{tabular}{cc}
  \includegraphics[width=0.48\textwidth]{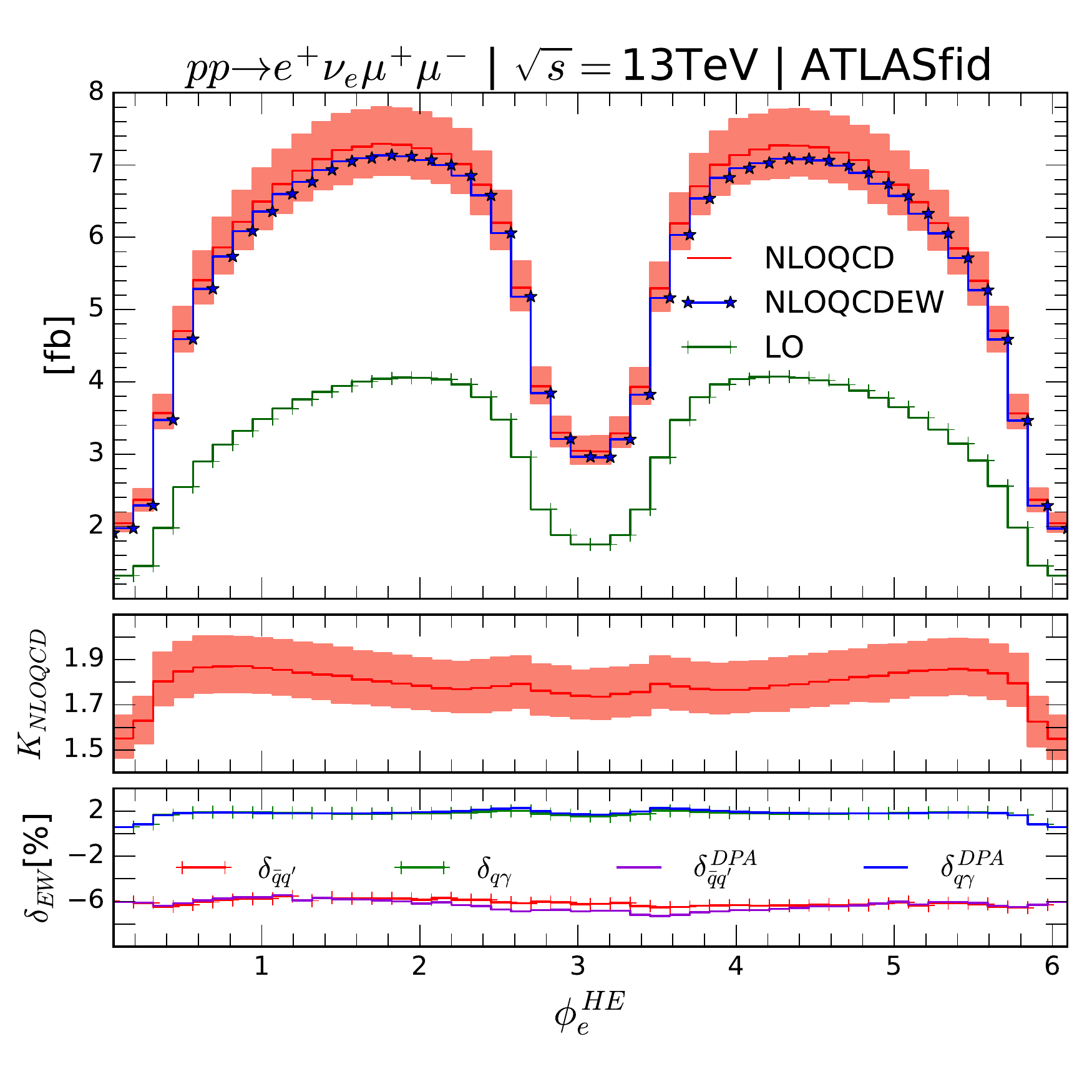}&
  \includegraphics[width=0.48\textwidth]{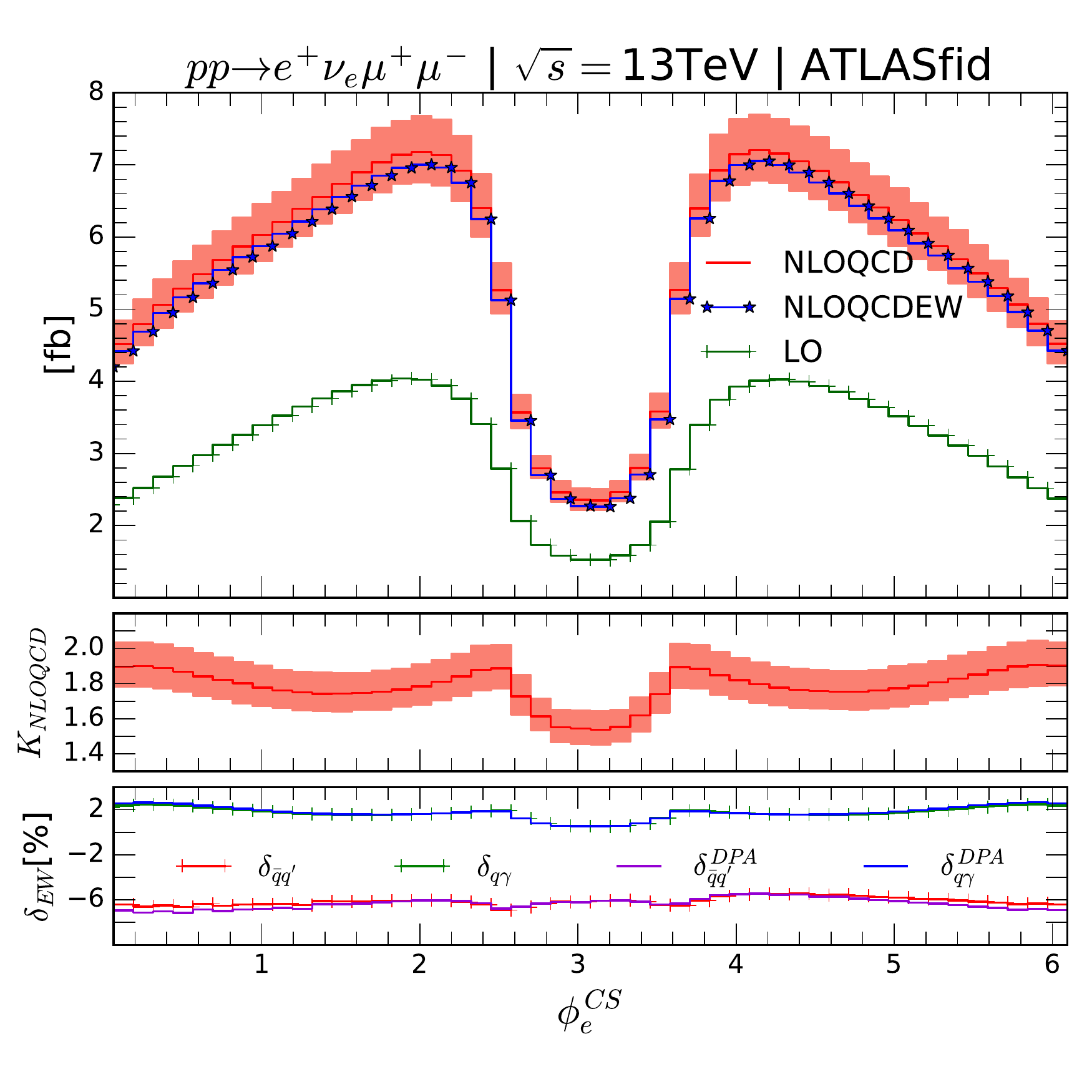}\\
  \includegraphics[width=0.48\textwidth]{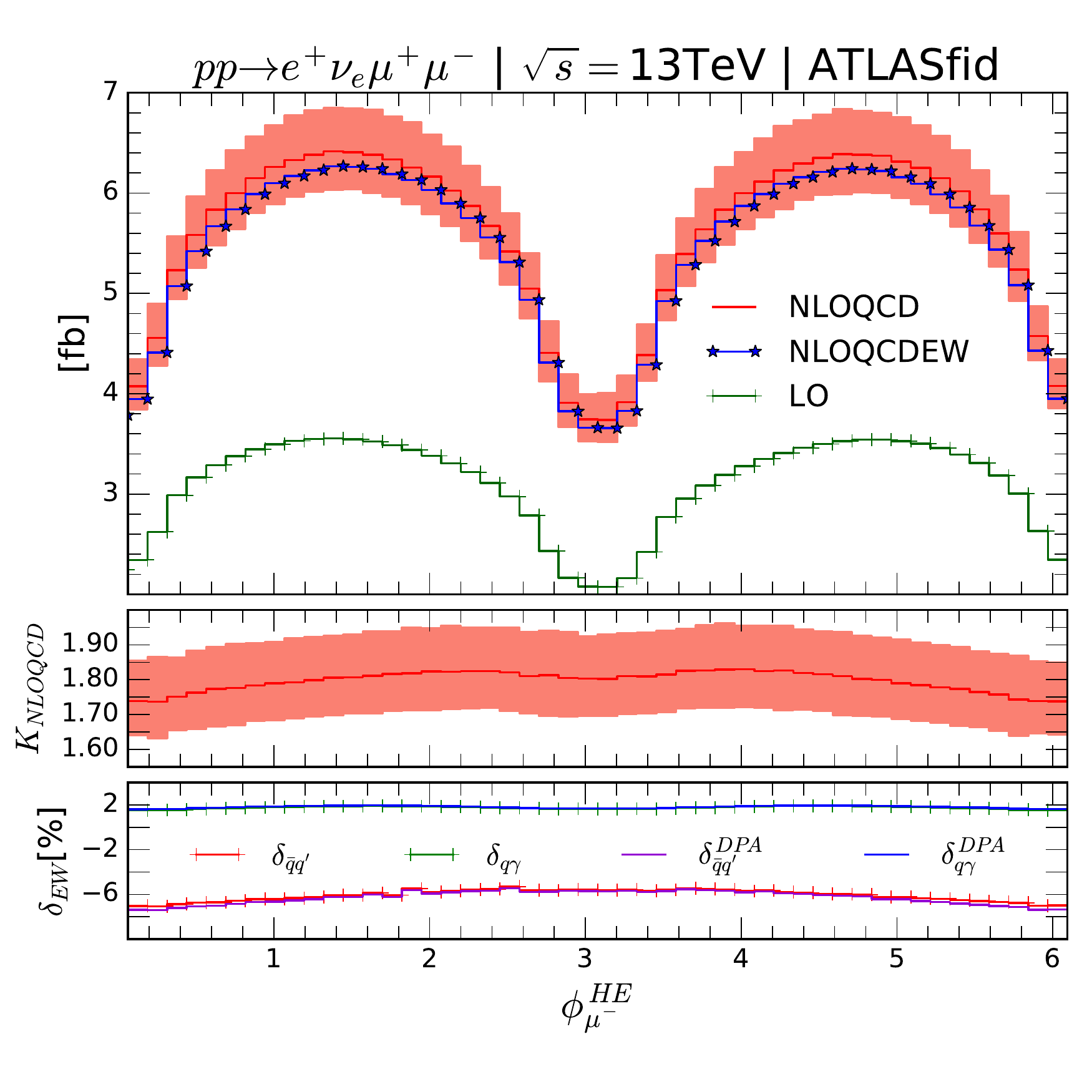}&
  \includegraphics[width=0.48\textwidth]{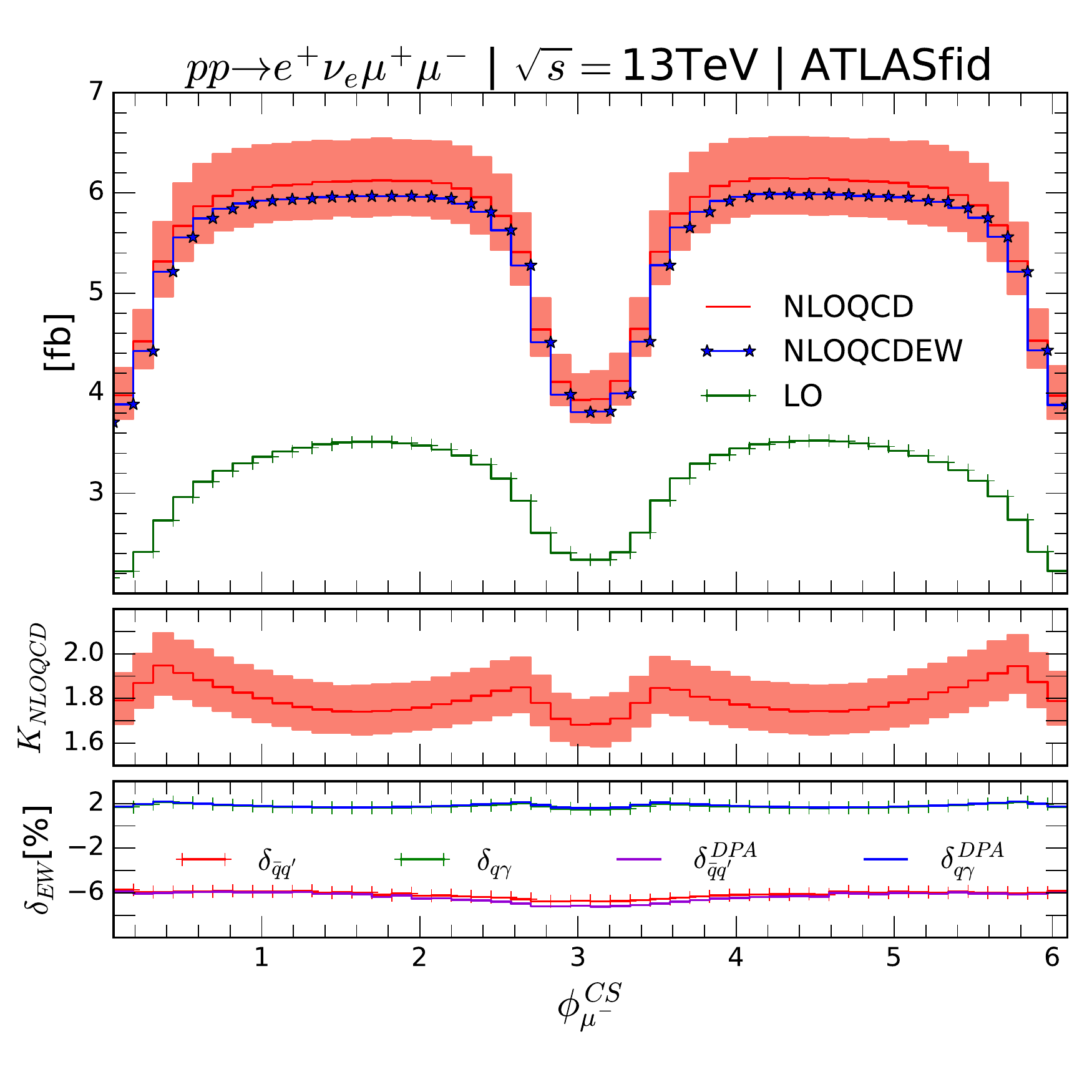}
  \end{tabular}
  \caption{Same as \fig{fig:dist_pT_W_Z_y_W_Z_Wp_atlas} but for the azimuthal-angle distributions of the electron (top row) 
   calculated in the Helicity (left) and Collins-Soper (right) coordinate systems. The same distributions for 
   the muon are shown in the bottom row.}
   \label{fig:dist_phi_HEL_CS_e_muon_Wp_atlas}
\end{figure}
\begin{figure}[ht!]
  \centering
  \begin{tabular}{cc}
  \includegraphics[width=0.48\textwidth]{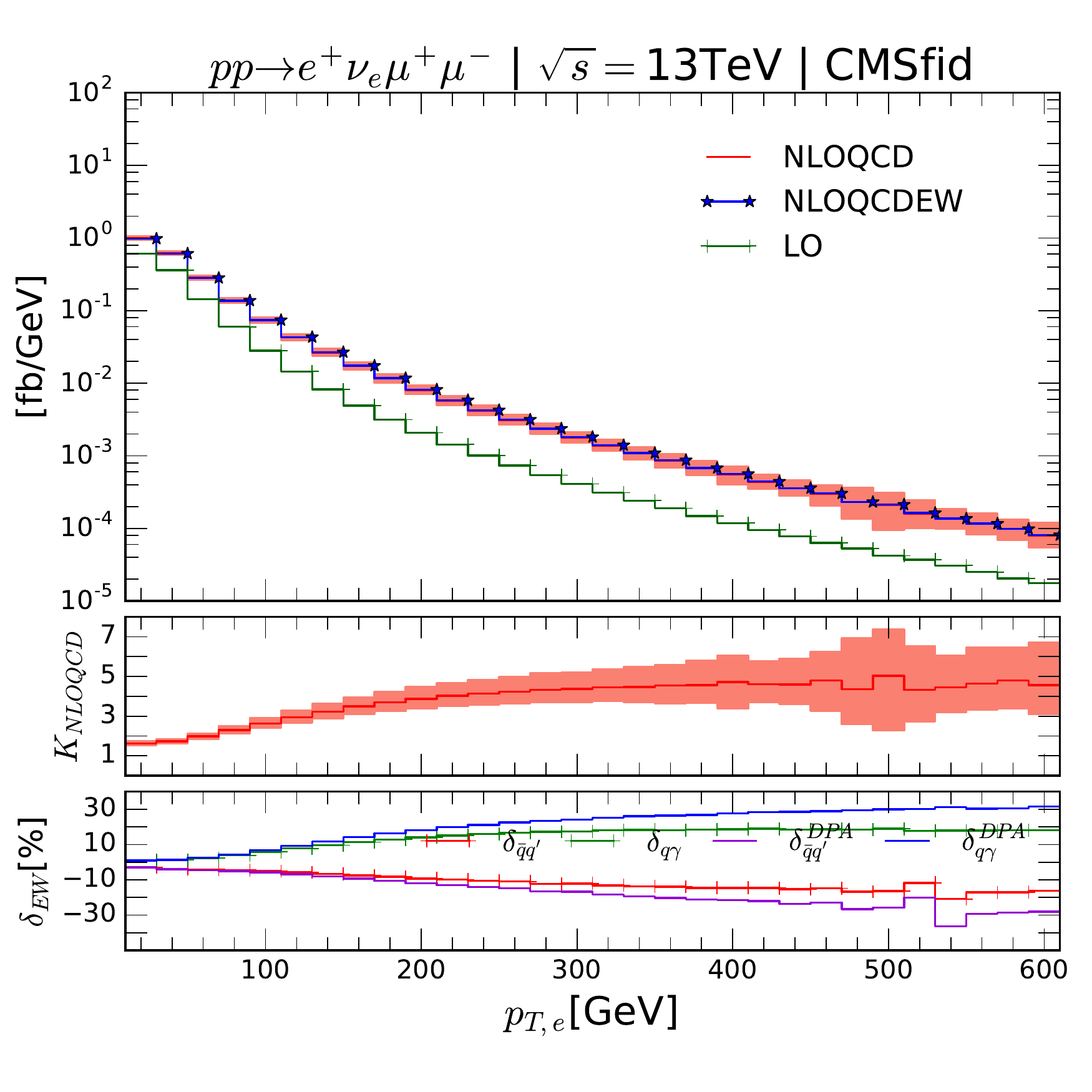}& 
  \includegraphics[width=0.48\textwidth]{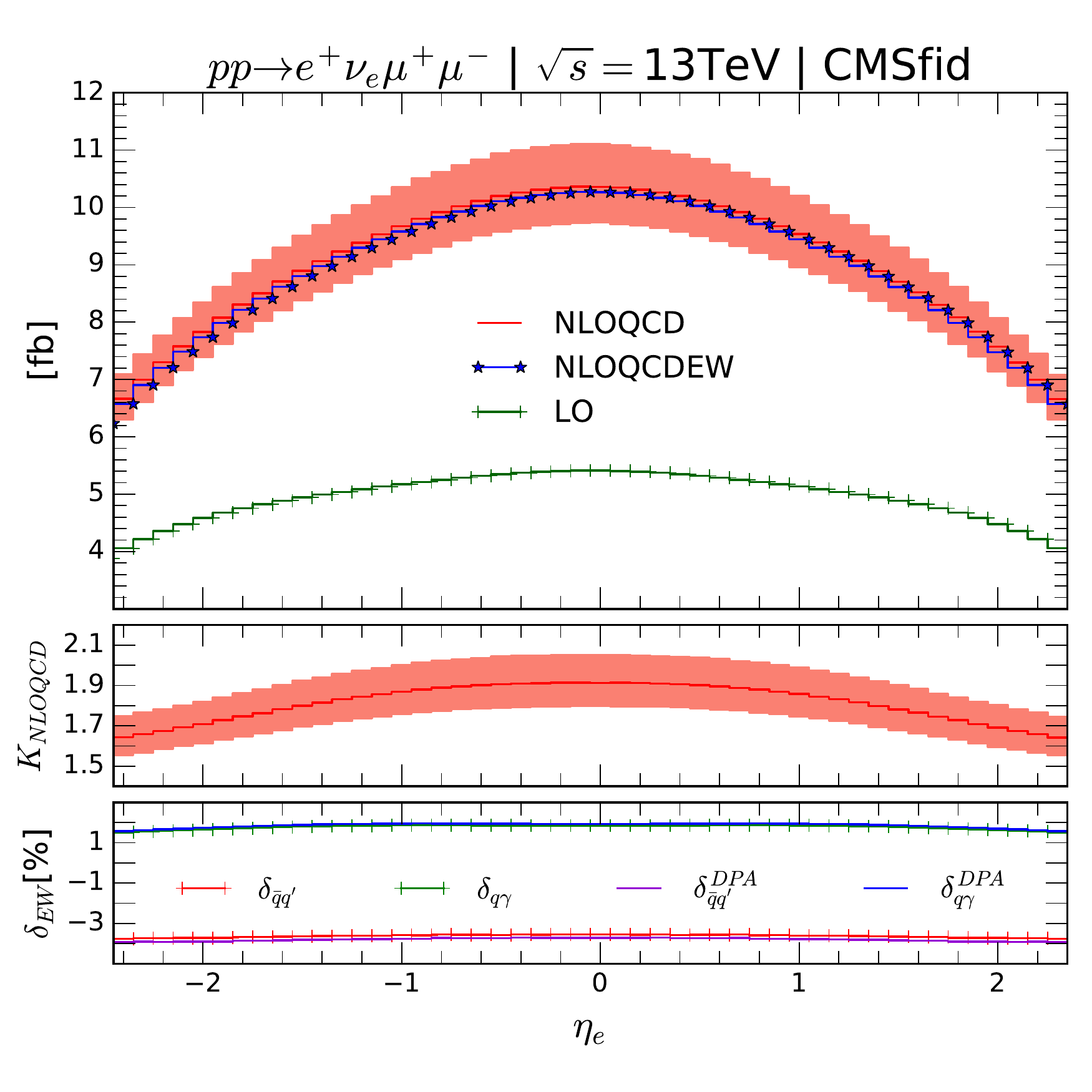}\\
  \includegraphics[width=0.48\textwidth]{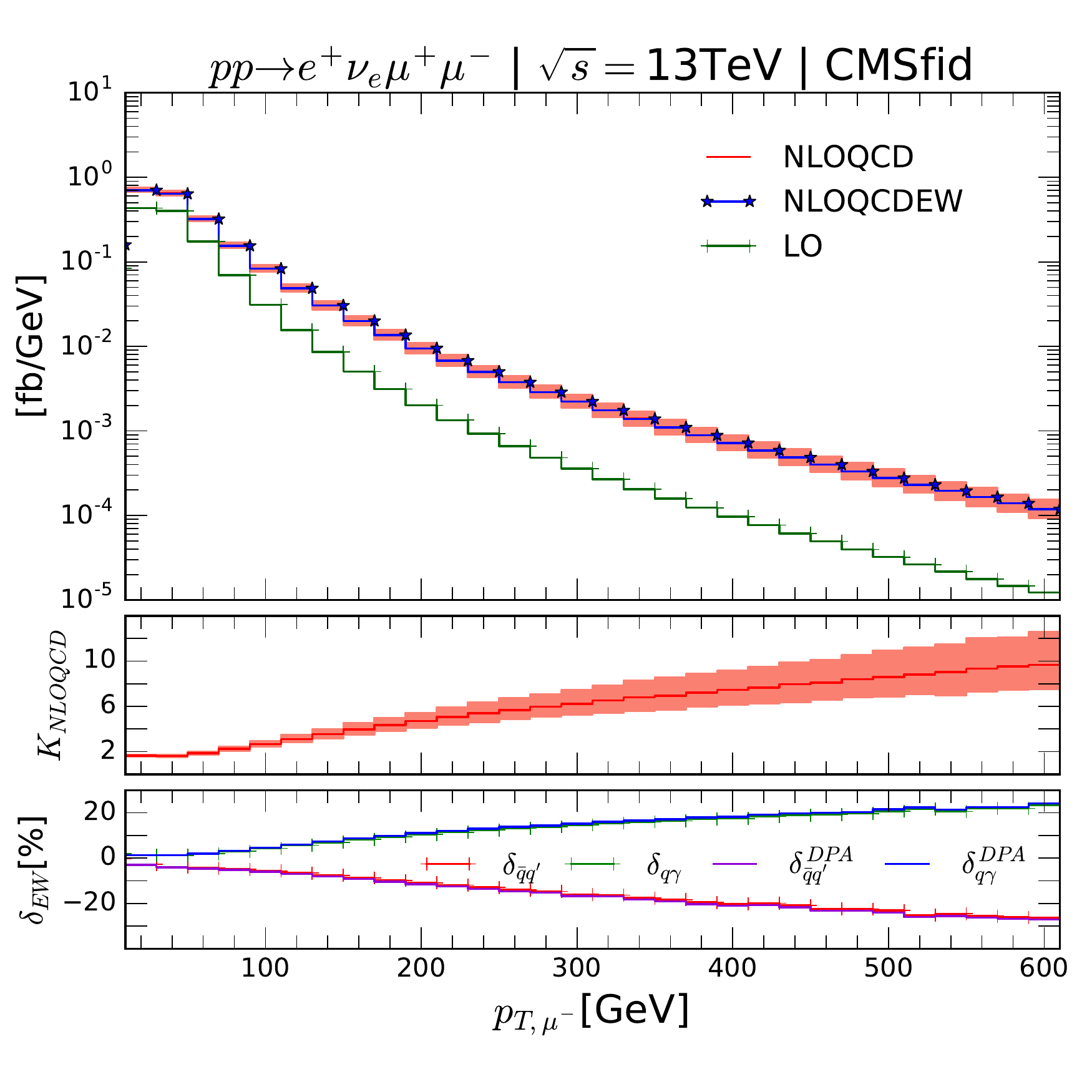}& 
  \includegraphics[width=0.48\textwidth]{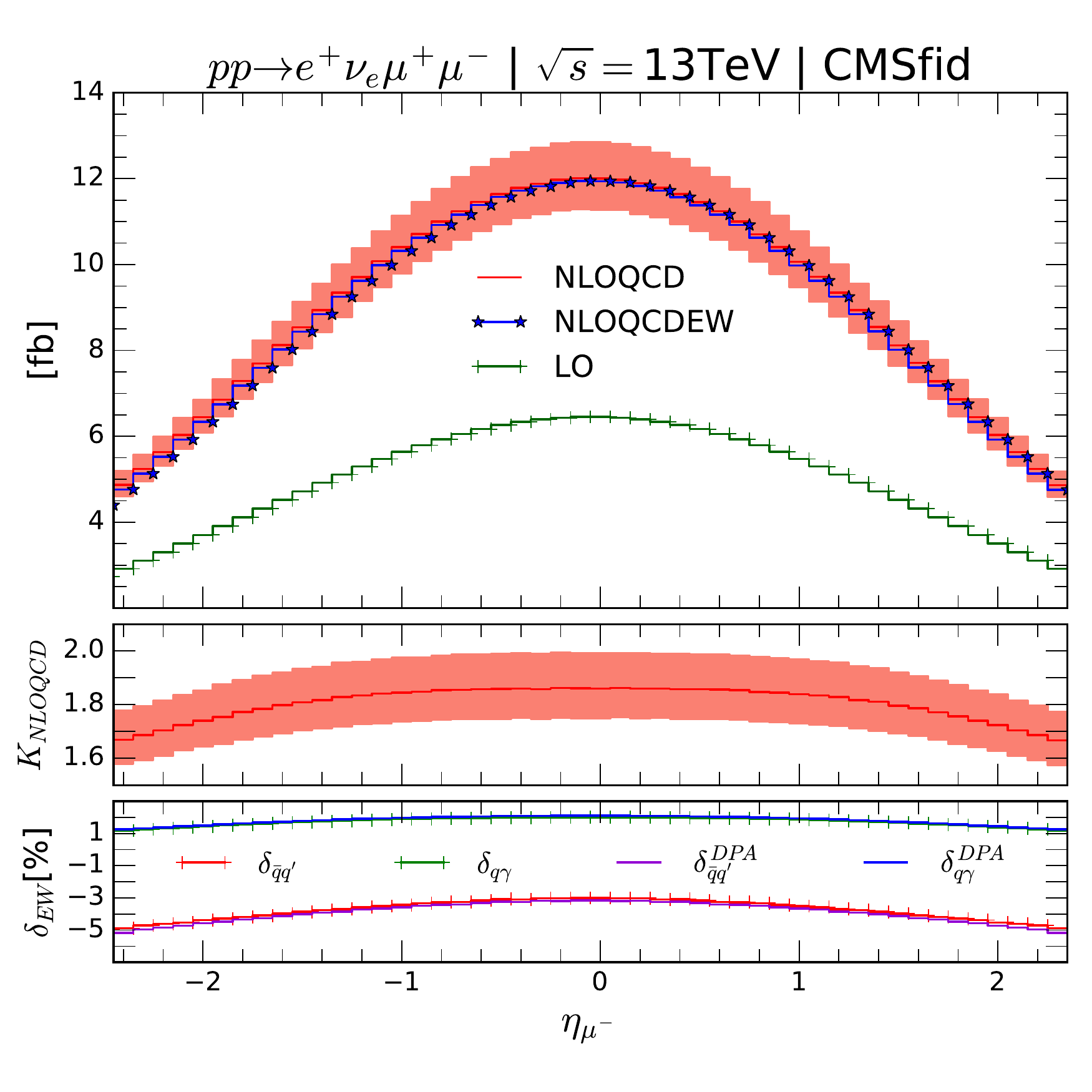}
  \end{tabular}
  \caption{Same as \fig{fig:dist_pT_e_mu_eta_e_mu_Wp_atlas} but with the CMS fiducial cuts.}
  \label{fig:dist_pT_e_mu_eta_e_mu_Wp_cms}
\end{figure}
\begin{figure}[ht!]
  \centering
  \begin{tabular}{cc}
  \includegraphics[width=0.48\textwidth]{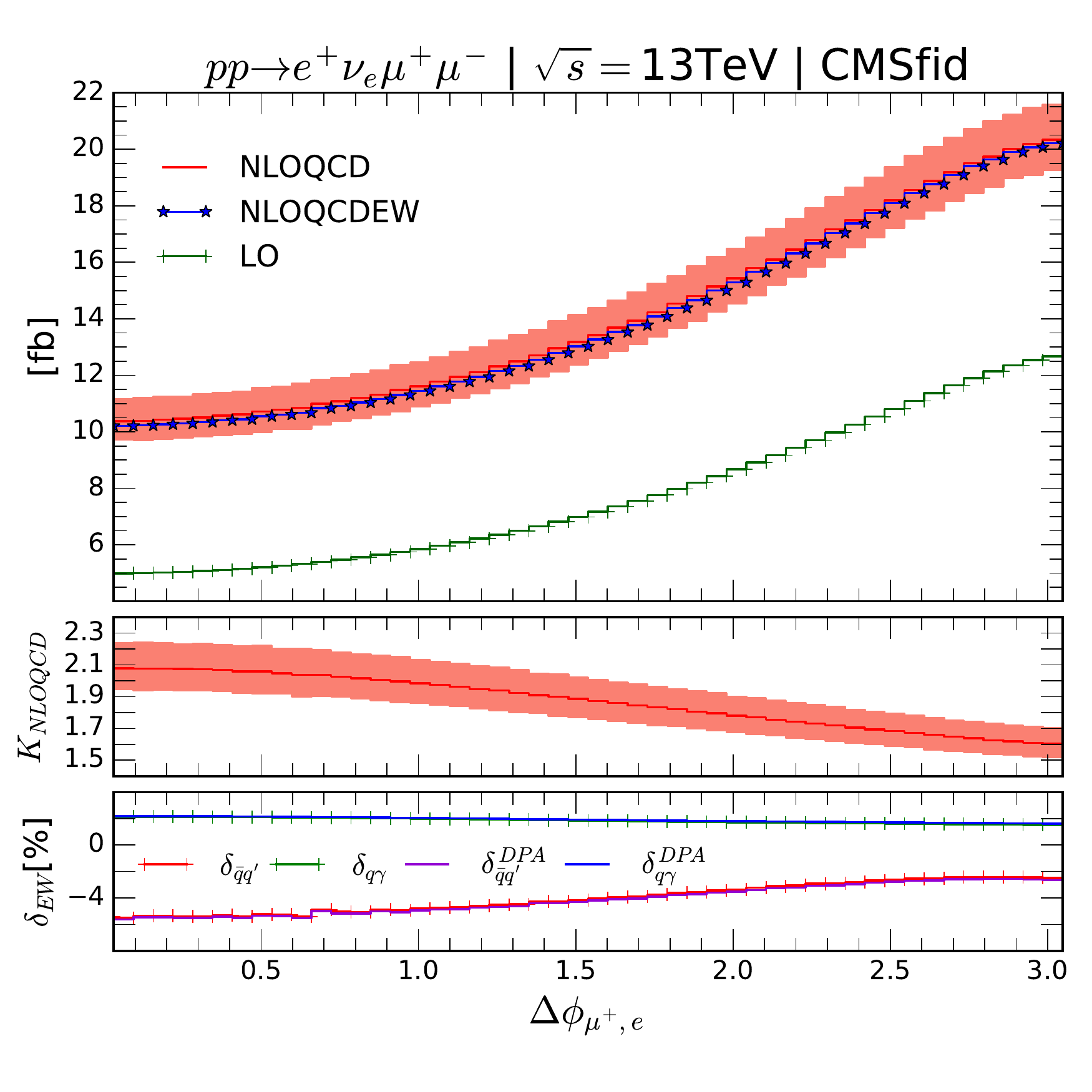}& 
  \includegraphics[width=0.48\textwidth]{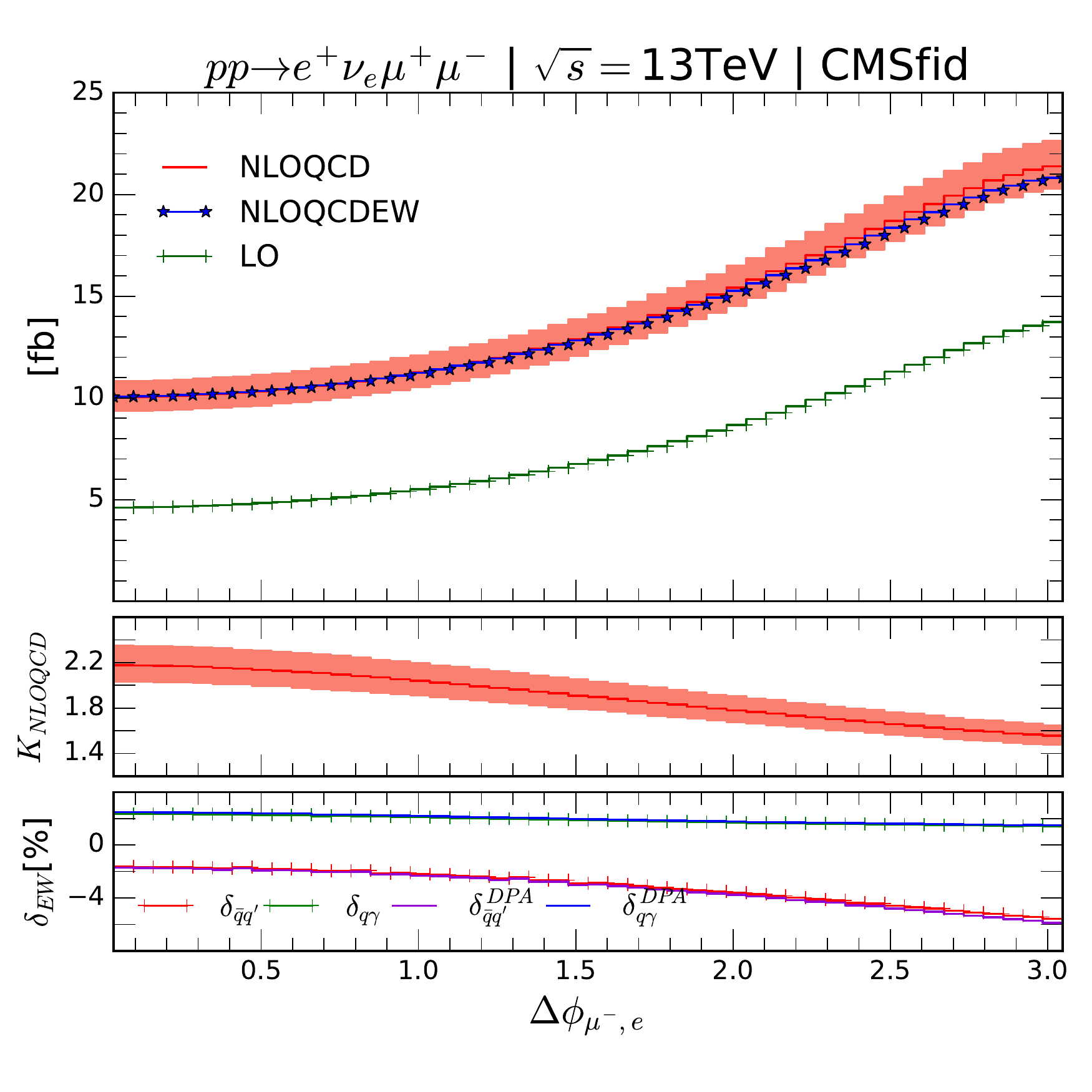}\\
  \includegraphics[width=0.48\textwidth]{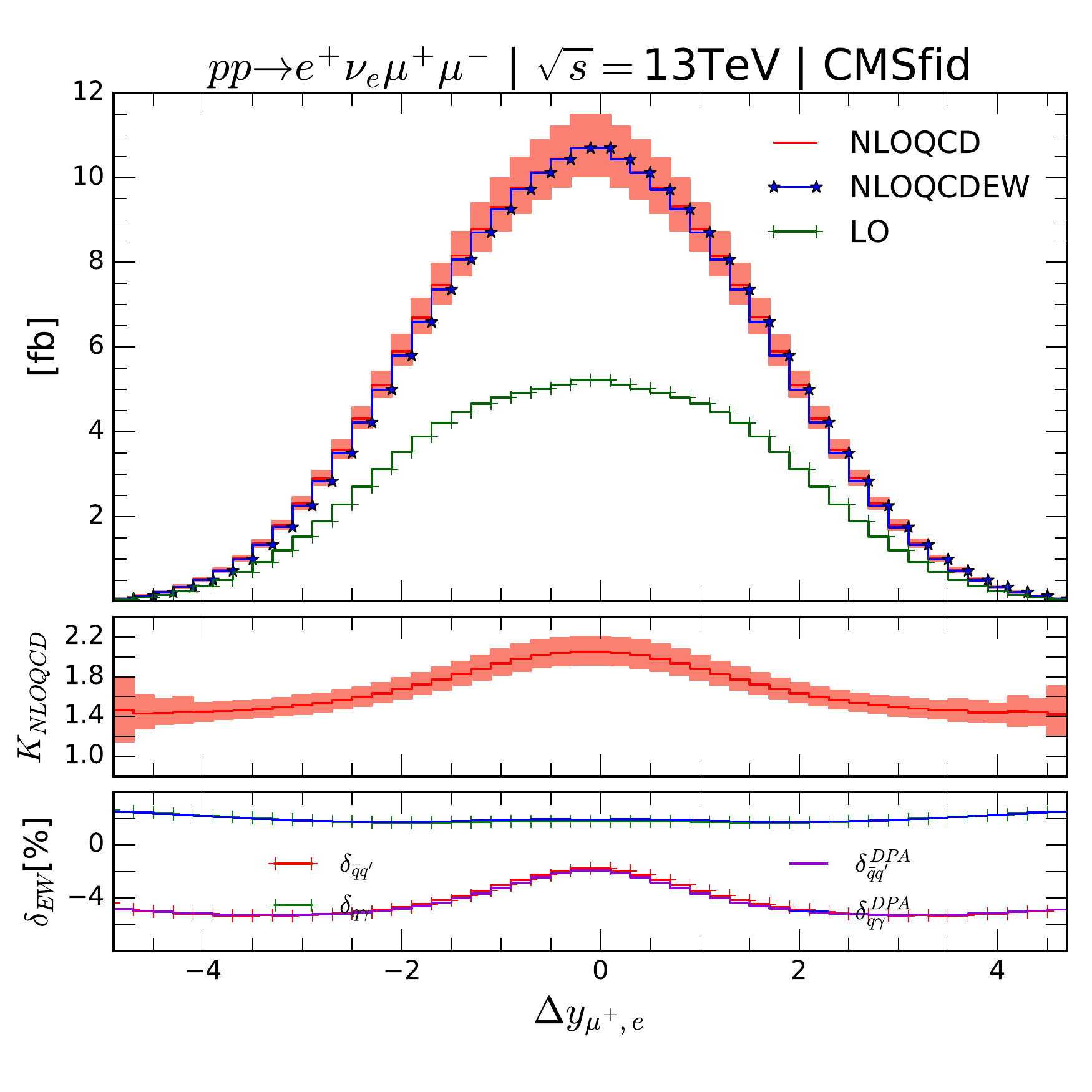}& 
  \includegraphics[width=0.48\textwidth]{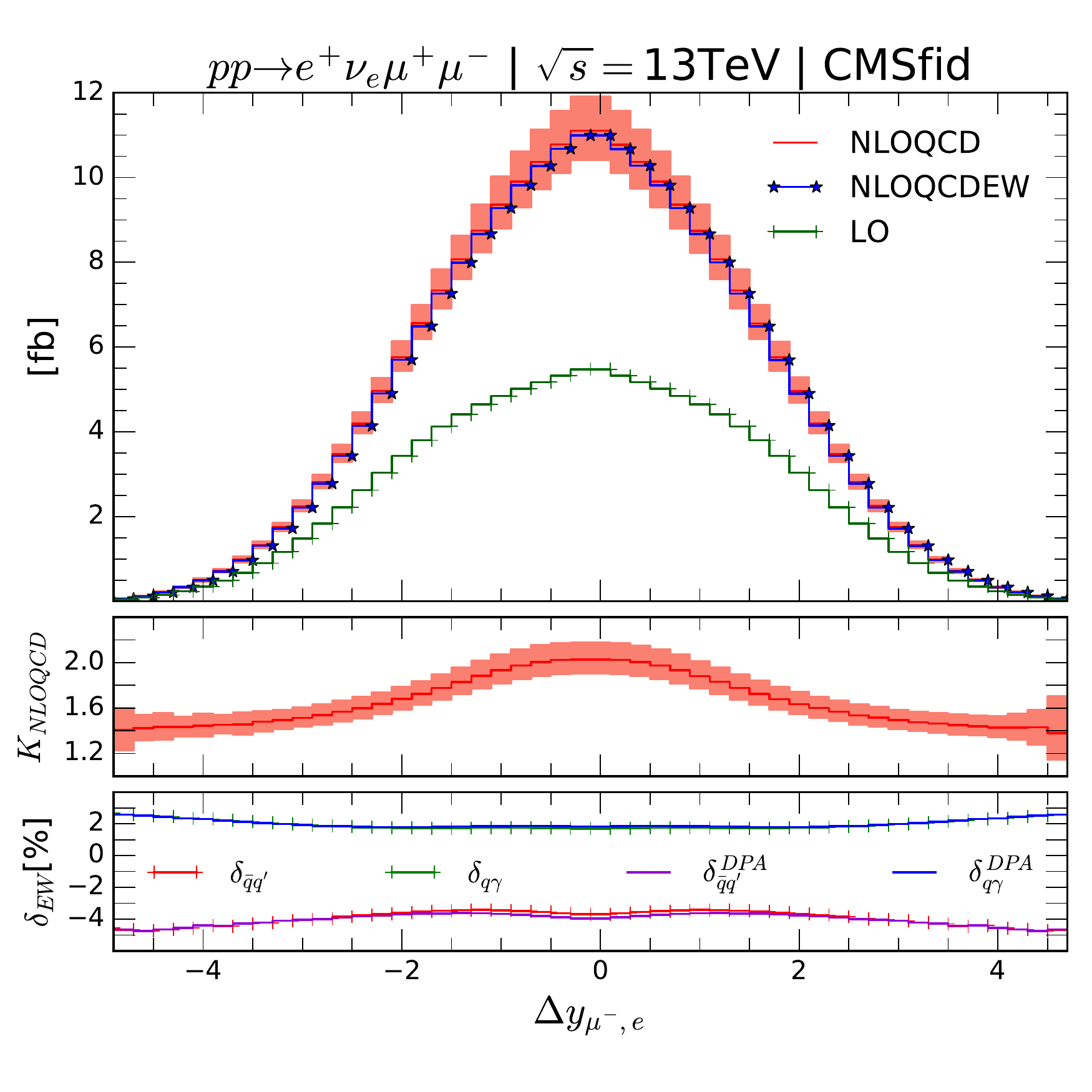}
  \end{tabular}
  \caption{Same as \fig{fig:dist_Delta_phi_y_Wp_atlas} but with the CMS fiducial cuts.}
  \label{fig:dist_Delta_phi_y_Wp_cms}
\end{figure}
\begin{figure}[ht!]
  \centering
  \begin{tabular}{cc}
  \includegraphics[width=0.48\textwidth]{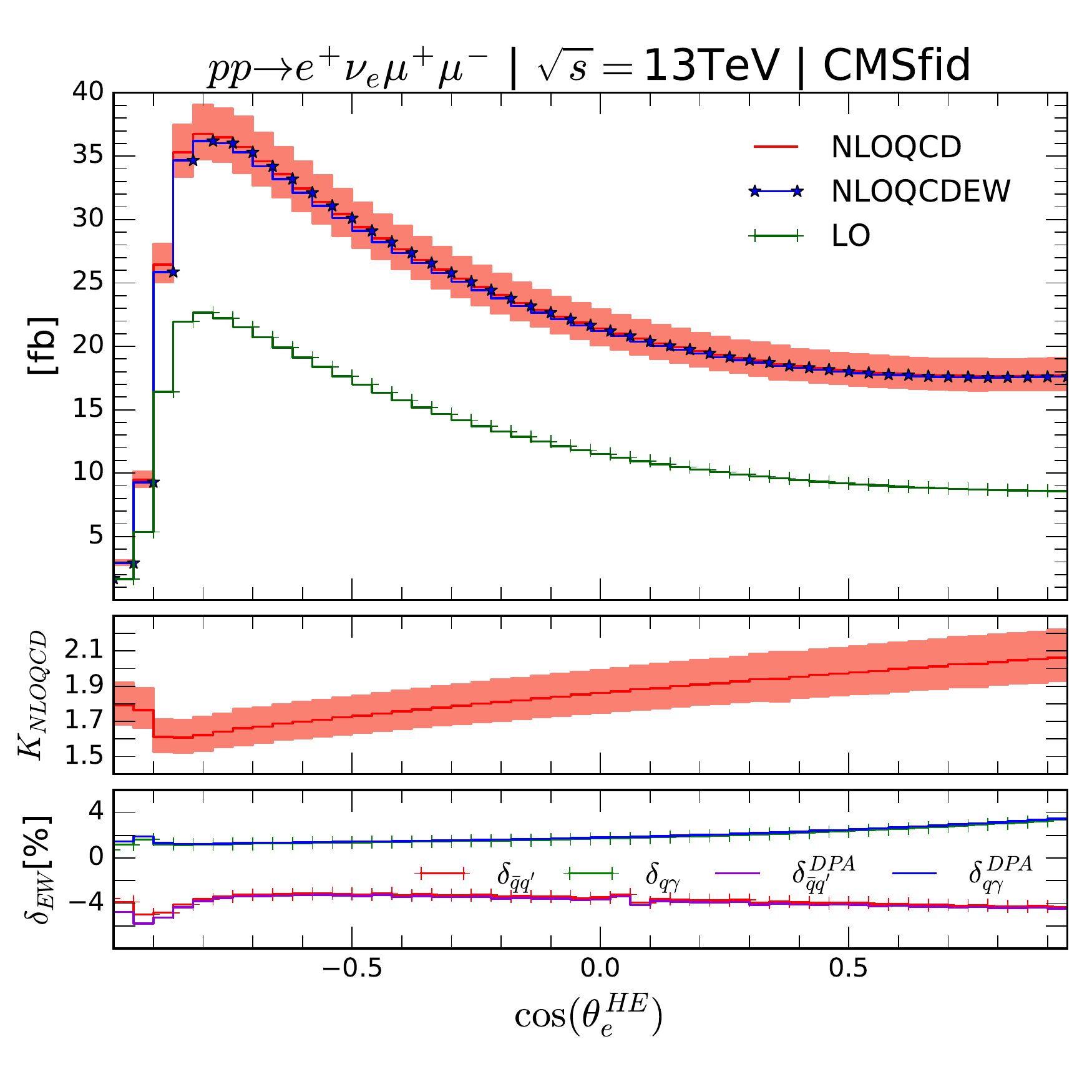}&
  \includegraphics[width=0.48\textwidth]{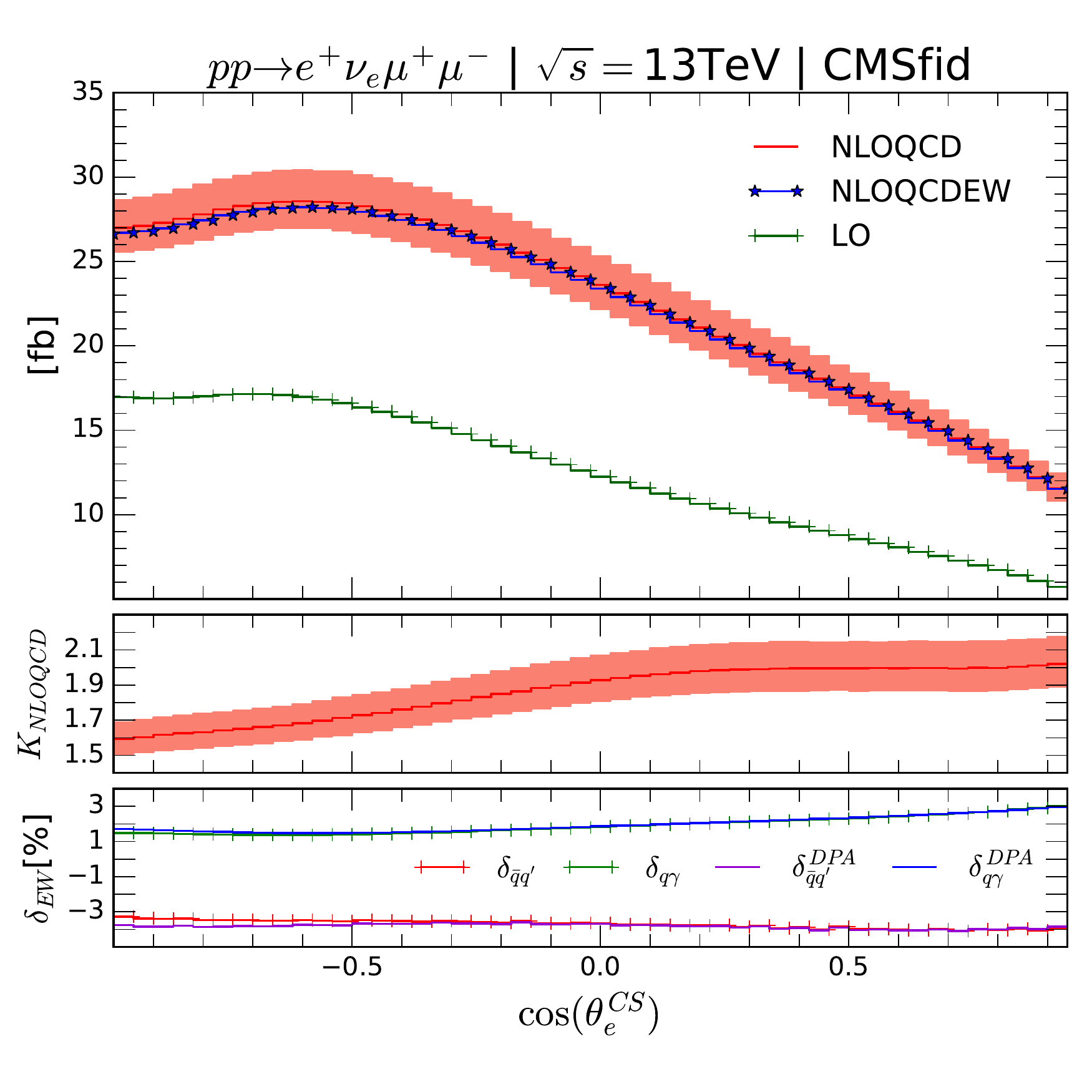}\\
  \includegraphics[width=0.48\textwidth]{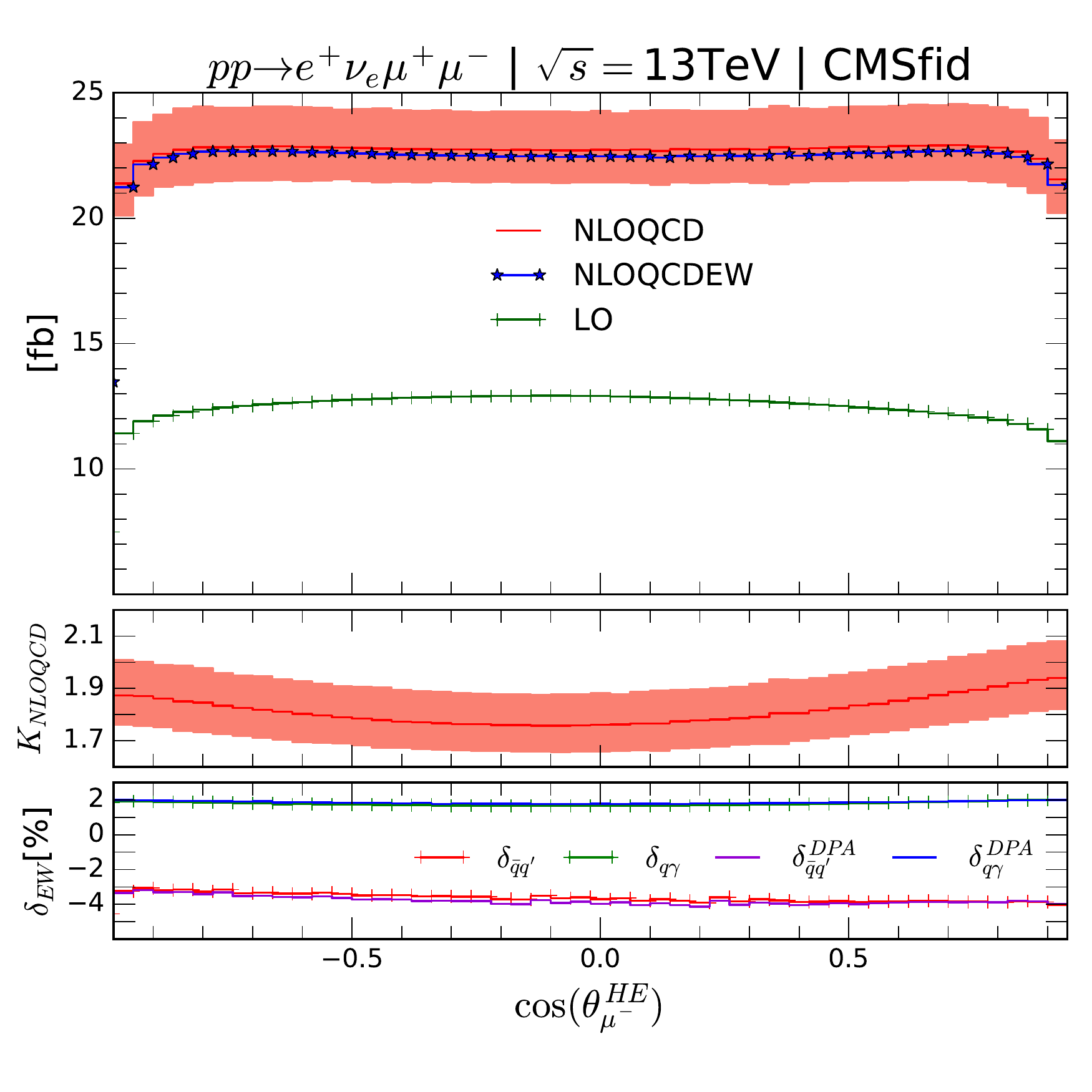}&
  \includegraphics[width=0.48\textwidth]{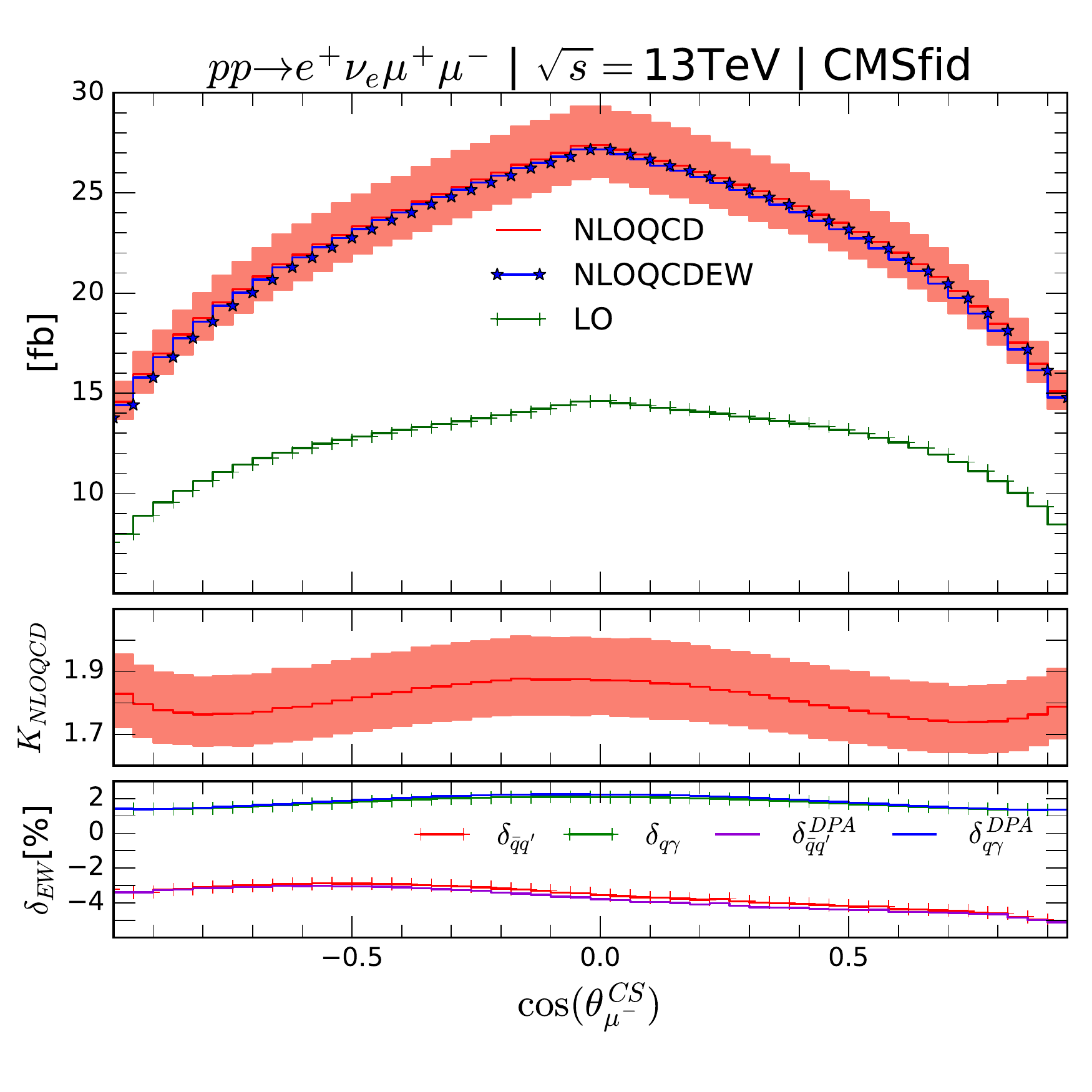}
  \end{tabular}
  \caption{Same as \fig{fig:dist_cos_theta_HEL_CS_e_muon_Wp_atlas} but with the CMS fiducial cuts.}
   \label{fig:dist_cos_theta_HEL_CS_e_muon_Wp_cms}
\end{figure}
\begin{figure}[ht!]
  \centering
  \begin{tabular}{cc}
  \includegraphics[width=0.48\textwidth]{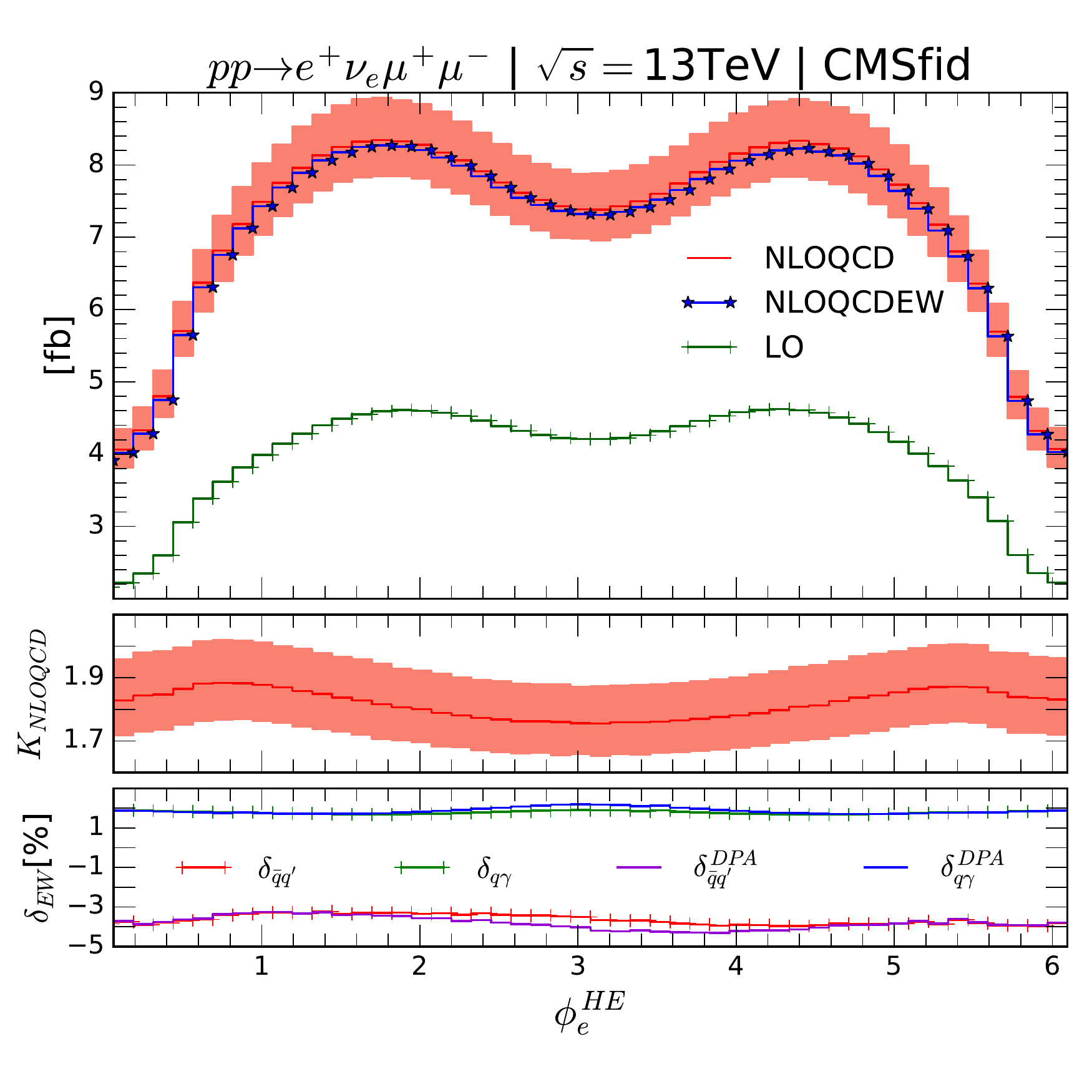}&
  \includegraphics[width=0.48\textwidth]{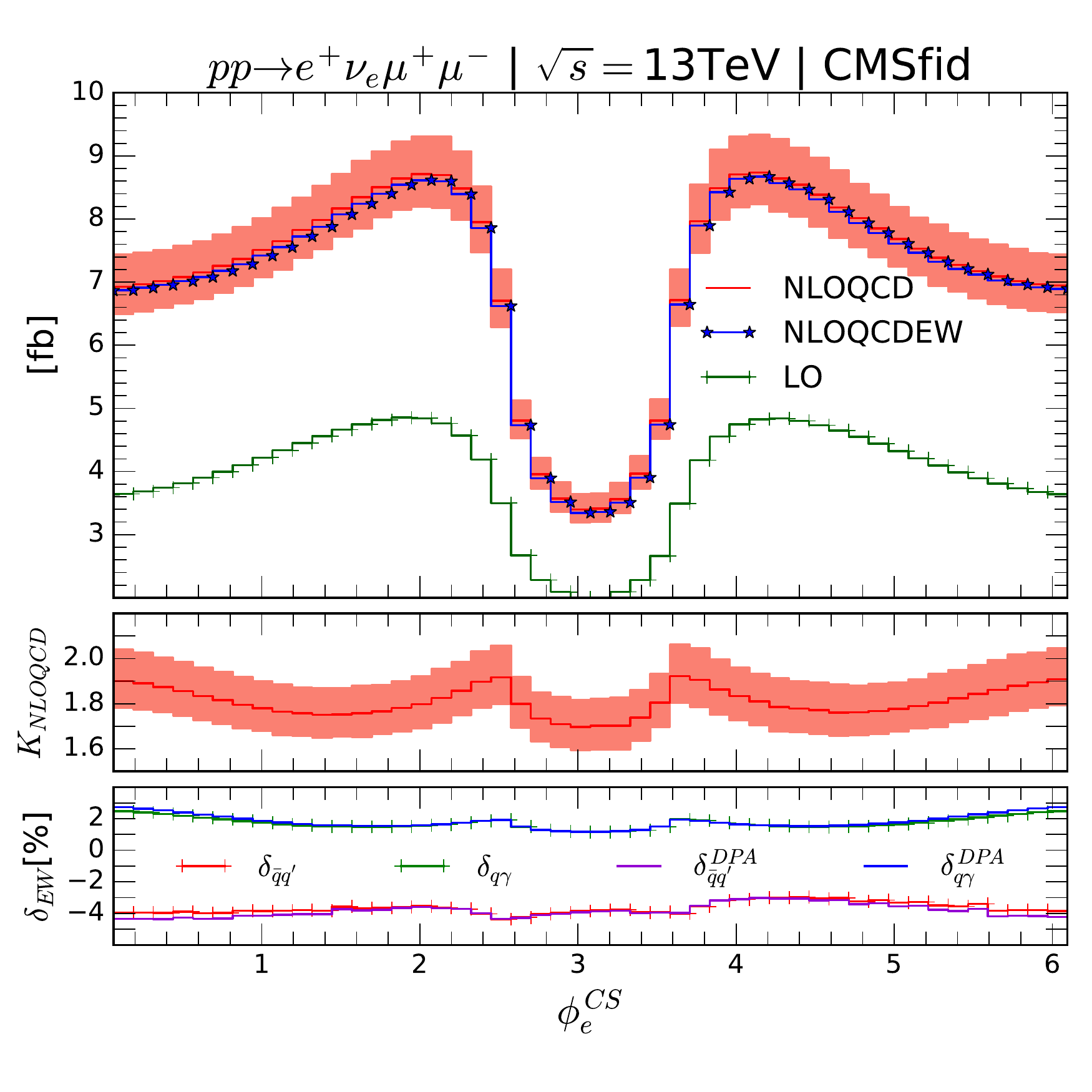}\\
  \includegraphics[width=0.48\textwidth]{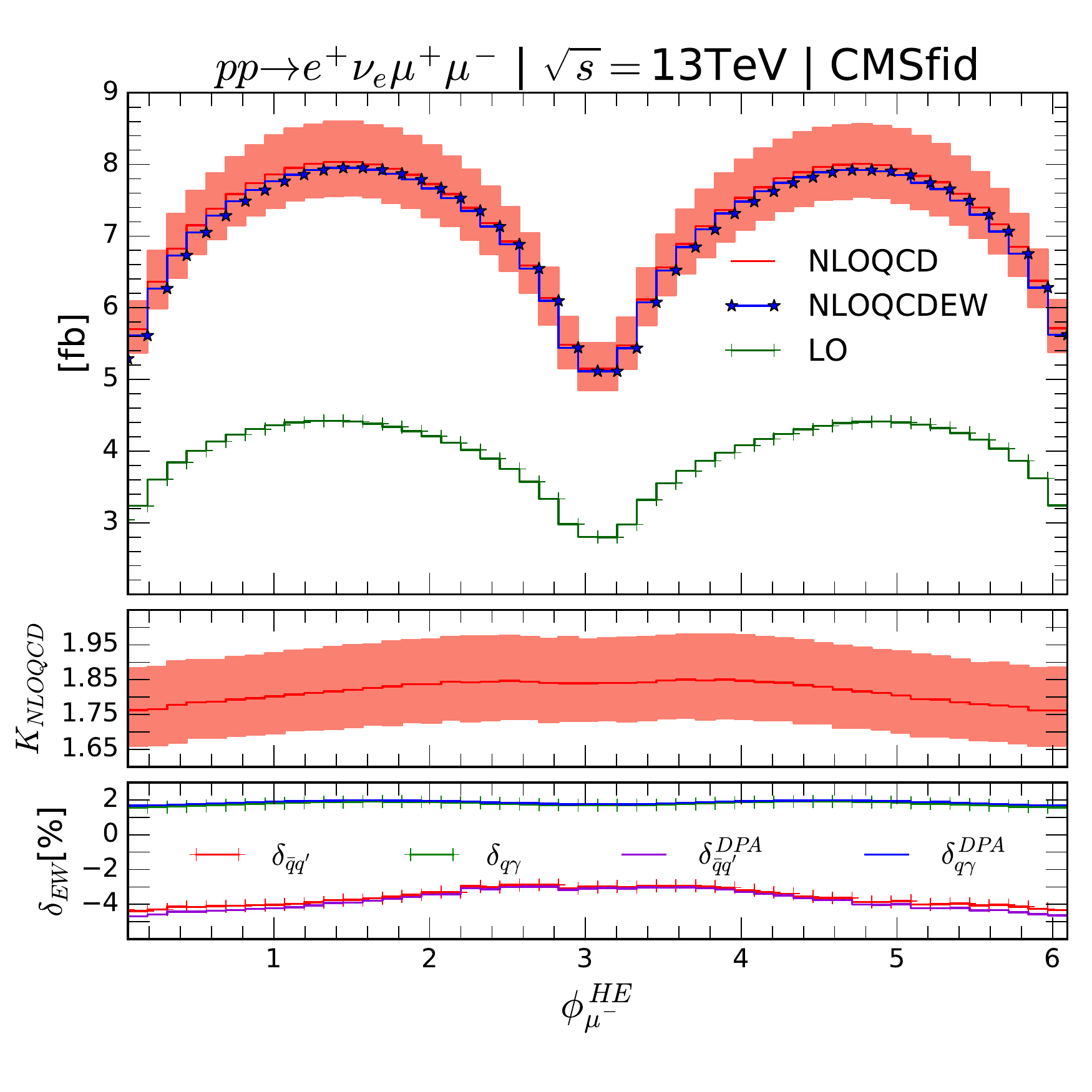}&
  \includegraphics[width=0.48\textwidth]{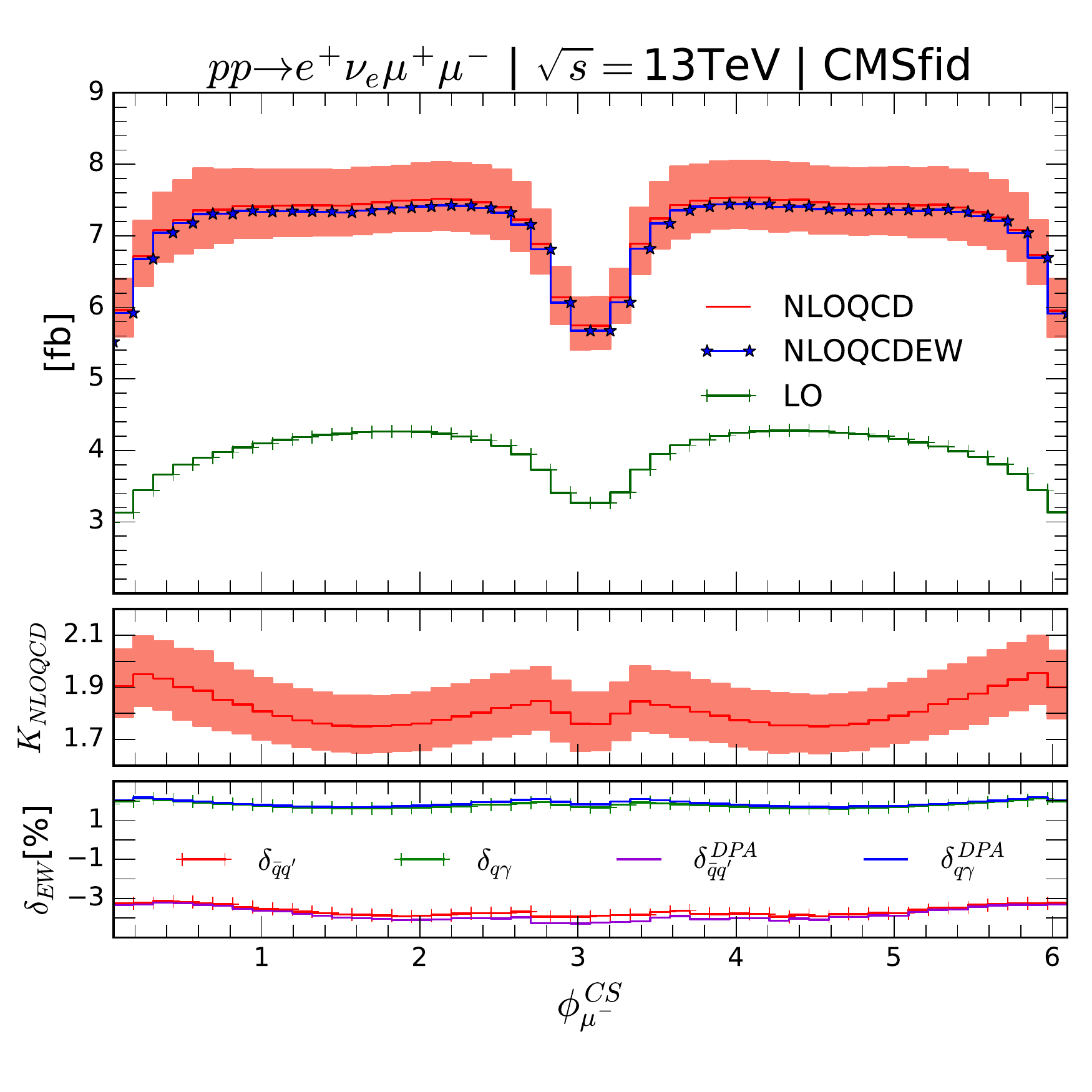}
  \end{tabular}
  \caption{Same as \fig{fig:dist_phi_HEL_CS_e_muon_Wp_atlas} but with the CMS fiducial cuts.}
   \label{fig:dist_phi_HEL_CS_e_muon_Wp_cms}
\end{figure}

In the middle panels we display the $K$-factor defined as 
\bea
K_\text{NLOQCD} = \fr{d\sigma_\text{NLOQCD}}{d\sigma_\text{LO}}.
\label{eq:K_factor}
\eea
Please note that we use the same NNLO PDF set in the numerator and the 
denominator.  
In the bottom panels we show the NLO EW corrections calculated using
DPA as explained in \sect{sect:cal:ew} relative to the full and DPA
Born cross sections. These corrections are defined as
\begin{align}
\delta_{\bar{q}q'} &= \fr{d\Delta\sigma^\text{NLOEW}_{\bar{q}q'}}{d\sigma^\text{LO}}, \quad 
\delta_{q\gamma} = \fr{d\Delta\sigma^\text{NLOEW}_{q\gamma}}{d\sigma^\text{LO}},\crn
\delta_{\bar{q}q'}^\text{DPA} &= \fr{d\Delta\sigma^\text{NLOEW}_{\bar{q}q'}}{d\sigma^\text{LO}_\text{DPA}}, \quad 
\delta_{q\gamma}^\text{DPA} = \fr{d\Delta\sigma^\text{NLOEW}_{q\gamma}}{d\sigma^\text{LO}_\text{DPA}}.\label{eq:defs_EW_cor}
\end{align} 
The total NLO EW correction is the sum of $\delta_{\bar{q}q'}$ and
$\delta_{q\gamma}$. Those EW corrections are defined in the same way
as in \bib{Biedermann:2017oae}, thereby enabling direct comparisons of
our DPA NLO EW corrections to the full results. 

As already observed in OS production \bibs{Baglio:2013toa}, the
$K$-factors in the $p_T$ distributions are increasing over the $p_T$
range to reach high values at high transverse momentum because of soft
weak boson emission in the quark-gluon and quark-photon induced
processes. For example, it reaches $K_\text{NLOQCD}=6.5$ at $p_{T,W} =
600$ GeV and $K_\text{NLOQCD}\simeq 8$ at $p_{T,Z}=600$ GeV. The same
observation is true for the transverse momentum distributions of the
individual final-state leptons displayed in
\fig{fig:dist_pT_W_Z_y_W_Z_Wp_atlas} and
\fig{fig:dist_pT_e_mu_eta_e_mu_Wp_atlas}. The uncertainties are also
increasing from close to $\pm 5\%$ at low $p_T$ up to $\simeq \pm
40\%$ at high $p_T$. Similar distributions can be obtained with the
CMS fiducial cuts, as displayed in \fig{fig:dist_pT_W_Z_y_W_Z_Wp_cms}
and \fig{fig:dist_pT_e_mu_eta_e_mu_Wp_cms}. In that case, the $K$
  factors are a bit smaller and the uncertainties are also smaller,
  reaching $\simeq \pm 20\%$ at high $p_T$, closer to the
  uncertainties that can be obtained in the OS production.

The EW corrections are nearly identical when using DPA or full LO
matrix elements for the transverse momenta of the $e^+\nu_e$ and
$\mu^+\mu^-$ systems, as well as for the individual muons. The Sudakov
regime of the $\bar{q}q'$ contribution is cancelled by the
quark-photon induced contributions that reach $+20\%$ at $p_{T,W/Z}=
600$ GeV. The distributions of the individual final-state leptons of
the $W$ boson system, however, display differences between the DPA and
the full LO matrix elements. This can be traced back to the
contribution of a $Z$ radiated off the electron or the neutrino and
splitting into the di-muon pair. This contribution, displayed in
\fig{fig:diags_pT_structure}, does not exist in the DPA and can lead to a
higher $p_{T,e/\nu}$ due to the hard $\mu^+\mu^-$ system that needs
to be balanced by the transverse momentum of the electron or the neutrino. 
The EW corrections calculated with the full LO matrix elements should
then be smaller in magnitude than the corresponding EW corrections
with the DPA matrix elements, as observed in
\fig{fig:dist_pT_W_Z_y_W_Z_Wp_atlas} (lower left) and
\fig{fig:dist_pT_e_mu_eta_e_mu_Wp_atlas} (upper left). The same is
true for the CMS fiducial cuts as displayed in
\fig{fig:dist_pT_W_Z_y_W_Z_Wp_cms} (lower left) and
\fig{fig:dist_pT_e_mu_eta_e_mu_Wp_cms} (upper left). 
Similar differences between the DPA and the full 
results for the case of $pp \to W^+ W^- \to \nu_\mu \mu^+ e^- \bar{\nu}_e + X$ 
have been discussed in \bib{Biedermann:2016guo}.

The $Z$ boson rapidity distribution displayed in
\fig{fig:dist_pT_W_Z_y_W_Z_Wp_atlas} (lower right) as well as the
pseudo-rapidity distributions of the charged leptons displayed in
\fig{fig:dist_pT_e_mu_eta_e_mu_Wp_atlas} (right-hand side) show
non-constant $K$-factors of order 2 and EW corrections close to
$-4\%$, summing the quark-photon induced
contributions and the $\bar{q}q'$ contributions.
 
The theoretical uncertainties at NLO  
are limited to $\simeq \pm
5\%$. 
We remind that the $\simeq +10\%$ correction stemming from the
NNLO QCD corrections~\cite{Grazzini:2017ckn} is not included there.

Similar distributions
are also obtained for the CMS fiducial cuts displayed in
\fig{fig:dist_pT_W_Z_y_W_Z_Wp_cms} (lower right) and
\fig{fig:dist_pT_e_mu_eta_e_mu_Wp_cms} (right-hand side).

We also display azimuthal-angle difference distributions in
\fig{fig:dist_Delta_phi_y_Wp_atlas} (ATLAS fiducial cuts) and
\fig{fig:dist_Delta_phi_y_Wp_cms} (CMS fiducial cuts), that 
can be directly compared to \bib{Biedermann:2017oae}. Other distributions 
including $p_{T,Z}$, $p_{T,\nu_e}$, $p_{T,e}$, $y_Z$, $\eta_\ell$, $\ldots$, 
can be compared as well. 
Comparing our
kinematical distributions to those of \bib{Biedermann:2017oae}, we
observe good agreement in shape and magnitude for 
$\delta_{\bar{q}q'}$ and 
$\delta_{q\gamma}$ corrections individually.

For completeness, we also display the $\cos\theta$ and $\phi$
distributions in \fig{fig:dist_cos_theta_HEL_CS_e_muon_Wp_atlas} and
\fig{fig:dist_phi_HEL_CS_e_muon_Wp_atlas} respectively, for ATLAS
fiducial cuts. The corresponding distributions for the CMS fiducial cuts
are given in \fig{fig:dist_cos_theta_HEL_CS_e_muon_Wp_cms} and
\fig{fig:dist_phi_HEL_CS_e_muon_Wp_cms}. The Helicity coordinate
system is displayed on the left-hand side while the Collins-Soper
coordinate system is displayed on the right-hand side, in all
plots. The $K$-factors are not constant and the theoretical
uncertainties are quite limited, ranging from $\simeq \pm 5\%$ to
$\simeq \pm 20\%$. In all distributions, the DPA and the full LO
matrix elements give the same EW corrections.


\clearpage
\pagebreak

\section{Numerical results for fiducial polarization observables: \boldmath $W^+Z$ channel}
\label{sect:numres}

\subsection{Fiducial angular coefficients and polarization fractions: $W^+Z$ channel}
\label{sect:numres:polar_observables_nloqcdew}

We start the discussion of the fiducial polarization observables with a
presentation of the NLO QCD and EW predictions for the angular
coefficients and the polarization fractions including PDF and scale
uncertainties. We will also display the LO results to get an insight
into the size of the NLO corrections. We calculate the polarization
coefficients via the numerical integration of the $\cos\theta-\phi$
distributions while the polarization fractions are calculated via the
numerical integration of $\cos\theta$ distributions. The statistical
uncertainty in the case of the NLO QCD predictions is found to be
negligible compared to the PDF or scale uncertainty and is not
given. We use the bin-averaging method for the numerical integration,
that gives for $n\times n$ bins $[a_i^{},b_i^{}] \times
[c_j^{},d_j^{}]$, 
\begin{align}
  \left\langle f(\theta)\right\rangle
  & = \int_{-1}^{1} d\!\cos\theta f(\theta) \frac{d\sigma}{\sigma d\!\cos\theta}
     \simeq \frac{1}{\sigma}
    \sum_{i=1}^n\left(\frac{d\sigma}{d\!\cos\theta}\right)_i^{}
    f_i^{},\\
\left\langle g(\theta,\phi)\right\rangle
  & = \int_{-1}^{1} d\!\cos\theta\int_0^{2\pi} d\phi g(\theta,\phi)
    \frac{d\sigma}{\sigma d\!\cos\theta d\phi} \simeq \frac{1}{\sigma}
    \sum_{1\leq i,j\leq n}^{}\left(\frac{d\sigma}{d\!\cos\theta
    d\phi}\right)_{ij}^{} g_{ij}^{},
\end{align}
with
\begin{align}
  f_i^{}
  = \frac{1}{b_i^{}-a_i^{}}\int_{a_i^{}}^{b_i^{}} f(\theta)
  d\!\cos\theta,\quad
  g_{ij}^{}
  = \frac{1}{\left(b_i^{}-a_i^{}\right)
  \left(d_j^{}-c_j^{}\right)}\int_{a_i^{}}^{b_i^{}} d\!\cos\theta
  \int_{c_j^{}}^{d_j^{}} d\phi g(\theta,\phi),
\label{eq:binavg}
\end{align}
where the integrals in \eq{eq:binavg} are analytically performed. 

Another obvious choice to calculate $f_i$ and $g_{ij}$ reads, for each bin, 
\begin{align}
  f_i^{} = f(\theta_i),\quad
  g_{ij}^{} = g(\theta_i,\phi_i),
\label{eq:bin_middle}
\end{align}
where $\theta_i$ and $\phi_i$ correspond to the middle of the bin. We have checked that 
results obtained from both methods are in good agreement. In the following, all numerical results 
are calculated using the bin-averaging method.   

The results for the fiducial coefficients $A_{0-7}$, depending on the choice of
the coordinate system (either HE or CS as defined in
\sect{sect:cal:polar_observables}), obtained using the full NLO QCD
matrix elements and the EW corrections in the DPA, are presented in
\tab{tab:coeff_Ai_Wp_ATLAS} (for the $W$ boson) and
\tab{tab:coeff_Ai_Z_ATLAS} (for the $Z$ boson) for the process $p p
\to e^+_{} \nu_e^{}\, \mu^+_{}\mu^-_{} + X$ using the ATLAS fiducial
cuts. The corresponding results using the CMS fiducial cuts are given in
\tab{tab:coeff_Ai_Wp_CMS} and \tab{tab:coeff_Ai_Z_CMS}. The QCD
corrections are always sizable in all coordinate systems, while the EW
corrections are more limited. As already seen in $W^+ + j$ production
at the LHC \bibs{Bern:2011ie}, the inclusive coefficients $A_5$, $A_6$, and $A_7$
are very small even after taking into account the QCD corrections. The
higher-order corrections can sometimes switch the sign of these
coefficients, see e.g. $A_7^{e^+}$ in the HE coordinate system. 
The smallness of these coefficients can be understood from the DPA LO results 
provided in \appen{appen:off_shell_NLOEW_effects}, where we can see that they are all 
consistently zero within the statistical errors, independent of the cuts or the coordinate system.   
This is also in line with the fact that those coefficients are, at the DPA LO and in the 
inclusive
 
phase-space limit, proportional 
to the imaginary parts of the spin-density matrix elements as shown in \eq{eq:relations_Ai_spin_matrix}, 
therefore expected to be vanishing. 
The scale and PDF uncertainties are very small, at maximum a few percents,
as expected from an observable built over a ratio of cross sections.

\begin{table}[ht!]
 \renewcommand{\arraystretch}{1.3}
\begin{center}
\setlength\tabcolsep{0.03cm}
\fontsize{5.0}{5.0}
\begin{tabular}{|c|c|c|c|c|c|c|c|c|}\hline
$\text{Method}$  & $A_0$ & $A_1$  & $A_2$ & $A_3$ & $A_4$ & $A_5$ & $A_6$ & $A_7$\\
\hline
{\fontsize{6.0}{6.0}$\text{HE LO}$} & $1.026(2)^{+5}_{-6}$ & $-0.286(2)^{+4}_{-3}$ & $-1.314(2)^{+3}_{-3}$ & $-0.251(2)^{+2}_{-2}$ & $-0.447(7)^{+3}_{-3}$ & $-0.002(0.2)^{+0.03}_{-0}$ & $-0.001(0.3)^{+0.1}_{-0.1}$ & $-0.004(0.2)^{+0.1}_{-0.02}$\\
\hline
{\fontsize{6.0}{6.0}$\text{HE NLOEW}$} & $1.028$ & $-0.284$ & $-1.324$ & $-0.252$ & $-0.438$ & $-0.004$ & $-0.004$ & $0.003$\\
\hline
{\fontsize{6.0}{6.0}$\text{HE NLOQCD}$} & $1.016(1)^{+3}_{-4}$ & $-0.326(2)^{+2}_{-3}$ & $-1.413(2)^{+10}_{-12}$ & $-0.229(1)^{+2}_{-1}$ & $-0.295(7)^{+11}_{-11}$ & $-0.001(1)^{+0.1}_{-0.2}$ & $-0.0002(6)^{+3}_{-2}$ & $0.003(1)^{+1}_{-0.5}$\\
\hline
{\fontsize{6.0}{6.0}$\text{HE NLOQCDEW}$} & $1.017$ & $-0.326$ & $-1.420$ & $-0.229$ & $-0.287$ & $-0.002$ & $-0.002$ & $0.007$\\
\hline\hline
{\fontsize{6.0}{6.0}$\text{CS LO}$} & $1.397(3)^{+4}_{-5}$ & $0.229(1)^{+3}_{-3}$ & $-0.945(1)^{+2}_{-2}$ & $0.003(2)^{+0.3}_{-1}$ & $-0.613(8)^{+4}_{-4}$ & $-0.0002(3)^{+0.1}_{-1}$ & $0.002(0.3)^{+0.1}_{-0.04}$ & $0.004(0.2)^{+0.01}_{-0.1}$\\
\hline
{\fontsize{6.0}{6.0}$\text{CS NLOEW}$} & $1.402$ & $0.225$ & $-0.952$ & $0.008$ & $-0.608$ & $0.001$ & $0.006$ & $-0.003$\\
\hline
{\fontsize{6.0}{6.0}$\text{CS NLOQCD}$} & $1.513(3)^{+7}_{-7}$ & $0.192(1)^{+2}_{-2}$ & $-0.918(3)^{+2}_{-2}$ & $0.061(4)^{+4}_{-4}$ & $-0.469(6)^{+10}_{-10}$ & $-0.0001(11)^{+0}_{-3}$ & $0.001(0.5)^{+0.3}_{-0.2}$ & $-0.003(0.4)^{+1}_{-1}$\\
\hline
{\fontsize{6.0}{6.0}$\text{CS NLOQCDEW}$} & $1.518$ & $0.189$ & $-0.921$ & $0.065$ & $-0.463$ & $0.0004$ & $0.003$ & $-0.007$\\
\hline
\end{tabular}
\caption{\small Fiducial angular coefficients of the $e^{+}$ distribution for
  the process $pp \to e^+ \nu_e\, \mu^+ \mu^- + X$ at LO, NLO EW, NLO QCD and
  NLO QCD+EW at the 13 TeV LHC with the ATLAS fiducial cuts. Results are
  presented for two coordinate systems: the helicity (HE) and
  Collins-Soper (CS) coordinate systems. The PDF uncertainties (in parenthesis) and the scale
  uncertainties are provided for the LO and NLO QCD results, all given on the last digit of the central prediction.}
\label{tab:coeff_Ai_Wp_ATLAS}
\end{center}
\end{table}

\begin{table}[ht!]
 \renewcommand{\arraystretch}{1.3}
\begin{center}
\setlength\tabcolsep{0.03cm}
\fontsize{5.0}{5.0}
\begin{tabular}{|c|c|c|c|c|c|c|c|c|}\hline
$\text{Method}$  & $A_0$ & $A_1$  & $A_2$ & $A_3$ & $A_4$ & $A_5$ & $A_6$ & $A_7$\\
\hline
{\fontsize{6.0}{6.0}$\text{HE LO}$} & $1.035(2)^{+2}_{-2}$ & $-0.304(1)^{+2}_{-1}$ & $-0.705(1)^{+0.3}_{-1}$ & $0.063(1)^{+0.04}_{-0.1}$ & $-0.017(1)^{+1}_{-1}$ & $-0.007(0.4)^{+0.1}_{-0}$ & $-0.007(0.2)^{+0}_{-0.2}$ & $0.003(0.2)^{+0}_{-0.1}$\\
\hline
{\fontsize{6.0}{6.0}$\text{HE NLOEW}$} & $1.039$ & $-0.307$ & $-0.717$ & $0.050$ & $-0.020$ & $-0.007$ & $-0.008$ & $0.003$\\
\hline
{\fontsize{6.0}{6.0}$\text{HE NLOQCD}$} & $0.985(2)^{+5}_{-6}$ & $-0.306(1)^{+4}_{-3}$ & $-0.734(1)^{+2}_{-2}$ & $0.031(1)^{+2}_{-2}$ & $0.003(1)^{+1}_{-1}$ & $-0.004(1)^{+0.3}_{-0.4}$ & $-0.004(1)^{+0.3}_{-0.2}$ & $0.003(1)^{+0.2}_{-0}$\\
\hline
{\fontsize{6.0}{6.0}$\text{HE NLOQCDEW}$} & $0.986$ & $-0.308$ & $-0.742$ & $0.023$ & $0.001$ & $-0.004$ & $-0.004$ & $0.003$\\
\hline\hline
{\fontsize{6.0}{6.0}$\text{CS LO}$} & $1.254(2)^{+2}_{-3}$ & $0.239(2)^{+2}_{-2}$ & $-0.488(1)^{+1}_{-1}$ & $-0.061(0.3)^{+0.3}_{-0.4}$ & $0.035(1)^{+1}_{-1}$ & $-0.0001(3)^{+1}_{-0}$ & $0.010(0.3)^{+0.1}_{-0.004}$ & $-0.003(0.2)^{+0.1}_{-0}$\\
\hline
{\fontsize{6.0}{6.0}$\text{CS NLOEW}$} & $1.266$ & $0.234$ & $-0.493$ & $-0.053$ & $0.023$ & $0.001$ & $0.012$ & $-0.003$\\
\hline
{\fontsize{6.0}{6.0}$\text{CS NLOQCD}$} & $1.267(2)^{+4}_{-4}$ & $0.221(1)^{+1}_{-1}$ & $-0.455(2)^{+2}_{-2}$ & $-0.021(1)^{+3}_{-3}$ & $0.023(1)^{+1}_{-1}$ & $0.0004(6)^{+2}_{-2}$ & $0.006(0.5)^{+0.2}_{-0.4}$ & $-0.003(1)^{+0}_{-0.1}$\\
\hline
{\fontsize{6.0}{6.0}$\text{CS NLOQCDEW}$} & $1.273$ & $0.218$ & $-0.457$ & $-0.016$ & $0.016$ & $0.001$ & $0.007$ & $-0.003$\\
\hline
\end{tabular}
\caption{\small Same as \tab{tab:coeff_Ai_Wp_ATLAS} but for the
  $\mu^{-}$ distribution.}
\label{tab:coeff_Ai_Z_ATLAS}
\end{center}
\end{table}

\begin{table}[ht!]
 \renewcommand{\arraystretch}{1.3}
\begin{center}
\setlength\tabcolsep{0.03cm}
\fontsize{5.0}{5.0}
\begin{tabular}{|c|c|c|c|c|c|c|c|c|}\hline
$\text{Method}$  & $A_0$ & $A_1$  & $A_2$ & $A_3$ & $A_4$ & $A_5$ & $A_6$ & $A_7$\\
\hline
{\fontsize{6.0}{6.0}$\text{HE LO}$} & $0.897(2)^{+5}_{-6}$ & $0.088(3)^{+6}_{-5}$ & $-0.626(2)^{+4}_{-3}$ & $-0.373(3)^{+3}_{-4}$ & $-0.488(8)^{+3}_{-4}$ & $-0.002(0.3)^{+0.1}_{-0}$ & $-0.001(0.3)^{+0.04}_{-0}$ & $-0.003(0.3)^{+0}_{-0.05}$\\
\hline
{\fontsize{6.0}{6.0}$\text{HE NLOEW}$} & $0.899$ & $0.092$ & $-0.625$ & $-0.374$ & $-0.480$ & $-0.004$ & $-0.004$ & $0.003$\\
\hline
{\fontsize{6.0}{6.0}$\text{HE NLOQCD}$} & $0.913(1)^{+2}_{-3}$ & $0.023(2)^{+4}_{-4}$ & $-0.672(1)^{+3}_{-3}$ & $-0.321(2)^{+3}_{-3}$ & $-0.325(7)^{+12}_{-11}$ & $-0.001(1)^{+0.2}_{-0.2}$ & $-0.0003(5)^{+3}_{-2}$ & $0.003(0.5)^{+0.5}_{-1}$\\
\hline
{\fontsize{6.0}{6.0}$\text{HE NLOQCDEW}$} & $0.915$ & $0.025$ & $-0.672$ & $-0.321$ & $-0.319$ & $-0.002$ & $-0.002$ & $0.006$\\
\hline\hline
{\fontsize{6.0}{6.0}$\text{CS LO}$} & $0.760(4)^{+8}_{-9}$ & $0.196(1)^{+3}_{-3}$ & $-0.764(1)^{+1}_{-1}$ & $0.052(2)^{+0.4}_{-1}$ & $-0.723(9)^{+4}_{-6}$ & $-0.00003(28)^{+2}_{-1}$ & $0.002(0.4)^{+0}_{-0.1}$ & $0.003(0.3)^{+0.05}_{-0}$\\
\hline
{\fontsize{6.0}{6.0}$\text{CS NLOEW}$} & $0.758$ & $0.192$ & $-0.767$ & $0.057$ & $-0.719$ & $0.001$ & $0.006$ & $-0.003$\\
\hline
{\fontsize{6.0}{6.0}$\text{CS NLOQCD}$} & $0.875(3)^{+7}_{-6}$ & $0.172(1)^{+1}_{-2}$ & $-0.711(3)^{+4}_{-4}$ & $0.097(4)^{+3}_{-3}$ & $-0.551(7)^{+12}_{-11}$ & $-0.0002(6)^{+1}_{-1}$ & $0.001(1)^{+0.2}_{-0.2}$ & $-0.003(0.5)^{+1}_{-0.5}$\\
\hline
{\fontsize{6.0}{6.0}$\text{CS NLOQCDEW}$} & $0.875$ & $0.170$ & $-0.712$ & $0.100$ & $-0.547$ & $0.0002$ & $0.003$ & $-0.007$\\
\hline
\end{tabular}
\caption{\small Same as \tab{tab:coeff_Ai_Wp_ATLAS} but with the CMS fiducial cuts.}
\label{tab:coeff_Ai_Wp_CMS}
\end{center}
\end{table}

\begin{table}[ht!]
 \renewcommand{\arraystretch}{1.3}
\begin{center}
\setlength\tabcolsep{0.03cm}
\fontsize{5.0}{5.0}
\begin{tabular}{|c|c|c|c|c|c|c|c|c|}\hline
$\text{Method}$  & $A_0$ & $A_1$  & $A_2$ & $A_3$ & $A_4$ & $A_5$ & $A_6$ & $A_7$\\
\hline
{\fontsize{6.0}{6.0}$\text{HE LO}$} & $0.858(3)^{+3}_{-4}$ & $-0.273(1)^{+1}_{-1}$ & $-0.570(1)^{+0.5}_{-1}$ & $0.068(1)^{+0.1}_{-0}$ & $-0.022(1)^{+1}_{-1}$ & $-0.007(0.2)^{+0.1}_{-0}$ & $-0.006(0.2)^{+0.1}_{-0.1}$ & $0.003(0.2)^{+0.1}_{-0.02}$\\
\hline
{\fontsize{6.0}{6.0}$\text{HE NLOEW}$} & $0.855$ & $-0.282$ & $-0.574$ & $0.055$ & $-0.026$ & $-0.007$ & $-0.008$ & $0.002$\\
\hline
{\fontsize{6.0}{6.0}$\text{HE NLOQCD}$} & $0.785(2)^{+8}_{-9}$ & $-0.300(1)^{+3}_{-2}$ & $-0.592(2)^{+1}_{-1}$ & $0.033(1)^{+2}_{-3}$ & $0.001(1)^{+2}_{-2}$ & $-0.004(1)^{+0.2}_{-0.2}$ & $-0.004(1)^{+0.2}_{-0.2}$ & $0.003(1)^{+0.02}_{-0.2}$\\
\hline
{\fontsize{6.0}{6.0}$\text{HE NLOQCDEW}$} & $0.782$ & $-0.304$ & $-0.594$ & $0.026$ & $-0.001$ & $-0.004$ & $-0.004$ & $0.003$\\
\hline\hline
{\fontsize{6.0}{6.0}$\text{CS LO}$} & $1.128(2)^{+1}_{-1}$ & $0.296(2)^{+2}_{-2}$ & $-0.303(1)^{+2}_{-1}$ & $-0.069(0.5)^{+0.4}_{-1}$ & $0.038(2)^{+1}_{-1}$ & $-0.0003(2)^{+1}_{-0}$ & $0.010(0.4)^{+0}_{-0.1}$ & $-0.003(0.2)^{+0.02}_{-0.1}$\\
\hline
{\fontsize{6.0}{6.0}$\text{CS NLOEW}$} & $1.141$ & $0.291$ & $-0.290$ & $-0.062$ & $0.025$ & $0.0005$ & $0.012$ & $-0.002$\\
\hline
{\fontsize{6.0}{6.0}$\text{CS NLOQCD}$} & $1.180(2)^{+3}_{-3}$ & $0.275(1)^{+1}_{-1}$ & $-0.200(3)^{+7}_{-8}$ & $-0.024(1)^{+3}_{-3}$ & $0.024(1)^{+1}_{-1}$ & $0.0003(5)^{+1}_{-1}$ & $0.006(0.5)^{+0.3}_{-0.3}$ & $-0.003(1)^{+0.2}_{-0.01}$\\
\hline
{\fontsize{6.0}{6.0}$\text{CS NLOQCDEW}$} & $1.188$ & $0.272$ & $-0.192$ & $-0.019$ & $0.017$ & $0.001$ & $0.006$ & $-0.003$\\
\hline
\end{tabular}
\caption{\small Same as \tab{tab:coeff_Ai_Z_ATLAS} but with the CMS fiducial cuts.}
\label{tab:coeff_Ai_Z_CMS}
\end{center}
\end{table}

The results for the fiducial polarization fractions $f_L$, $f_R$, and $f_0$ are
given in \tab{tab:coeff_fL0R_WpZ_ATLAS} for the case of the ATLAS fiducial
cuts, while the results for the CMS fiducial cuts are displayed in
\tab{tab:coeff_fL0R_WpZ_CMS}. It is noted that these fractions can be 
also calculated from the angular coefficients provided in \tab{tab:coeff_Ai_Wp_ATLAS} and \tab{tab:coeff_Ai_Z_ATLAS}.
The results of the ATLAS and CMS fiducial
cuts are similar in the HE coordinate system, except that the L
and R fractions are higher with the CMS fiducial cuts than with the ATLAS
fiducial cuts. In the CS coordinate system, however, there
is a sizable difference for $f_R^{W^+}$, that is close to zero and
negative at LO for the ATLAS
fiducial cuts while being positive and of the order of $0.13$ for the CMS
fiducial cuts. We note that having a negative fraction is 
possible when looking at \eq{eq:def_fLR0}. This happens because 
we are considering fiducial fractions and are using the Collins-Soper coordinate system. 
The fiducial fractions are all positive in the helicity system. 
This can be understood because the
$z$ axis is aligned along the vector-boson direction of flight in the helicity system, 
but not in the Collins-Soper one. In the limit of LO DPA and of 
inclusive cut, the polarization fractions in the helicity system 
can be proven to be positive because they are truly fractions. 
 
The QCD corrections are sizable in all polarization
fractions. The EW corrections, however, are negligible for the $W^+$
polarization fractions and small but noticeable for the $Z$
polarization fraction, in particular reaching 
$+4\%$ ($-5\%$) 
of the NLO QCD results for the
$f_R^Z$ ($f_L^Z$) fractions  
in the $W^+Z$ channel in the CS coordinate system with 
either the ATLAS or CMS fiducial cuts, see \tab{tab:coeff_fL0R_WpZ_ATLAS} and \tab{tab:coeff_fL0R_WpZ_CMS}. 
The EW corrections are a bit higher, reaching $+7\%$ ($-4\%$) for the 
$f_R^Z$ ($f_L^Z$) fractions  
in the $W^-Z$ channel in the HE coordinate system with 
the ATLAS fiducial cuts, see \tab{tab:coeff_fL0R_WmZ_ATLAS}. 

We trace back the origin of the large EW correction to the $f_R^Z$ fraction 
to the angular coefficient $A_4$. We see that the EW correction to the $A_4^Z$ in the 
$W^+Z$ channel in the CS coordinate system with the ATLAS fiducial cuts 
is $-30\%$ compared to the NLO QCD prediction, see \tab{tab:coeff_Ai_Z_ATLAS}. 
This can be further understood by inspecting
\tab{tab:coeff_Ai_full_DPA_Z_processWp_Atlas} in \appen{appen:off_shell_NLOEW_effects},
where we observe two things: (i) The $q\gamma$ correction is negligible, 
while the $\bar{q}q'$ one is large. (ii) The origin of this large EW correction 
comes from the radiative decay of the $Z$ boson. The EW correction to the $Z$ decay, including both 
the virtual and real photon emission contributions, 
induces $-36\%$  
correction to the DPA LO result. 
It would be interesting to see if these large effects are still present when considering 
the inclusive polarization observables $A_i$, because, as discussed 
in \sect{sect:cal:polar_observables}, the 
spin-density matrix defined in \eq{eq:spin_matrix_DPA}
is independent of the decay mode.  
Similar large EW corrections are also seen in the coefficient $A_3^Z$, where it is 
also due to the radiative corrections to the $Z$ decay.   

It is worth noting that $A_3^Z$ and $A_4^Z$ are proportional to the EW parameter $c$, which 
is very sensitive to the value of $s^2_W$, see \eq{eq:relations_Ai_spin_matrix}, while it 
is not the case for the $W$ bosons. So, it may be not so surprising after all that they are 
sensitive to the EW corrections to the $Z$ decay.

The PDF uncertainty is very limited, of the order of $\pm 1\%$ at
maximum, and the scale uncertainty is also very small, of the order of
a few percent. This is expected as the polarization fractions are
built from ratios of cross sections. The only exception is for
$f_R^{W^+}$ in the CS coordinate system, as a result of the smallness of this
coefficient. It is worth mentioning that the combined NLO QCD+EW
results are not simply the sum of the NLO EW corrections on the
polarization fraction and of the NLO QCD results. 
For example, the NLO
EW corrections on $f_L^Z$ with the CMS fiducial cuts and in the CS coordinate system 
are $\delta_{\rm
  EW} = -0.069$. Naively summed to the NLO QCD result $f_L^{Z,{\rm
    QCD}}=0.234$ this would give $f_L^Z = 0.165$, instead of the true
NLO QCD+EW result $f_L^Z=0.223$.  
This demonstrates the usefulness of a
fully combined analysis of the QCD and EW corrections for the
calculation of the polarization observables.

\begin{table}[ht!]
  \renewcommand{\arraystretch}{1.3}
\begin{center}
    \fontsize{8}{8}
\begin{tabular}{|c|c|c|c||c|c|c|}\hline
$\text{Method}$  & $f^{W^+}_L$ & $f^{W^+}_0$ & $f^{W^+}_R$ & $f^Z_L$ & $f^Z_0$ & $f^Z_R$\\
\hline
$\text{HE LO}$ & $0.355(2)^{+2}_{-2}$ & $0.513(1)^{+2}_{-3}$ & $0.132(2)^{+1}_{-1}$ & $0.222(1)^{+0.4}_{-1}$ & $0.518(1)^{+1}_{-1}$ & $0.261(1)^{+2}_{-1}$\\
\hline
$\text{HE NLOEW}$ & $0.352$ & $0.514$ & $0.134$ & $0.216$ & $0.519$ & $0.264$\\
\hline
$\text{HE NLOQCD}$ & $0.320(2)^{+2}_{-2}$ & $0.508(1)^{+2}_{-2}$ & $0.172(2)^{+4}_{-3}$ & $0.257(1)^{+3}_{-3}$ & $0.493(1)^{+2}_{-3}$ & $0.251(1)^{+1}_{-0.5}$\\
\hline
$\text{HE NLOQCDEW}$ & $0.317$ & $0.509$ & $0.174$ & $0.255$ & $0.493$ & $0.252$\\
\hline\hline
$\text{CS LO}$ & $0.304(3)^{+2}_{-2}$ & $0.698(1)^{+2}_{-2}$ & $-0.002(1)^{+0.1}_{-0.1}$ & $0.228(2)^{+0.03}_{-0.2}$ & $0.627(1)^{+1}_{-1}$ & $0.145(2)^{+1}_{-1}$\\
\hline
$\text{CS NLOEW}$ & $0.302$ & $0.701$ & $-0.003$ & $0.210$ & $0.633$ & $0.157$\\
\hline
$\text{CS NLOQCD}$ & $0.239(2)^{+4}_{-4}$ & $0.757(1)^{+4}_{-3}$ & $0.004(1)^{+1}_{-1}$ & $0.210(1)^{+1}_{-1}$ & $0.634(1)^{+2}_{-2}$ & $0.156(1)^{+2}_{-2}$\\
\hline
$\text{CS NLOQCDEW}$ & $0.236$ & $0.759$ & $0.005$ & $0.200$ & $0.637$ & $0.163$\\
\hline
\end{tabular}
    \caption{\small $W^+_{}$ and $Z$ fiducial polarization fractions in the
      process $pp \to e^+ \nu_e\, \mu^+ \mu^- + X$ at LO, NLO EW, NLO QCD and
  NLO QCD+EW at the 13 TeV LHC with the ATLAS fiducial cuts. Results are
  presented for two coordinate systems: the helicity (HE) and
  Collins-Soper (CS) coordinate systems. The PDF uncertainties (in parenthesis) and the scale
  uncertainties are provided for the LO and NLO QCD results, all given on the last digit of the central prediction.}
    \label{tab:coeff_fL0R_WpZ_ATLAS}
\end{center}
\end{table}

\begin{table}[ht!]
  \renewcommand{\arraystretch}{1.3}
  \begin{center}
    \fontsize{8}{8}
\begin{tabular}{|c|c|c|c||c|c|c|}\hline
$\text{Method}$  & $f^{W^+}_L$ & $f^{W^+}_0$ & $f^{W^+}_R$ & $f^Z_L$ & $f^Z_0$ & $f^Z_R$\\
\hline
$\text{HE LO}$ & $0.398(2)^{+3}_{-2}$ & $0.448(1)^{+3}_{-3}$ & $0.154(2)^{+1}_{-1}$ & $0.260(1)^{+0.2}_{-0.4}$ & $0.429(1)^{+2}_{-2}$ & $0.312(2)^{+2}_{-2}$\\
\hline
$\text{HE NLOEW}$ & $0.395$ & $0.450$ & $0.155$ & $0.256$ & $0.427$ & $0.317$\\
\hline
$\text{HE NLOQCD}$ & $0.353(2)^{+2}_{-3}$ & $0.457(1)^{+1}_{-1}$ & $0.190(2)^{+3}_{-3}$ & $0.305(1)^{+4}_{-4}$ & $0.392(1)^{+4}_{-4}$ & $0.302(2)^{+1}_{-1}$\\
\hline
$\text{HE NLOQCDEW}$ & $0.351$ & $0.457$ & $0.191$ & $0.304$ & $0.391$ & $0.305$\\
\hline\hline
$\text{CS LO}$ & $0.491(3)^{+4}_{-3}$ & $0.380(2)^{+4}_{-5}$ & $0.129(1)^{+1}_{-1}$ & $0.262(2)^{+0.3}_{-0.3}$ & $0.564(1)^{+1}_{-1}$ & $0.174(2)^{+1}_{-1}$\\
\hline
$\text{CS NLOEW}$ & $0.490$ & $0.379$ & $0.131$ & $0.244$ & $0.571$ & $0.185$\\
\hline
$\text{CS NLOQCD}$ & $0.419(2)^{+4}_{-4}$ & $0.438(2)^{+4}_{-3}$ & $0.143(1)^{+2}_{-2}$ & $0.234(1)^{+2}_{-2}$ & $0.590(1)^{+2}_{-1}$ & $0.176(1)^{+1}_{-1}$\\
\hline
$\text{CS NLOQCDEW}$ & $0.418$ & $0.438$ & $0.144$ & $0.223$ & $0.594$ & $0.183$\\
\hline
\end{tabular}
    \caption{\small Same as \tab{tab:coeff_fL0R_WpZ_ATLAS} but with
      CMS fiducial cuts.}
    \label{tab:coeff_fL0R_WpZ_CMS}
  \end{center}
\end{table}

\clearpage
\pagebreak

\subsection{Distributions of fiducial polarization fractions: $W^+Z$ channel}
\label{sect:numres:polar_observables_dist}

We finish our presentation of the numerical results by the
  discussion of a few differential distributions of the fiducial polarization
  fractions. We include the QCD and EW corrections and display the
  combined scale+PDF uncertainty on the NLO QCD predictions.

\begin{figure}[ht!]
  \centering
  \begin{tabular}{cc}
    \includegraphics[width=0.48\textwidth]{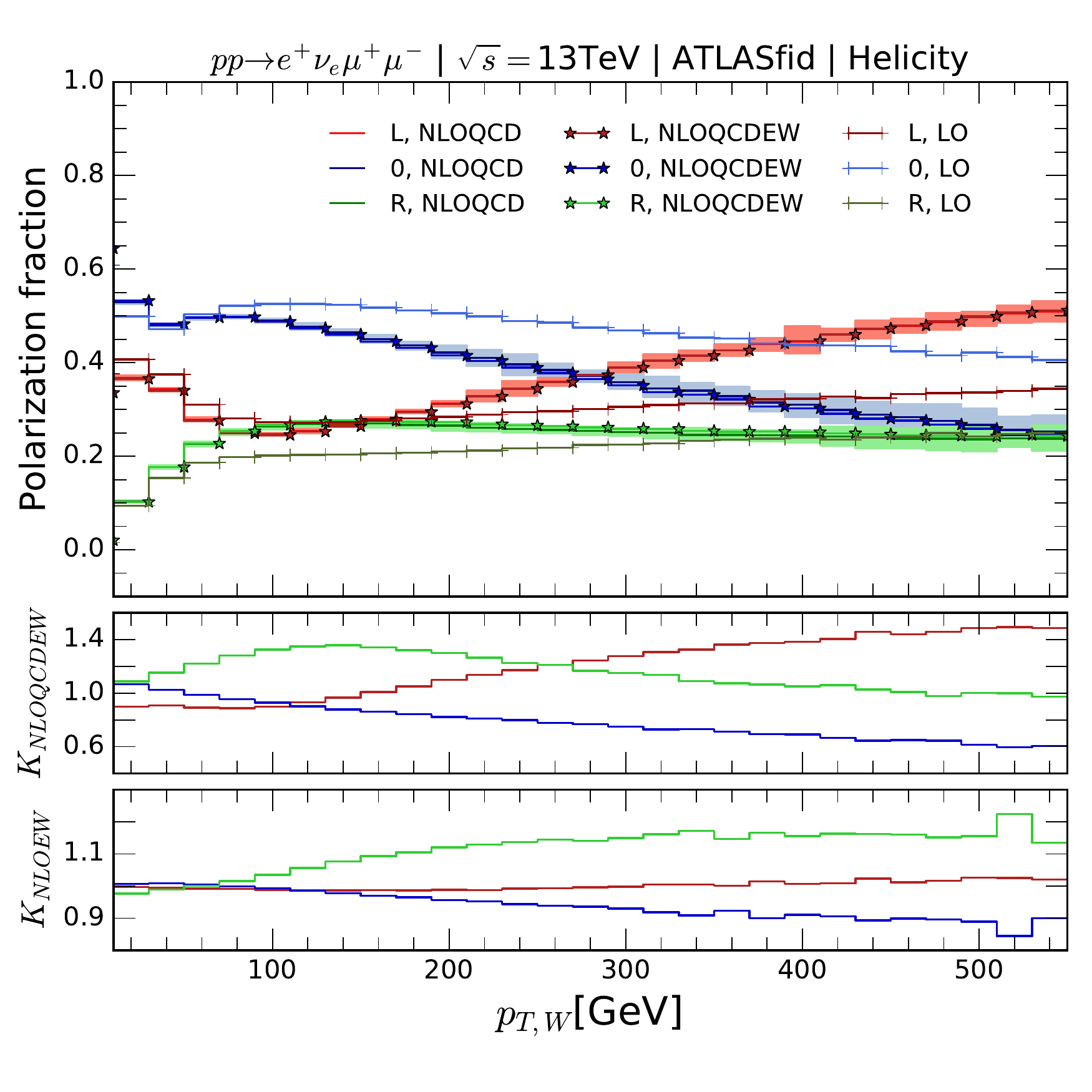}& 
    \includegraphics[width=0.48\textwidth]{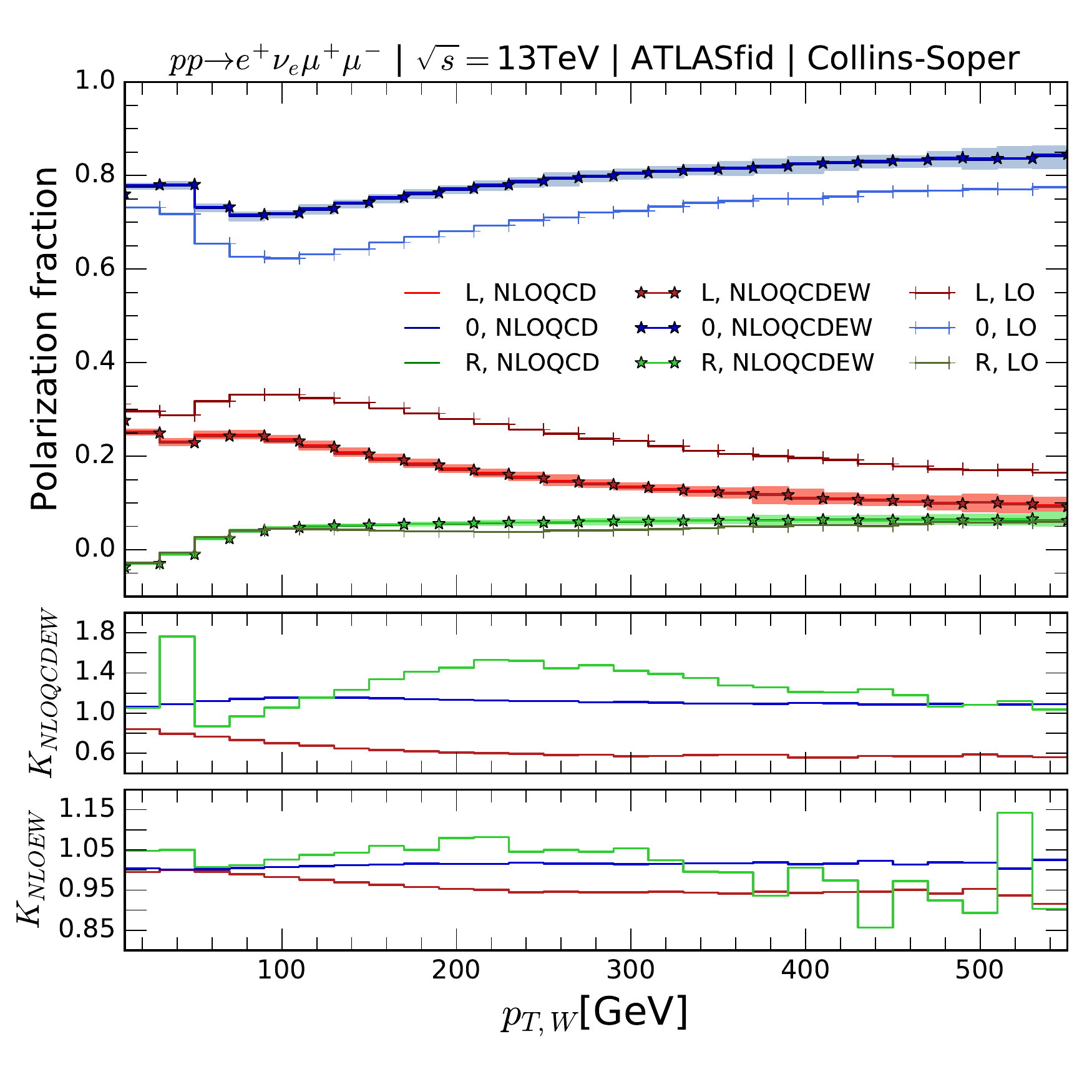}
  \end{tabular}
  \caption{Transverse momentum distributions of the $W^+$ boson fiducial polarization fractions
    for the process $pp \to e^+ \nu_e\, \mu^+ \mu^-
    + X$ at the 13 TeV LHC with the ATLAS fiducial cuts. The left-hand-side
    plot is for the HE coordinate system, while the right-hand-side
    plot is for the CS coordinate system. The bands include PDF and
    scale uncertainties calculated at NLOQCD.}
  \label{fig:dist_ptWp_ATLAS_NLOQCDEW}
\end{figure}

\begin{figure}[ht!]
  \centering
  \begin{tabular}{cc}
    \includegraphics[width=0.48\textwidth]{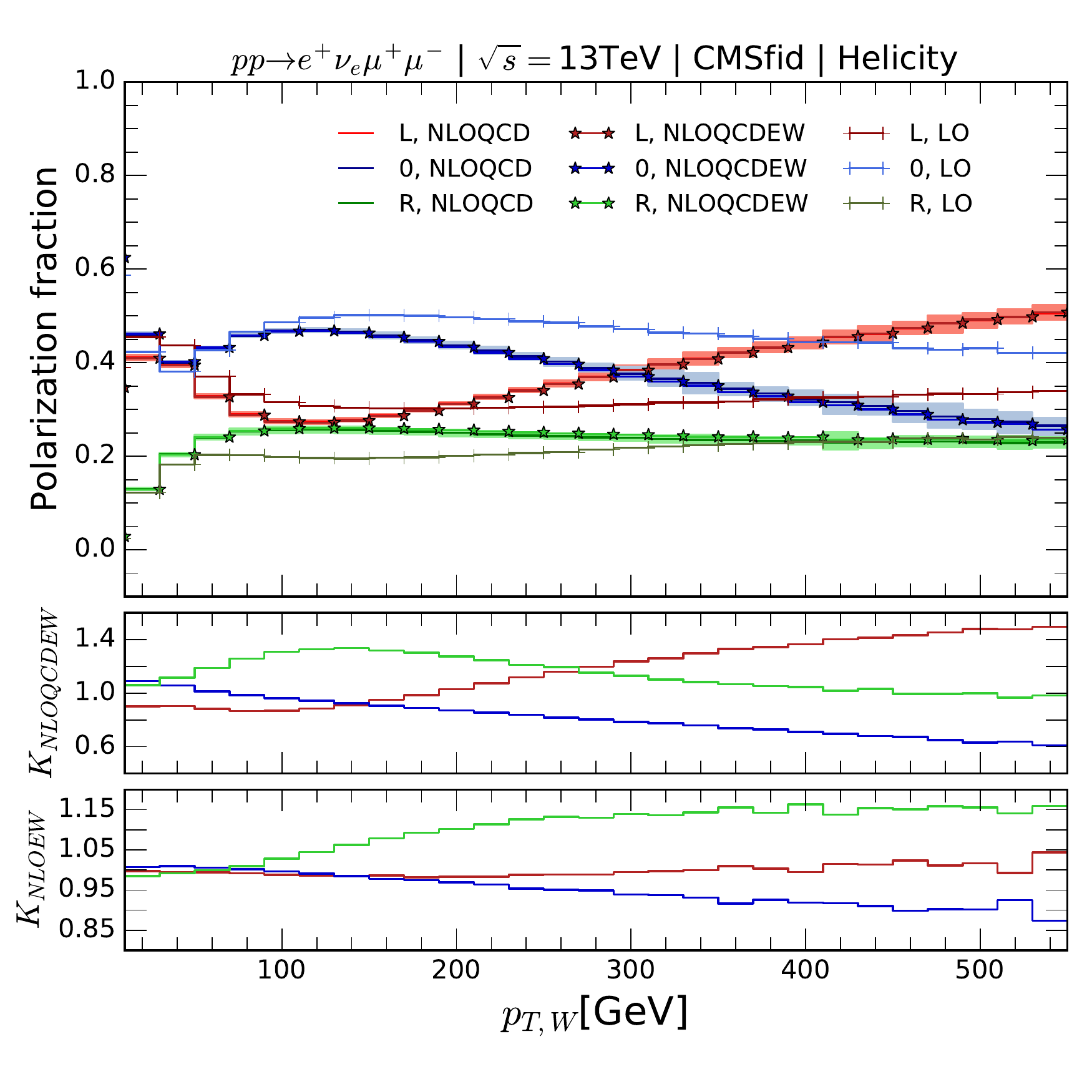}& 
    \includegraphics[width=0.48\textwidth]{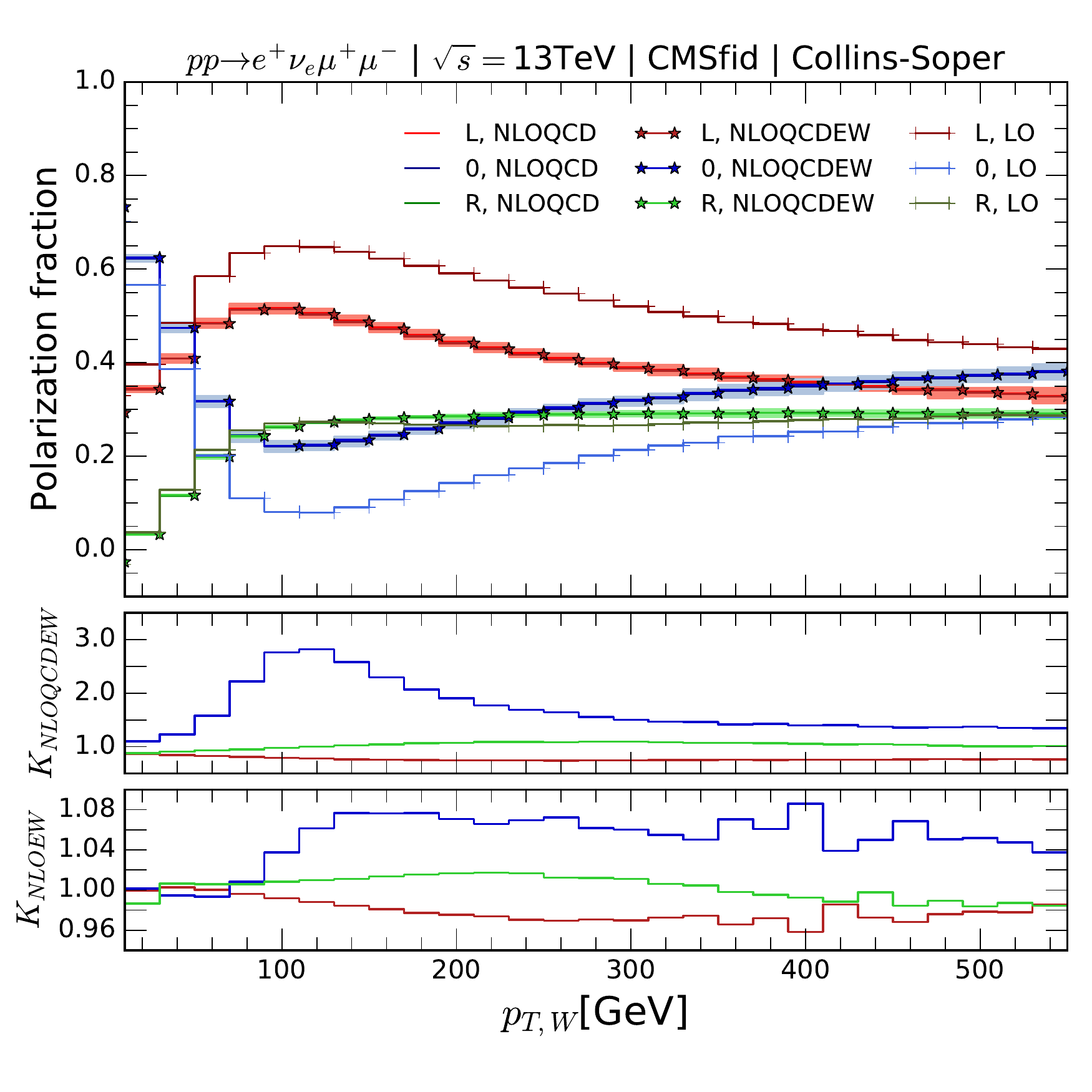}
  \end{tabular}
  \caption{Same as \fig{fig:dist_ptWp_ATLAS_NLOQCDEW} but for
    CMS fiducial cuts.}
  \label{fig:dist_ptWp_CMS_NLOQCDEW}
\end{figure}

\begin{figure}[ht!]
  \centering
 \begin{tabular}{cc}
   \includegraphics[width=0.48\textwidth]{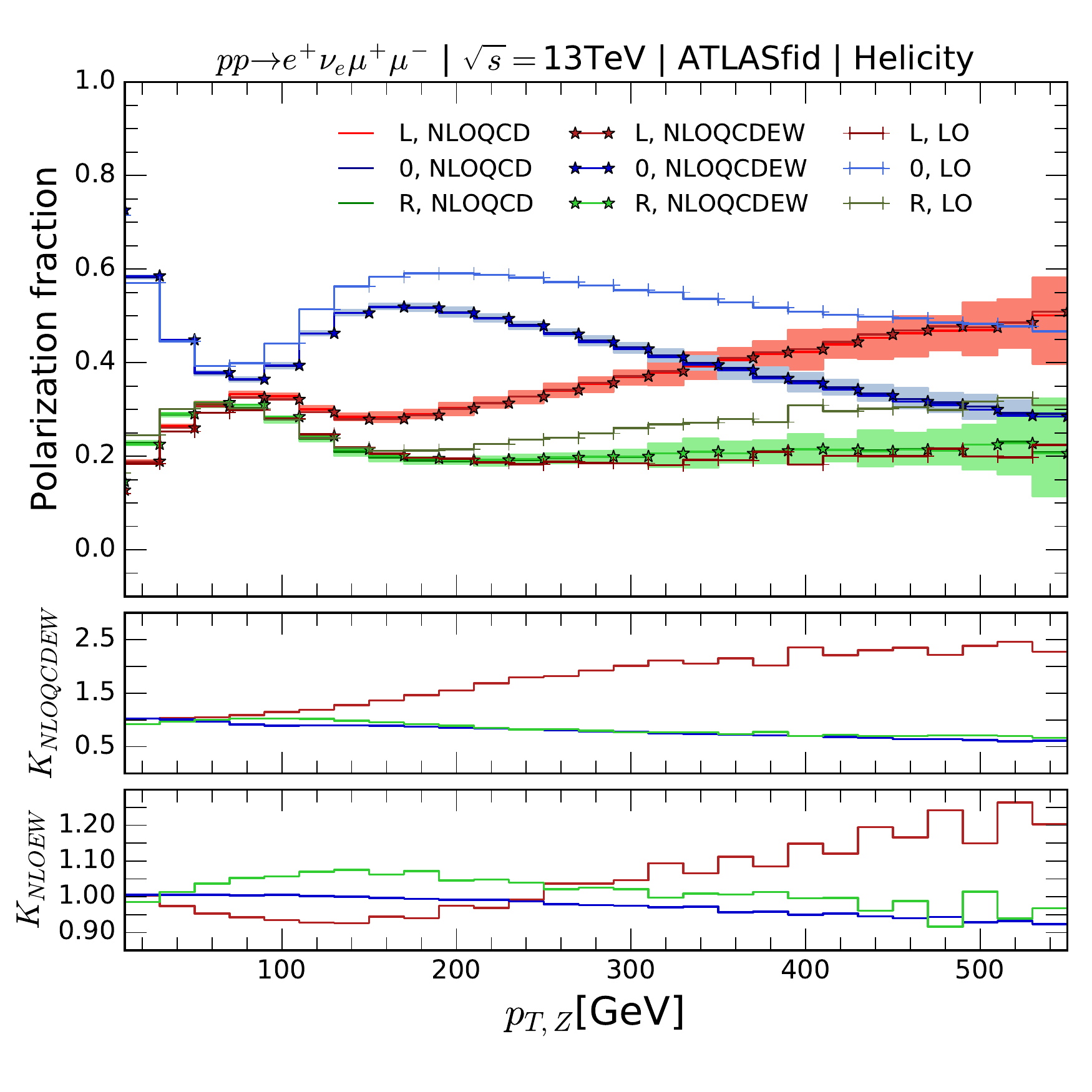}& 
   \includegraphics[width=0.48\textwidth]{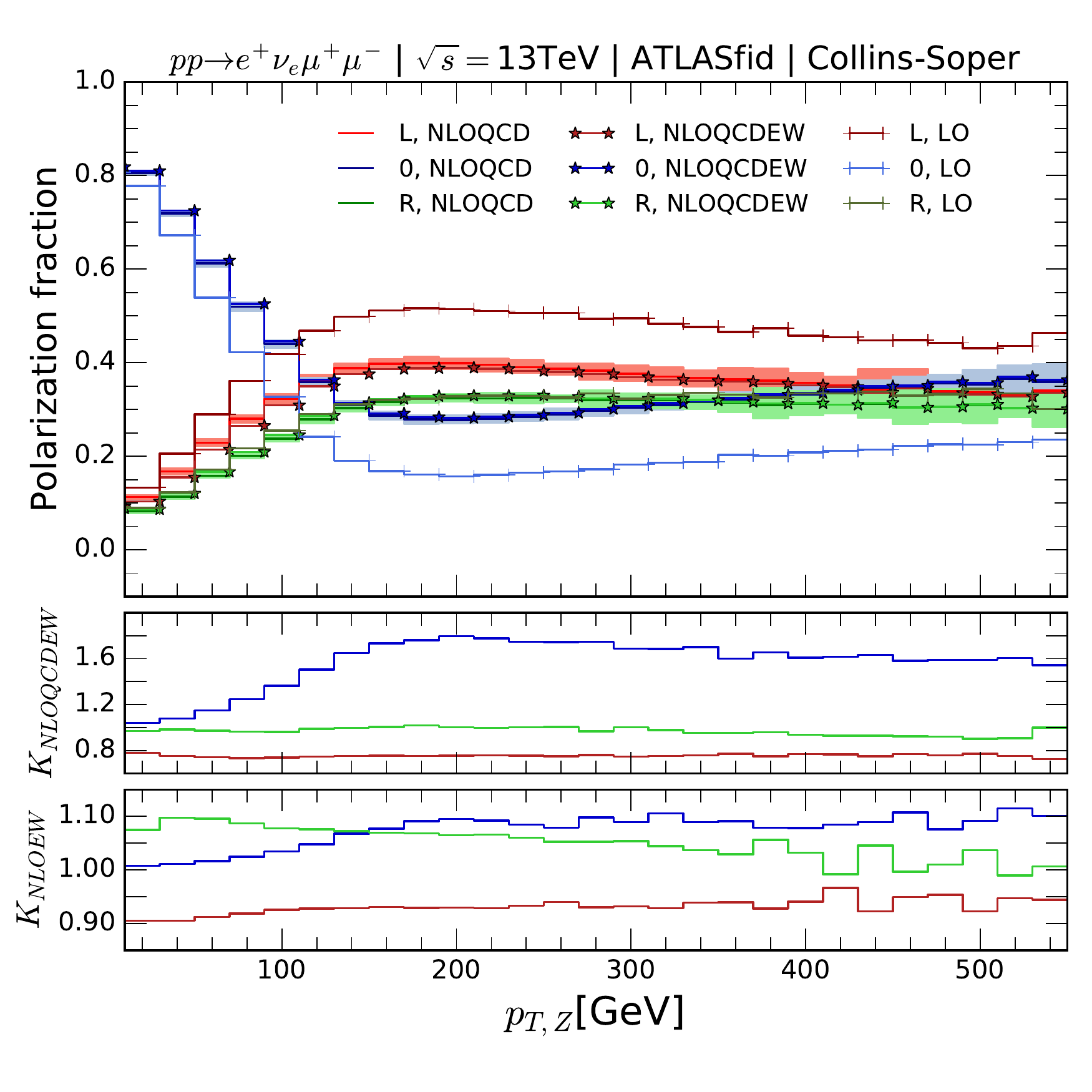}
 \end{tabular}
 \caption{Same as \fig{fig:dist_ptWp_ATLAS_NLOQCDEW} but for the $Z$ boson.}
 \label{fig:dist_pTZp_ATLAS_NLOQCDEW}
\end{figure}

\begin{figure}[ht!]
  \centering
  \begin{tabular}{cc}
    \includegraphics[width=0.48\textwidth]{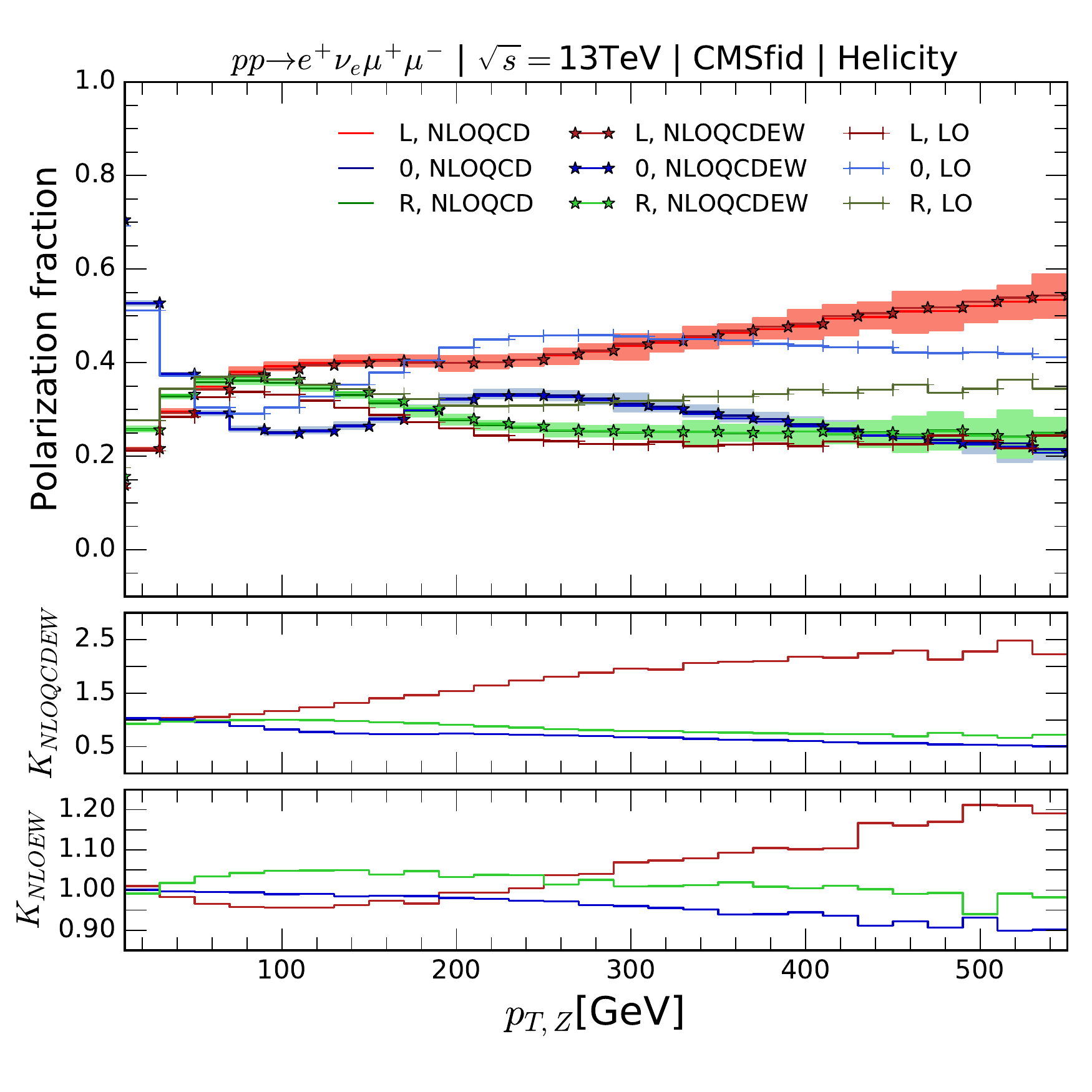}& 
    \includegraphics[width=0.48\textwidth]{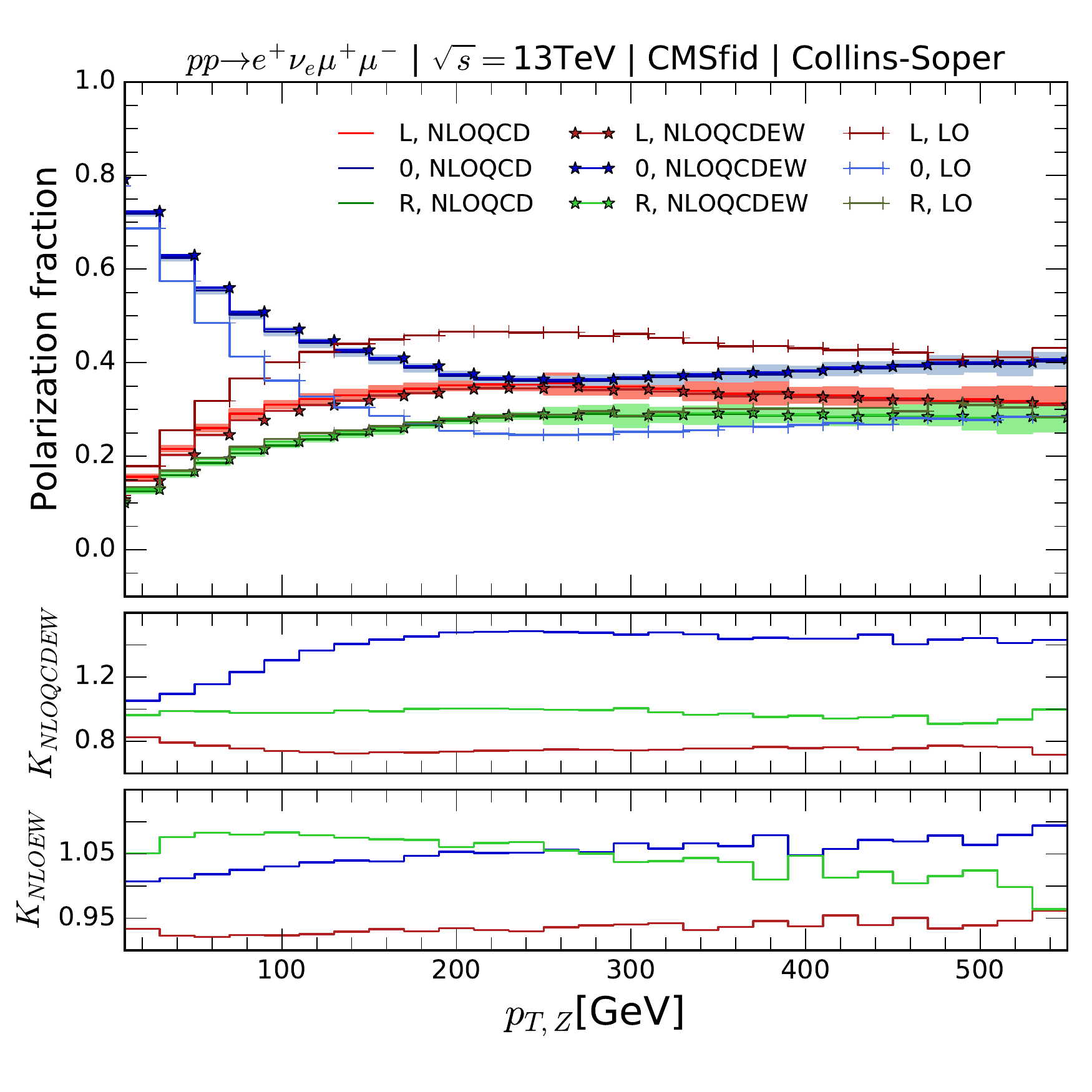}
  \end{tabular}
  \caption{Same as \fig{fig:dist_pTZp_ATLAS_NLOQCDEW} but for
   the CMS fiducial cuts.}
  \label{fig:dist_pTZp_CMS_NLOQCDEW}
\end{figure}

We display in \fig{fig:dist_ptWp_ATLAS_NLOQCDEW} the $p_{T,W}$
distribution of the fiducial polarization fractions when using the ATLAS fiducial
cuts and in \fig{fig:dist_ptWp_CMS_NLOQCDEW} when using the CMS fiducial
cuts. The corresponding distributions for $p_{T,Z}$ can be found in
\fig{fig:dist_pTZp_ATLAS_NLOQCDEW} and
\fig{fig:dist_pTZp_CMS_NLOQCDEW} respectively. The left-hand side
shows the results in the helicity (HE) coordinate system while the
right-hand side shows the results in the Collins-Soper (CS) coordinate
system. The fractions are very different from one coordinate system to
the other and in both cases the NLO corrections are sizable. For the
longitudinal polarization fraction $f_0$ (displayed in blue) the NLO
corrections are decreasing in the HE coordinate system for both $W$ and $Z$
polarization fractions, from $\simeq +10\%$ down to $-40\%$ at
$p_{T,W/Z}=540$ GeV. The NLO EW corrections are in particular not
negligible: they reach by themselves $\simeq -10\%$ at large
$p_T$. The left-handed polarization fraction $f_L$ (displayed in red)
shows a different behavior, with increasing NLO corrections driven by
the QCD corrections. They reach $\simeq +40\%$ at large $p_{T,W}$ and
$+150\%$ at large $p_{T,Z}$ for the HE coordinate system with the ATLAS or CMS fiducial cuts. 
The right-handed polarization fraction
$f_R$ (displayed in green) starts at $\simeq +10\%$, reaches a peak of
$\simeq +35\%$ at $p_T = 150$ GeV and then decreases down to zero at
large $p_T$. Out of these combined EW+QCD corrections the NLO EW
corrections can reach $+10\%$ at large $p_{T,W}$, signaling their
importance. 
For example, for the $p_{T,W}$ distribution with the ATLAS cuts and the 
HE coordinate system, the QCD+EW correction to $f_R$ is almost zero at high energies, 
but the EW correction alone is about $+10\%$. Same is true with the CMS cuts. 
For $f_R^Z$ fraction, the EW corrections are even larger at high $p_{T,Z}$ but are 
buried in the QCD corrections that are much larger in that case.

The CS coordinate system displays a complete different
behavior. Except at some specific locations where the LO predictions
are close to zero, the NLO $K$-factors are close to one for the
right-handed polarization fraction $f_R$. The NLO corrections are
constant at high $p_T$ for the longitudinal polarization fractions as
well as for the left-handed polarization fractions. Again the NLO EW
corrections can be sizable, e.g. close to $\pm 10\%$ for the $p_{T,Z}$
distribution of the polarization fractions using the ATLAS fiducial
cuts.

\begin{figure}[ht!]
  \centering
  \begin{tabular}{cc}
    \includegraphics[width=0.48\textwidth]{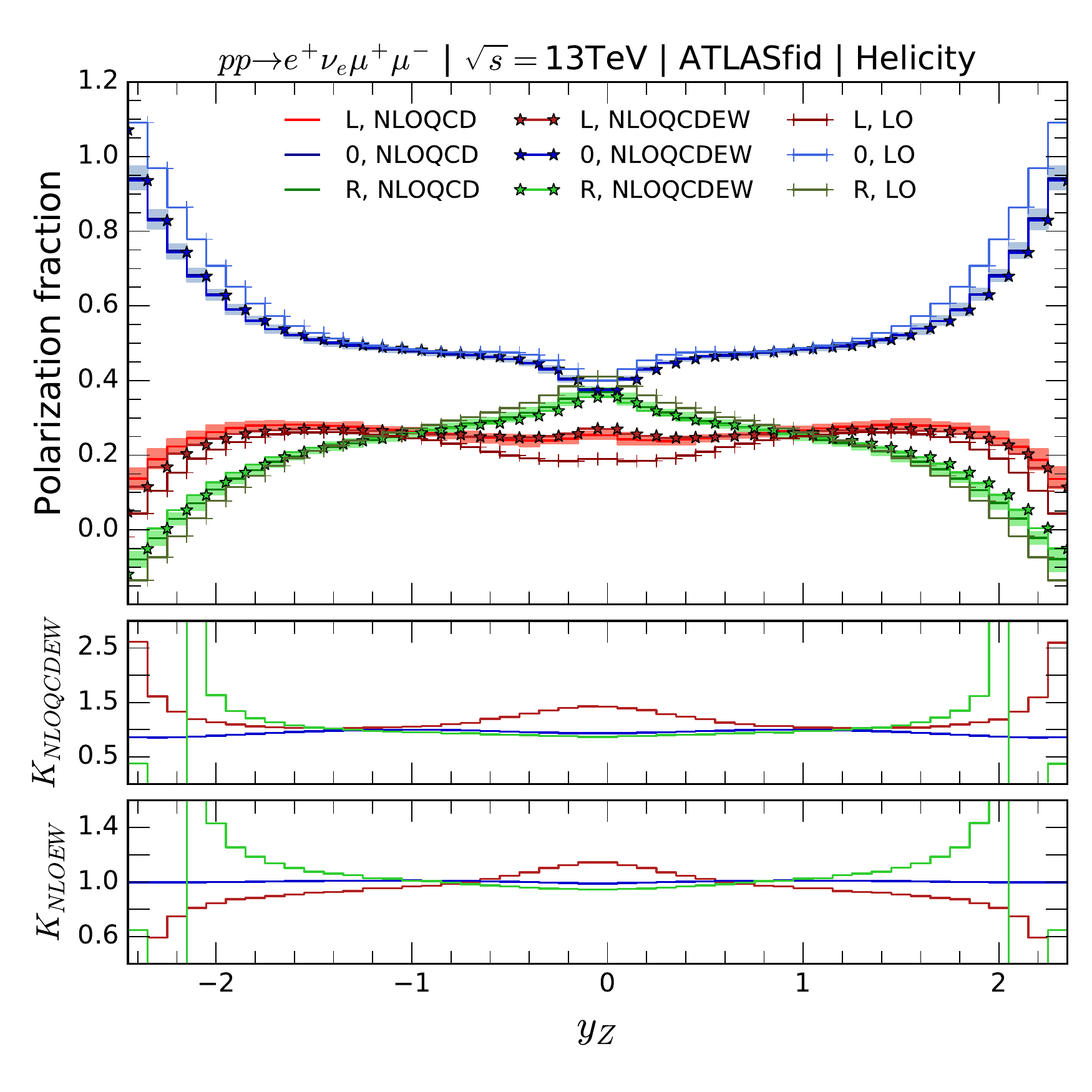}& 
    \includegraphics[width=0.48\textwidth]{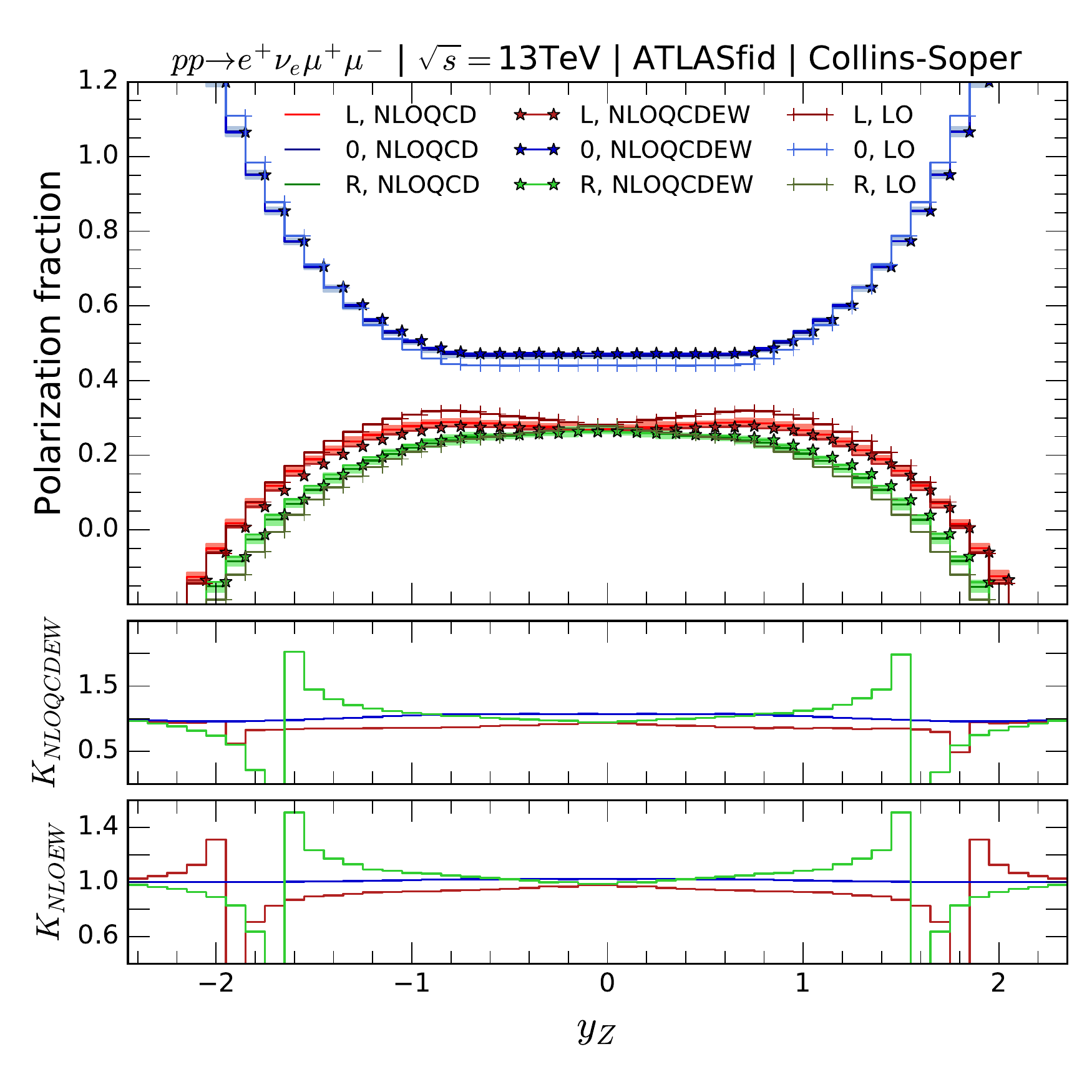}
  \end{tabular}
  \caption{Same as \fig{fig:dist_ptWp_ATLAS_NLOQCDEW} but for
the rapidity distributions of the $Z$ polarization fractions.}
  \label{fig:dist_yZp_ATLAS_NLOQCDEW}
\end{figure}

\begin{figure}[ht!]
  \centering
  \begin{tabular}{cc}
    \includegraphics[width=0.48\textwidth]{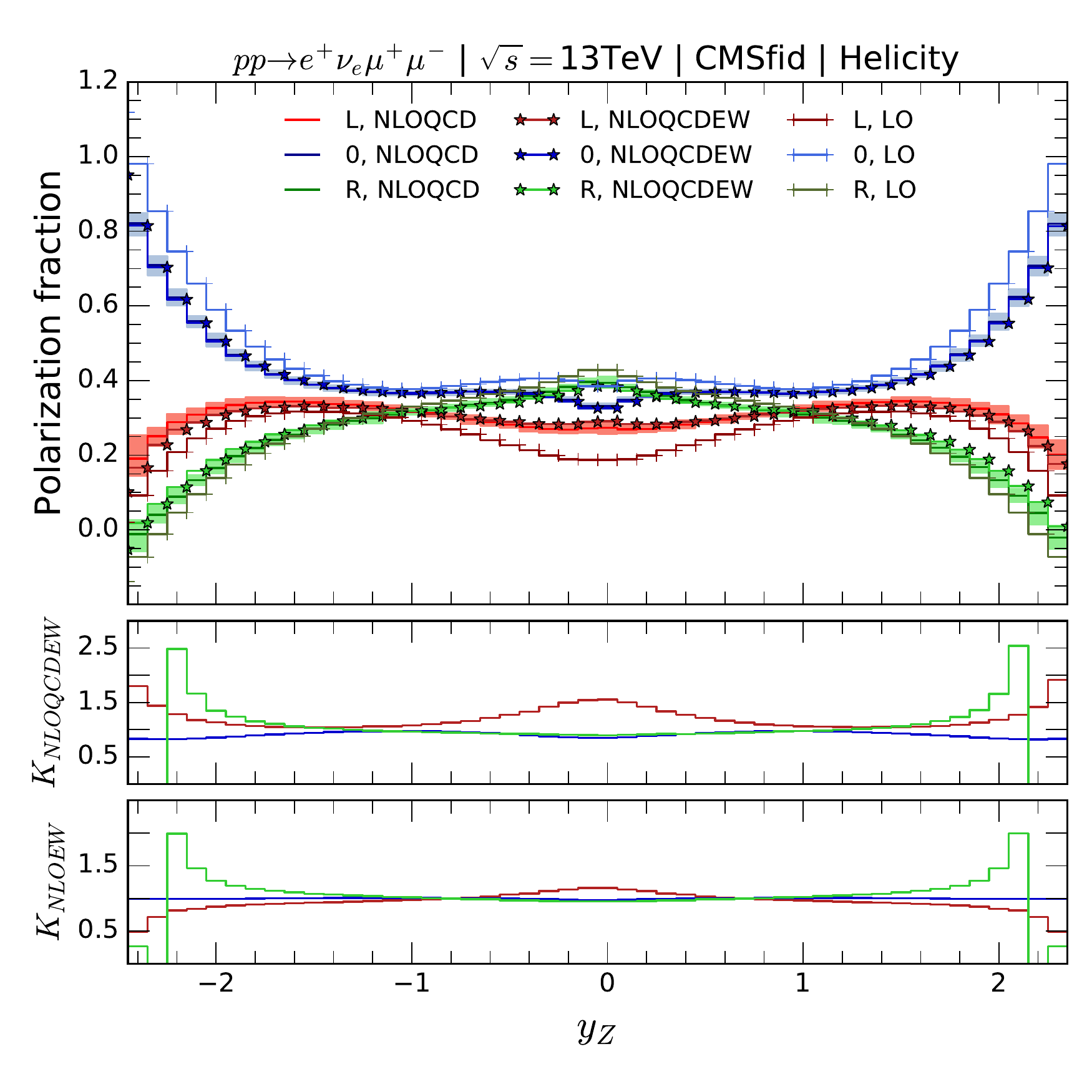}& 
    \includegraphics[width=0.48\textwidth]{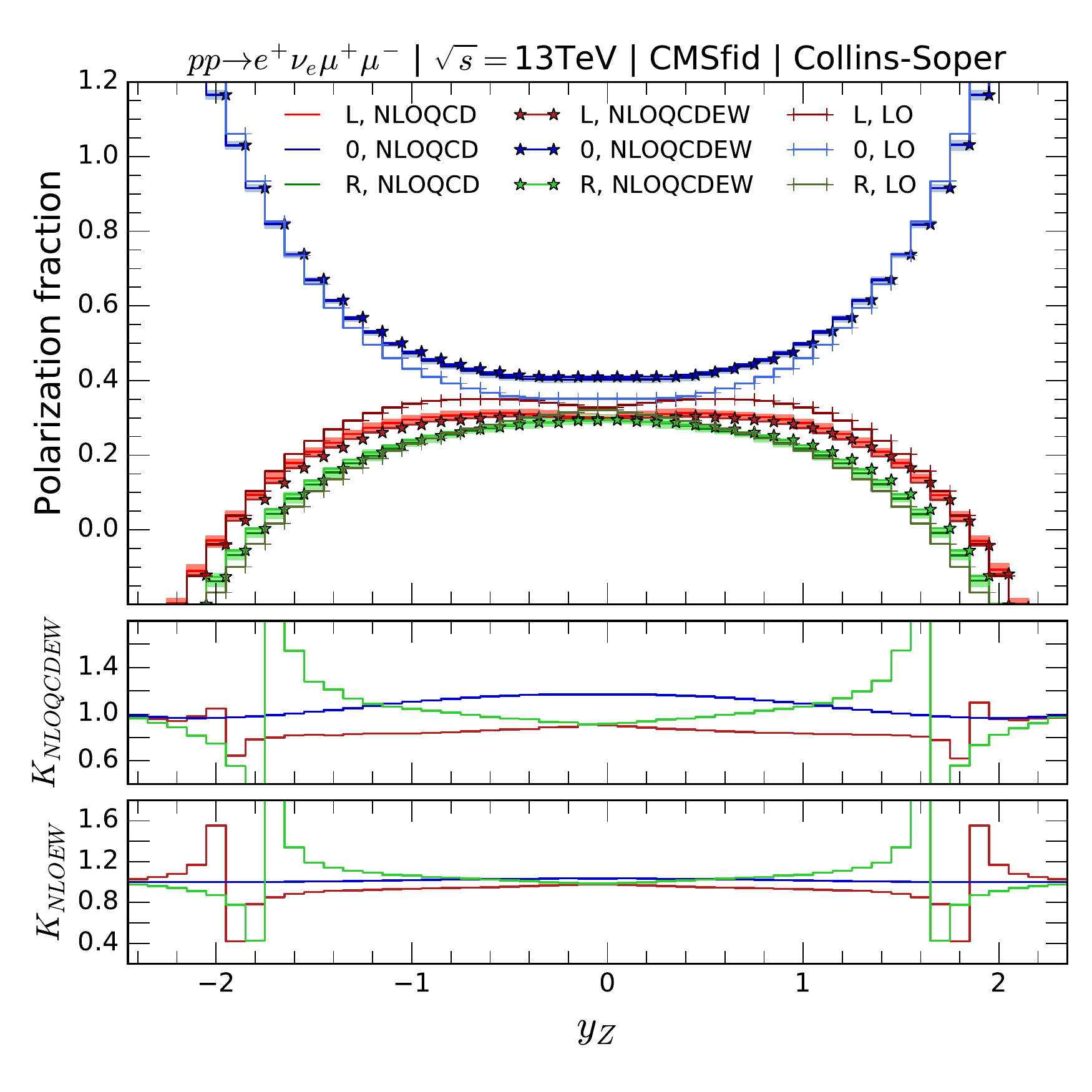}
  \end{tabular}
  \caption{Same as \fig{fig:dist_yZp_ATLAS_NLOQCDEW} but for
   the CMS fiducial cuts.}
  \label{fig:dist_yZp_CMS_NLOQCDEW}
\end{figure}

\begin{figure}[ht!]
  \centering
  \begin{tabular}{cc}
    \includegraphics[width=0.48\textwidth]{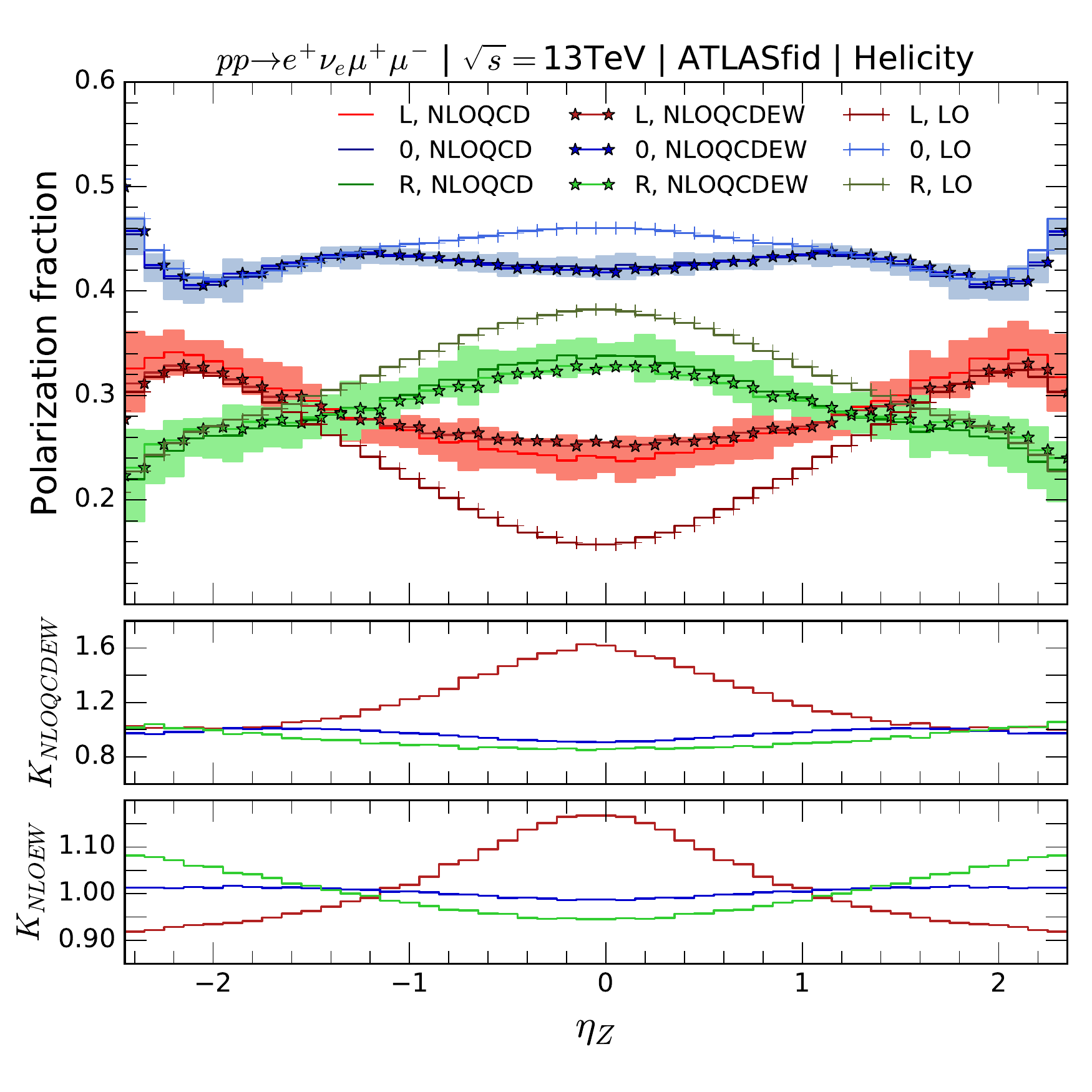}& 
    \includegraphics[width=0.48\textwidth]{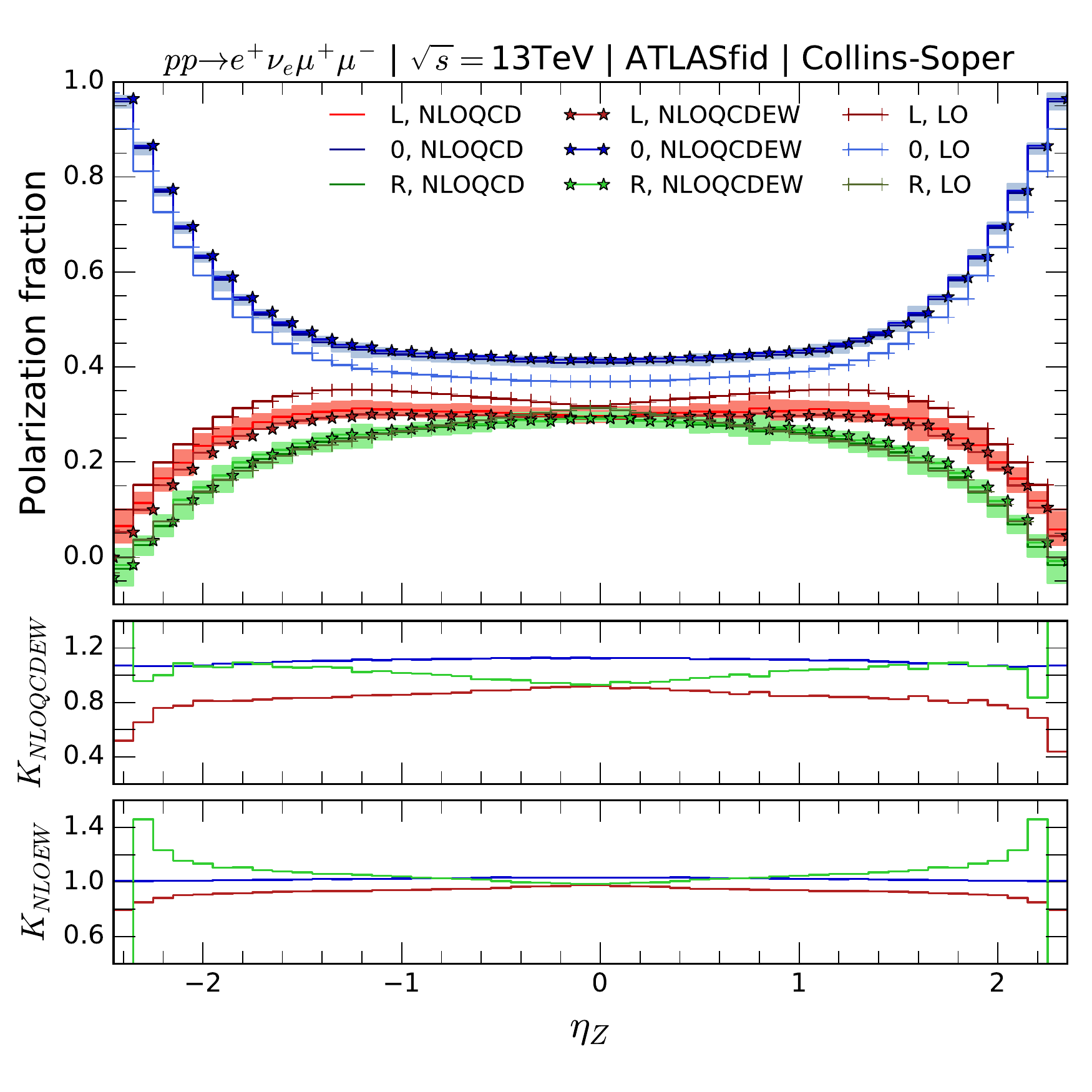}
  \end{tabular}
  \caption{Same as \fig{fig:dist_ptWp_ATLAS_NLOQCDEW} but for
the pseudo-rapidity distributions of the $Z$ fiducial polarization fractions.}
  \label{fig:dist_etaZp_ATLAS_NLOQCDEW}
\end{figure}

\begin{figure}[ht!]
  \centering
  \begin{tabular}{cc}
    \includegraphics[width=0.48\textwidth]{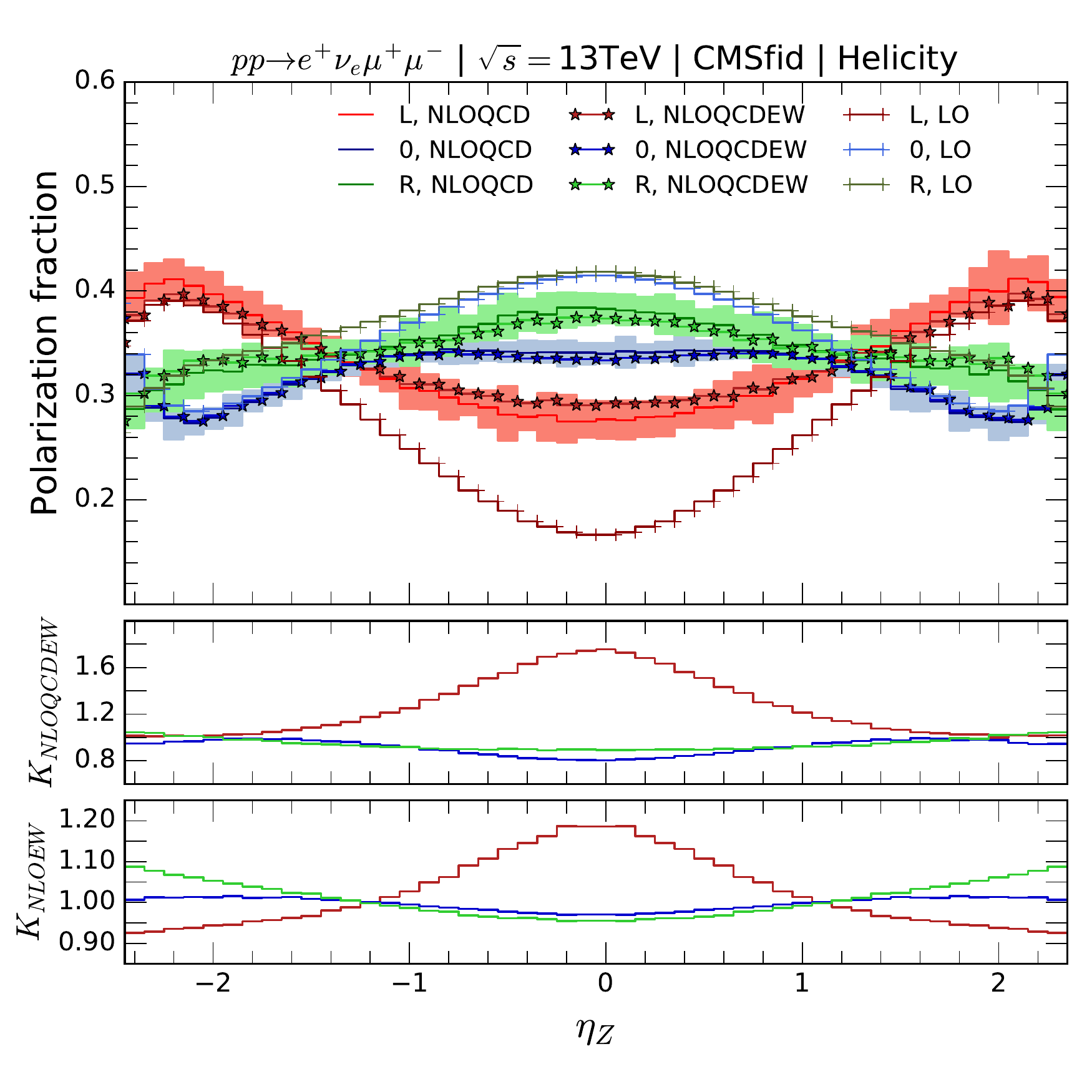}& 
    \includegraphics[width=0.48\textwidth]{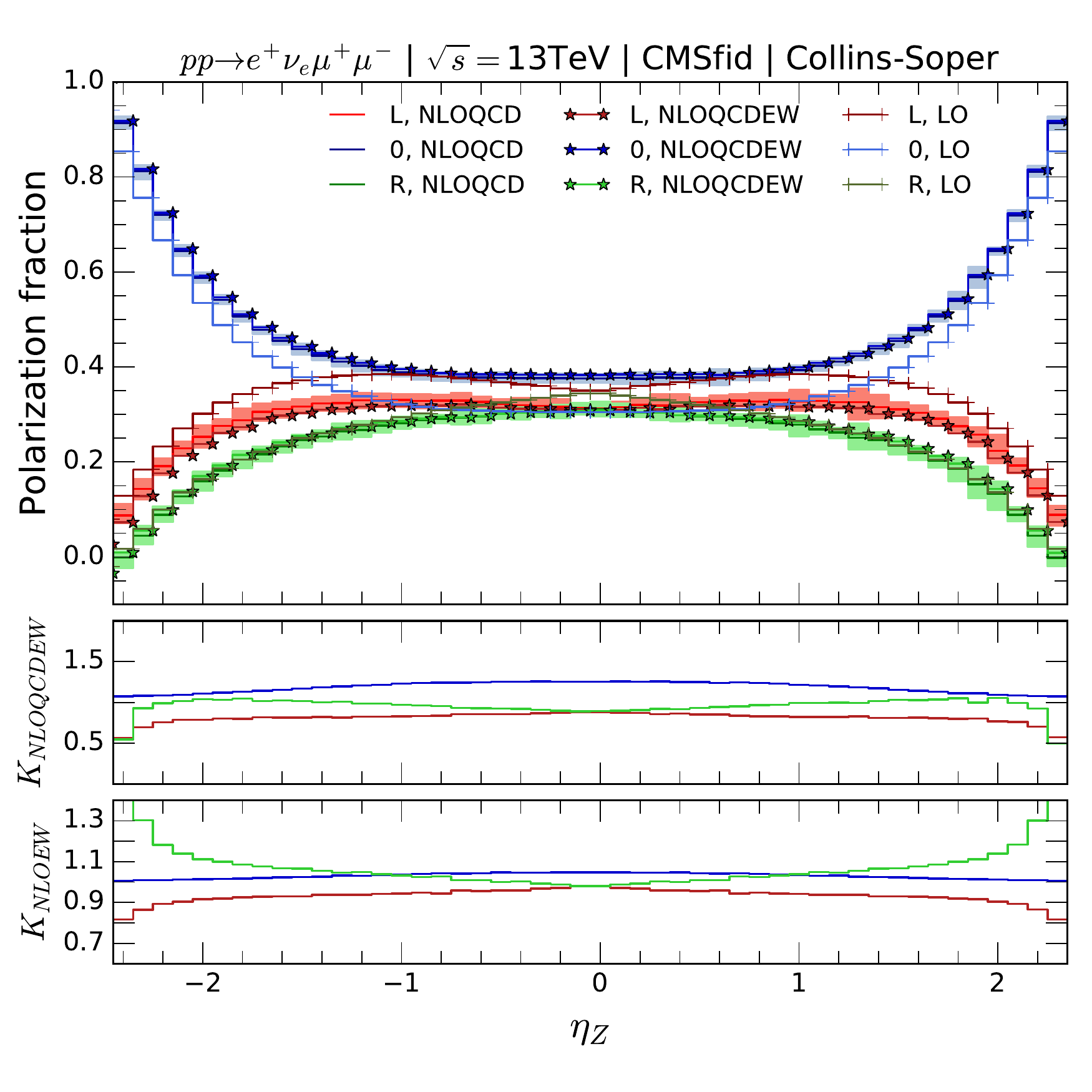}
  \end{tabular}
  \caption{Same as \fig{fig:dist_etaZp_ATLAS_NLOQCDEW} but for
   the CMS fiducial cuts.}
  \label{fig:dist_etaZp_CMS_NLOQCDEW}
\end{figure}

The rapidity and pseudo-rapidity distributions of the $Z$ boson,
shown in \fig{fig:dist_yZp_ATLAS_NLOQCDEW},
\fig{fig:dist_yZp_CMS_NLOQCDEW}, \fig{fig:dist_etaZp_ATLAS_NLOQCDEW},
and \fig{fig:dist_etaZp_CMS_NLOQCDEW} display also the importance of the 
higher-order corrections. Except at the edges of each distribution, where the 
LO results are close to zero, the
bulk of the NLO EW corrections is between $-20\%$ and $+10\%$. The
total NLO corrections, including QCD effects, are about 
$+50\%$ for the left-handed polarization fraction in the bulk. On the
edges of the distribution the $K$-factors can reach values of $2$ or
$3$, again due to the smallness of the LO results. In all
distributions the combined scale+PDF uncertainty is very small, as
seen by the bands in all figures. They do not exceed $\simeq +5\%$. 

Finally, it is important to note that the longitudinal  
fraction of a massive gauge boson decreases at large $p_{T,V}$ according 
to the equivalence theorem. This feature is seen for the 
fiducial longitudinal fraction $f_0$ in the helicity coordinate system, 
but not in the Collins-Soper system. 
As above mentioned, this is because 
the $z$ axis is aligned along the vector-boson direction of flight in the 
helicity system, but not in the Collins-Soper one. 
Therefore, the $f_0$ fraction in the Collins-Soper coordinate system 
is not the longitudinal fraction.   

\clearpage
\pagebreak

\section{Conclusions}
\label{sect:conclusion}
We have presented in this paper an NLO analysis 
, including both
QCD and EW corrections, of the fiducial angular coefficients and of the
fiducial polarization fractions of the $W^\pm$ and $Z$ bosons in the process $p
p \to e \nu \mu^+\mu^-$ at the 13 TeV LHC, using the fiducial
phase space provided by the ATLAS and CMS experiments and in two different
coordinate systems. The LO and NLO QCD predictions include off-shell
effects, while the EW corrections have been calculated in the
DPA. Comparing our predictions for the cross sections as well as for
the kinematical distributions to the full NLO EW results, we find that the DPA
predicts the quark-photon induced and the 
quark-antiquark corrections correctly. Very good agreement has been found 
for many kinematical distributions. In particular, the shape of
the kinematical distributions is well reproduced by the DPA.

We have included the scale and PDF uncertainties, added linearly, in
our predictions for the fiducial angular coefficients and for the fiducial polarization
fractions. They have been found to be very small. 
The EW corrections are found to be
significant, of about $-30\%$ to the NLO QCD predictions, 
in two angular coefficients $A_3$ and $A_4$ 
for the $Z$ boson, and they are mainly due to the EW
corrections to the $Z$ decay into charged leptons. As $f_L$ and $f_R$
can be built out of $A_4$, significant EW corrections, of about $\pm 5\%$ to the 
NLO QCD results, are also found
for these polarization fractions. 
Meanwhile, 
those EW corrections have been found very small for the $W^\pm$ bosons. 
We have also studied the transverse
momenta, rapidity, and pseudo-rapidity distributions of the fiducial 
polarization fractions and we have found that the EW corrections can
also be significant over the whole range of transverse momenta for
$f_L^Z$ and $f_R^Z$. This happens for both the ATLAS and CMS fiducial cuts
and in the two coordinate systems we have considered, namely the
helicity and Collins-Soper coordinate systems.  

We have observed that the fiducial polarization observables in the Collins-Soper coordinate system 
have unexpected behaviors such as negative fractions and not-decreasing longitudinal fraction 
at large transverse momentum values. Meanwhile, in the helicity system, 
the fiducial fractions are all positive and the longitudinal fraction decreases 
with large $p_{T,V}$ for both the $W$ and $Z$ bosons in accordance with the equivalence theorem. 
This can be understood 
because the $z$ axis is aligned along the vector-boson direction of flight in the helicity system, 
but not in the Collins-Soper one. Therefore, the fractions calculated in the helicity system 
are closer to the on-shell values.

This study also shows that it is easy to calculate the fiducial polarization observables. 
They can also be viewed as a simple way to characterize 
the three-dimensional polar-azimuthal 
angular distributions in terms of eight parameters, where three of them are very small and can be 
neglected. They share some common properties with 
the inclusive polarization observables and 
would enable theorists to perform precise comparisons with measurements 
without doing the template fitting step. 

We therefore recommend that experimentalists provide measurements for 
these fiducial coefficients in the helicity coordinate system for both the $W$ and $Z$ bosons.

\appendix

\section{NLO EW corrections in the DPA}
\label{appen:DPA_details} 

We spell out here all necessary details of our NLO EW calculation in
the DPA in such a way that the reader will have all information needed
to reproduce our results. In order to achieve this, the on-shell
mappings and the method for calculating NLO EW corrections have to be
specified.

For the DPA LO, the OS mapping is done as follows. We first generate
the exact kinematics, i.e. the momenta $(p_1,p_2,k_i)$ with $i=1,4$ as
defined in \sect{sect:cal:ew}. The exact momenta of the two gauge
bosons are $q_1$ and $q_2$ as defined in \eq{eq:mom1}. The
corresponding OS momenta $\hat{q}_j$ are calculated as follows. In the
$WZ$ center-of-mass system, we choose
\bea
\vec{\hat{q}}_1 = b \vec{q}_1.
\eea 
We note that this choice is the same as the one used in
\bib{Denner:2000bj}. We then obtain, with $p_1^\mu = (E,0,0,E)$,
\bea
\hat{q}_{10} &=& \fr{M^2_{V_1} - M^2_{V_2}}{4E} + E,\quad 
\hat{q}_{20} = \fr{M^2_{V_2} - M^2_{V_1}}{4E} + E,\crn
q_{10} &=& \fr{q_1^2 - q_2^2}{4E} + E,\quad
q_{20} = \fr{q_2^2 - q_1^2}{4E} + E,\crn
b &=& \sqrt{(\hat{q}^2_{10} - M^2_{V_1})/(q^2_{10} - q^2_1)}.
\label{eq:map1}
\eea
Since the parameter $b$ must be real (otherwise we get complex
momenta), we obtain the condition for the partonic energy
\bea
2E > M_{V_1} + M_{V_2},
\label{eq:cut_E}
\eea
which is equivalent to \eq{eq:cond_DPA}. The next step is to perform
the decays $\hat{q}_j \to \hat{k}_{i_1}+\hat{k}_{i_2}$ to obtain the
OS-projected momenta $\hat{k}_{i}$. They are first calculated in the
rest frame of $V_j$ using two random numbers for each gauge boson and
then boost back. This Monte-Carlo method is described in
\bib{Berends:1994pv} and is also implemented in the {\tt VBFNLO}
program, see the subroutine {\tt TwoBodyDecay0($R_1$, $R_2$,
  $\ldots$)} there. In our code, the same random numbers $R_1$, $R_2$
used in generating $k_i$ are used for $\hat{k}_i$. The momenta of the
initial partons are unchanged, i.e. $\hat{p}_i = p_i$, $i = 1,2$. We
call this method {\it OS mapping R} \footnote{As a way to remember it,
  R stands here for random numbers.} to distinguish with the {\it OS
  mapping D} \footnote{ stands here for the authors of
  \bib{Denner:2000bj}.} method defined as follows. Following
\bib{Denner:2000bj}, the lepton momenta are calculated as
\begin{align}
\hat{k}^\mu_1 &= c k^\mu_1, \quad \hat{k}^\mu_3 = d k^\mu_3,\quad
\hat{k}_2 = \hat{q}_1 - ck_1, \quad \hat{k}_4 = \hat{q}_2 - dk_3,\crn
c &= \fr{M^2_{V_1}}{2\hat{q}_1k_1}, \quad d =
\fr{M^2_{V_2}}{2\hat{q}_2k_3}.
\label{eq:OS_lep_mom_D}
\end{align}
A comparison of these two methods for the cross section at DPA LO is
presented in \tab{tab:compare_OSmap_LO}. The results presented in this
paper are obtained using the OS mapping R.
\begin{table}[ht!]
 \renewcommand{\arraystretch}{1.3}
  \begin{center}
   \small
\begin{tabular}{|c|c|c|c|c|}\hline
Cut  &  Process  & LO~[fb] & DPA mapping R~[fb] & DPA mapping D~[fb] 
\\\hline
ATLAS fid.  &  $e^+\nu_e\mu^+\mu^-$ & $19.344[2]$ & $18.740[2]$ & $18.960[2]$ \\
\hline
ATLAS fid.  &  $e^-\bar{\nu}_e\mu^+\mu^-$ & $13.001[1]$ & $12.987[1]$ & $12.763[1]$ \\
\hline
CMS fid.  &  $e^+\nu_e\mu^+\mu^-$ & $24.6225[4]$ & $23.510[2]$ & $23.922[2]$ \\
\hline
CMS fid.  &  $e^-\bar{\nu}_e\mu^+\mu^-$ & $16.3205[2]$ & $16.157[1]$ & $15.847[1]$ \\
\hline
\end{tabular}
\caption{\small Born cross sections in fb obtained using the full
  matrix elements and also the DPA matrix elements with the OS mapping
  R and OS mapping D methods (see text). The numbers in square
  brackets represent the statistical error.}
\label{tab:compare_OSmap_LO}
\end{center}
\end{table} 

The OS-projected momenta are used to calculate the DPA matrix
elements. However, for kinematical cuts and distributions, we use the
exact kinematics for the LO-like phase space. The OS-projected momenta can also be used here but
we have checked that using exact momenta for cuts gives better
agreement with the full LO results. 
For a real-emission phase space with one extra
particle, it is more complicated because there are tilde kinematics
introduced by the dipole-subtraction method. Here we have to use 
the OS-projected momenta everywhere, i.e. for the matrix elements, cuts, and distributions. 
This point will be
discussed in detail below.

The virtual EW corrections are calculated as in \bib{Baglio:2013toa}
and hence there is no need to repeat it here. The kinematics 
are the same as for the DPA LO. As done in \bib{Baglio:2013toa}, we have 
checked that the virtual corrections including the endpoint contribution 
(also called I operator) defined in \refs{Dittmaier:1999mb,Basso:2015gca} to the decays are UV and IR finite. 
Mass regularization has been used and checks have been performed to make sure that 
the results are independent of the masses of the light fermions, i.e. all but the top quark.

For the $\delta\mathcal{A}_\text{rad,DPA}^{\bar{q}q'\to V_1V_2 \to
  4l\gamma}$ contribution in \eq{eq:rad_DPA}, there are three
terms. For the first term, the photon is radiated from the process
$\bar{q}q'\to V_1V_2$. This contains IR divergences needed to cancel
with the corresponding term in the virtual corrections. The
OS-projected momenta are calculated as follows. First, the exact
kinematics for the process $\bar{q}q'\to 4l + \gamma(k_\gamma)$ are
generated and the momenta $q_1 = k_1 + k_2$, $q_2 = k_3 + k_4$ for the
intermediate gauge bosons are calculated. We then boost to the center-of-mass system of 
the two gauge bosons and calculate $\hat{q}_j$ as above. The
OS-projected momenta for the leptons are calculated from $\hat{q}_j$
using the OS mapping D method. Using the OS mapping R method should
work as well, but we have not tried to do this. After this, all
momenta can be boosted back to the partonic or hadronic CMS as
needed. The momenta of the initial partons and of the photon are
unchanged. Since we use the dipole-subtraction method, there are
subtraction terms involved. The amplitude in each of these terms is
written in a factorized form similar to the DPA LO one. The
OS-projected momenta for those amplitudes are calculated as
follows. First, the tilde momenta, defined in \bib{Dittmaier:1999mb},
corresponding to the reduced process $\bar{q}q'\to V_1V_2$ are
calculated as in \bib{Baglio:2013toa} using the OS momenta
$\hat{q}_j$. These tilde momenta are OS by construction,
i.e. $\tilde{q}_j^2 = M_{V_j}^2$. We then generate the OS momenta for
the leptons from $\tilde{q}_j$ using \eq{eq:OS_lep_mom_D} with $k_1$
and $k_3$ now chosen to be $\hat{k}_1$ and $\hat{k}_3$. We have
checked that this choice gives an IR-safe result. It is noted that we
use the same factor $1/(Q_1Q_2)$ calculated from the exact momenta for
all subtraction terms. For the kinematical cuts and distributions of
the subtraction terms, we use the OS-projected momenta that enter the
reduced amplitudes there. To cancel the IR divergences, we have to use 
the OS-projected momenta for cuts and distributions of the parent $N+1$ 
contribution. The same choice is used for the subtraction
terms of the radiative decays discussed next.

For the second (or third) term in the RHS of \eq{eq:rad_DPA} we have
to focus on the radiative decay $V_1 \to l_1 l_2 \gamma$. The
phase space is generated in the same way as for the DPA LO, but we
have to replace the LO decay $V_1 \to l_1 l_2$ by the radiative
one. In our code, we have to call the subroutine {\tt
  ThreeBodyDecay0()} of the {\tt VBFNLO} program, which uses five
random numbers to generate three Euler angles and two final-state
particle energies. This routine is called two times: First to
calculate the exact momenta and the phase-space Jacobian, 
second to calculate the OS-projected
momenta using the $\hat{q}_j$ as described above. The same set of
random numbers is used in both times. The other decay $V_2 \to l_3
l_4$ is calculated exactly as for the DPA LO. From this one can see
that the OS mapping R method is very general and can be easily
generalized for more complicated processes. We have also found another
method to calculate the OS-projected momenta, similar to the OS
mapping D method but for $1\to 3$ decays this time. This can be done
as follows. We choose
\begin{align}
\hat{k}_i^\mu &= c k_i^\mu, \quad i=1,2, \quad \hat{k}_\gamma = \hat{q}_1 - c(\hat{k}_1 + \hat{k}_2),\crn  
c &= \fr{2(k_1 + k_2) \hat{q}_1 - 2\sqrt{[(k_1 + k_2) \hat{q}_1]^2 - 2(k_1k_2)M^2_{V_1}}   
}{4k_1k_2},
\end{align}
where we have to take the minus sign solution for $c$ because the plus
sign solution makes $(\hat{k}_1 + \hat{k}_\gamma)^2 + (\hat{k}_2 +
\hat{k}_\gamma)^2$ negative. We have also proved that the argument of
the square root function is always positive when the DPA condition
$(q_1 + q_2)^2 > (M_{V_1} + M_{V_2})^2$ is satisfied. We call this OS
mapping T method \footnote{T stands here for three-body decays.}. We
have checked that both choices give similar results and a comparison
is presented in \tab{tab:compare_decayV_NLOEW}. The final NLO EW
results of this paper are calculated using the OS mapping R method.
\begin{table}[ht!]
 \renewcommand{\arraystretch}{1.3}
  \begin{center}
\setlength\tabcolsep{0.03cm}
\fontsize{10.0}{10.0}
\begin{tabular}{|c|c|c|c||c|c|}\hline
Cut  &  Process & $W$ (map. R)~[fb] & $W$ (map. T)~[fb] & $Z$ (map. R)~[fb] & $Z$ (map. T)~[fb]  
\\\hline
ATLAS fid.  &  $e^+\nu_e\mu^+\mu^-$ & $-0.1326[3]$ & $-0.1283[8]$ & $-0.7803[9]$ & $-0.7651[10]$ \\
\hline
ATLAS fid.  &  $e^-\bar{\nu}_e\mu^+\mu^-$ & $-0.0813[2]$ & $-0.0796[4]$ & $-0.5262[5]$ & $-0.5163[6]$ \\
\hline
CMS fid.  &  $e^+\nu_e\mu^+\mu^-$ & $-0.1451[5]$ & $-0.1388[6]$ & $-0.4090[6]$ & $-0.4032[9]$ \\
\hline
CMS fid.  &  $e^-\bar{\nu}_e\mu^+\mu^-$ & $-0.0869[3]$ & $-0.0843[7]$ & $-0.2699[7]$ & $-0.2693[7]$ \\
\hline
\end{tabular}
\caption{\small Cross sections in fb of the second ($W$) and third ($Z$) terms in
  the RHS of \eq{eq:rad_DPA} with subtraction terms obtained using the
  DPA matrix elements with the OS mapping R and OS mapping T methods
  (see text). The numbers in square brackets represent the statistical
  error.}
\label{tab:compare_decayV_NLOEW}
\end{center}
\end{table}  

We now describe how the corresponding subtraction terms are
calculated. For $Z(\hat{q}_2) \to \mu^+(\hat{k}_3) + \mu^-(\hat{k}_4)
+ \gamma(\hat{k}_\gamma)$ decay, the photon can only be radiated off
the final state leptons. The subtraction term is calculated in a
straightforward way using the method of \bib{Dittmaier:1999mb}. For
$W^+(\hat{q}_2) \to e^+(\hat{k}_1) + \nu_e(\hat{k}_2) +
\gamma(\hat{k}_\gamma)$ decay, it is more complicated because the
photon can be radiated from the initial-state $W$ boson. The method of
\bib{Dittmaier:1999mb} cannot be applied here because it is for $2\to
n$ processes. Fortunately, the subtraction method for $1\to n$ decay
processes has been worked out in \bib{Basso:2015gca} and can be
directly used here. The tilde momenta calculated from the hat ones are
all on-shell by construction and hence no further OS projection is
needed. As before, the same factor $1/(Q_1Q_2)$ calculated from the
exact momenta is used for all subtraction terms here as
well. Concerning cross check, we have performed two independent
calculations and checked that the soft and collinear limits work.

For the last term defined in \eq{eq:ind_DPA}, the central piece is the
factor $\delta A_\text{ind}^{q\gamma \to V_1 V_2q'}$ which has been
calculated in \bib{Baglio:2013toa} and is therefore taken over. For
the LO decay factors, they are calculated as for the DPA LO term, using
the OS mapping R method to get the momenta. 
Same as for the above photon-emission corrections 
involving subtraction terms, we use 
the OS-projected momenta for the kinematical cuts and distributions of
the subtraction terms and of the parent $N+1$ 
contribution.

\section{Kinematical distributions: \boldmath $W^-Z$ channel}
\label{appen:kin_dist_Wm}

\begin{figure}[ht!]
  \centering
  \begin{tabular}{cc}
  \includegraphics[width=0.48\textwidth]{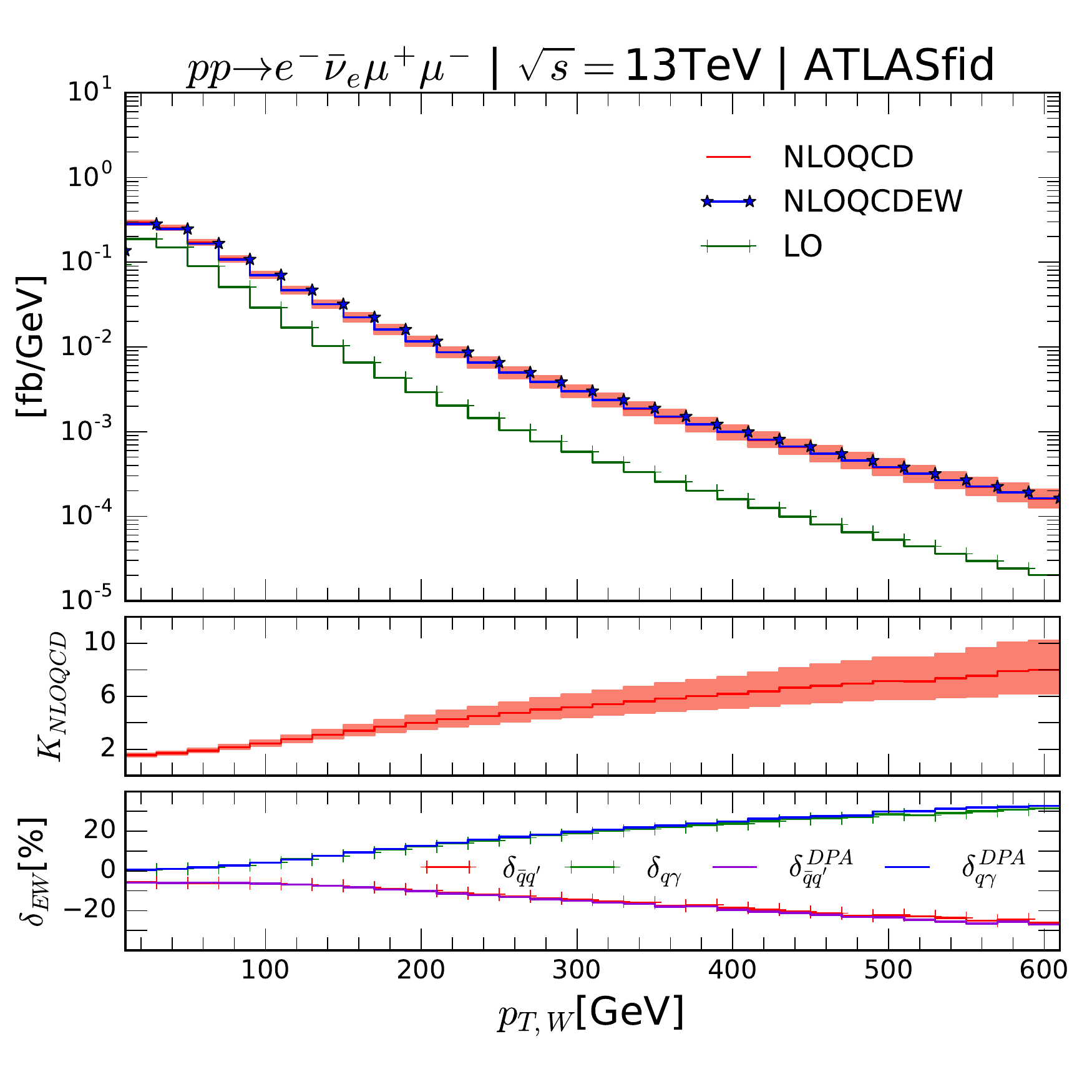}& 
  \includegraphics[width=0.48\textwidth]{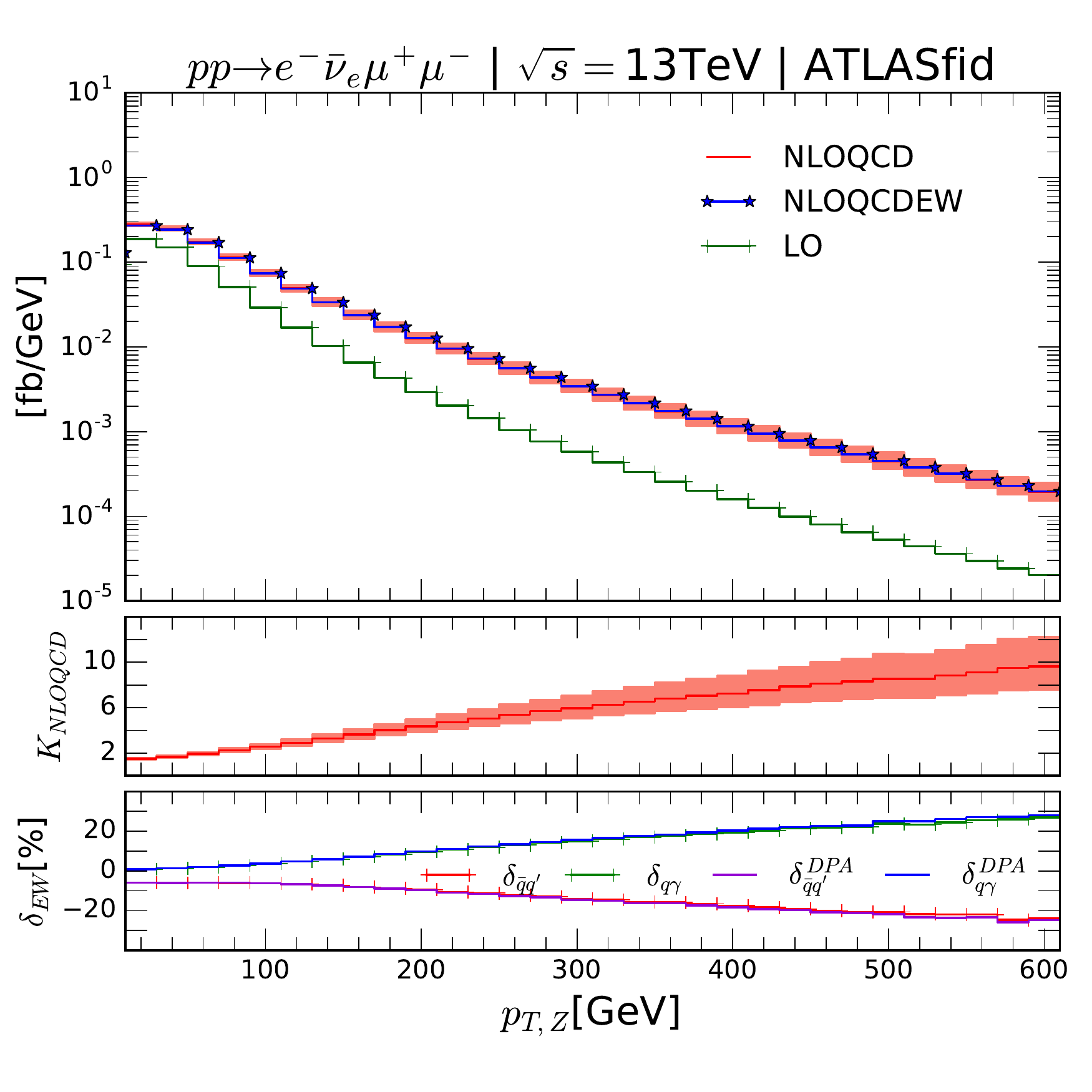}\\
  \includegraphics[width=0.48\textwidth]{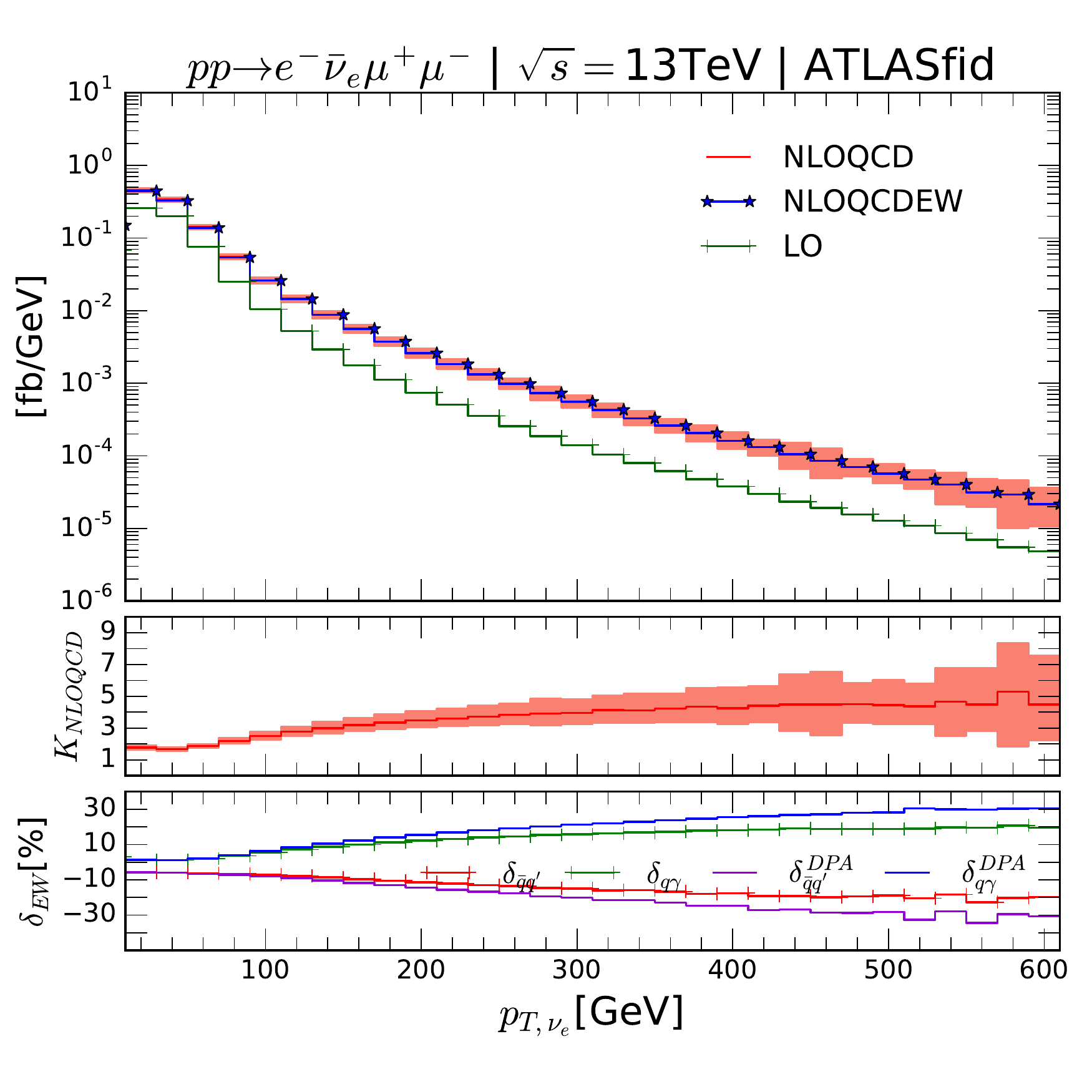}& 
  \includegraphics[width=0.48\textwidth]{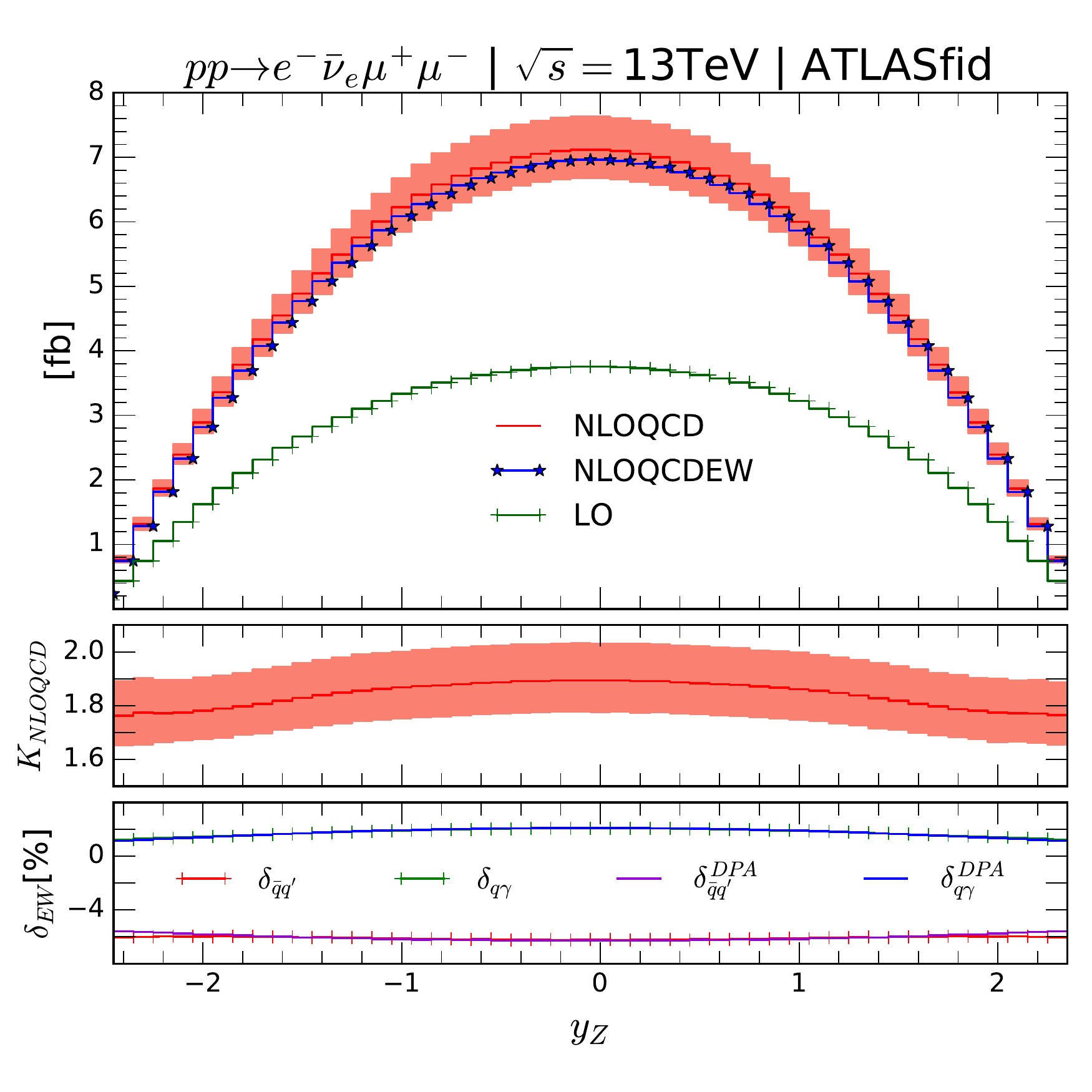}
  \end{tabular}
  \caption{Distributions of the transverse momentum of the $e^-\nu_e$ (top left) and 
    $\mu^+\mu^-$ (top right) systems and of the neutrino (bottom left)
    in the processes $p p\to e^-_{} \nu_e^{}\,
    \mu^+_{}\mu^-_{} +X$ at the 13 TeV LHC with the
    ATLAS fiducial cuts. Rapidity distribution of the $\mu^+\mu^-$ system 
    is also displayed at the bottom right corner. 
    The upper panels show the absolute values of 
    the cross sections at LO (in green), NLO QCD (red), and NLO QCD+EW (blue). 
    The middle panels display the ratio of the NLO QCD cross sections to the
    corresponding LO ones. The bands indicate the total theoretical uncertainty calculated 
    as a linear sum of PDF and scale uncertainties at NLO QCD. The bottom panels show the 
    NLO EW corrections (see text) calculated using DPA relative to the full LO (marked with plus signs) and DPA LO cross sections.}
  \label{fig:dist_pT_W_Z_y_W_Z_Wm_atlas}
\end{figure}

\begin{figure}[ht!]
  \centering
  \begin{tabular}{cc}
  \includegraphics[width=0.48\textwidth]{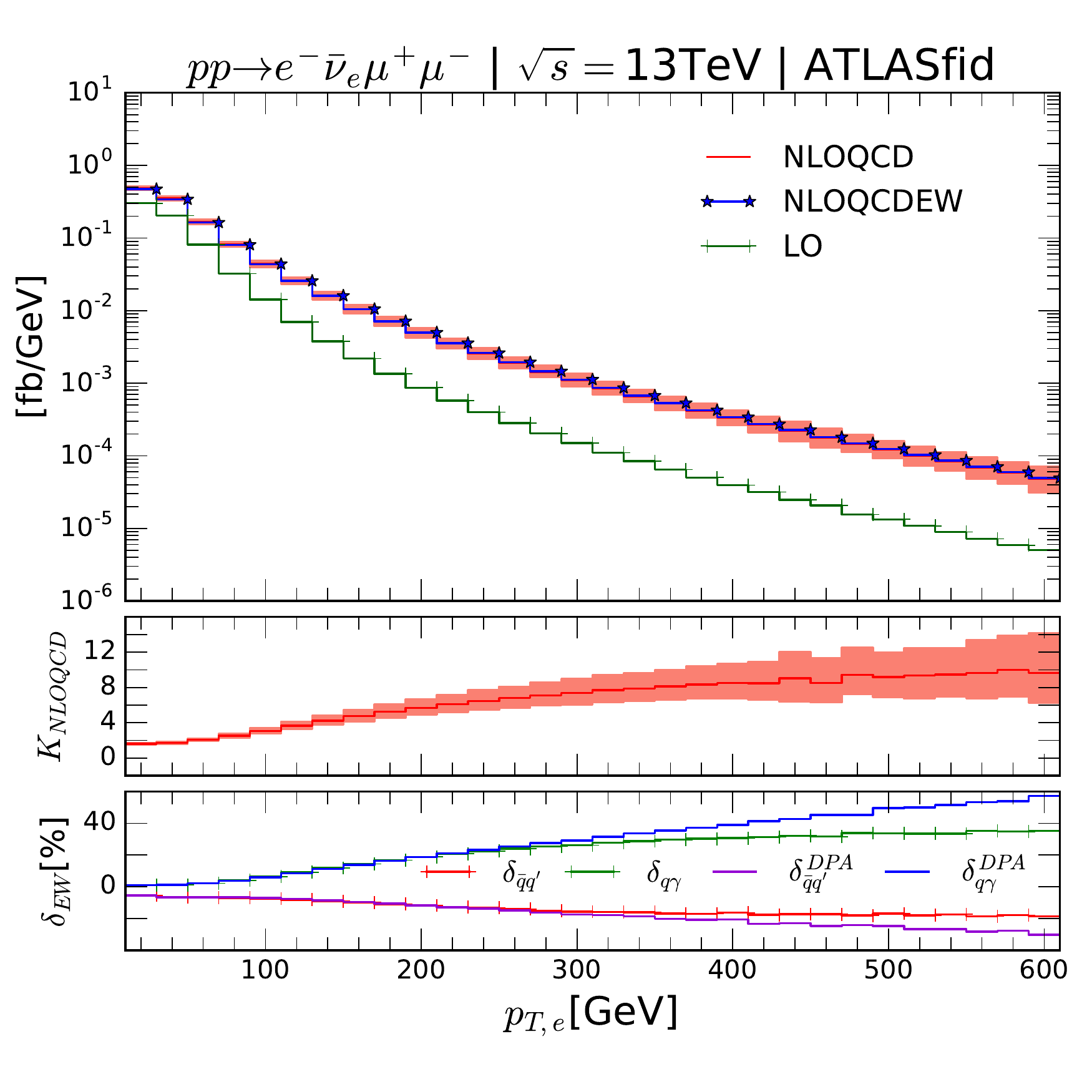}& 
  \includegraphics[width=0.48\textwidth]{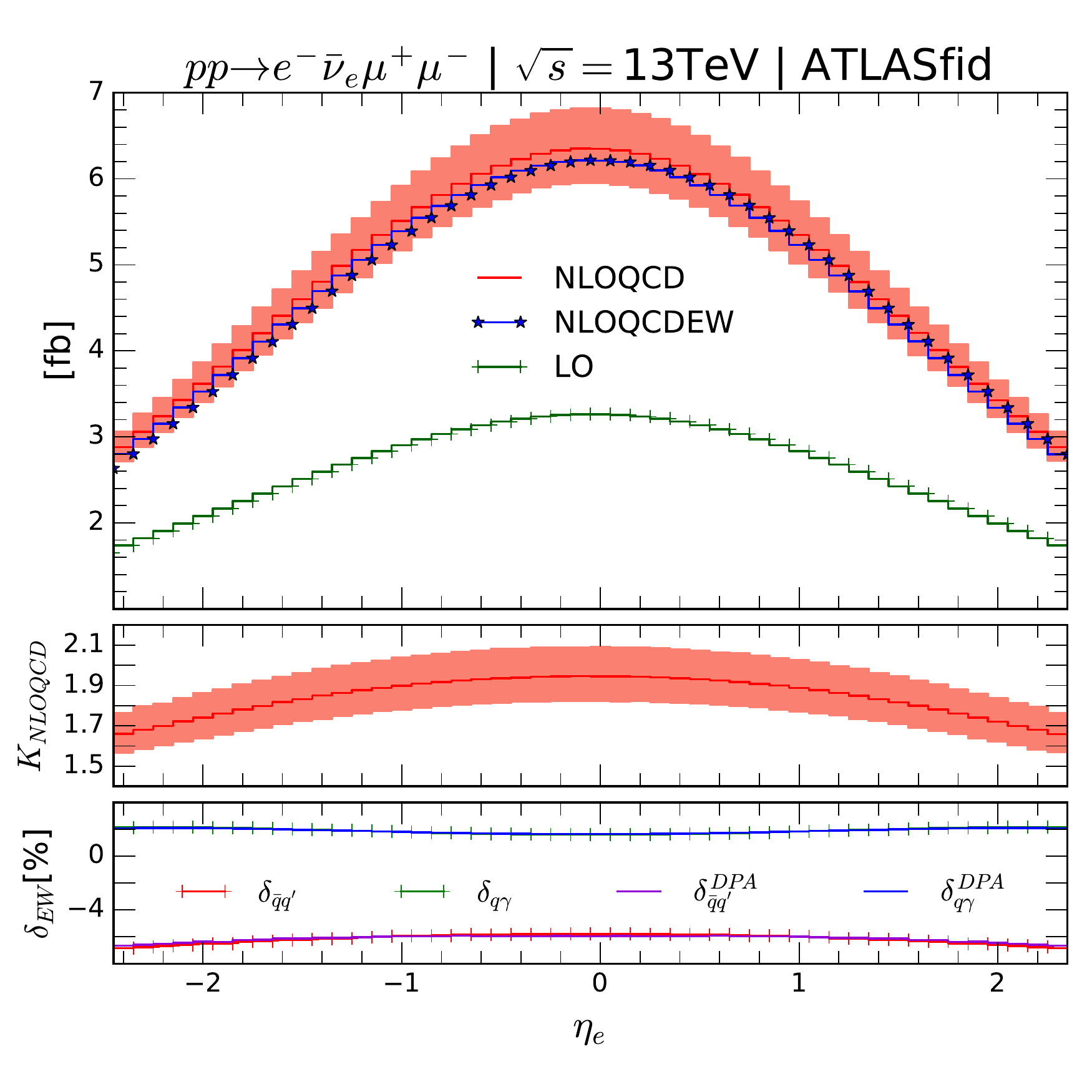}\\
  \includegraphics[width=0.48\textwidth]{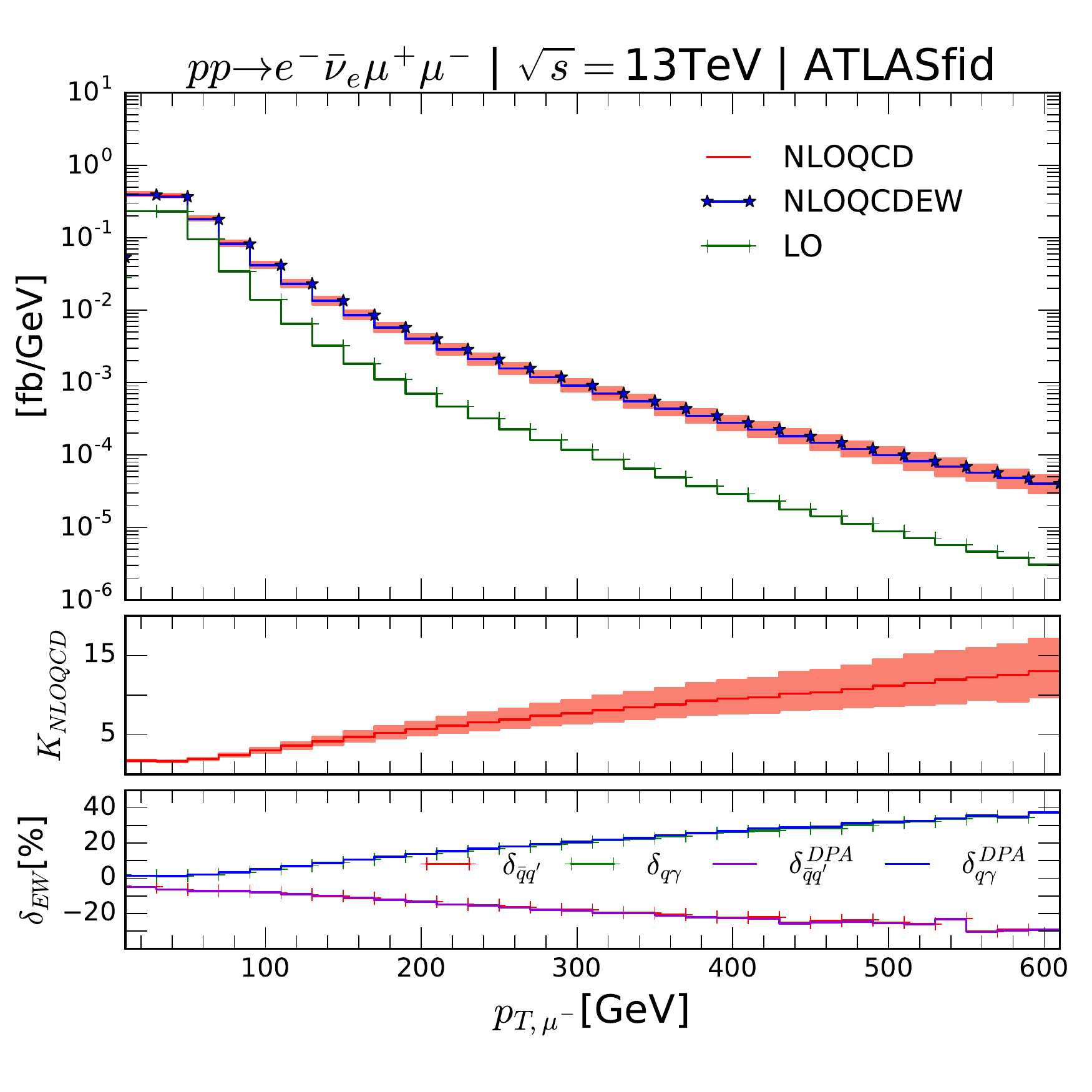}& 
  \includegraphics[width=0.48\textwidth]{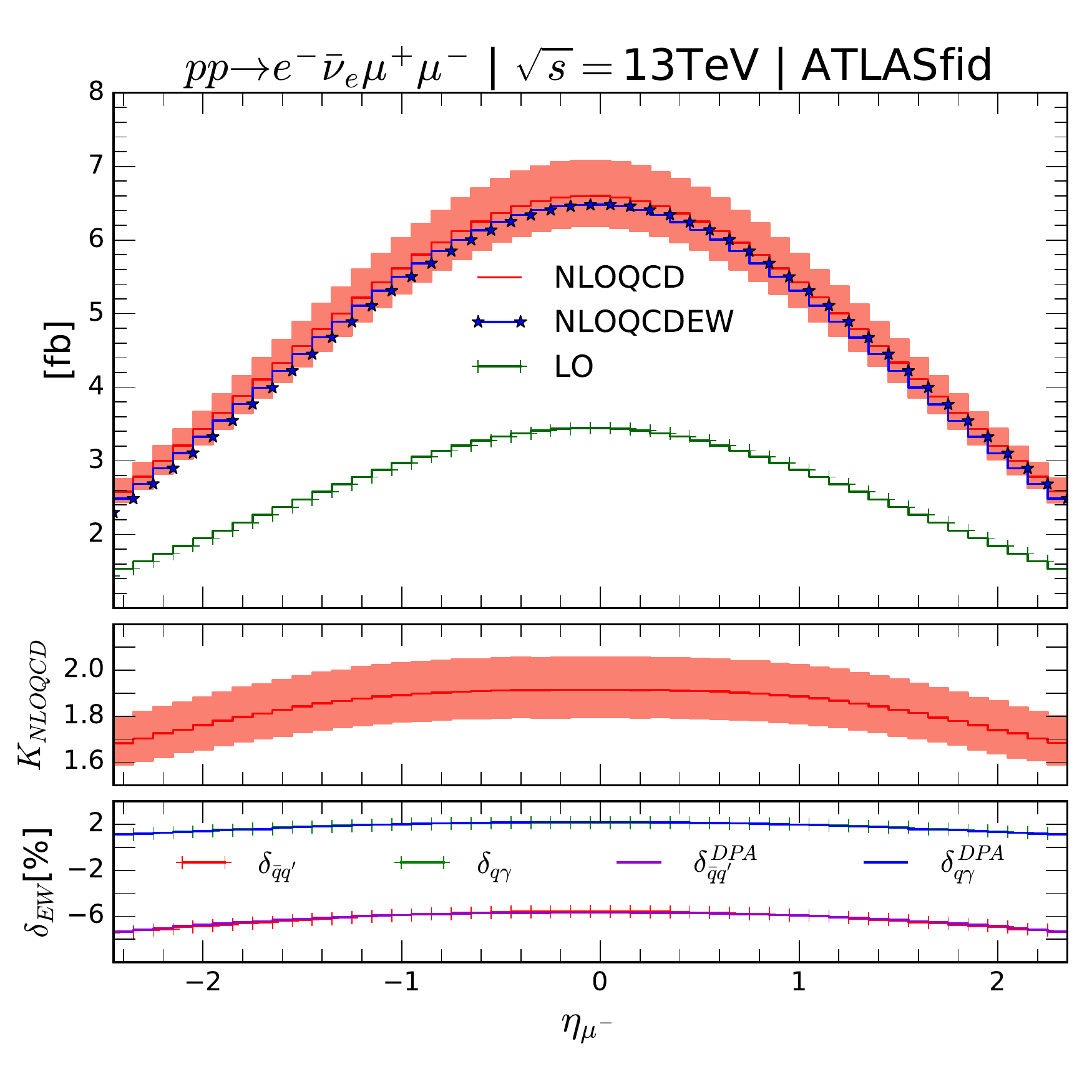}
  \end{tabular}
  \caption{Same as \fig{fig:dist_pT_W_Z_y_W_Z_Wm_atlas} but for the transverse momentum and pseudo-rapidity 
  distributions of the positron (top row) and the muon (bottom row).}
  \label{fig:dist_pT_e_mu_eta_e_mu_Wm_atlas}
\end{figure}

\begin{figure}[ht!]
  \centering
  \begin{tabular}{cc}
  \includegraphics[width=0.48\textwidth]{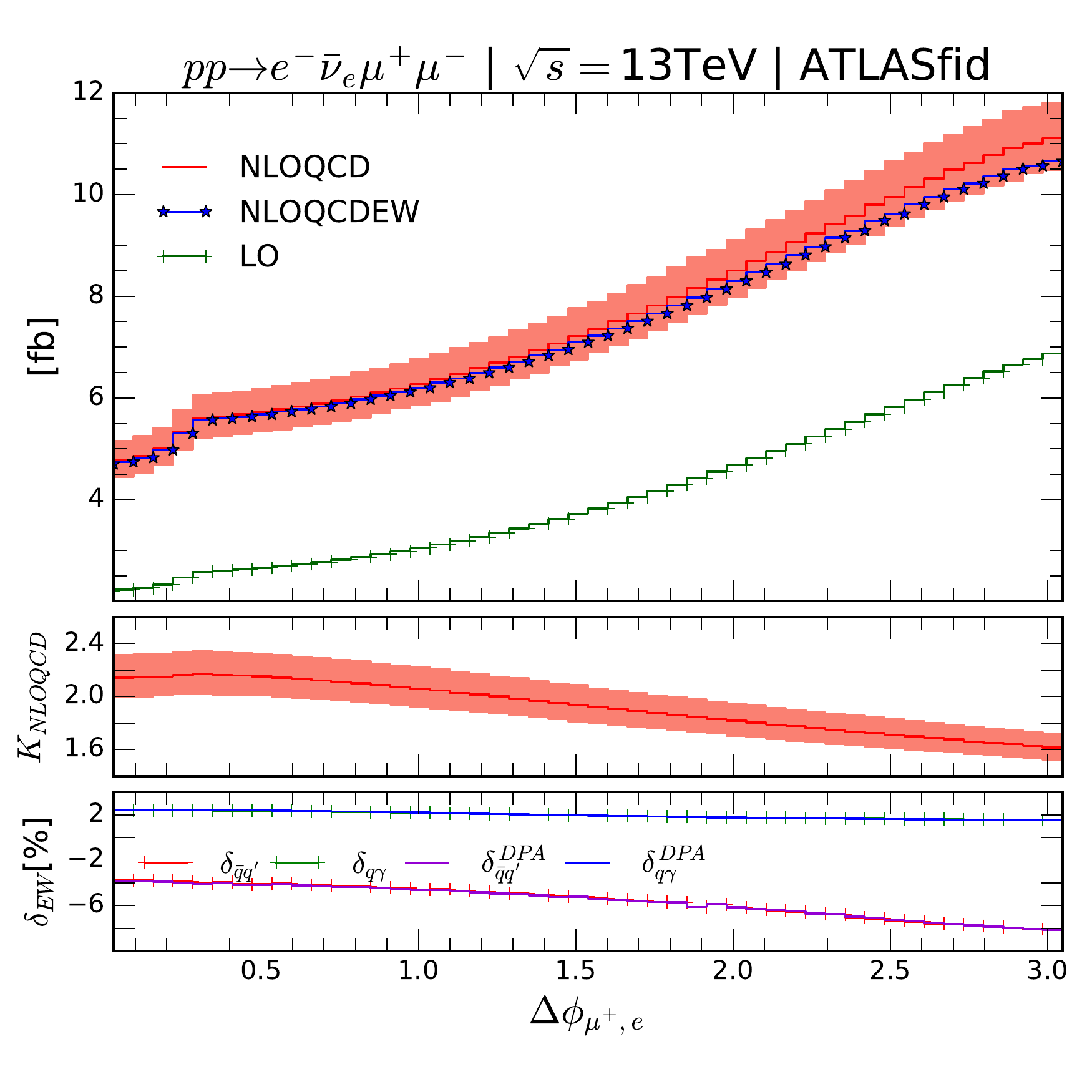}& 
  \includegraphics[width=0.48\textwidth]{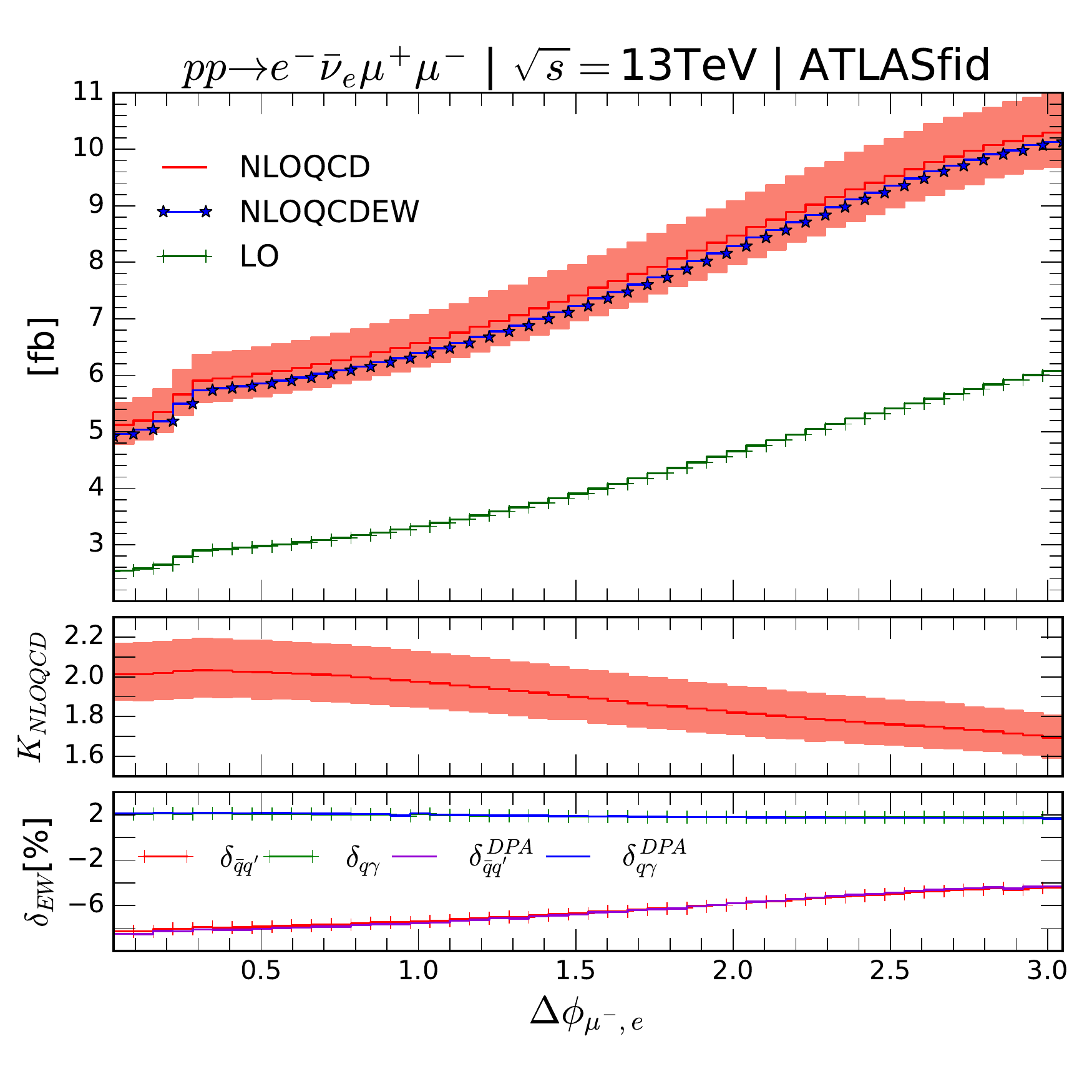}\\
  \includegraphics[width=0.48\textwidth]{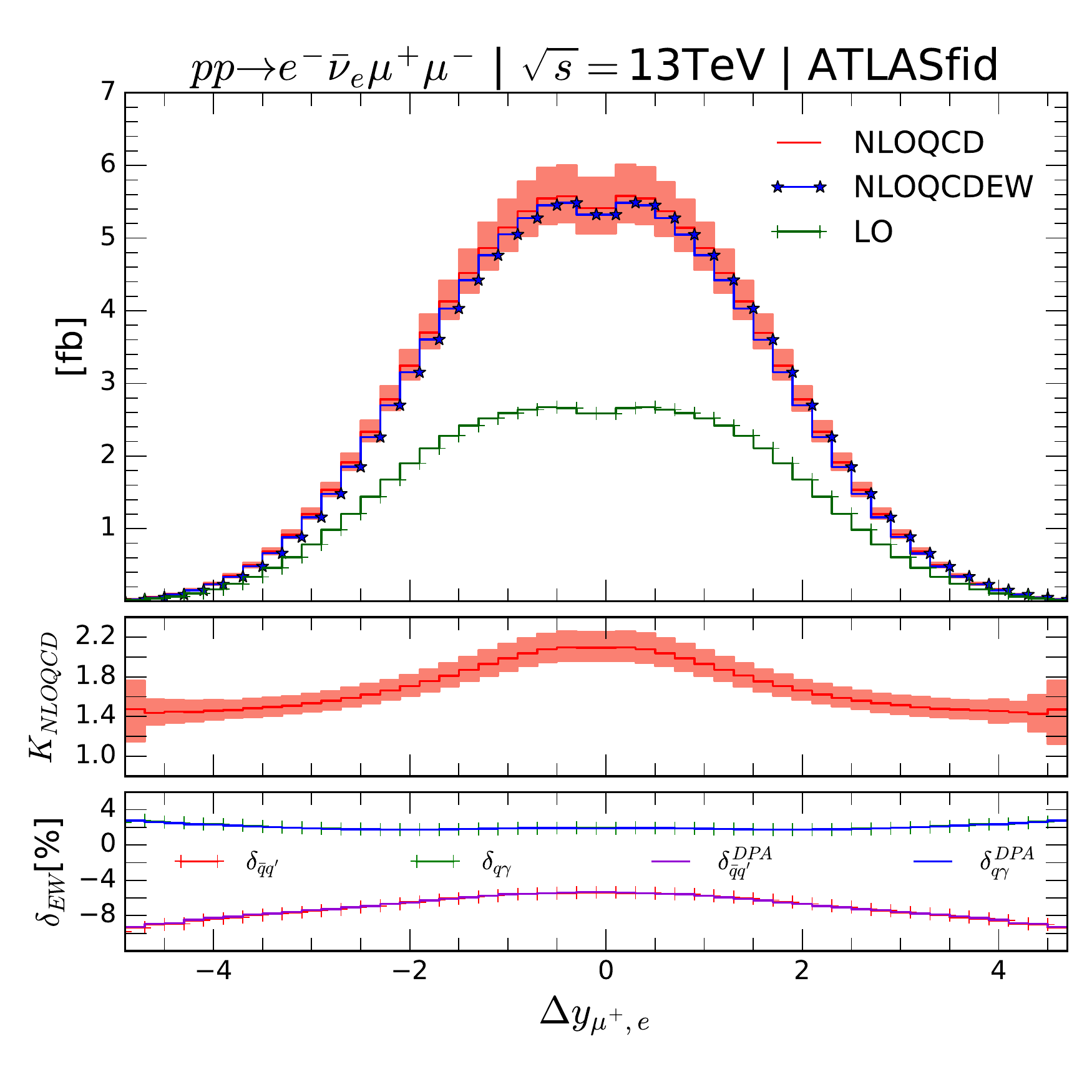}& 
  \includegraphics[width=0.48\textwidth]{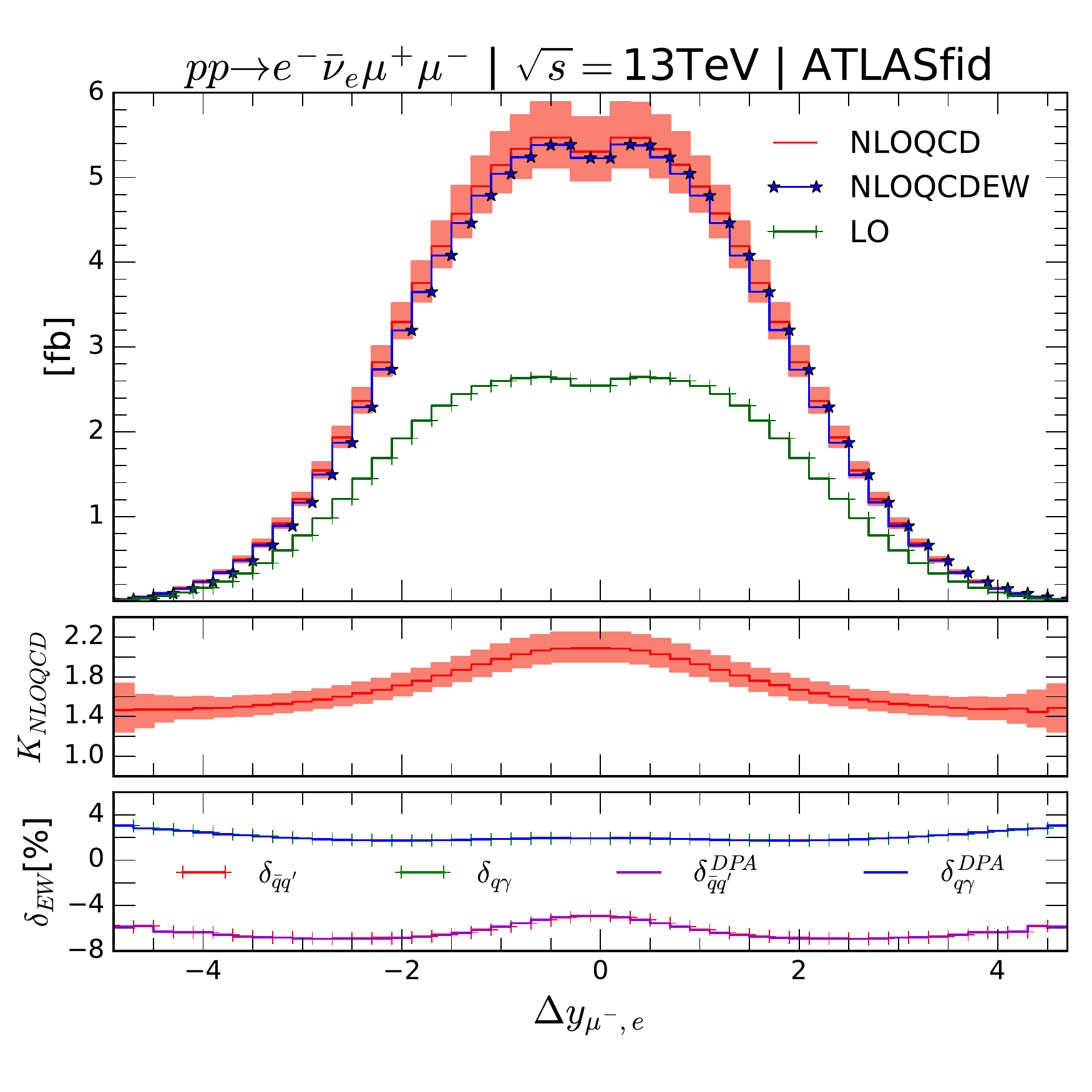}
  \end{tabular}
  \caption{Same as \fig{fig:dist_pT_W_Z_y_W_Z_Wm_atlas} but for the azimuthal-angle difference (top row) 
   and the rapidity difference (bottom row) between the $\mu^+$ and $e^-$ (left column) and between the 
   $\mu^-$ and $e^-$ (right column).}
  \label{fig:dist_Delta_phi_y_Wm_atlas}
\end{figure}

\begin{figure}[ht!]
  \centering
  \begin{tabular}{cc}
  \includegraphics[width=0.48\textwidth]{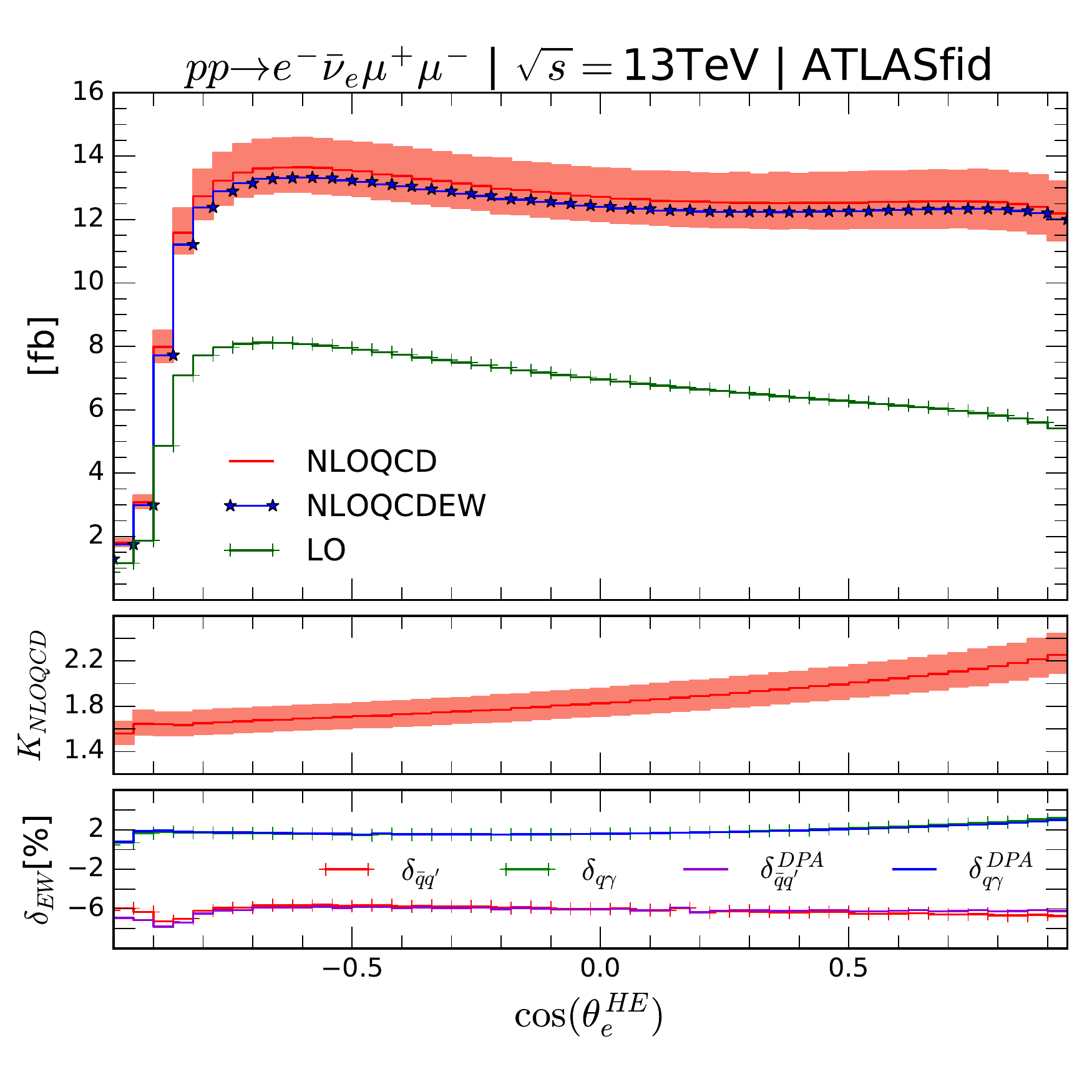}&
  \includegraphics[width=0.48\textwidth]{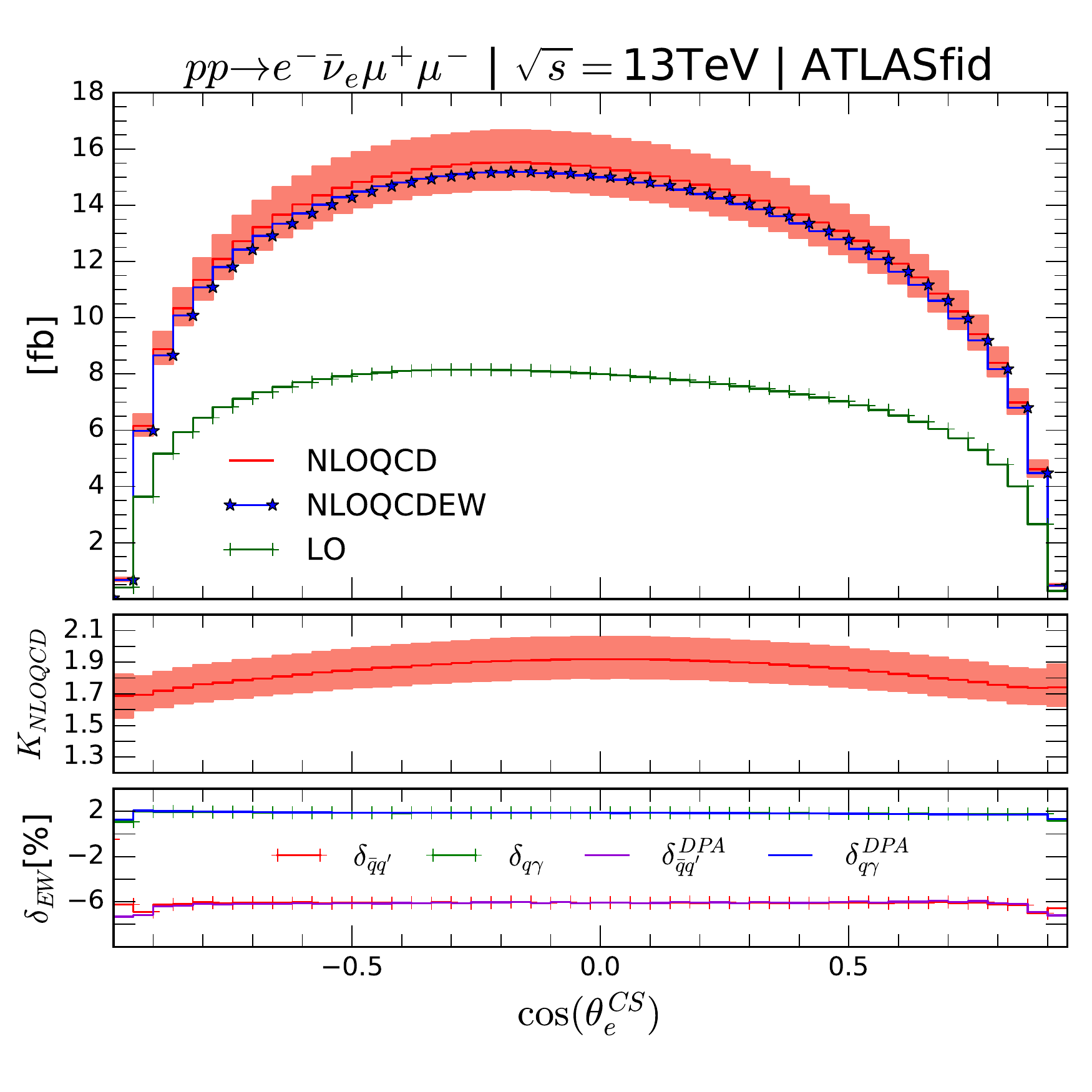}\\
  \includegraphics[width=0.48\textwidth]{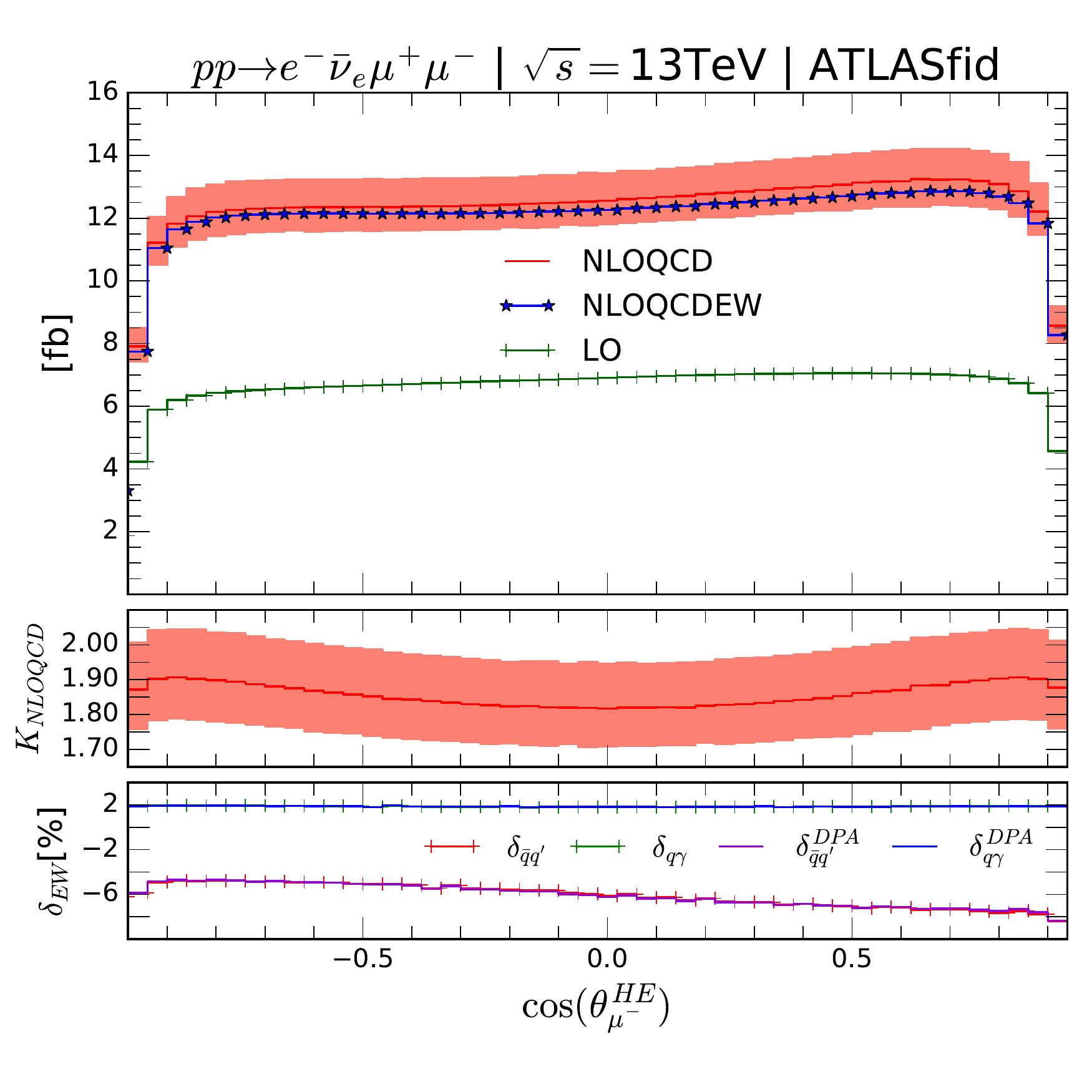}&
  \includegraphics[width=0.48\textwidth]{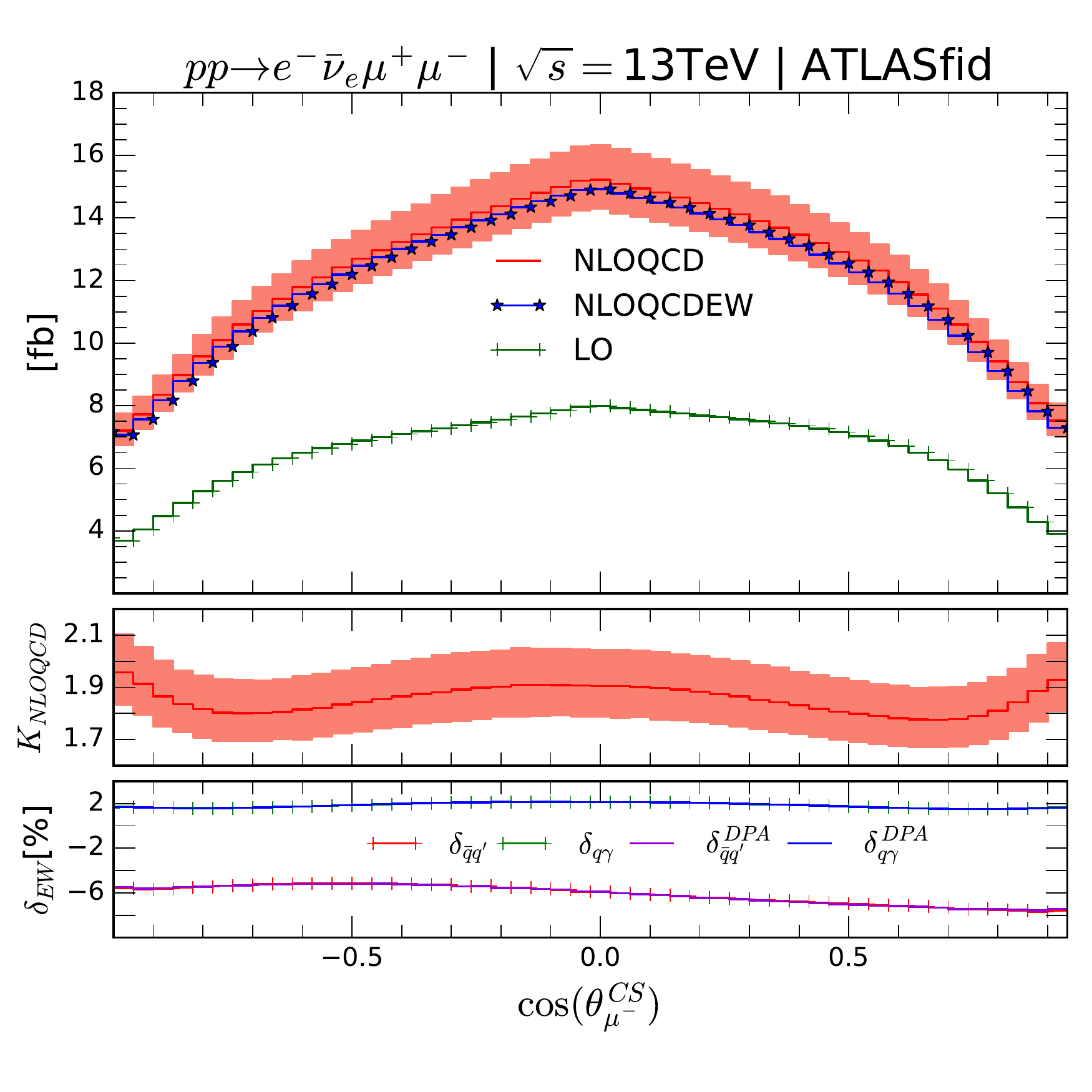}
  \end{tabular}
  \caption{Same as \fig{fig:dist_pT_W_Z_y_W_Z_Wm_atlas} but for the $\cos\theta$ distributions of the electron (top row) 
   calculated in the Helicity (left) and Collins-Soper (right) coordinate systems. The same distributions for 
   the muon are shown in the bottom row.}
   \label{fig:dist_cos_theta_HEL_CS_e_muon_Wm_atlas}
\end{figure}

\begin{figure}[ht!]
  \centering
  \begin{tabular}{cc}
  \includegraphics[width=0.48\textwidth]{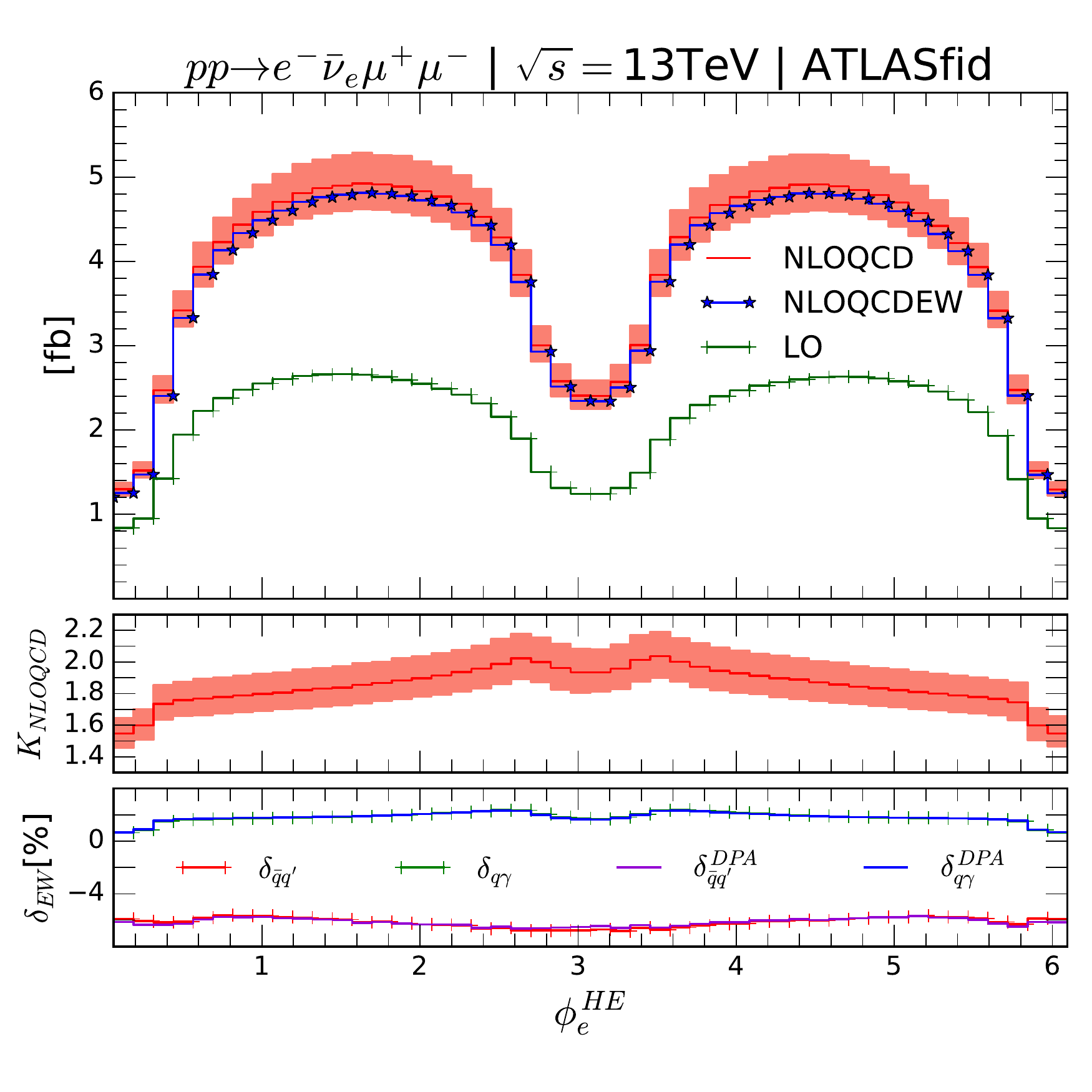}&
  \includegraphics[width=0.48\textwidth]{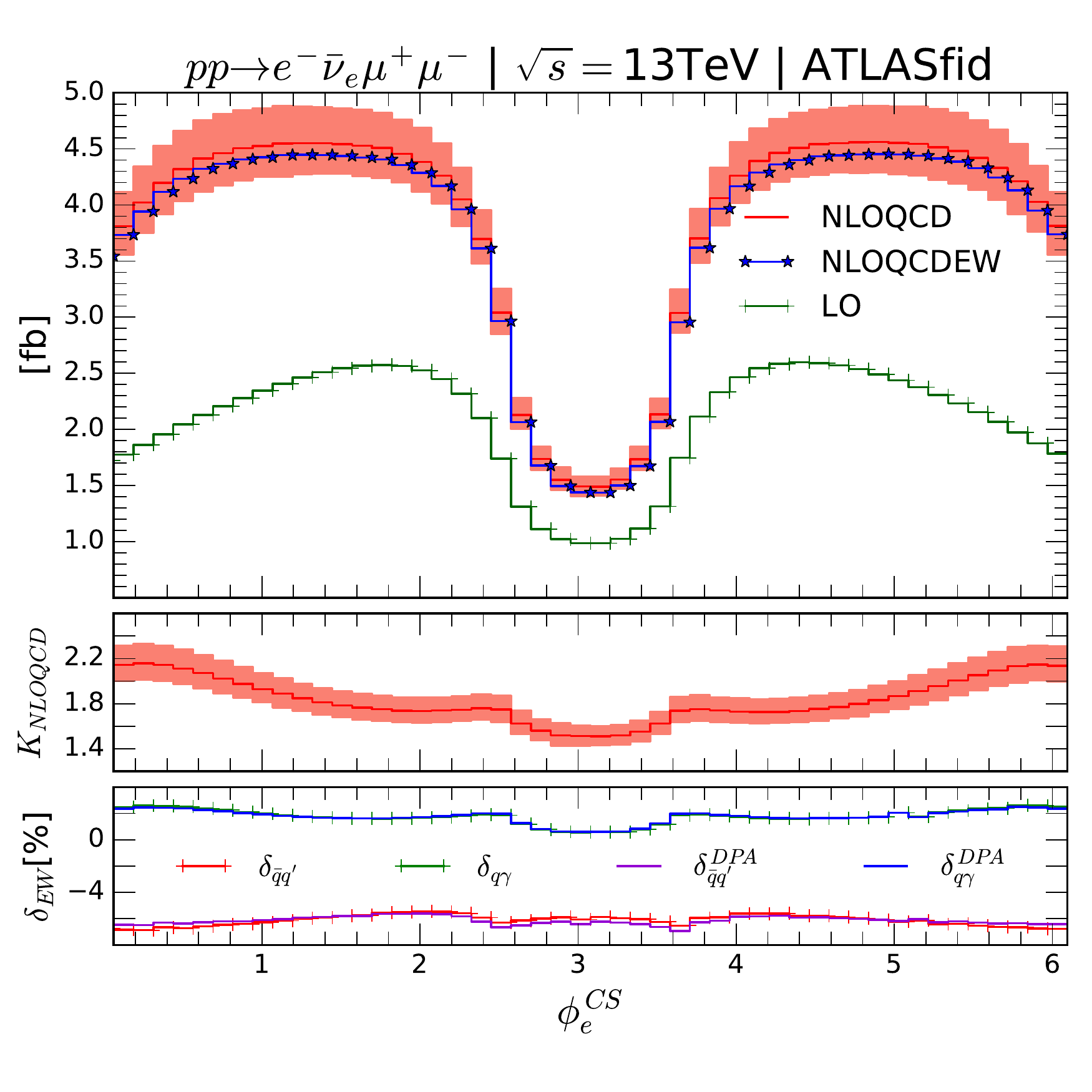}\\
  \includegraphics[width=0.48\textwidth]{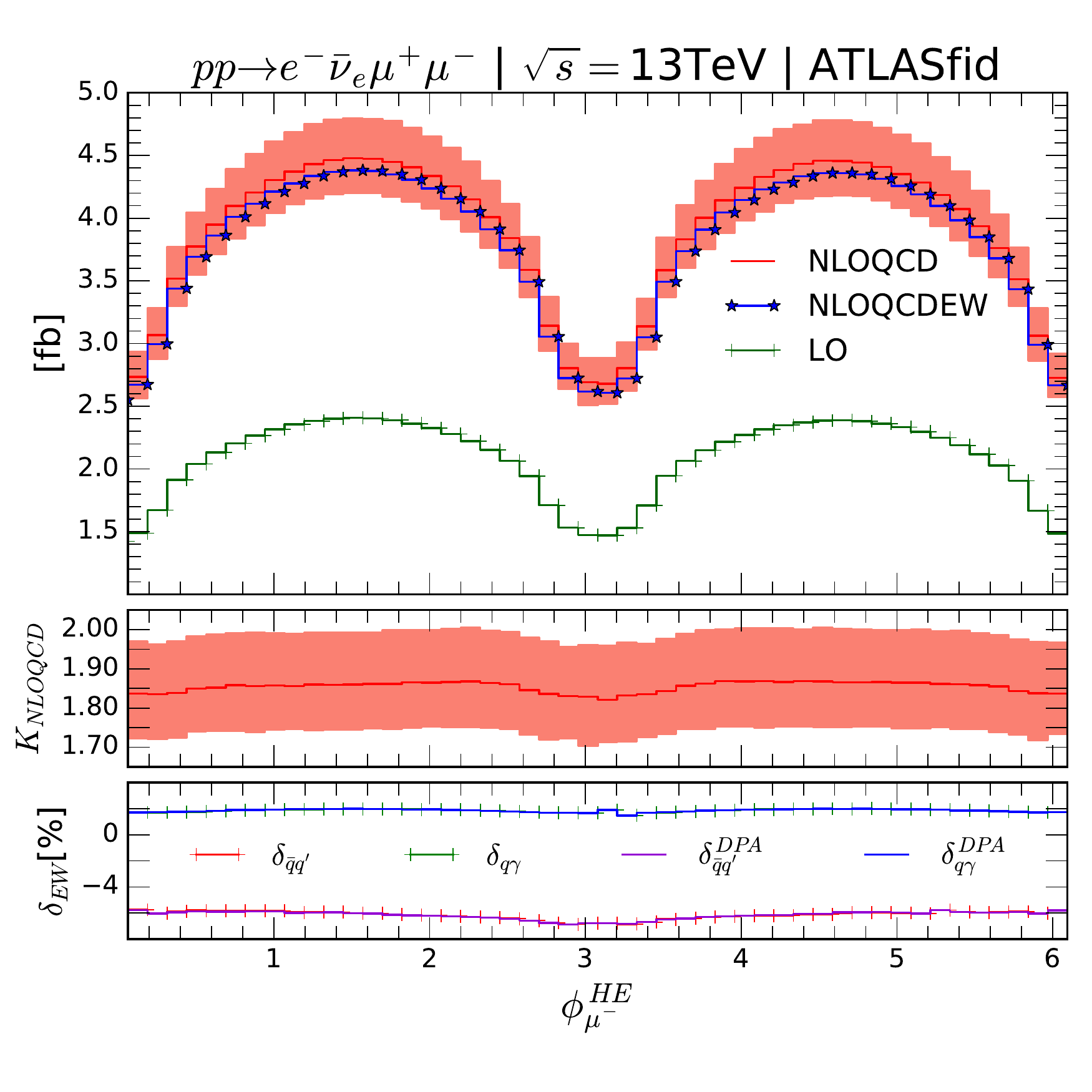}&
  \includegraphics[width=0.48\textwidth]{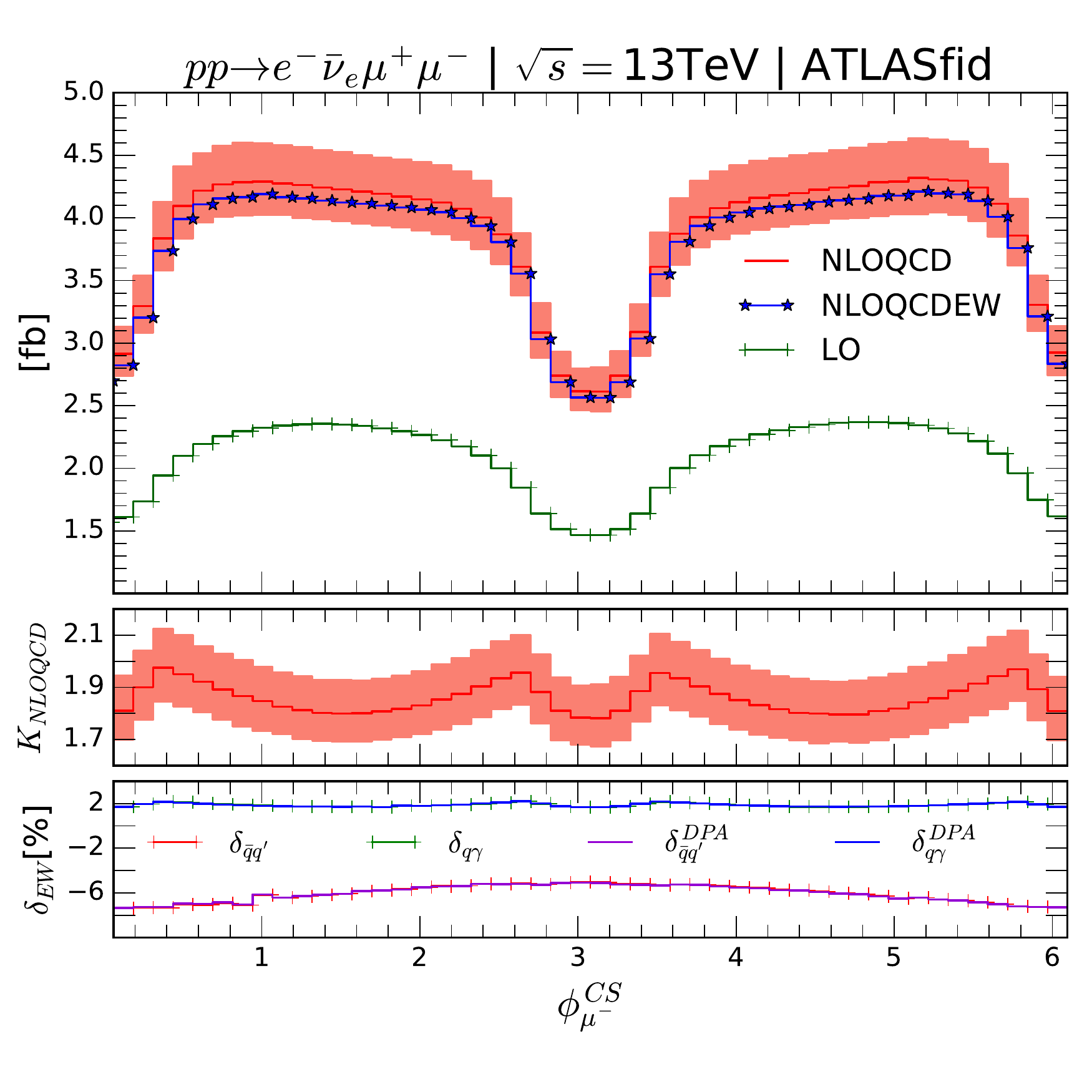}
  \end{tabular}
  \caption{Same as \fig{fig:dist_pT_W_Z_y_W_Z_Wm_atlas} but for the azimuthal-angle distributions of the electron (top row) 
   calculated in the Helicity (left) and Collins-Soper (right) coordinate systems. The same distributions for 
   the muon are shown in the bottom row.}
   \label{fig:dist_phi_HEL_CS_e_muon_Wm_atlas}
\end{figure}

\begin{figure}[ht!]
  \centering
  \begin{tabular}{cc}
  \includegraphics[width=0.48\textwidth]{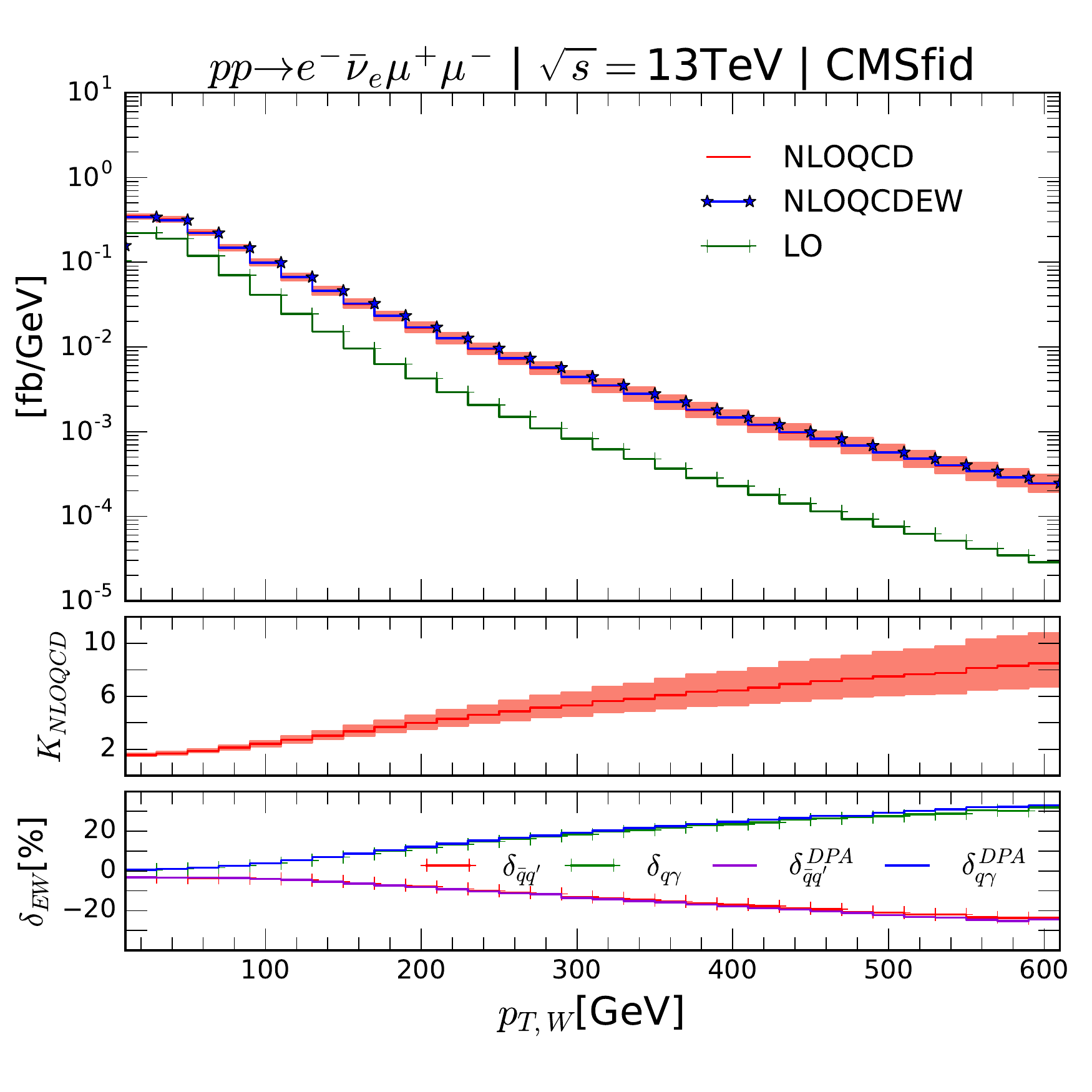}& 
  \includegraphics[width=0.48\textwidth]{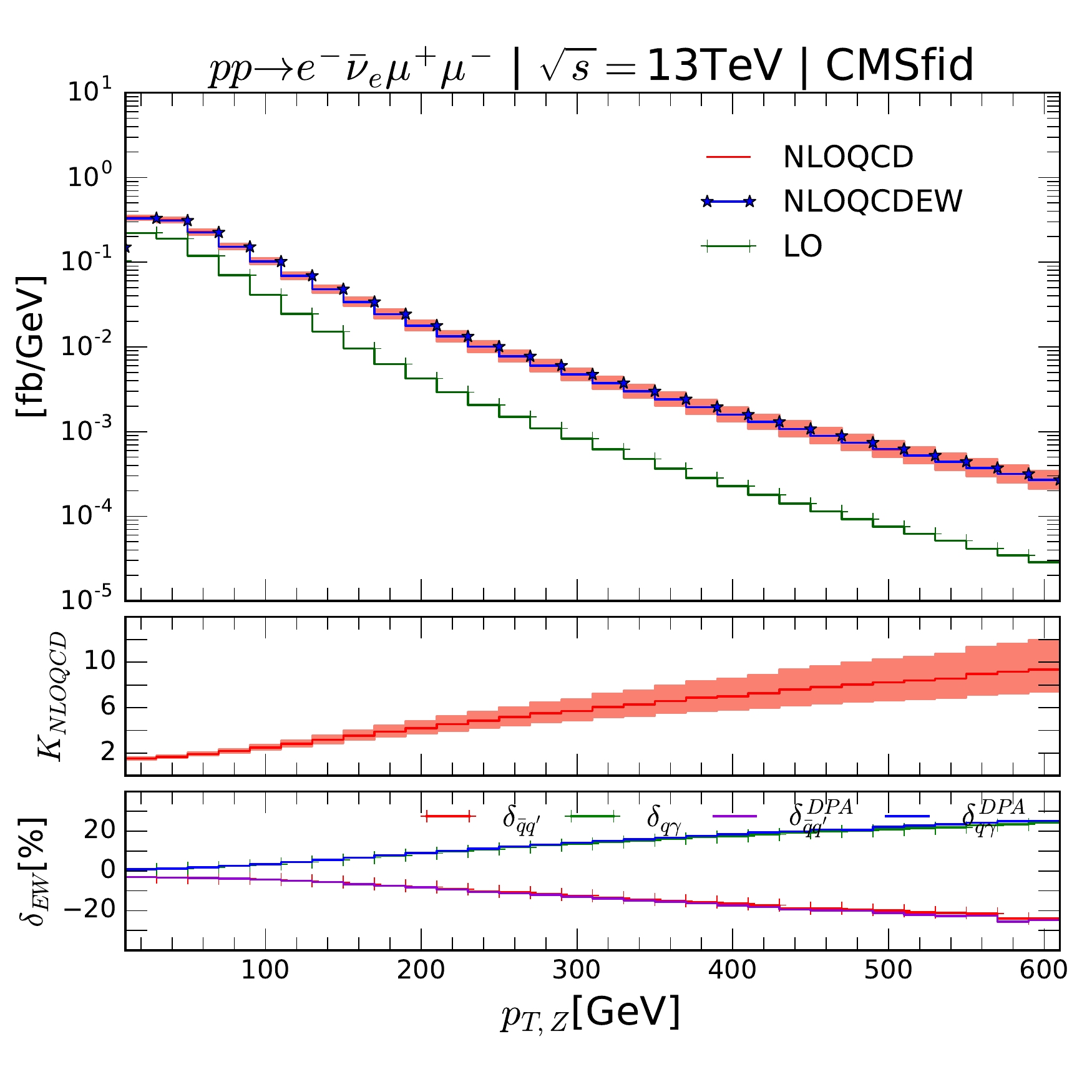}\\
  \includegraphics[width=0.48\textwidth]{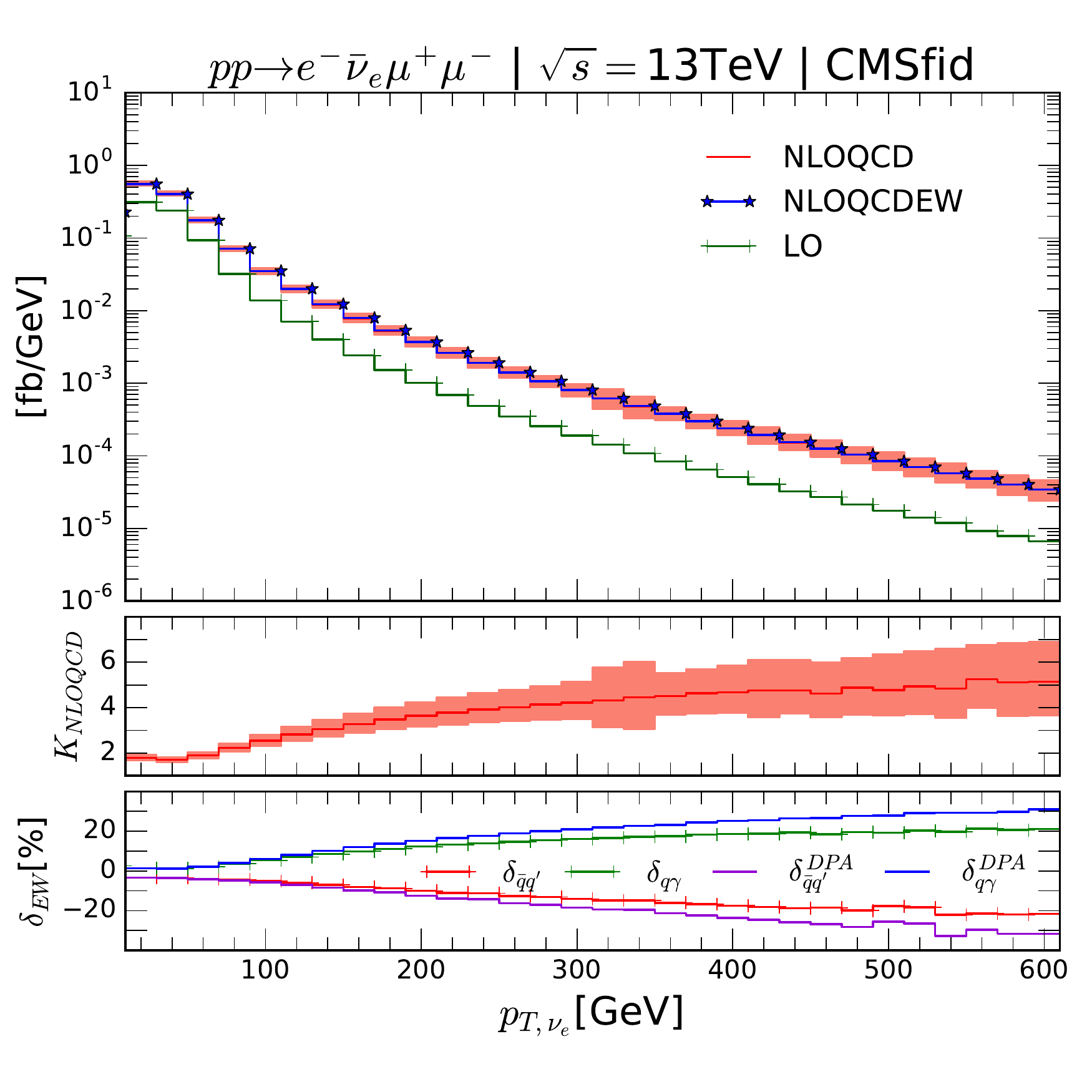}& 
  \includegraphics[width=0.48\textwidth]{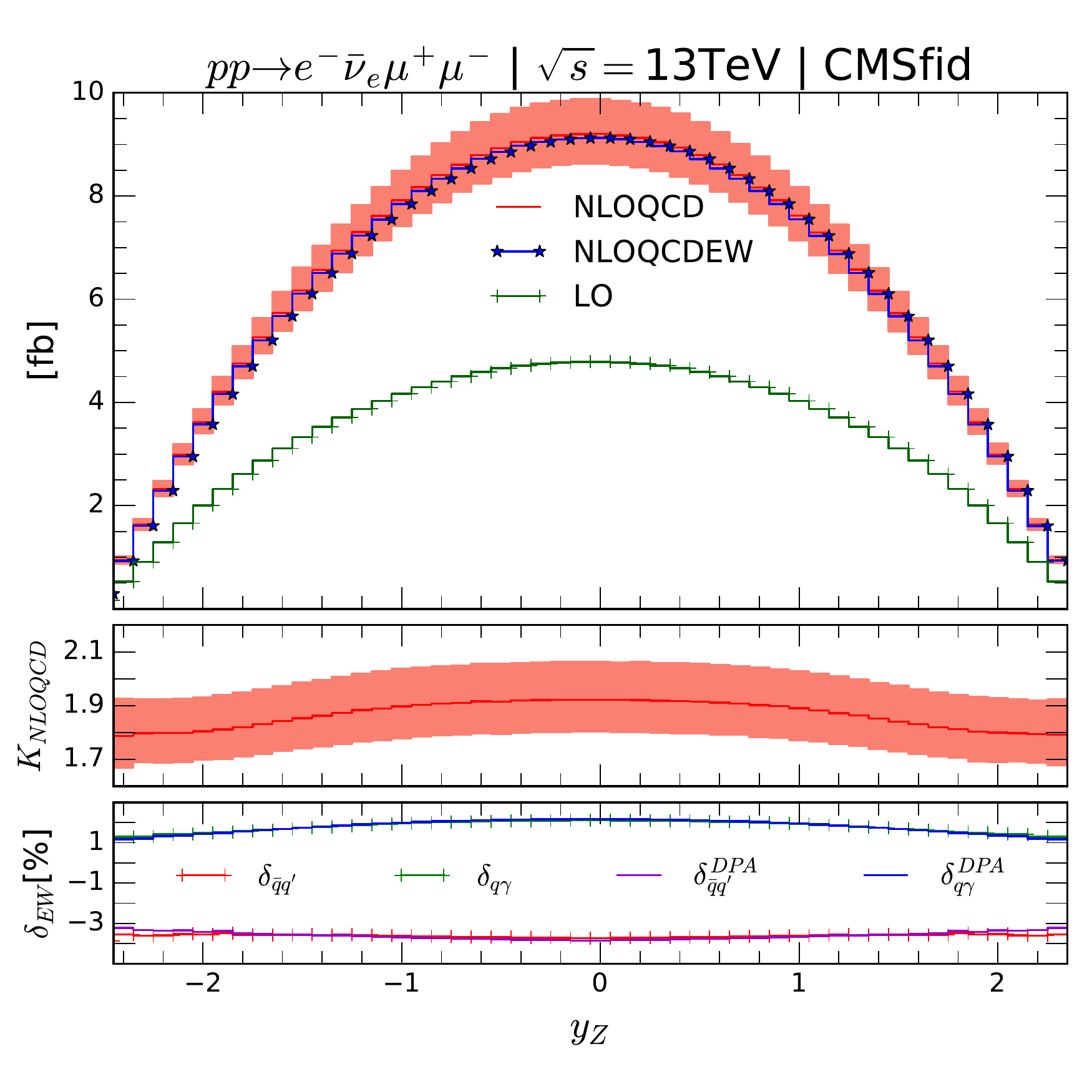}
  \end{tabular}
  \caption{Same as \fig{fig:dist_pT_W_Z_y_W_Z_Wm_atlas} but with the CMS fiducial cuts.}
  \label{fig:dist_pT_W_Z_y_W_Z_Wm_cms}
\end{figure}

\begin{figure}[ht!]
  \centering
  \begin{tabular}{cc}
  \includegraphics[width=0.48\textwidth]{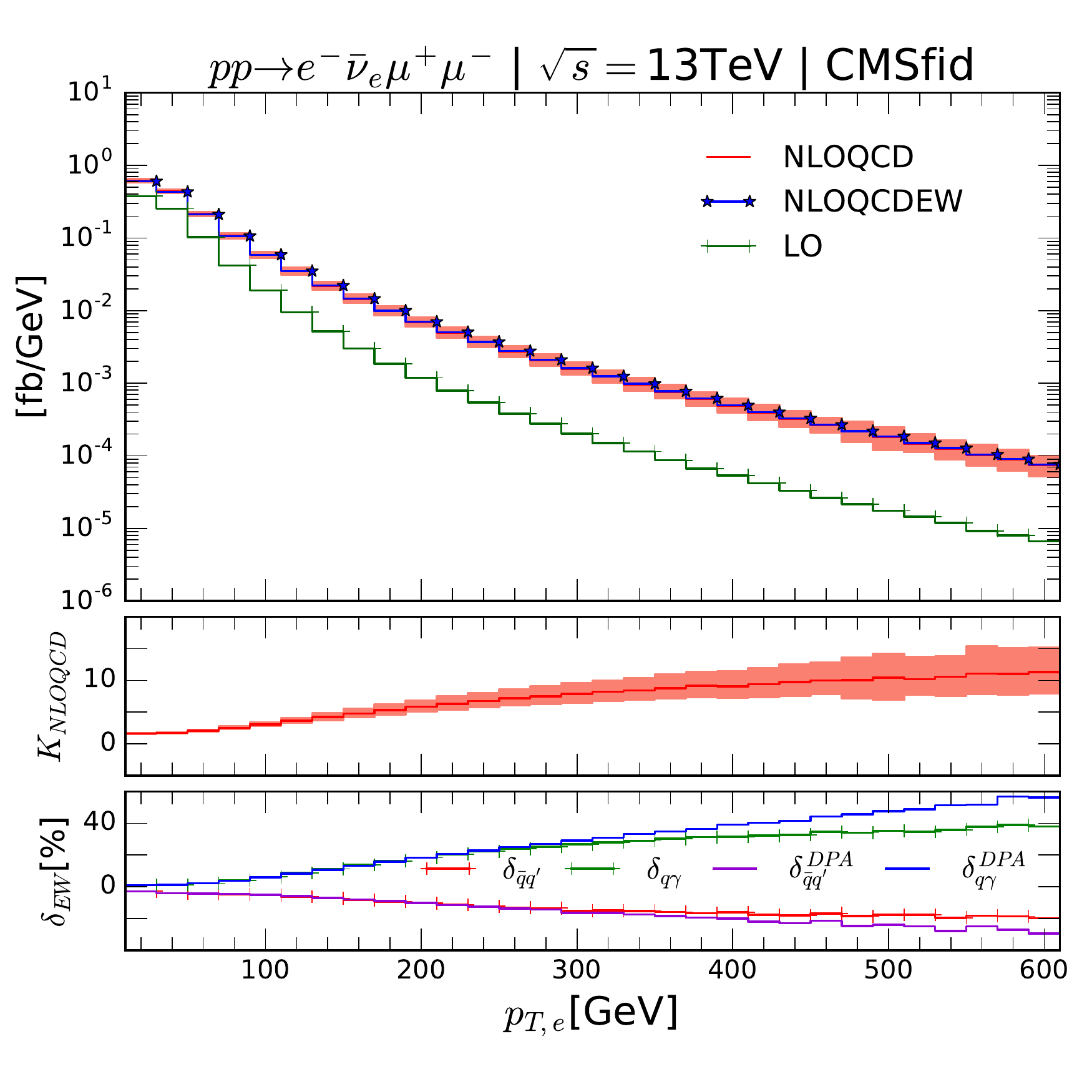}& 
  \includegraphics[width=0.48\textwidth]{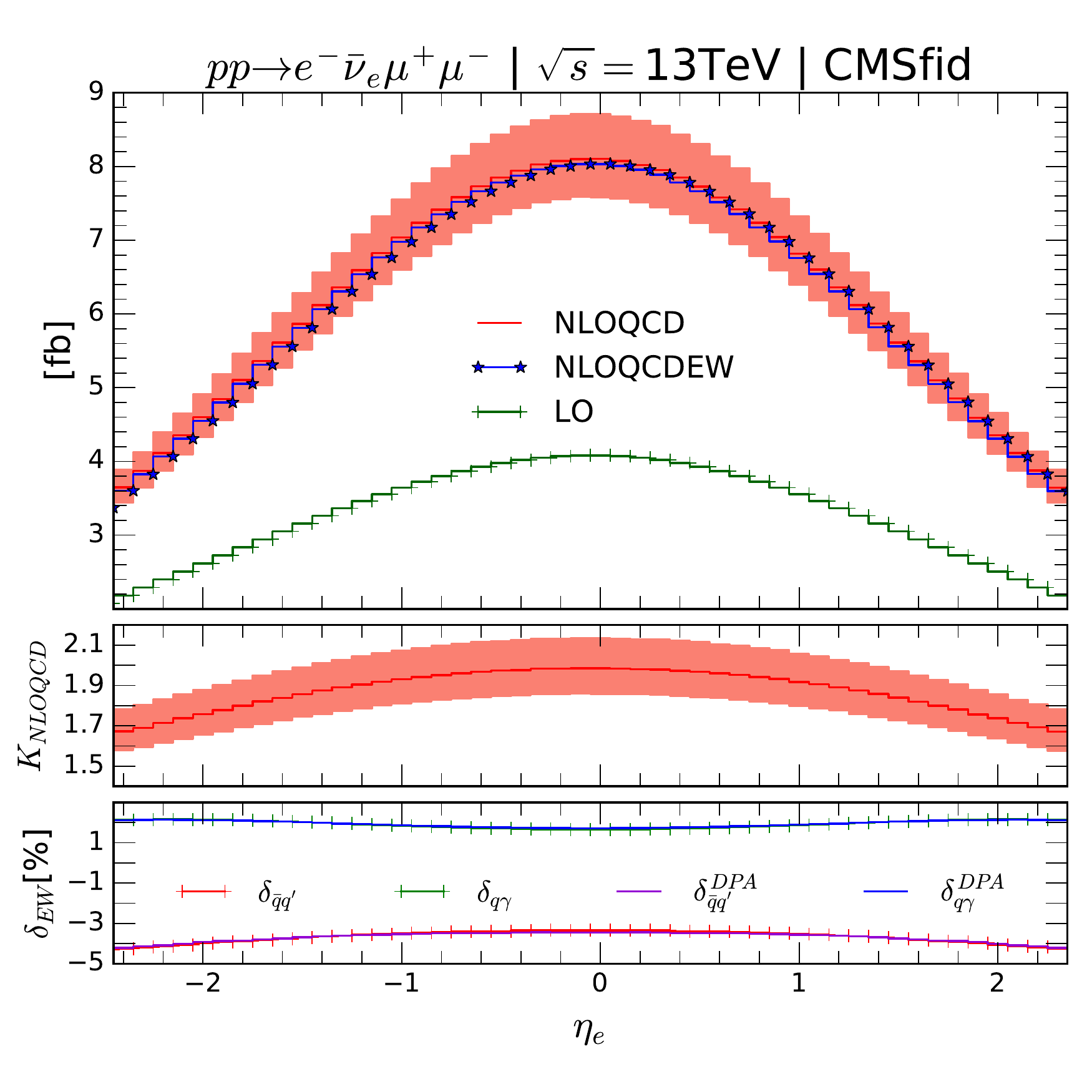}\\
  \includegraphics[width=0.48\textwidth]{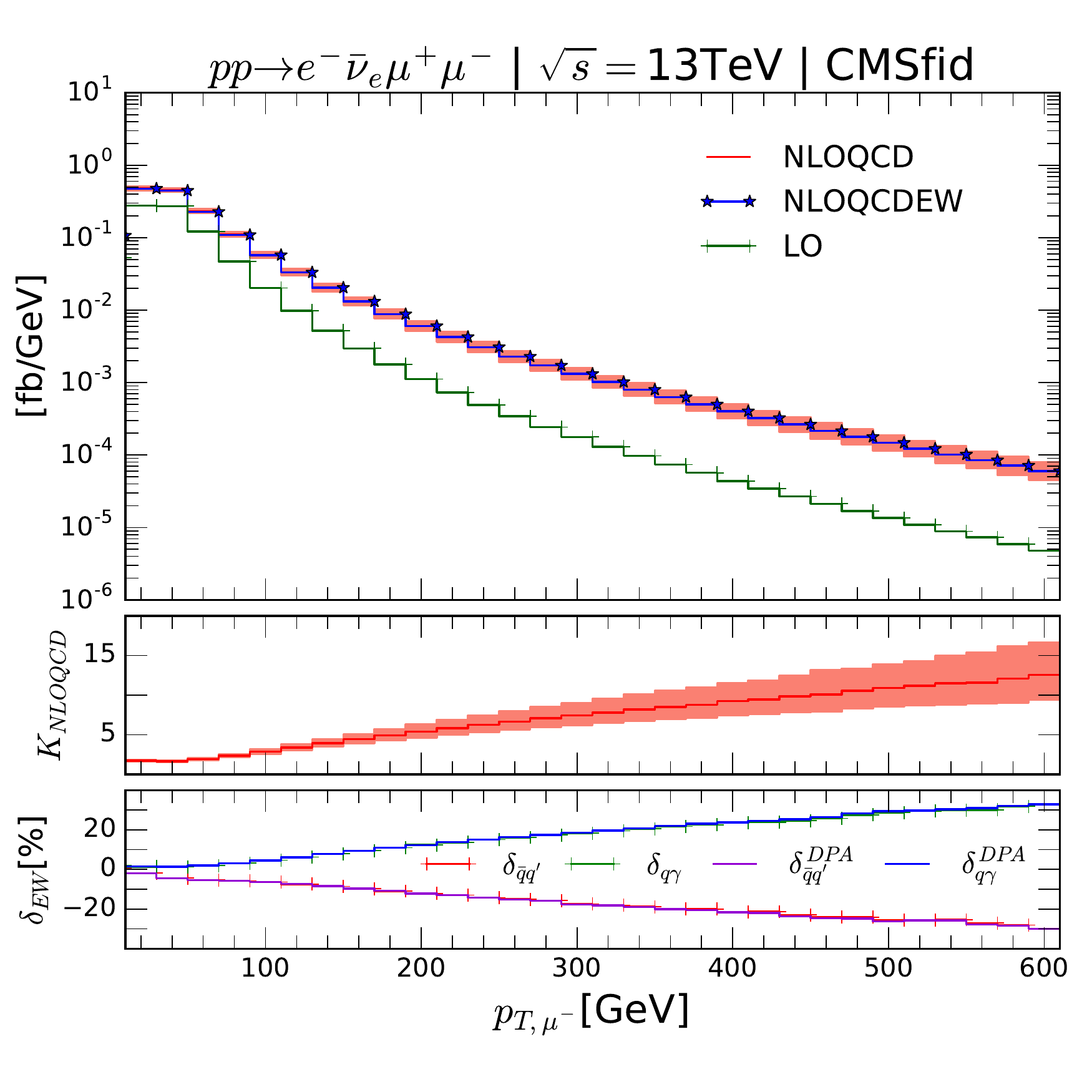}& 
  \includegraphics[width=0.48\textwidth]{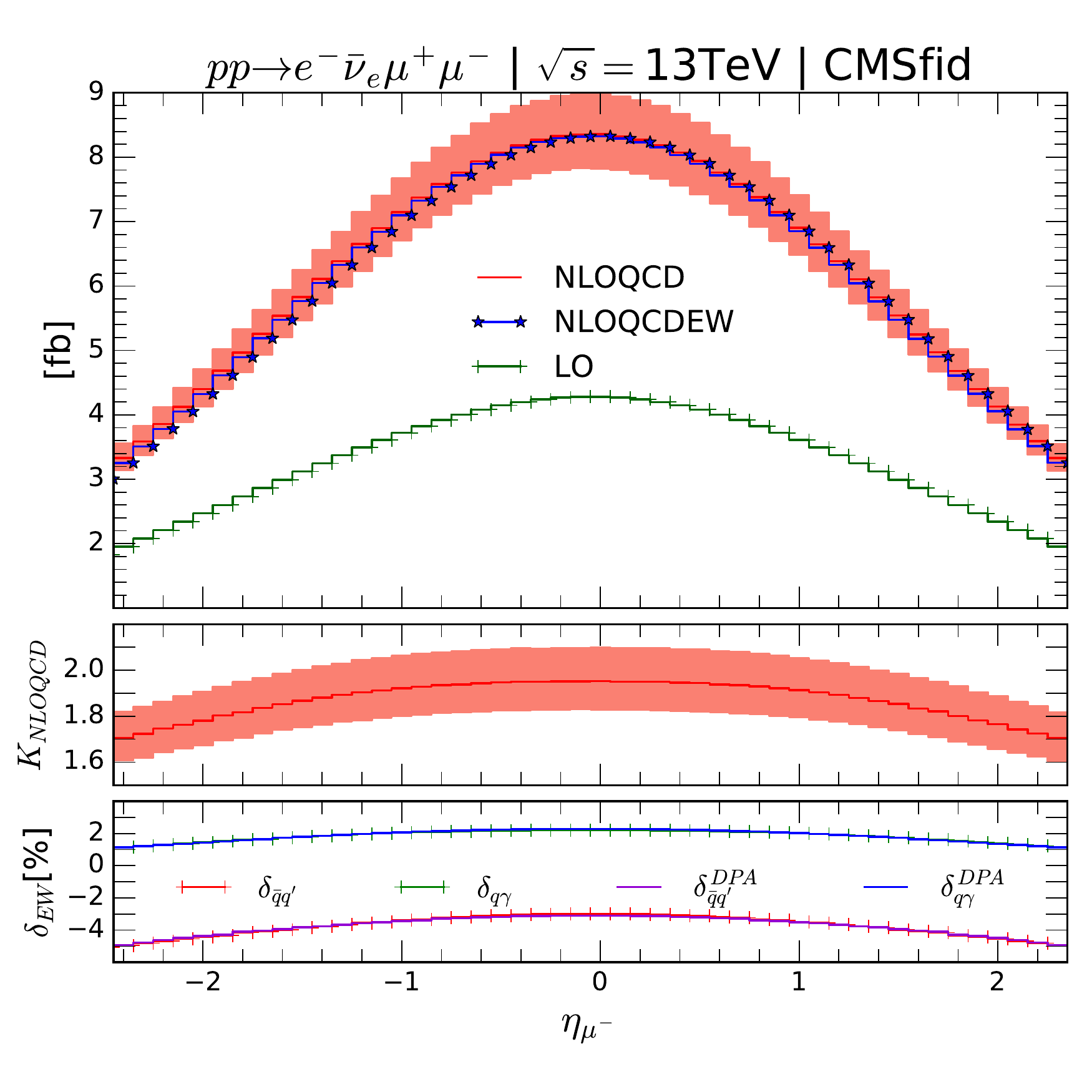}
  \end{tabular}
  \caption{Same as \fig{fig:dist_pT_e_mu_eta_e_mu_Wm_atlas} but with the CMS fiducial cuts.}
  \label{fig:dist_pT_e_mu_eta_e_mu_Wm_cms}
\end{figure}

\begin{figure}[ht!]
  \centering
  \begin{tabular}{cc}
  \includegraphics[width=0.48\textwidth]{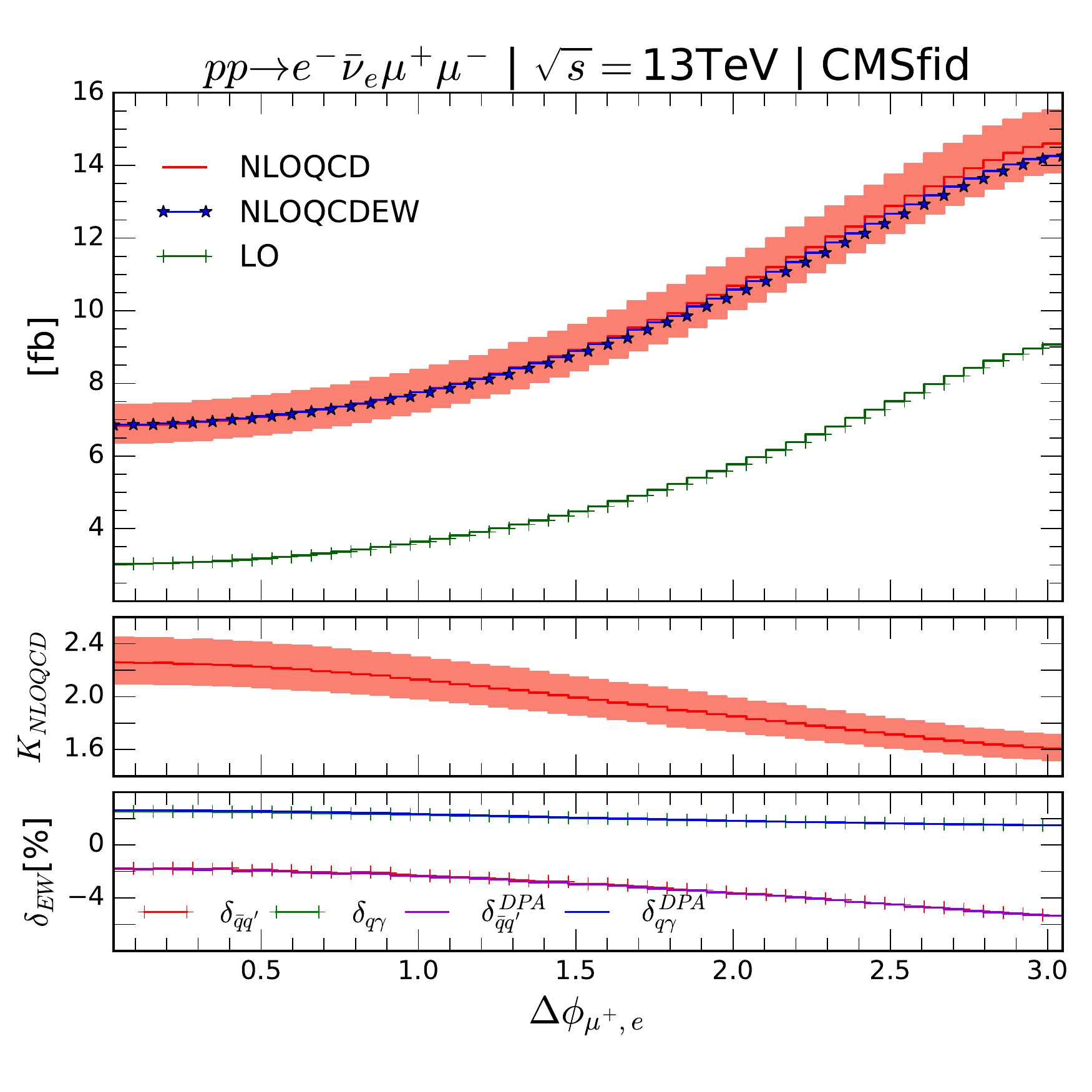}& 
  \includegraphics[width=0.48\textwidth]{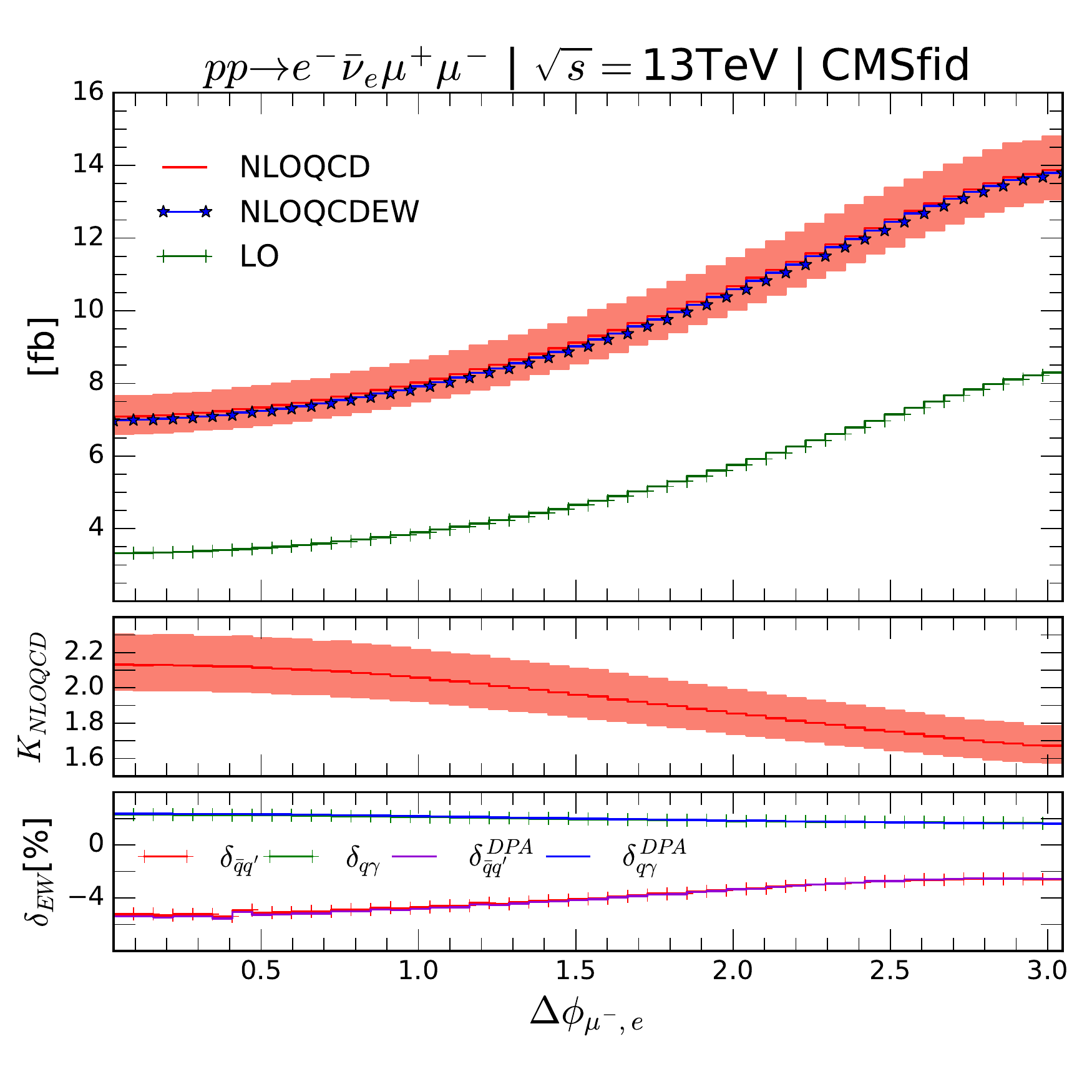}\\
  \includegraphics[width=0.48\textwidth]{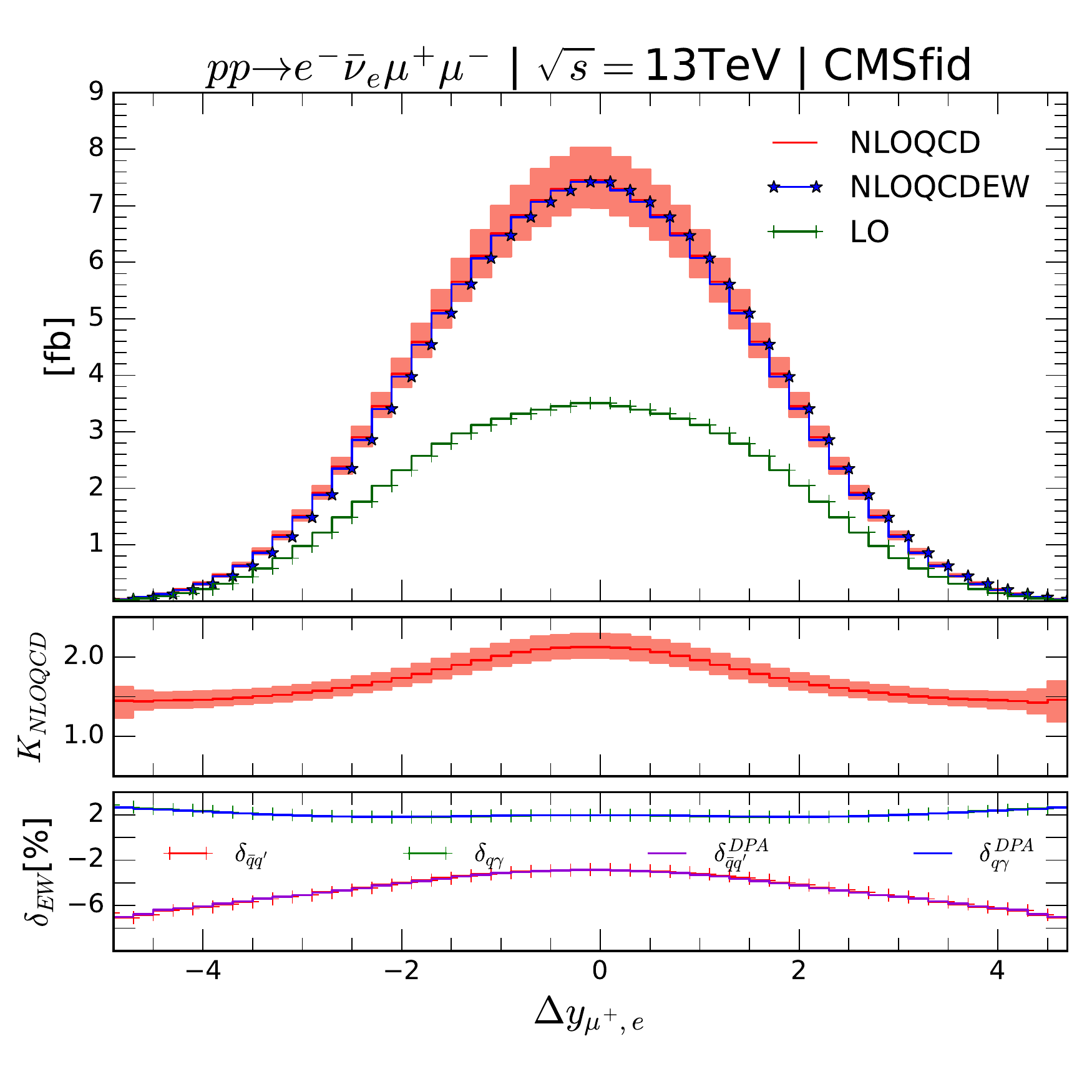}& 
  \includegraphics[width=0.48\textwidth]{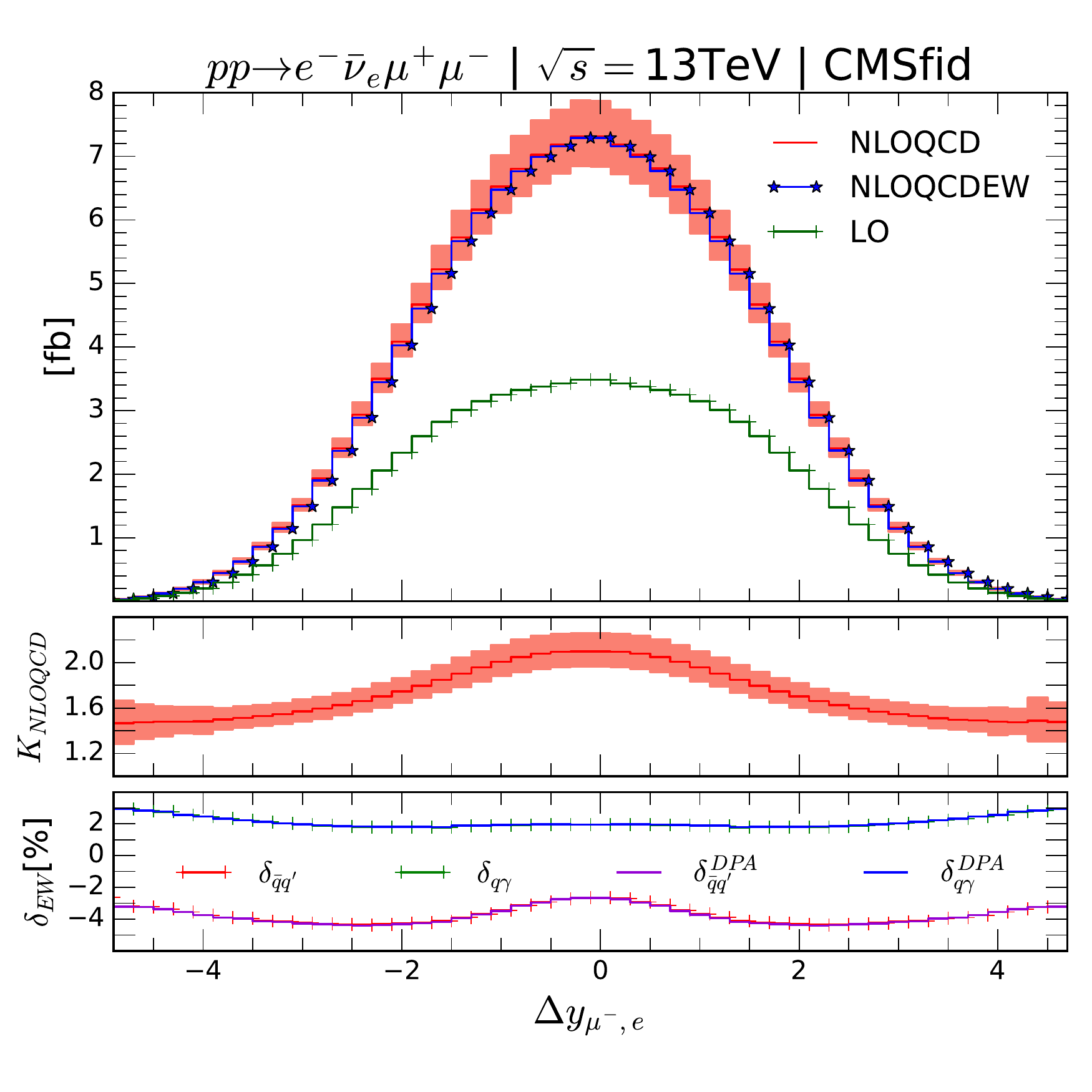}
  \end{tabular}
  \caption{Same as \fig{fig:dist_Delta_phi_y_Wm_atlas} but with the CMS fiducial cuts.}
  \label{fig:dist_Delta_phi_y_Wm_cms}
\end{figure}

\begin{figure}[ht!]
  \centering
  \begin{tabular}{cc}
  \includegraphics[width=0.48\textwidth]{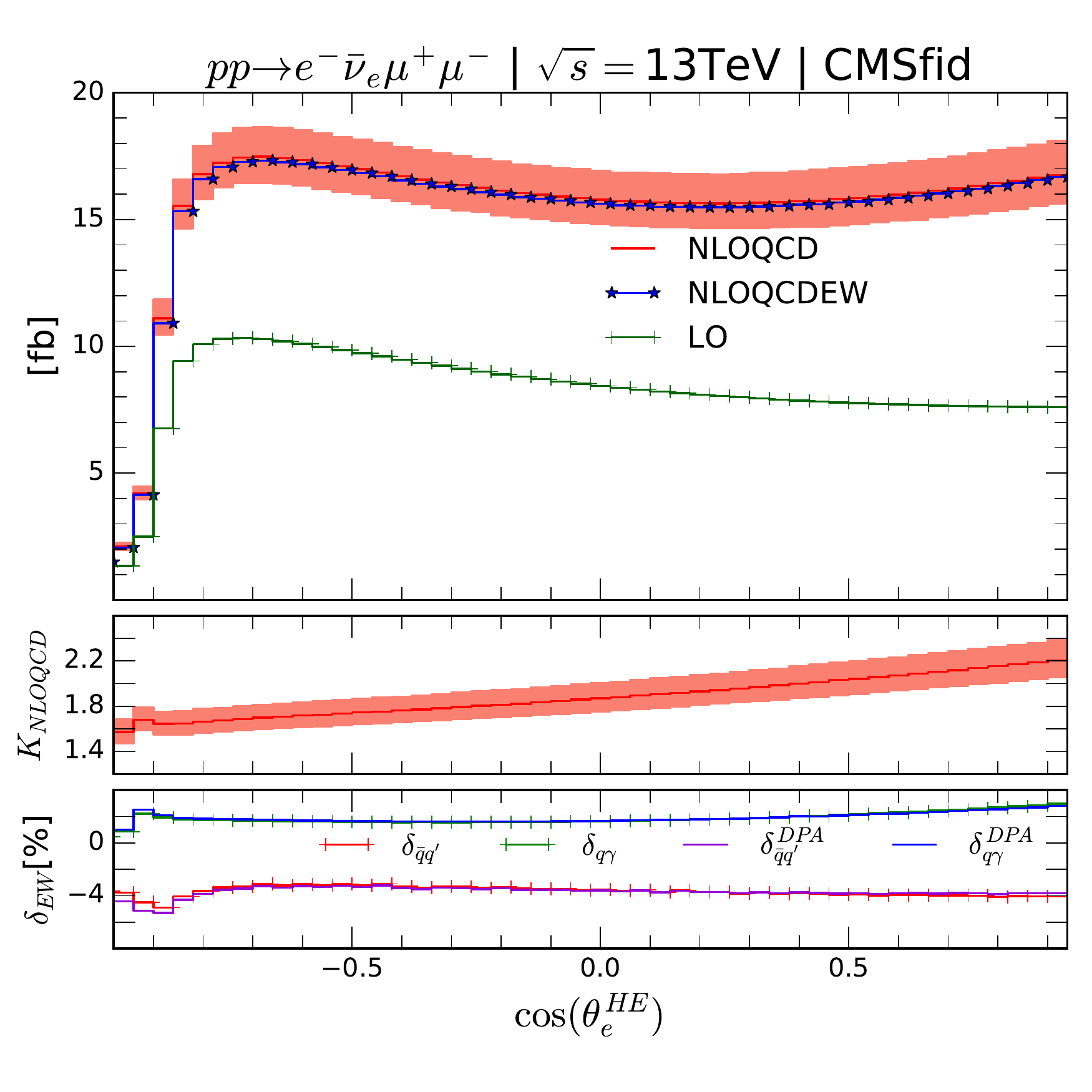}&
  \includegraphics[width=0.48\textwidth]{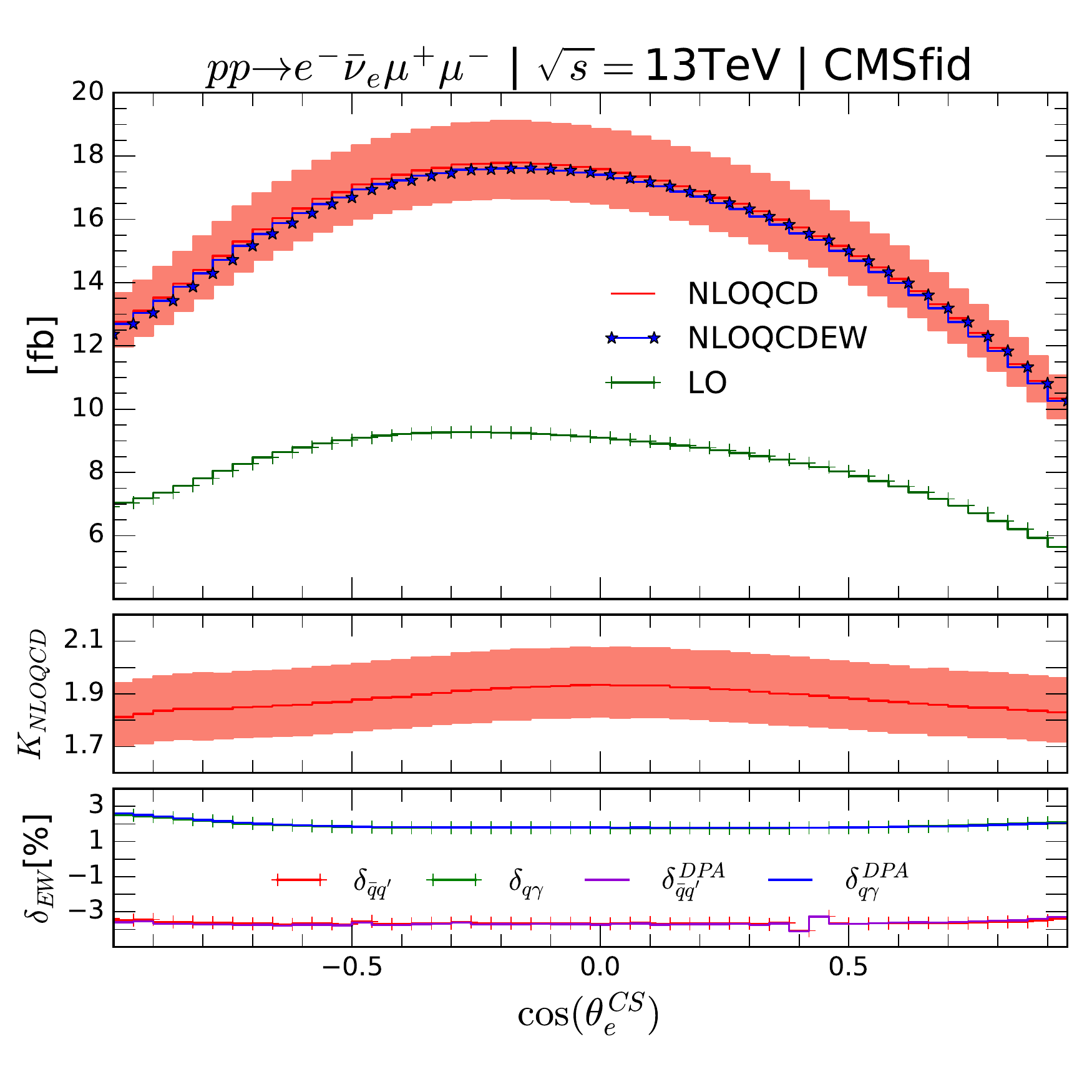}\\
  \includegraphics[width=0.48\textwidth]{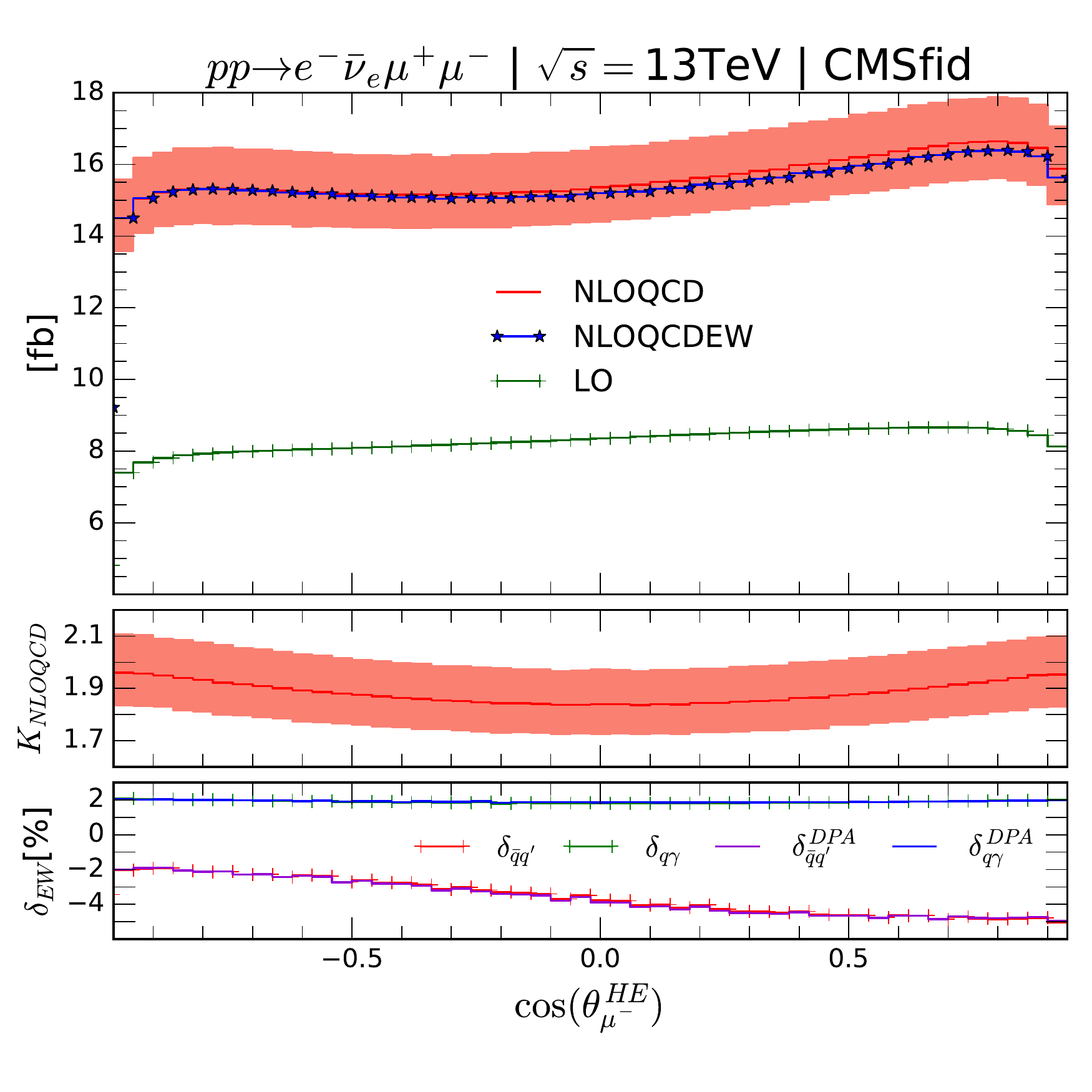}&
  \includegraphics[width=0.48\textwidth]{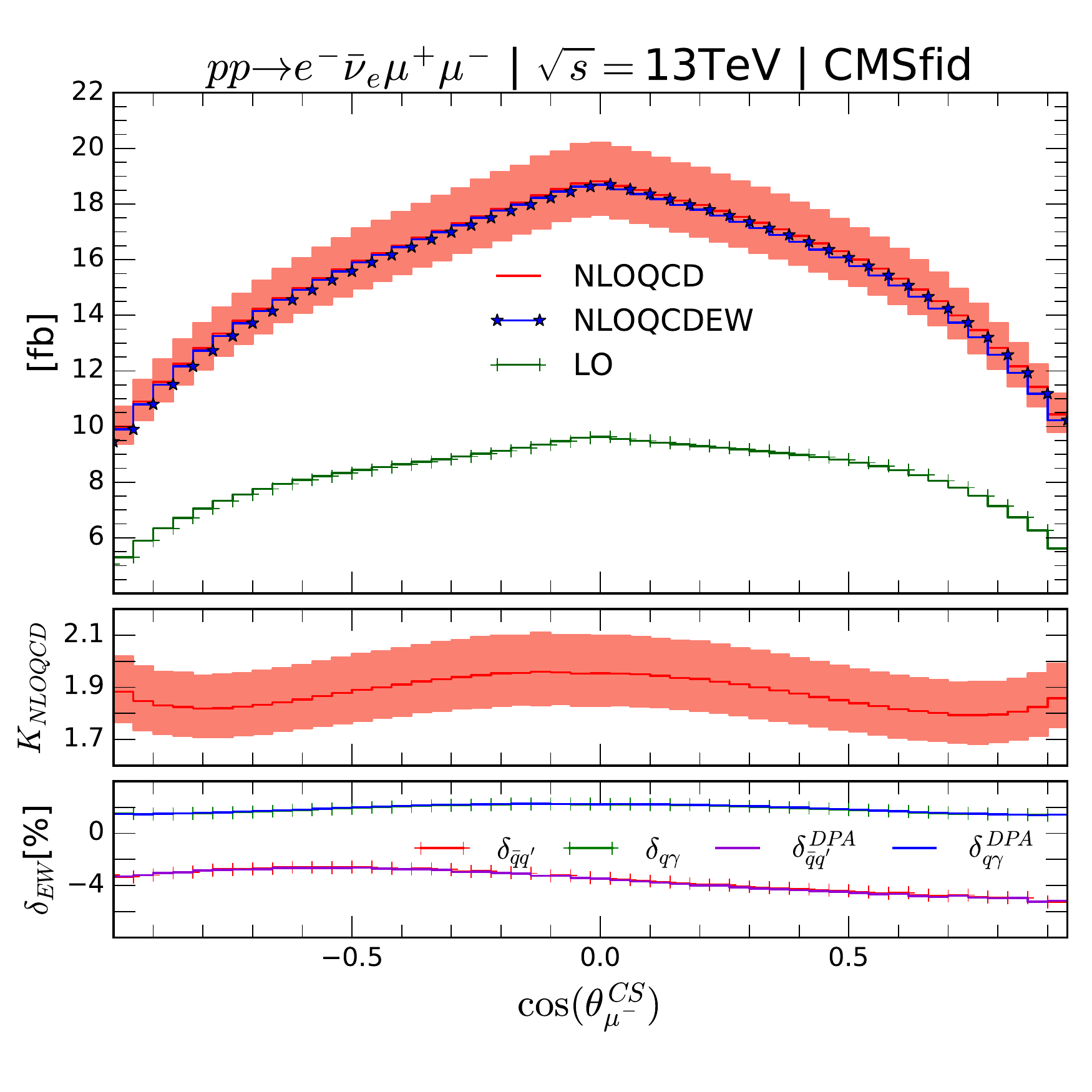}
  \end{tabular}
  \caption{Same as \fig{fig:dist_cos_theta_HEL_CS_e_muon_Wm_atlas} but with the CMS fiducial cuts.}
   \label{fig:dist_cos_theta_HEL_CS_e_muon_Wm_cms}
\end{figure}

\begin{figure}[ht!]
  \centering
  \begin{tabular}{cc}
  \includegraphics[width=0.48\textwidth]{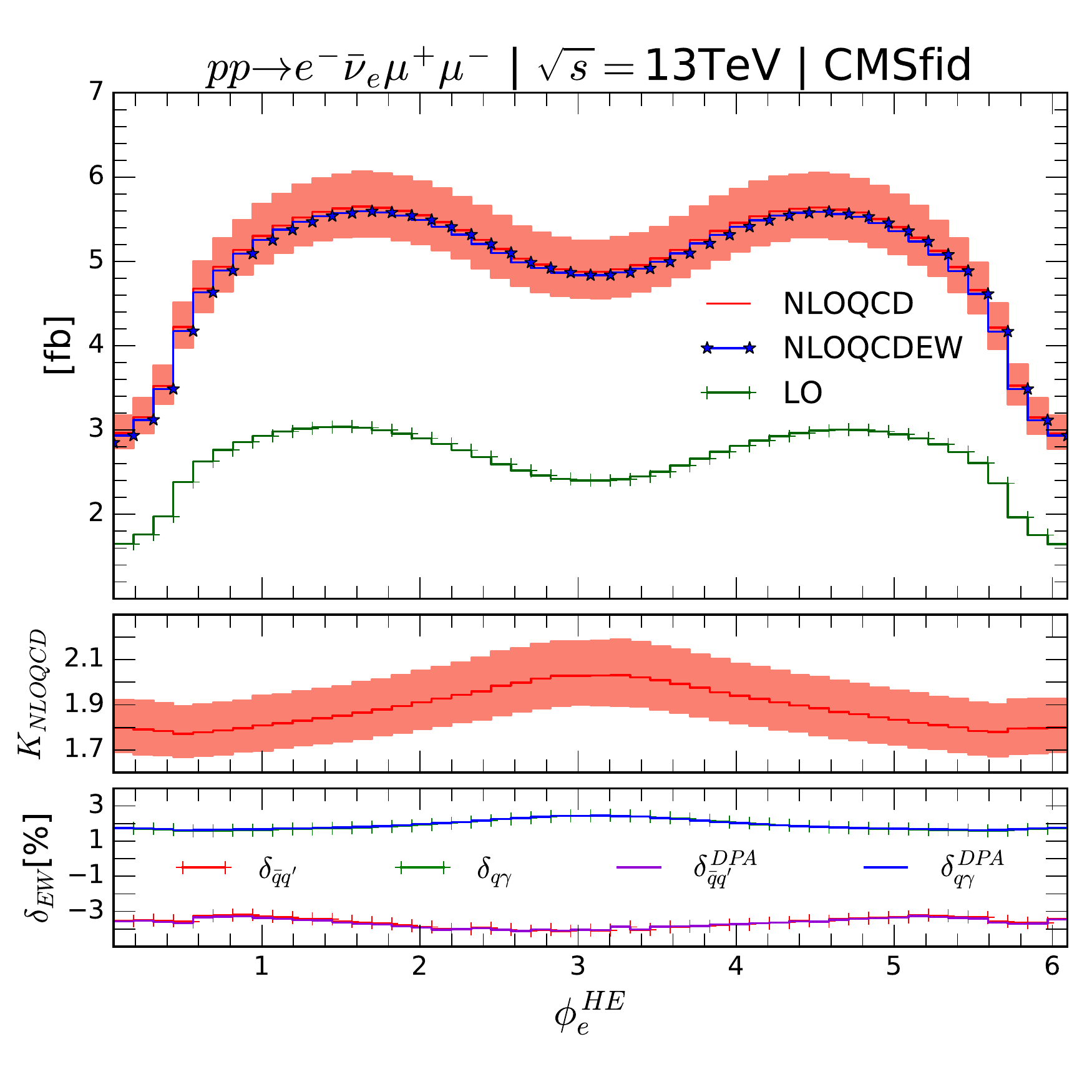}&
  \includegraphics[width=0.48\textwidth]{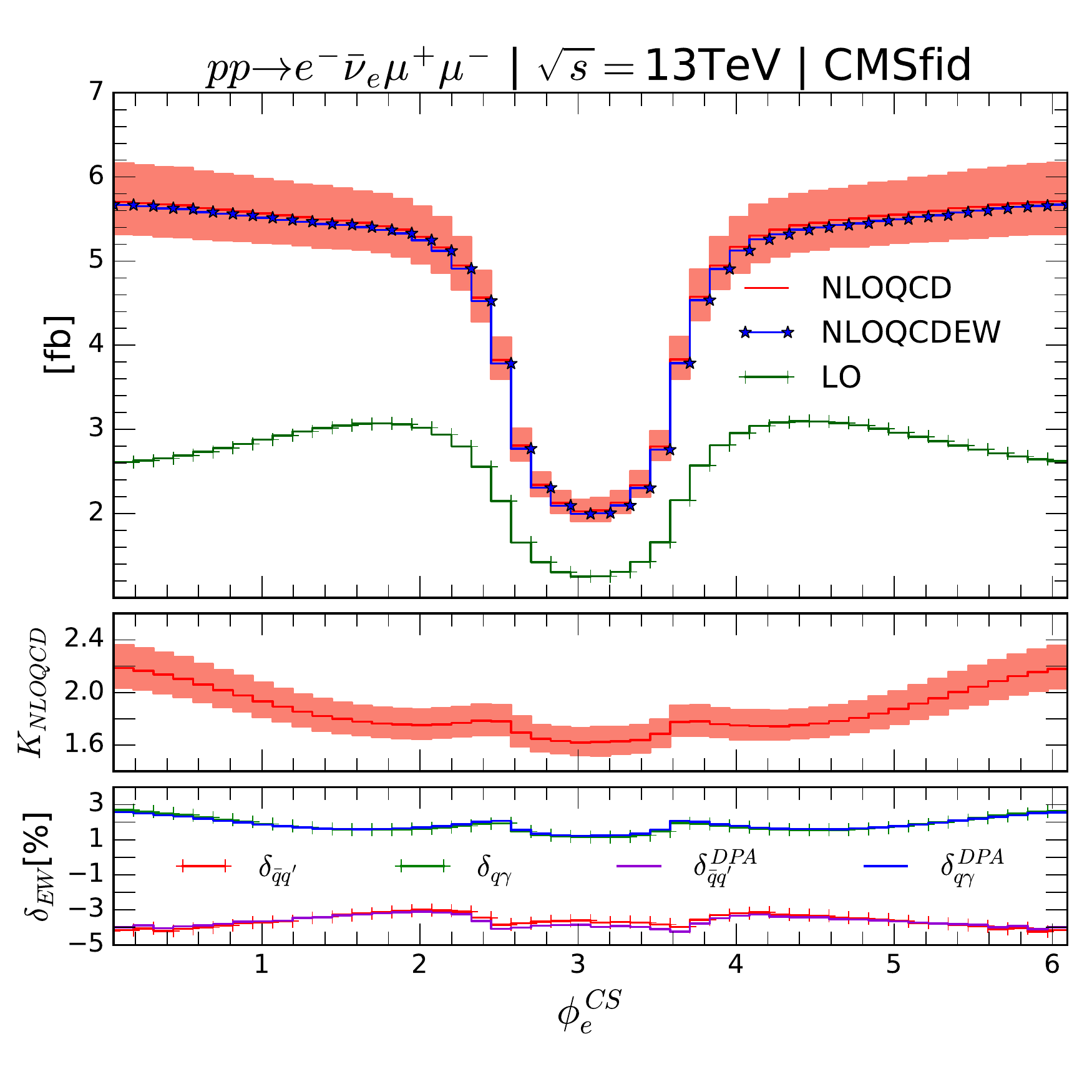}\\
  \includegraphics[width=0.48\textwidth]{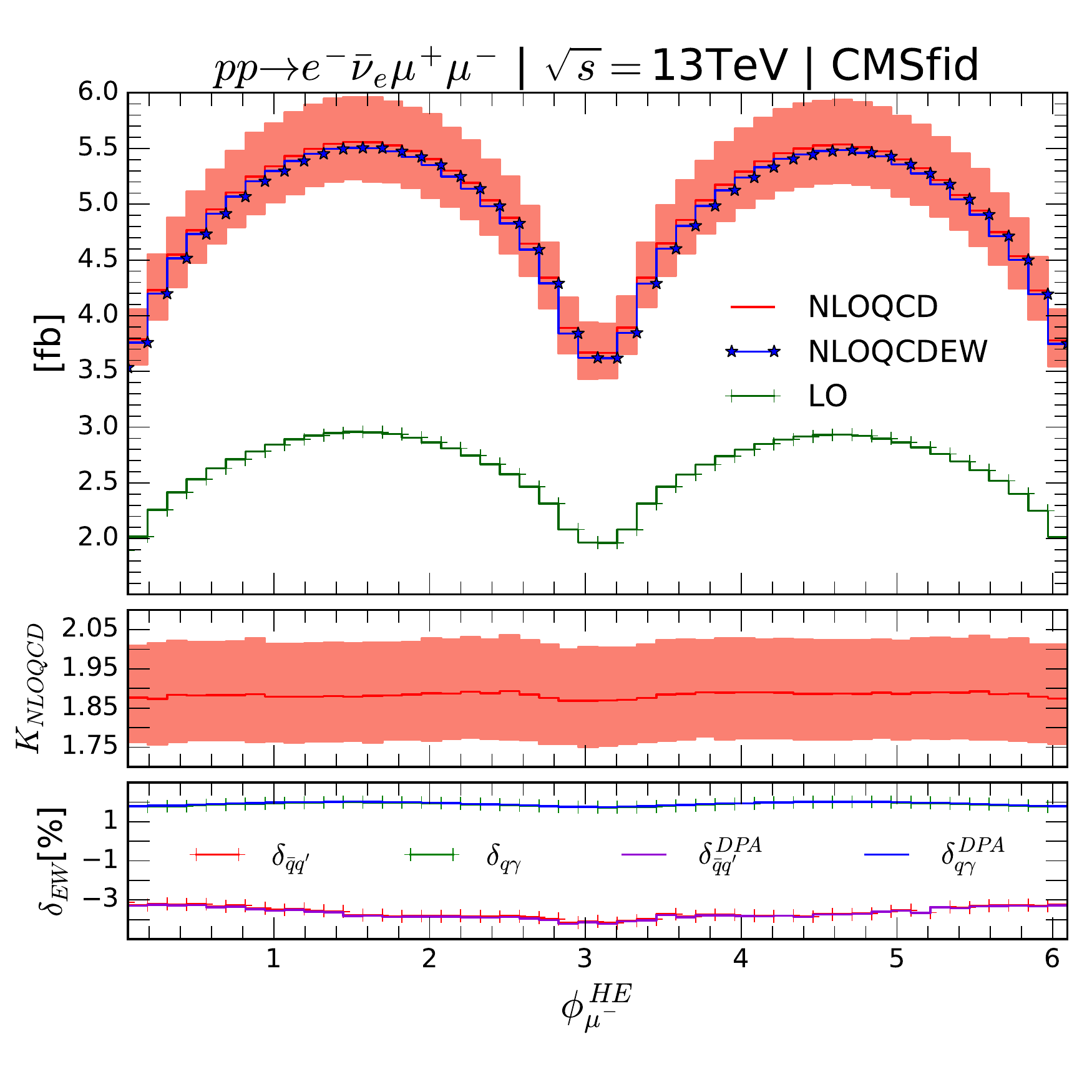}&
  \includegraphics[width=0.48\textwidth]{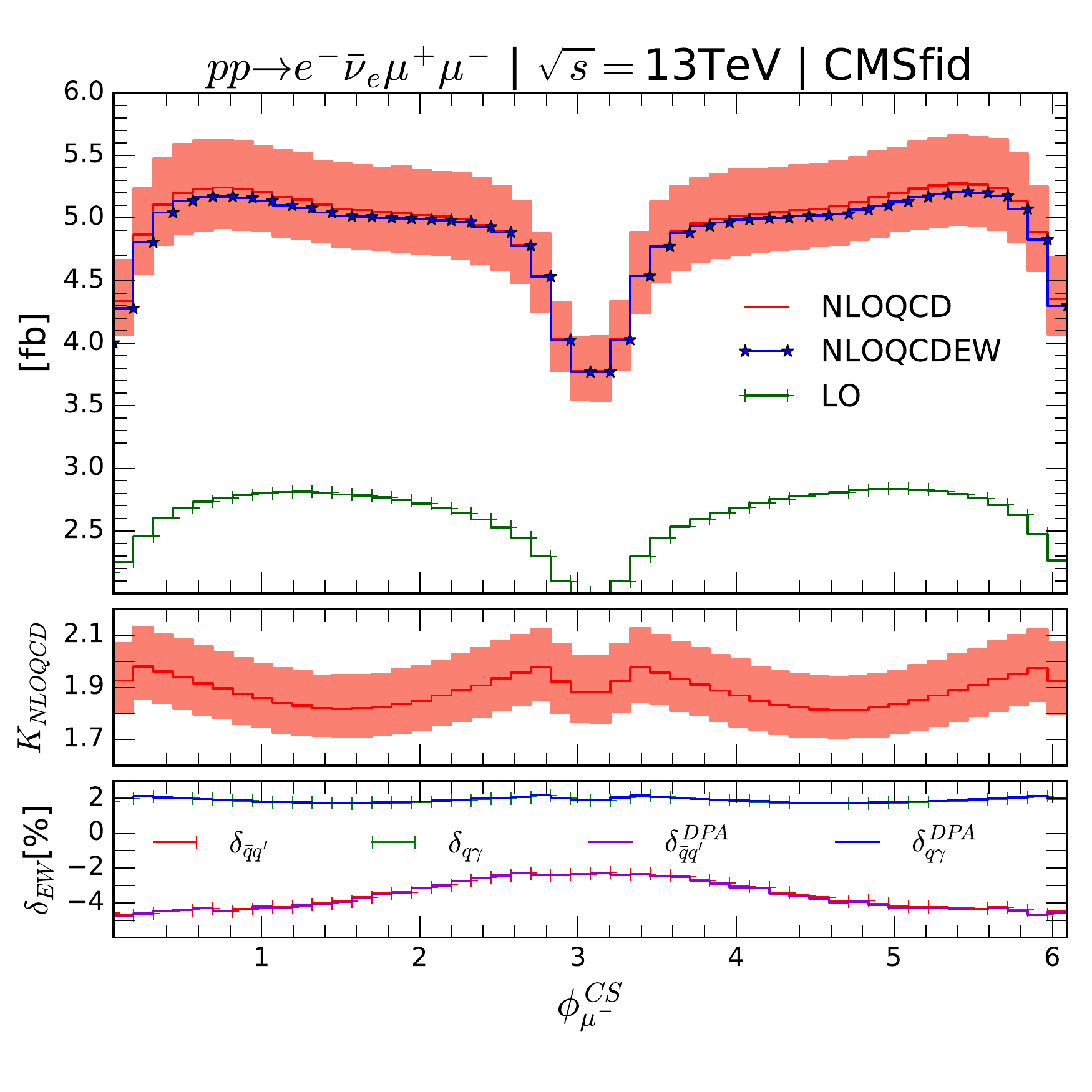}
  \end{tabular}
  \caption{Same as \fig{fig:dist_phi_HEL_CS_e_muon_Wm_atlas} but with the CMS fiducial cuts.}
   \label{fig:dist_phi_HEL_CS_e_muon_Wm_cms}
\end{figure}

We present in this appendix the kinematical distributions for the
process $p p\to e^-\bar{\nu}_e \mu^+\mu^- +X$. They are very similar to
the kinematical distributions of the process $p p\to
e^+\nu_e\mu^+\mu^-+X$ presented in \sect{sect:numset:kinematicdist}, hence we do
not repeat our analysis and just display the plots. Plots for the ATLAS
fiducial cuts are displayed in \fig{fig:dist_pT_W_Z_y_W_Z_Wm_atlas},
\fig{fig:dist_pT_e_mu_eta_e_mu_Wm_atlas},
\fig{fig:dist_Delta_phi_y_Wm_atlas},
\fig{fig:dist_cos_theta_HEL_CS_e_muon_Wm_atlas}, and
\fig{fig:dist_phi_HEL_CS_e_muon_Wm_atlas}. Plots for the CMS fiducial cuts
are presented in \fig{fig:dist_pT_W_Z_y_W_Z_Wm_cms},
\fig{fig:dist_pT_e_mu_eta_e_mu_Wm_cms},
\fig{fig:dist_Delta_phi_y_Wm_cms},
\fig{fig:dist_cos_theta_HEL_CS_e_muon_Wm_cms}, and
\fig{fig:dist_phi_HEL_CS_e_muon_Wm_cms}.


\clearpage
\pagebreak

\section{Numerical results for fiducial polarization observables: \boldmath
  $W^-Z$ channel}
\label{appen:numres_Wm}

We present in this appendix the numerical results for the fiducial polarization
observables in the process $pp \to e^-\bar{\nu}_e\mu^+\mu^-+X$. The
analysis is identical to that of the process $pp \to
e^+\nu_e\mu^+\mu^-+X$ carried out in \sect{sect:numres}, hence we do
not repeat ourselves and just present the numbers for the angular
coefficients in \tab{tab:coeff_Ai_Wm_ATLAS},
\tab{tab:coeff_Ai_Z_2_ATLAS}, \tab{tab:coeff_Ai_Wm_CMS}, and
\tab{tab:coeff_Ai_Z_2_CMS}; and for the polarization fractions in
\tab{tab:coeff_fL0R_WmZ_ATLAS} and \tab{tab:coeff_fL0R_WmZ_CMS}.

\begin{figure}[hb!]
  \centering
  \begin{tabular}{cc}
  \includegraphics[width=0.48\textwidth]{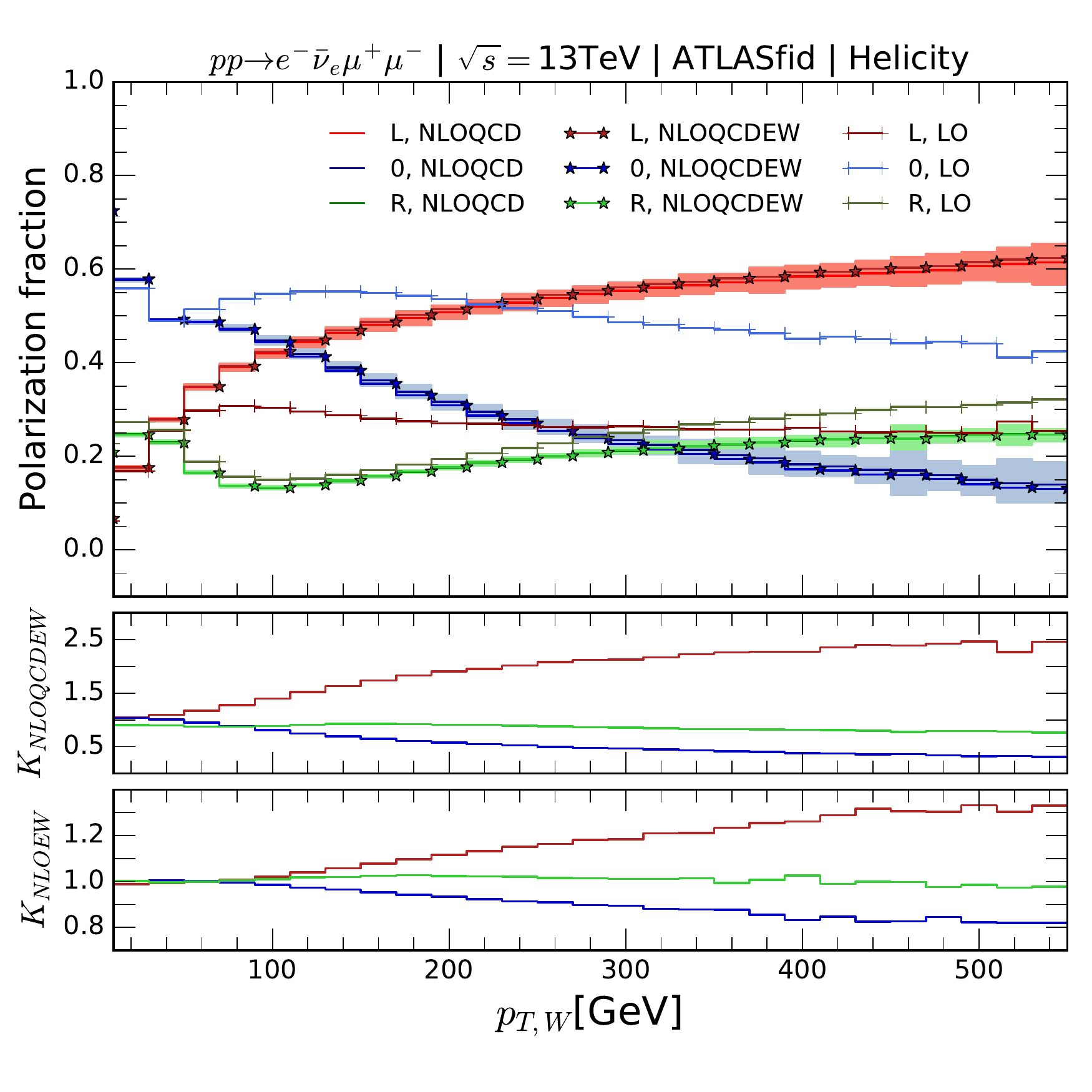}& 
  \includegraphics[width=0.48\textwidth]{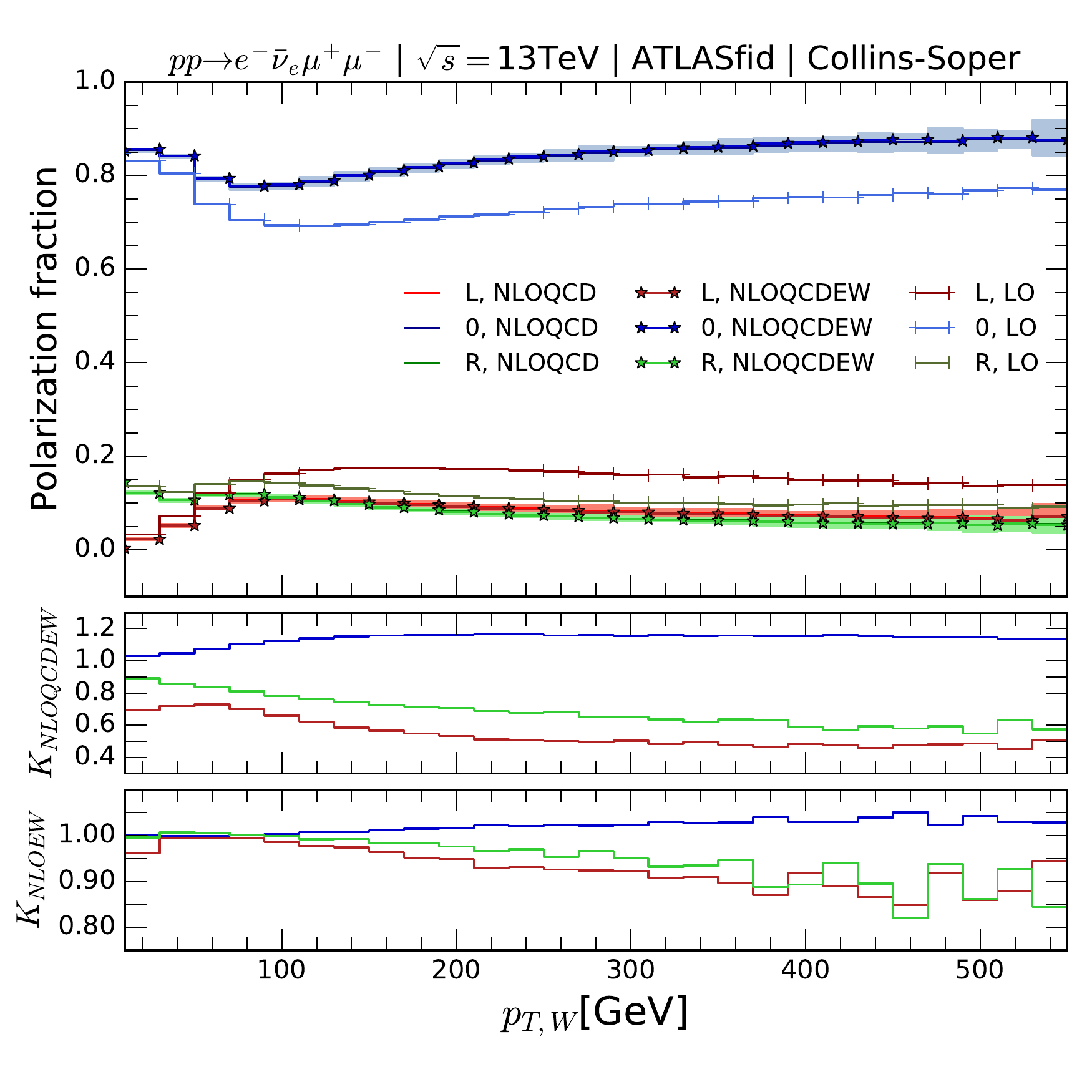}
  \end{tabular}
  \caption{Transverse momentum distributions of the $W^-$ boson fiducial polarization fractions
    for the process $pp \to e^- \bar{\nu}_e\, \mu^+ \mu^- + X$ at the 13 TeV LHC with the ATLAS fiducial cuts. 
    The left-hand-side
    plot is for the HE coordinate system, while the right-hand-side
    plot is for the CS coordinate system. The bands include PDF and
    scale uncertainties calculated at NLOQCD.}
  \label{fig:dist_ptWm_ATLAS_NLOQCDEW}
\end{figure}

\begin{figure}[hb!]
  \centering
 \begin{tabular}{cc}
  \includegraphics[width=0.48\textwidth]{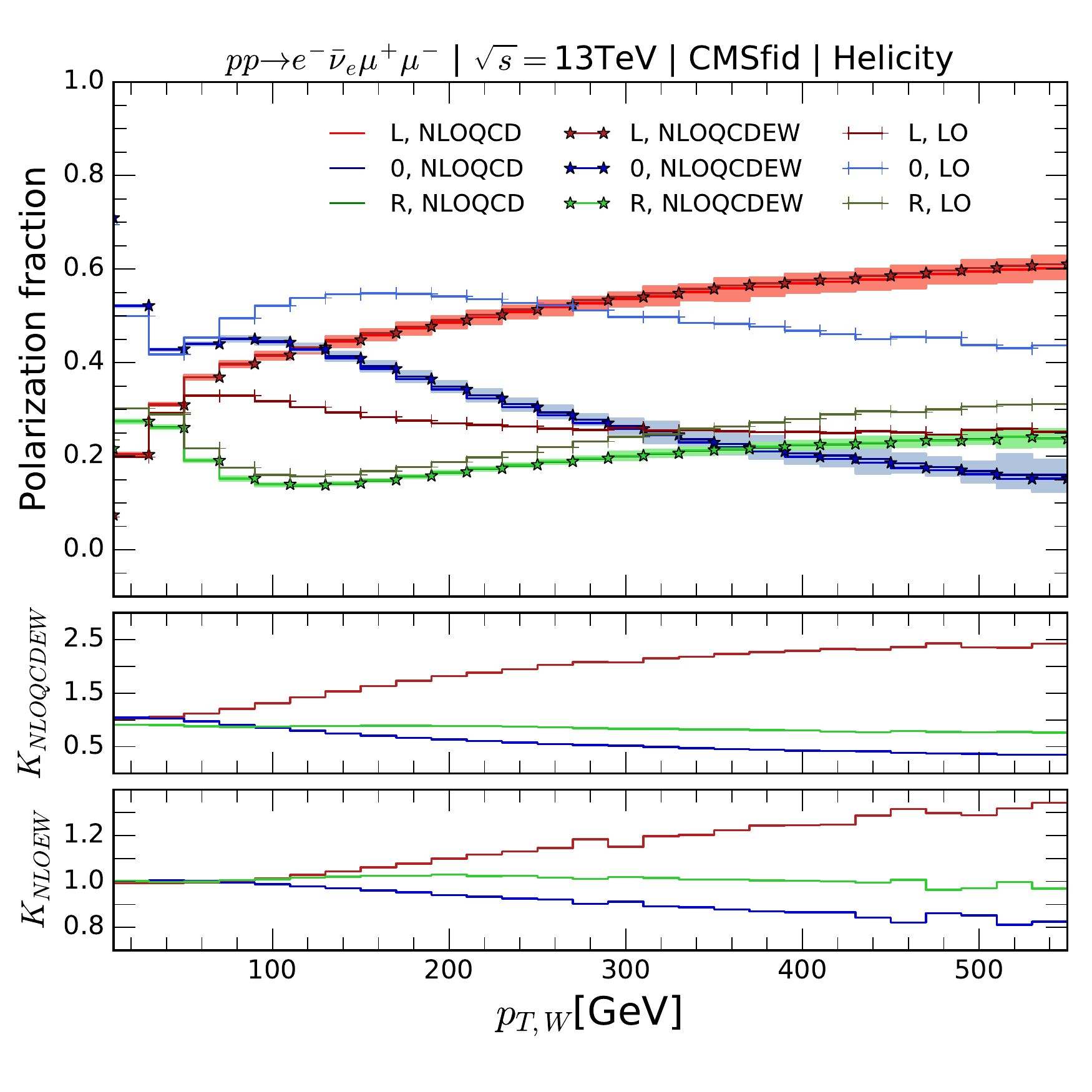}& 
  \includegraphics[width=0.48\textwidth]{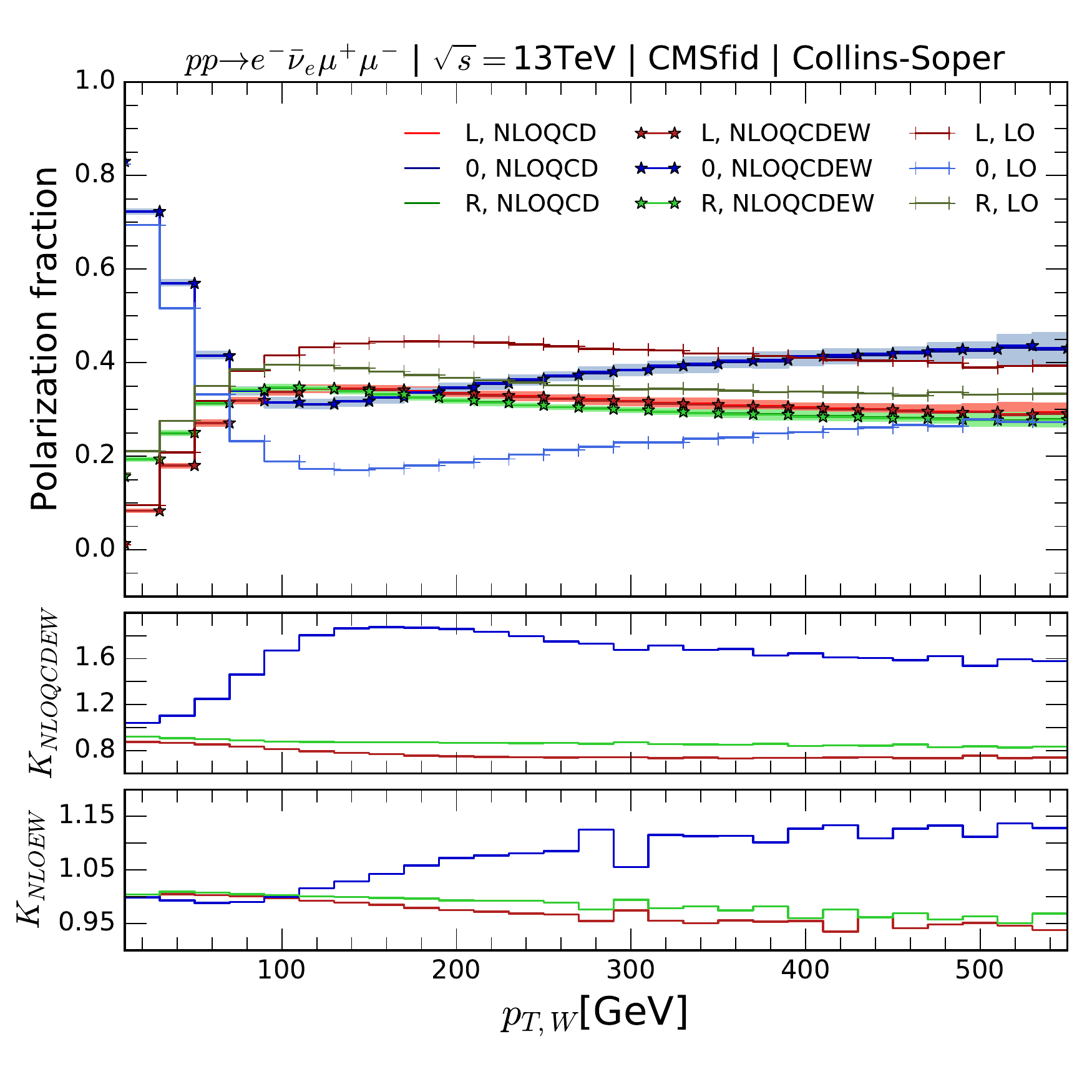}
  \end{tabular}
  \caption{Same as \fig{fig:dist_ptWm_ATLAS_NLOQCDEW} but with the CMS fiducial cuts.}
  \label{fig:dist_ptWm_CMS_NLOQCDEW}
\end{figure}

\begin{figure}[hb!]
  \centering
 \begin{tabular}{cc}
  \includegraphics[width=0.48\textwidth]{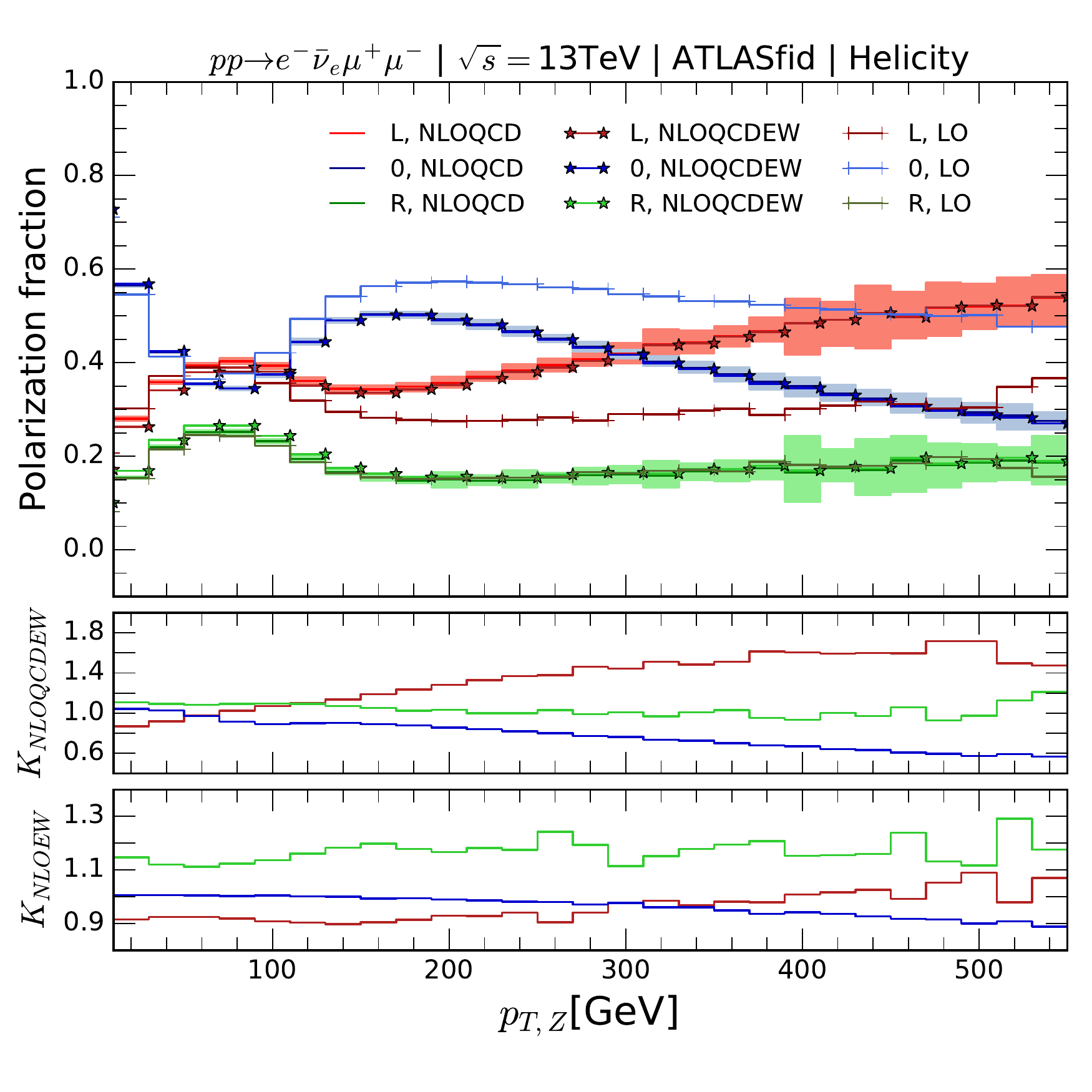}& 
  \includegraphics[width=0.48\textwidth]{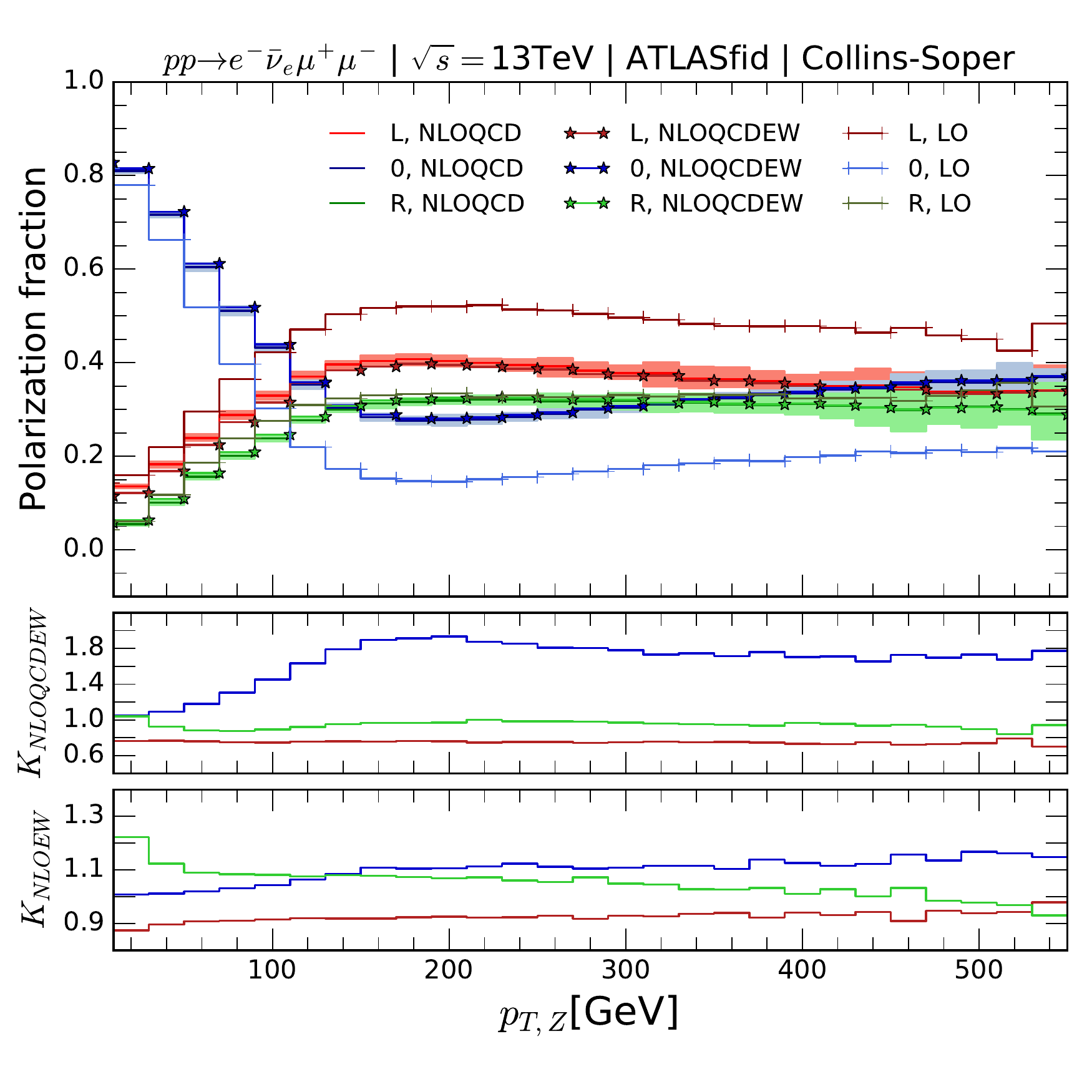}
  \end{tabular}
  \caption{Same as \fig{fig:dist_ptWm_ATLAS_NLOQCDEW} but for the $Z$ boson.}
  \label{fig:dist_pTZm_ATLAS_NLOQCDEW}
\end{figure}

\begin{figure}[hb!]
  \centering
 \begin{tabular}{cc}
  \includegraphics[width=0.48\textwidth]{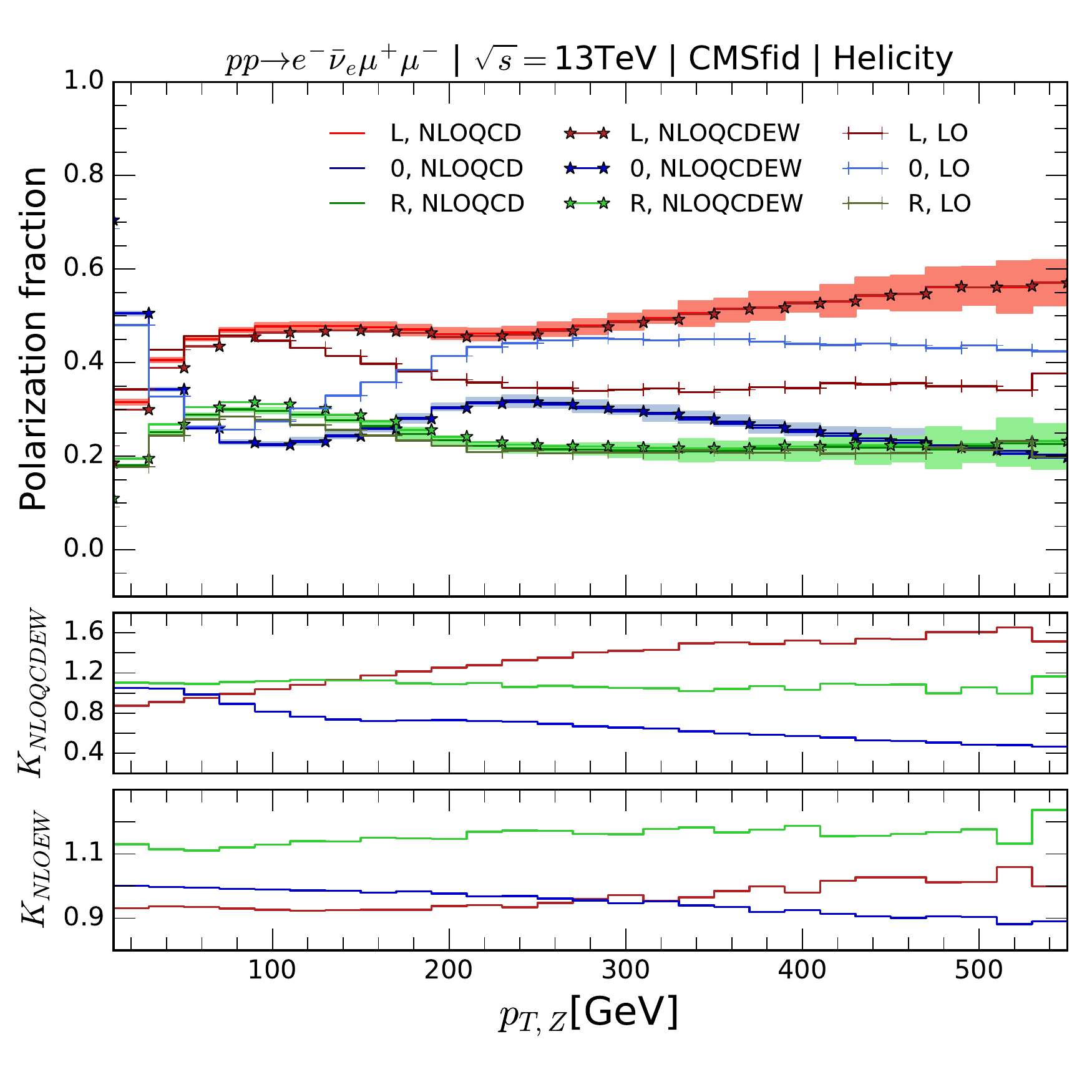}& 
  \includegraphics[width=0.48\textwidth]{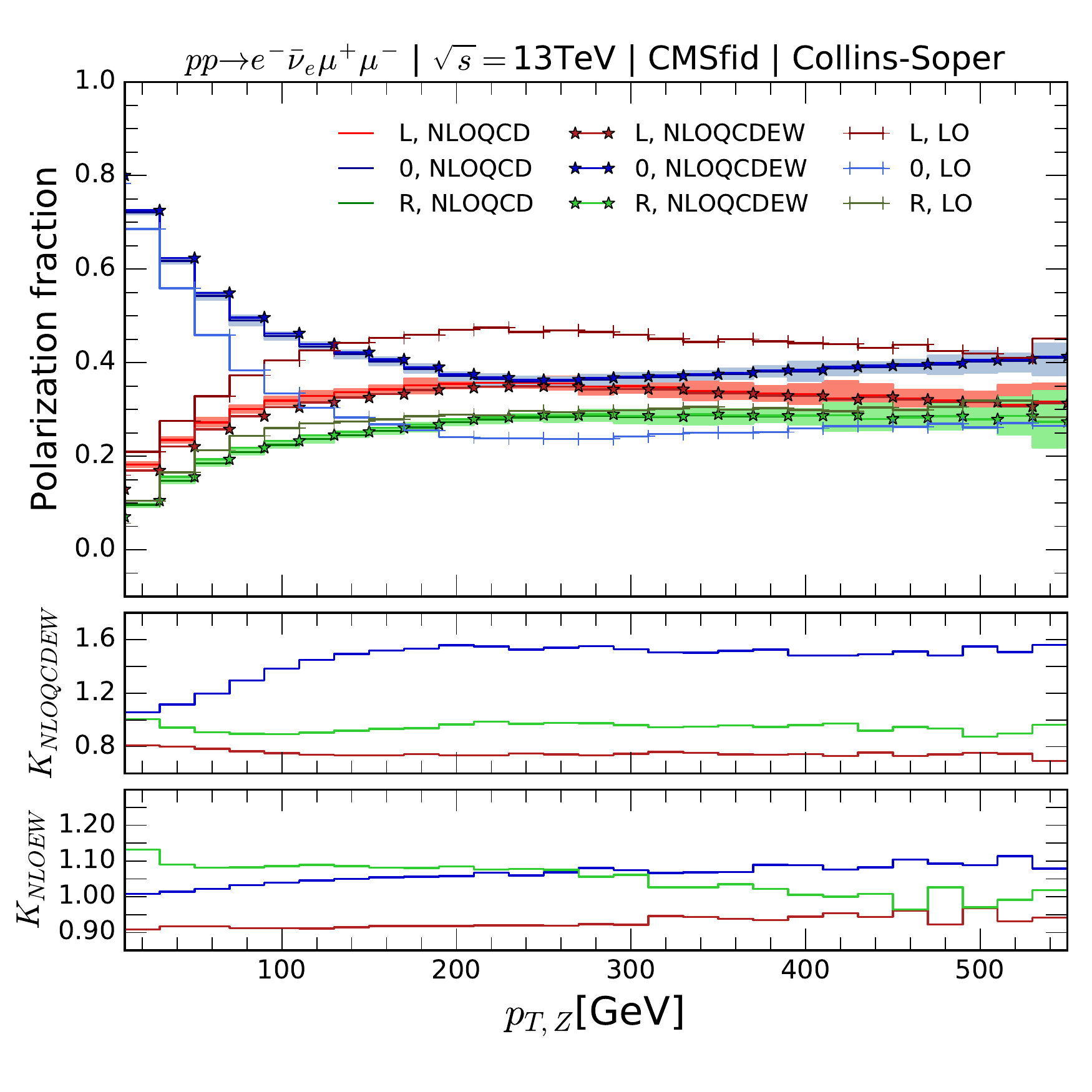}
  \end{tabular}
  \caption{Same as \fig{fig:dist_pTZm_ATLAS_NLOQCDEW} but with the CMS cuts.}
  \label{fig:dist_pTZm_CMS_NLOQCDEW}
\end{figure}

\begin{figure}[hb!]
  \centering
 \begin{tabular}{cc}
  \includegraphics[width=0.48\textwidth]{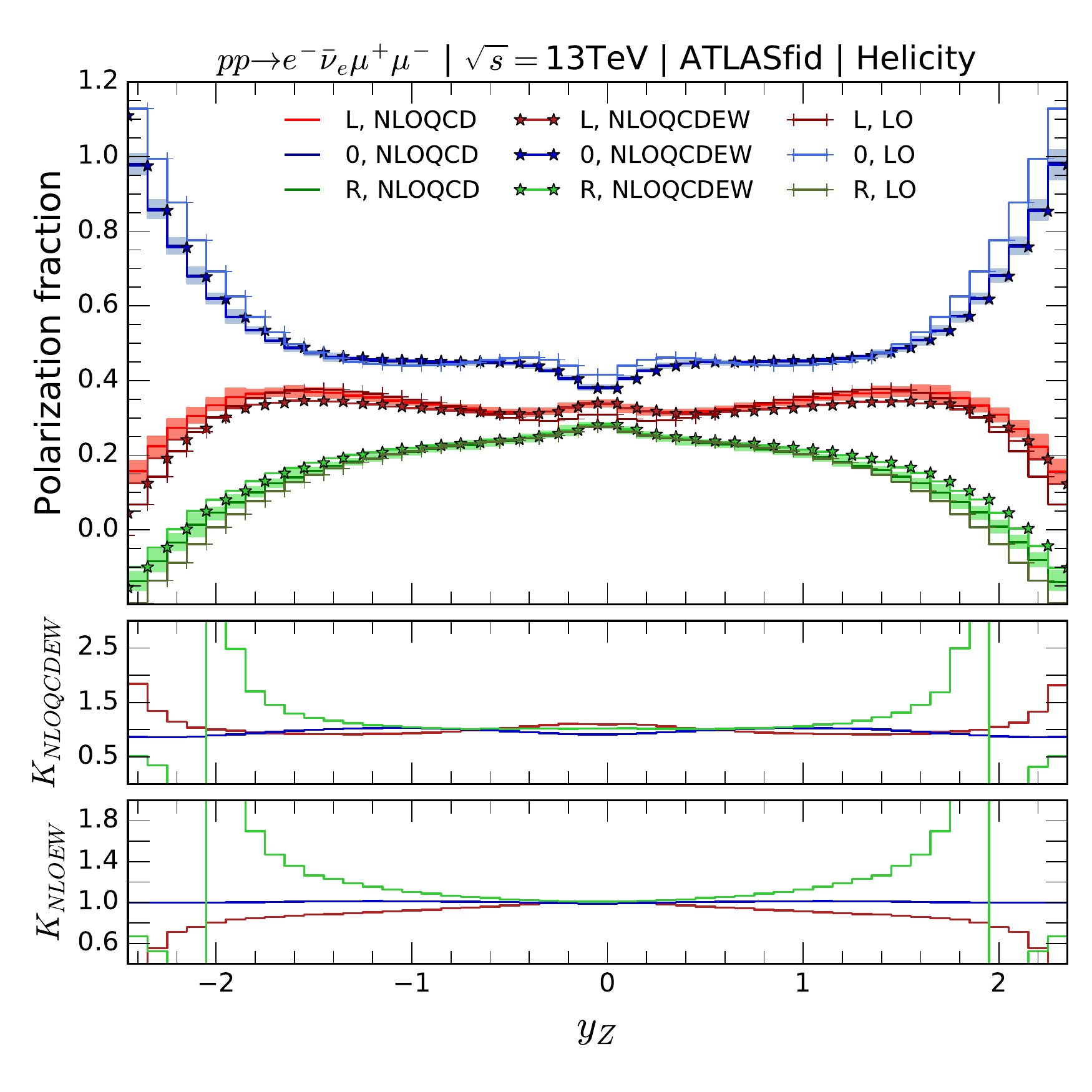}& 
  \includegraphics[width=0.48\textwidth]{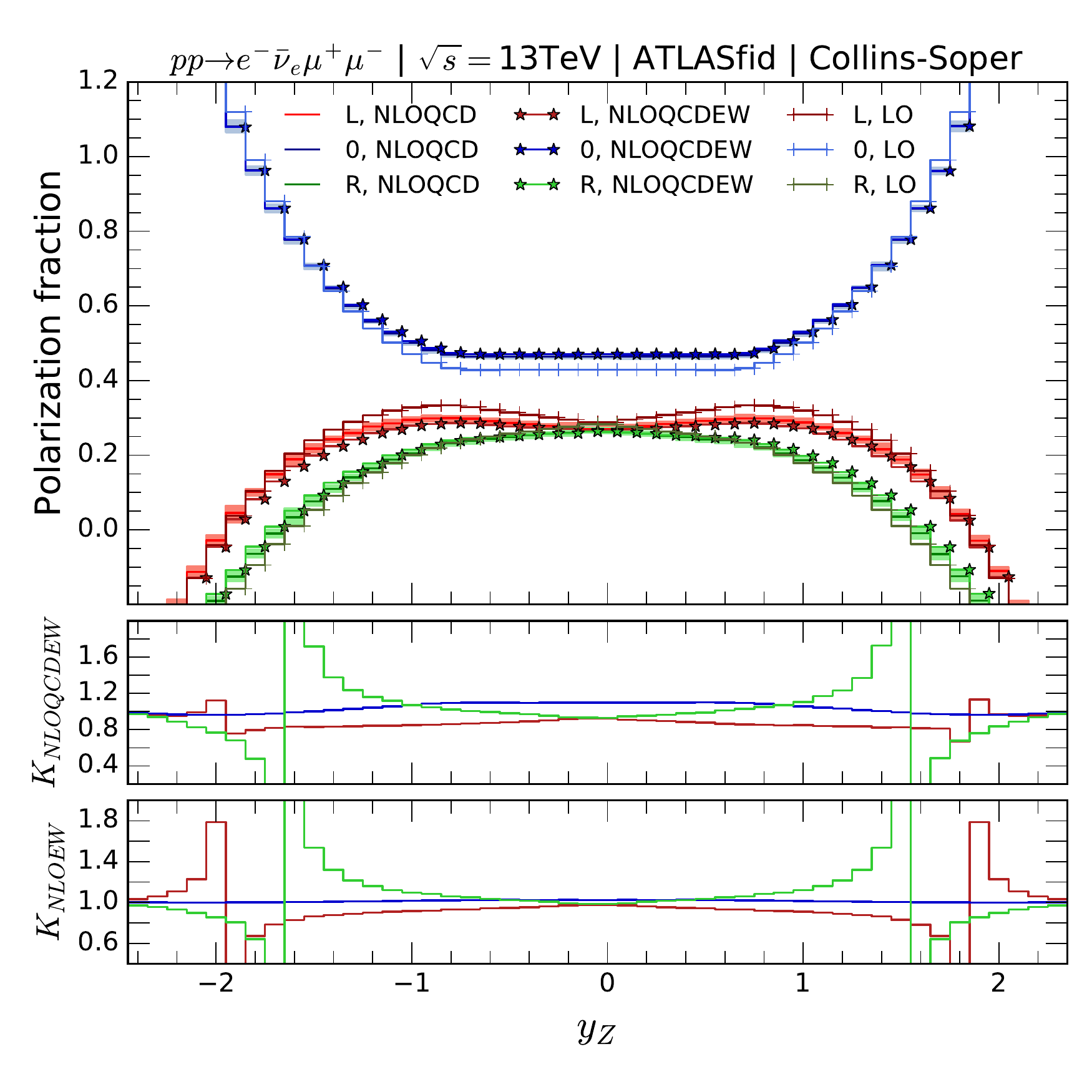}
  \end{tabular}
  \caption{Same as \fig{fig:dist_ptWm_ATLAS_NLOQCDEW} but for
the rapidity distributions of the $Z$ fiducial polarization fractions.}
  \label{fig:dist_yZm_ATLAS_NLOQCDEW}
\end{figure}

\begin{figure}[hb!]
  \centering
 \begin{tabular}{cc}
  \includegraphics[width=0.48\textwidth]{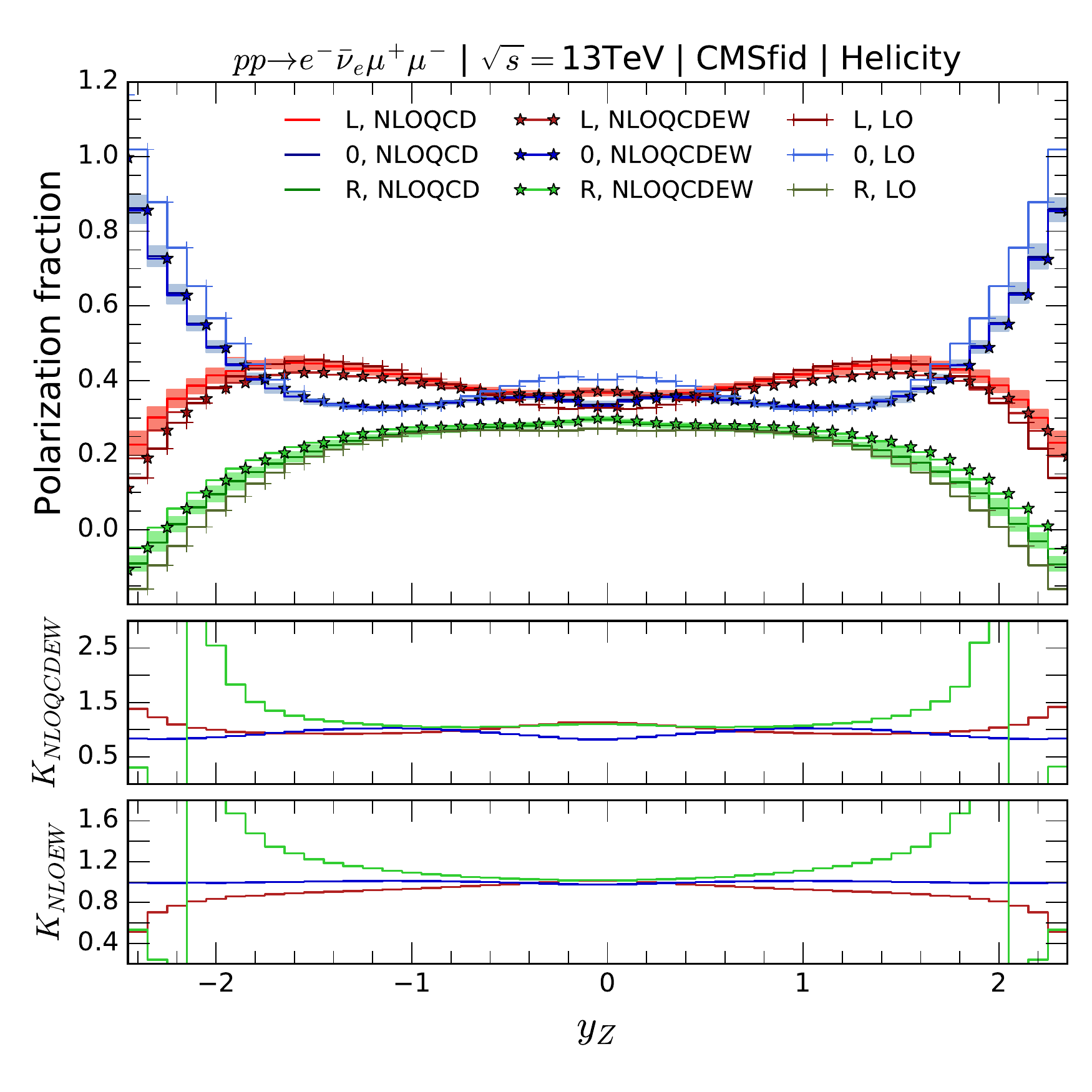}& 
  \includegraphics[width=0.48\textwidth]{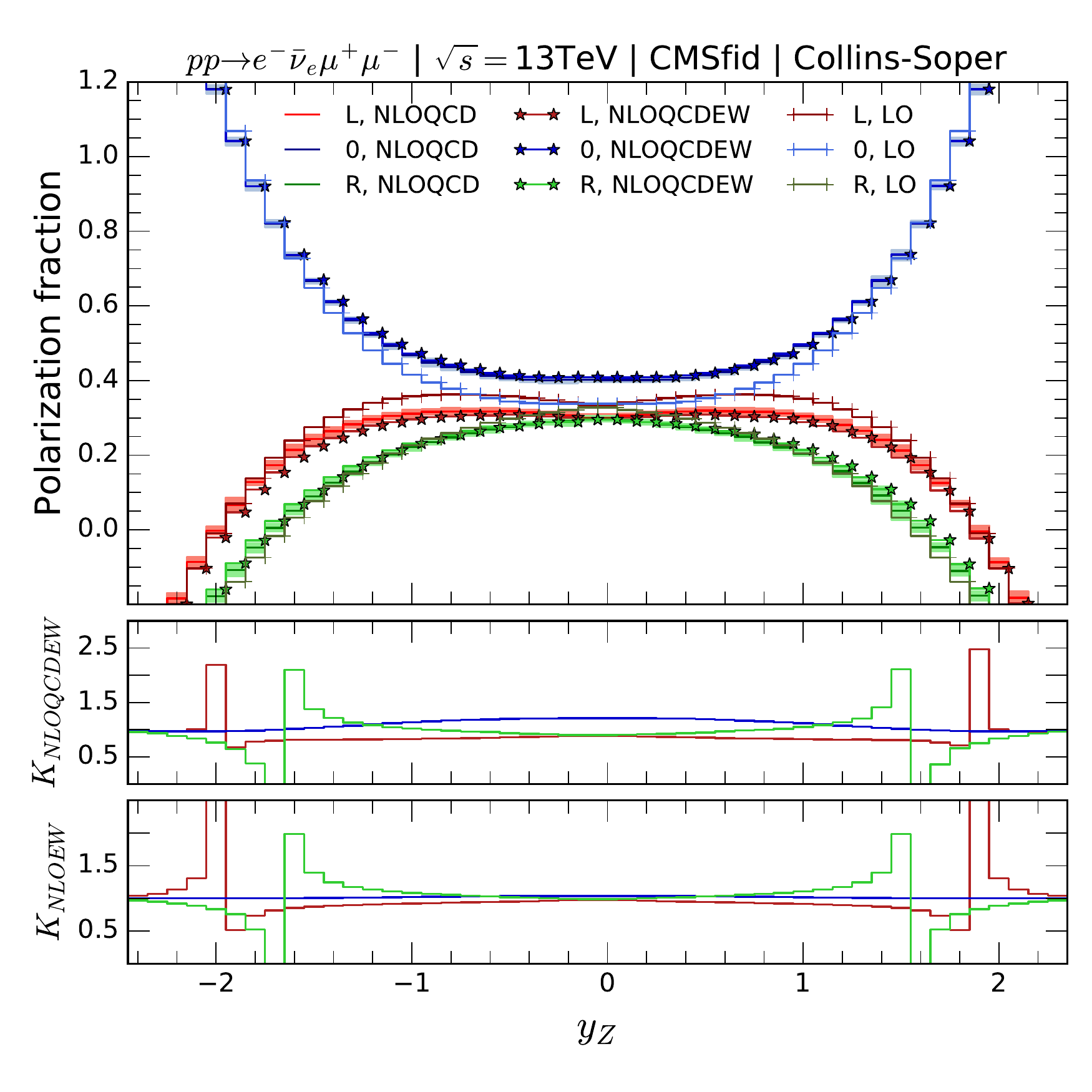}
  \end{tabular}
  \caption{Same as \fig{fig:dist_yZm_ATLAS_NLOQCDEW} but with the CMS fiducial cuts.}
  \label{fig:dist_yZm_CMS_NLOQCDEW}
\end{figure}

\begin{figure}[hb!]
  \centering
 \begin{tabular}{cc}
  \includegraphics[width=0.48\textwidth]{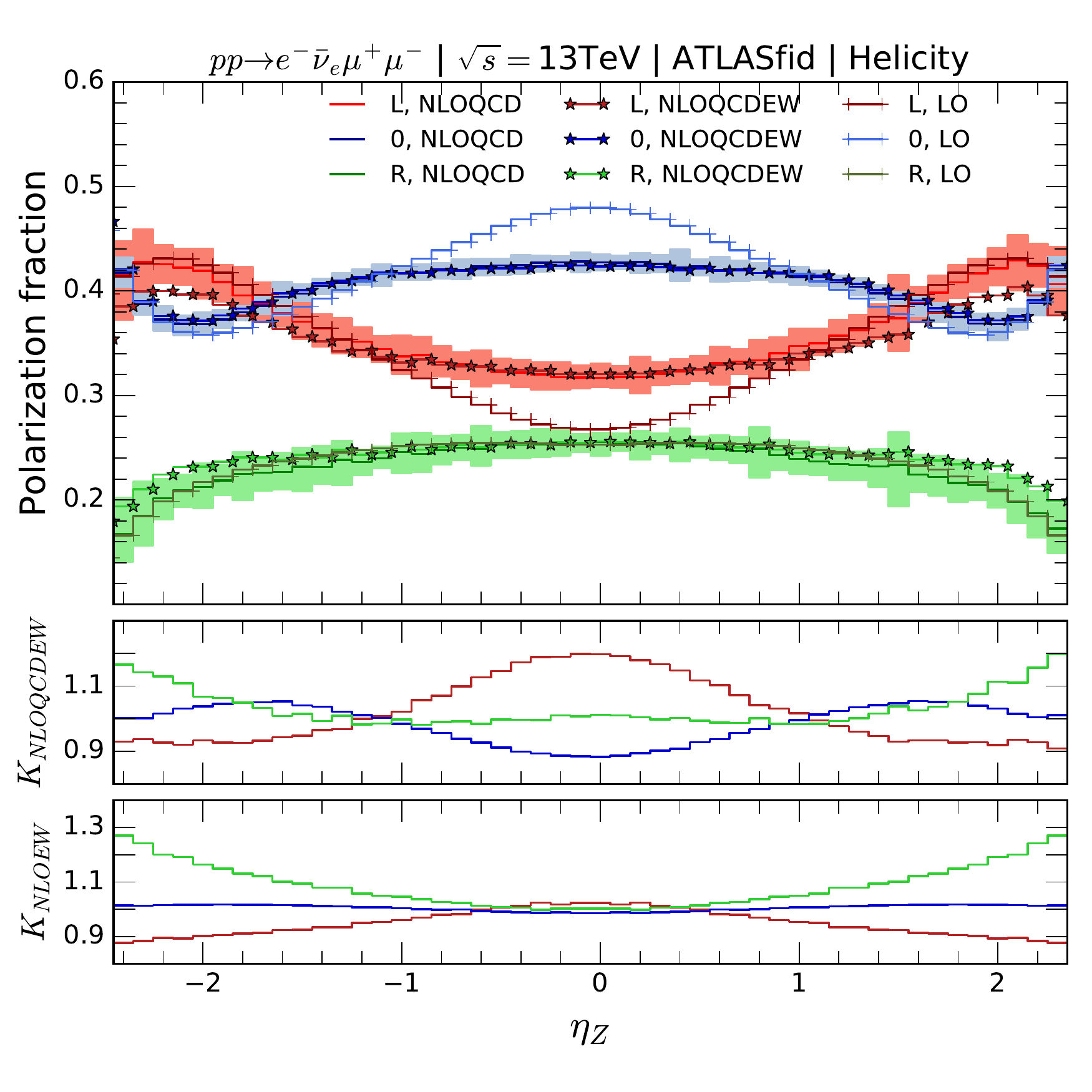}& 
  \includegraphics[width=0.48\textwidth]{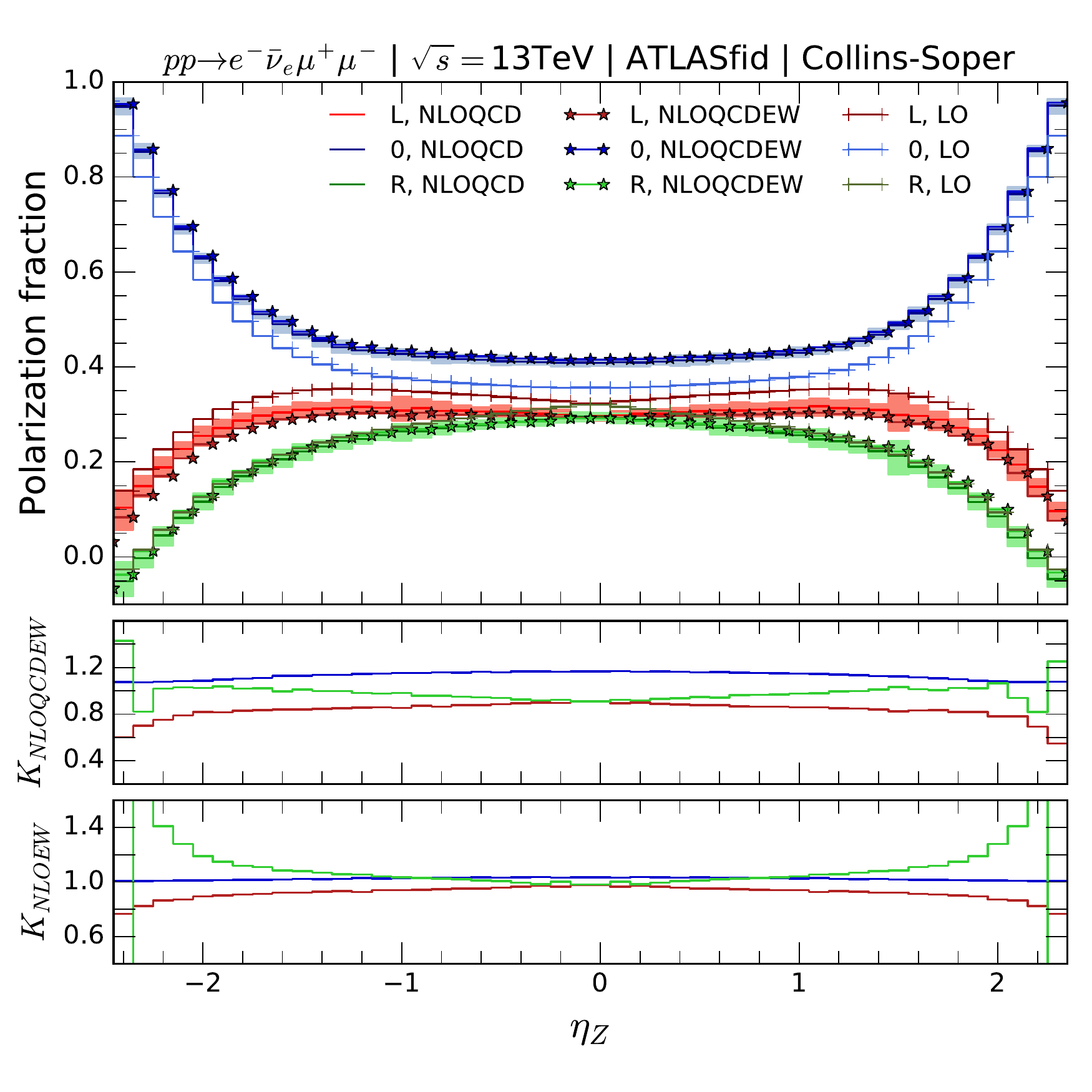}
  \end{tabular}
  \caption{Same as \fig{fig:dist_ptWm_ATLAS_NLOQCDEW} but for
the pseudo-rapidity distributions of the $Z$ fiducial polarization fractions.}
  \label{fig:dist_etaZm_ATLAS_NLOQCDEW}
\end{figure}

\begin{figure}[hb!]
  \centering
 \begin{tabular}{cc}
  \includegraphics[width=0.48\textwidth]{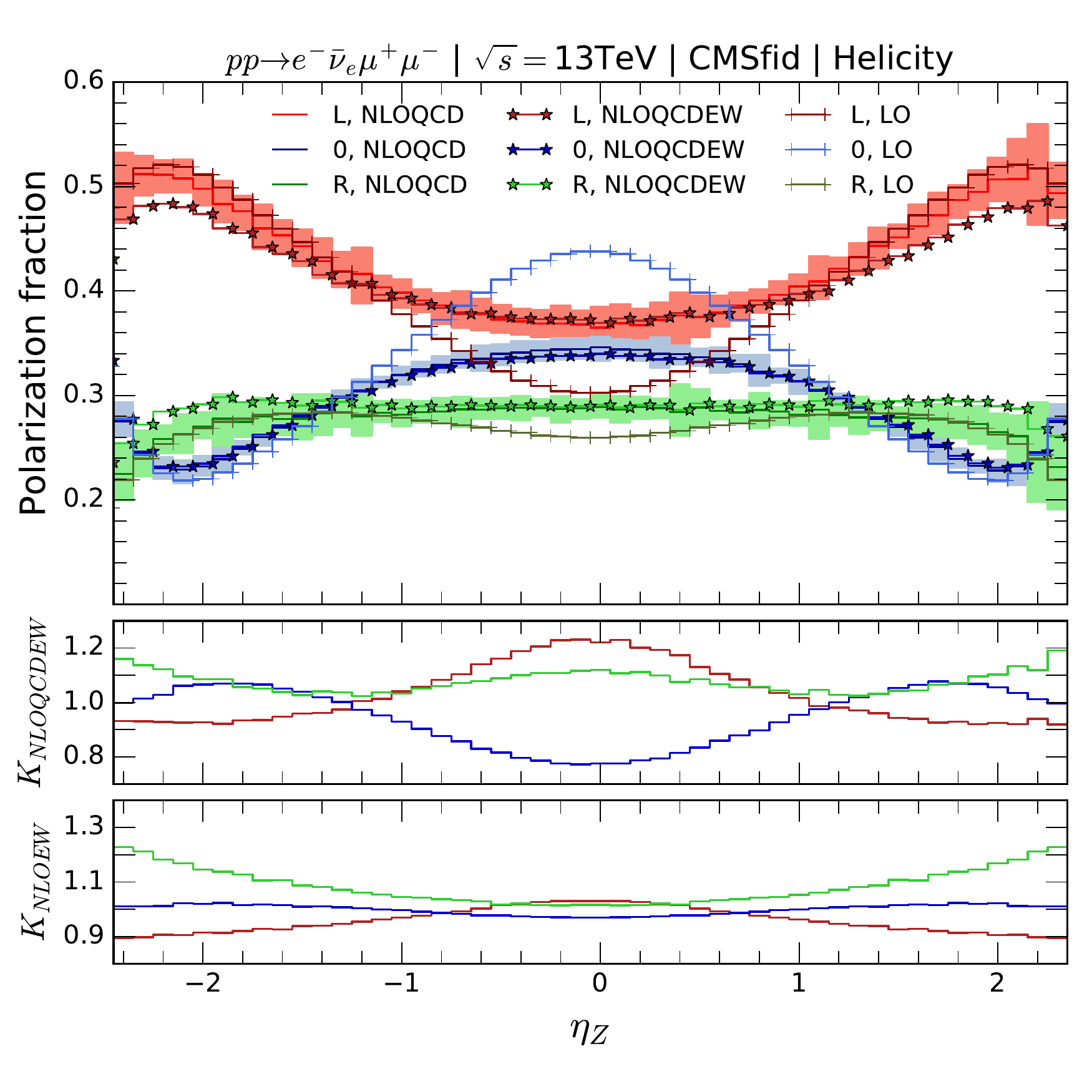}& 
  \includegraphics[width=0.48\textwidth]{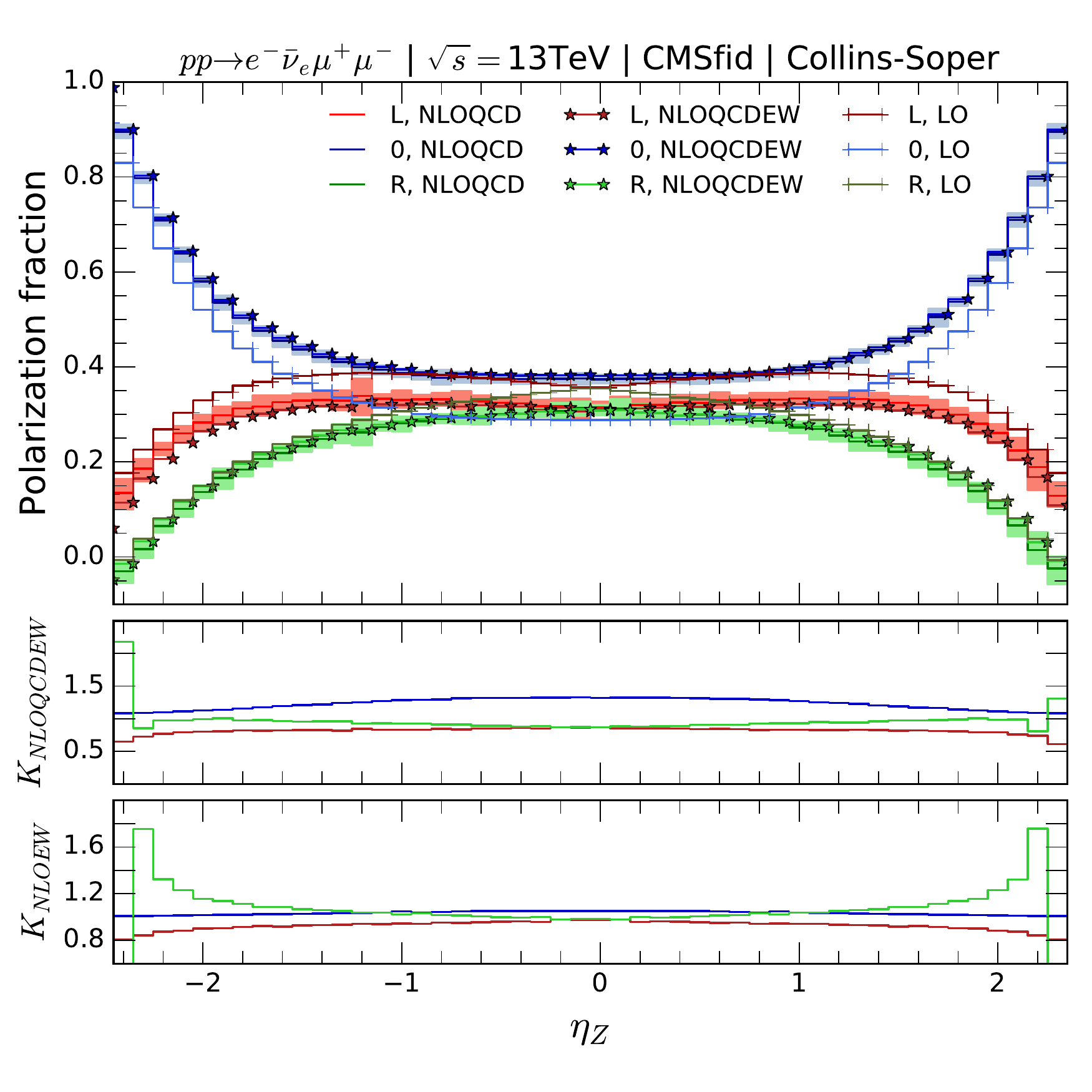}
  \end{tabular}
  \caption{Same as \fig{fig:dist_etaZm_ATLAS_NLOQCDEW} but with the CMS fiducial cuts.}
  \label{fig:dist_etaZm_CMS_NLOQCDEW}
\end{figure}

The corresponding transverse momentum distributions of the fiducial 
polarization fractions can be found in
\fig{fig:dist_ptWm_ATLAS_NLOQCDEW}, \fig{fig:dist_ptWm_CMS_NLOQCDEW},
\fig{fig:dist_pTZm_ATLAS_NLOQCDEW}, and
\fig{fig:dist_pTZm_CMS_NLOQCDEW}. Rapidity and pseudo-rapidity
distributions are displayed in \fig{fig:dist_yZm_ATLAS_NLOQCDEW},
\fig{fig:dist_yZm_CMS_NLOQCDEW}, \fig{fig:dist_etaZm_ATLAS_NLOQCDEW},
and \fig{fig:dist_etaZm_CMS_NLOQCDEW}.

\subsection{Fiducial angular coefficients and polarization fractions: $W^-Z$ channel}
\label{appen:numres:polar_observables_nloqcdew_Wm}
\begin{table}[ht!]
 \renewcommand{\arraystretch}{1.3}
\begin{center}
\setlength\tabcolsep{0.03cm}
\fontsize{5.0}{5.0}
\begin{tabular}{|c|c|c|c|c|c|c|c|c|}\hline
$\text{Method}$  & $A_0$ & $A_1$  & $A_2$ & $A_3$ & $A_4$ & $A_5$ & $A_6$ & $A_7$\\
\hline
{\fontsize{6.0}{6.0}$\text{HE LO}$} & $1.110(1)^{+2}_{-2}$ & $-0.388(1)^{+2}_{-1}$ & $-1.359(1)^{+2}_{-2}$ & $-0.127(2)^{+2}_{-3}$ & $-0.025(6)^{+2}_{-2}$ & $0.0003(3)^{+1}_{-1}$ & $0.003(0.2)^{+0.01}_{-0.1}$ & $0.011(0.3)^{+0.05}_{-0}$\\
\hline
{\fontsize{6.0}{6.0}$\text{HE NLOEW}$} & $1.107$ & $-0.385$ & $-1.372$ & $-0.126$ & $-0.021$ & $0.002$ & $0.006$ & $0.010$\\
\hline
{\fontsize{6.0}{6.0}$\text{HE NLOQCD}$} & $1.029(2)^{+9}_{-9}$ & $-0.425(1)^{+2}_{-2}$ & $-1.361(2)^{+3}_{-5}$ & $-0.227(2)^{+7}_{-7}$ & $0.175(6)^{+17}_{-16}$ & $0.001(1)^{+0.04}_{-0.1}$ & $0.002(1)^{+0.1}_{-0.3}$ & $0.002(1)^{+1}_{-0.5}$\\
\hline
{\fontsize{6.0}{6.0}$\text{HE NLOQCDEW}$} & $1.026$ & $-0.424$ & $-1.368$ & $-0.228$ & $0.181$ & $0.001$ & $0.003$ & $0.001$\\
\hline\hline
{\fontsize{6.0}{6.0}$\text{CS LO}$} & $1.578(2)^{+1}_{-1}$ & $0.214(1)^{+1}_{-1}$ & $-0.893(1)^{+2}_{-2}$ & $0.139(2)^{+2}_{-1}$ & $-0.123(8)^{+3}_{-3}$ & $-0.003(0.2)^{+0.1}_{-0.1}$ & $-0.003(0.2)^{+0.1}_{-0.1}$ & $-0.011(0.3)^{+0}_{-0.05}$\\
\hline
{\fontsize{6.0}{6.0}$\text{CS NLOEW}$} & $1.580$ & $0.213$ & $-0.902$ & $0.141$ & $-0.124$ & $-0.004$ & $-0.005$ & $-0.010$\\
\hline
{\fontsize{6.0}{6.0}$\text{CS NLOQCD}$} & $1.650(2)^{+4}_{-4}$ & $0.156(1)^{+4}_{-4}$ & $-0.743(3)^{+9}_{-9}$ & $0.356(5)^{+18}_{-17}$ & $-0.112(6)^{+1}_{-1}$ & $-0.001(0.5)^{+0.1}_{-0.1}$ & $-0.002(0.5)^{+0.2}_{-0.2}$ & $-0.002(1)^{+0.5}_{-1}$\\
\hline
{\fontsize{6.0}{6.0}$\text{CS NLOQCDEW}$} & $1.653$ & $0.154$ & $-0.744$ & $0.362$ & $-0.112$ & $-0.002$ & $-0.003$ & $-0.001$\\
\hline
\end{tabular}
\caption{\small Fiducial angular coefficients of the $e^{-}$ distribution for
  the process $pp \to e^- \bar{\nu}_e\, \mu^+ \mu^- + X$ at LO, NLO EW, NLO QCD and
  NLO QCD+EW at the 13 TeV LHC with the ATLAS fiducial cuts. Results are
  presented for two coordinate systems: the helicity (HE) and
  Collins-Soper (CS) coordinate systems. The PDF uncertainties (in parenthesis) and the scale
  uncertainties are provided for the LO and NLO QCD results, all given on the last digit of the central prediction.}
\label{tab:coeff_Ai_Wm_ATLAS}
\end{center}
\end{table} 

\begin{table}[ht!]
 \renewcommand{\arraystretch}{1.3}
\begin{center}
\setlength\tabcolsep{0.03cm}
\fontsize{5.0}{5.0}
\begin{tabular}{|c|c|c|c|c|c|c|c|c|}\hline
$\text{Method}$  & $A_0$ & $A_1$  & $A_2$ & $A_3$ & $A_4$ & $A_5$ & $A_6$ & $A_7$\\
\hline
{\fontsize{6.0}{6.0}$\text{HE LO}$} & $0.989(1)^{+1}_{-1}$ & $-0.326(1)^{+1}_{-1}$ & $-0.736(1)^{+1}_{-1}$ & $-0.013(1)^{+0.1}_{-0.1}$ & $0.061(1)^{+0.03}_{-0}$ & $0.003(0.2)^{+0.1}_{-0.04}$ & $0.009(0.2)^{+0}_{-0.1}$ & $0.006(0.2)^{+0}_{-0.1}$\\
\hline
{\fontsize{6.0}{6.0}$\text{HE NLOEW}$} & $0.992$ & $-0.330$ & $-0.748$ & $-0.008$ & $0.039$ & $0.004$ & $0.011$ & $0.006$\\
\hline
{\fontsize{6.0}{6.0}$\text{HE NLOQCD}$} & $0.950(1)^{+4}_{-5}$ & $-0.323(1)^{+4}_{-3}$ & $-0.754(1)^{+1}_{-1}$ & $-0.016(1)^{+0.2}_{-0.2}$ & $0.062(1)^{+1}_{-1}$ & $0.001(0.5)^{+0.1}_{-0.05}$ & $0.005(1)^{+0.3}_{-0.2}$ & $0.004(0.4)^{+0.3}_{-0.2}$\\
\hline
{\fontsize{6.0}{6.0}$\text{HE NLOQCDEW}$} & $0.951$ & $-0.326$ & $-0.760$ & $-0.013$ & $0.050$ & $0.002$ & $0.006$ & $0.004$\\
\hline\hline
{\fontsize{6.0}{6.0}$\text{CS LO}$} & $1.251(1)^{+1}_{-2}$ & $0.313(1)^{+2}_{-2}$ & $-0.477(1)^{+1}_{-1}$ & $0.050(0.2)^{+0.1}_{-0.1}$ & $0.048(2)^{+0.1}_{-0}$ & $-0.004(0.3)^{+0.1}_{-0}$ & $-0.010(0.2)^{+0.1}_{-0.1}$ & $-0.006(0.2)^{+0.1}_{-0}$\\
\hline
{\fontsize{6.0}{6.0}$\text{CS NLOEW}$} & $1.263$ & $0.308$ & $-0.480$ & $0.032$ & $0.030$ & $-0.005$ & $-0.011$ & $-0.006$\\
\hline
{\fontsize{6.0}{6.0}$\text{CS NLOQCD}$} & $1.271(1)^{+3}_{-4}$ & $0.268(1)^{+3}_{-3}$ & $-0.436(2)^{+3}_{-3}$ & $0.055(1)^{+1}_{-1}$ & $0.036(1)^{+1}_{-1}$ & $-0.003(0.5)^{+0.2}_{-0.1}$ & $-0.005(1)^{+0.1}_{-0.4}$ & $-0.004(0.4)^{+0.2}_{-0.2}$\\
\hline
{\fontsize{6.0}{6.0}$\text{CS NLOQCDEW}$} & $1.278$ & $0.265$ & $-0.436$ & $0.046$ & $0.026$ & $-0.003$ & $-0.005$ & $-0.004$\\
\hline
\end{tabular}
\caption{\small Same as \tab{tab:coeff_Ai_Wm_ATLAS} but for the
  $\mu^{-}$ distribution.}
\label{tab:coeff_Ai_Z_2_ATLAS}
\end{center}
\end{table}

\begin{table}[ht!]
 \renewcommand{\arraystretch}{1.3}
\begin{center}
\setlength\tabcolsep{0.03cm}
\fontsize{5.0}{5.0}
\begin{tabular}{|c|c|c|c|c|c|c|c|c|}\hline
$\text{Method}$  & $A_0$ & $A_1$  & $A_2$ & $A_3$ & $A_4$ & $A_5$ & $A_6$ & $A_7$\\
\hline
{\fontsize{6.0}{6.0}$\text{HE LO}$} & $1.006(1)^{+2}_{-2}$ & $-0.079(2)^{+3}_{-3}$ & $-0.742(2)^{+4}_{-3}$ & $-0.156(3)^{+3}_{-3}$ & $-0.002(7)^{+2}_{-2}$ & $0.0005(3)^{+0}_{-1}$ & $0.003(0.2)^{+0}_{-0.03}$ & $0.010(0.2)^{+0.03}_{-0.02}$\\
\hline
{\fontsize{6.0}{6.0}$\text{HE NLOEW}$} & $1.003$ & $-0.072$ & $-0.742$ & $-0.156$ & $0.001$ & $0.002$ & $0.006$ & $0.009$\\
\hline
{\fontsize{6.0}{6.0}$\text{HE NLOQCD}$} & $0.949(1)^{+6}_{-7}$ & $-0.126(2)^{+3}_{-3}$ & $-0.671(2)^{+7}_{-7}$ & $-0.249(3)^{+7}_{-6}$ & $0.192(7)^{+17}_{-16}$ & $0.001(0.4)^{+0.1}_{-0.1}$ & $0.002(0.4)^{+0.1}_{-0.2}$ & $0.002(0.5)^{+0.5}_{-0.5}$\\
\hline
{\fontsize{6.0}{6.0}$\text{HE NLOQCDEW}$} & $0.947$ & $-0.122$ & $-0.671$ & $-0.250$ & $0.195$ & $0.001$ & $0.003$ & $0.001$\\
\hline\hline
{\fontsize{6.0}{6.0}$\text{CS LO}$} & $1.013(3)^{+4}_{-5}$ & $0.216(1)^{+1}_{-1}$ & $-0.735(1)^{+1}_{-1}$ & $0.182(2)^{+1}_{-1}$ & $-0.131(9)^{+4}_{-4}$ & $-0.002(0.2)^{+0.01}_{-0.1}$ & $-0.003(0.2)^{+0.1}_{-0}$ & $-0.010(0.2)^{+0.02}_{-0.03}$\\
\hline
{\fontsize{6.0}{6.0}$\text{CS NLOEW}$} & $1.007$ & $0.214$ & $-0.739$ & $0.184$ & $-0.133$ & $-0.003$ & $-0.005$ & $-0.009$\\
\hline
{\fontsize{6.0}{6.0}$\text{CS NLOQCD}$} & $1.067(2)^{+3}_{-4}$ & $0.157(1)^{+4}_{-4}$ & $-0.554(4)^{+14}_{-14}$ & $0.388(5)^{+17}_{-16}$ & $-0.123(6)^{+1}_{-0.4}$ & $-0.001(0.5)^{+0.1}_{-0.1}$ & $-0.002(0.4)^{+0.1}_{-0.1}$ & $-0.002(0.5)^{+0.5}_{-0.5}$\\
\hline
{\fontsize{6.0}{6.0}$\text{CS NLOQCDEW}$} & $1.064$ & $0.155$ & $-0.554$ & $0.391$ & $-0.124$ & $-0.001$ & $-0.003$ & $-0.001$\\
\hline
\end{tabular}
\caption{\small Same as \tab{tab:coeff_Ai_Wm_ATLAS} but with the CMS fiducial cuts.}
\label{tab:coeff_Ai_Wm_CMS}
\end{center}
\end{table} 

\begin{table}[ht!]
 \renewcommand{\arraystretch}{1.3}
\begin{center}
\setlength\tabcolsep{0.03cm}
\fontsize{4.0}{4.0}
\begin{tabular}{|c|c|c|c|c|c|c|c|c|}\hline
$\text{Method}$  & $A_0$ & $A_1$  & $A_2$ & $A_3$ & $A_4$ & $A_5$ & $A_6$ & $A_7$\\
\hline
{\fontsize{6.0}{6.0}$\text{HE LO}$} & $0.798(1)^{+2}_{-3}$ & $-0.288(1)^{+1}_{-1}$ & $-0.603(1)^{+1}_{-2}$ & $-0.017(1)^{+0.1}_{-0.1}$ & $0.072(1)^{+0.04}_{-0}$ & $0.003(0.2)^{+0.1}_{-0.04}$ & $0.009(0.4)^{+0}_{-0.05}$ & $0.006(0.2)^{+0.04}_{-0.01}$\\
\hline
{\fontsize{6.0}{6.0}$\text{HE NLOEW}$} & $0.795$ & $-0.297$ & $-0.606$ & $-0.012$ & $0.048$ & $0.004$ & $0.011$ & $0.006$\\
\hline
{\fontsize{6.0}{6.0}$\text{HE NLOQCD}$} & $0.739(1)^{+7}_{-7}$ & $-0.312(1)^{+3}_{-2}$ & $-0.609(1)^{+0.4}_{-0.5}$ & $-0.018(1)^{+0.2}_{-0.1}$ & $0.069(1)^{+1}_{-1}$ & $0.001(1)^{+0.2}_{-0.1}$ & $0.005(0.5)^{+0.3}_{-0.3}$ & $0.004(0.4)^{+0.3}_{-0.1}$\\
\hline
{\fontsize{6.0}{6.0}$\text{HE NLOQCDEW}$} & $0.737$ & $-0.317$ & $-0.611$ & $-0.015$ & $0.057$ & $0.001$ & $0.006$ & $0.004$\\
\hline\hline
{\fontsize{6.0}{6.0}$\text{CS LO}$} & $1.113(1)^{+0.4}_{-0.03}$ & $0.382(1)^{+3}_{-2}$ & $-0.291(1)^{+1}_{-1}$ & $0.061(0.2)^{+0.1}_{-0.03}$ & $0.050(2)^{+0.1}_{-0.01}$ & $-0.005(0.3)^{+0.1}_{-0}$ & $-0.010(0.3)^{+0.1}_{-0.1}$ & $-0.006(0.2)^{+0.01}_{-0.04}$\\
\hline
{\fontsize{6.0}{6.0}$\text{CS NLOEW}$} & $1.128$ & $0.377$ & $-0.276$ & $0.042$ & $0.032$ & $-0.005$ & $-0.011$ & $-0.006$\\
\hline
{\fontsize{6.0}{6.0}$\text{CS NLOQCD}$} & $1.176(1)^{+4}_{-3}$ & $0.329(1)^{+3}_{-3}$ & $-0.176(2)^{+8}_{-8}$ & $0.063(1)^{+1}_{-1}$ & $0.037(1)^{+1}_{-1}$ & $-0.003(1)^{+0.3}_{-0.02}$ & $-0.005(0.4)^{+0.3}_{-0.3}$ & $-0.004(0.4)^{+0.2}_{-0.2}$\\
\hline
{\fontsize{6.0}{6.0}$\text{CS NLOQCDEW}$} & $1.184$ & $0.326$ & $-0.167$ & $0.053$ & $0.028$ & $-0.003$ & $-0.005$ & $-0.004$\\
\hline
\end{tabular}
\caption{\small Same as \tab{tab:coeff_Ai_Z_2_ATLAS} but with the CMS fiducial cuts.}
\label{tab:coeff_Ai_Z_2_CMS}
\end{center}
\end{table}
\begin{table}[ht!]
  \renewcommand{\arraystretch}{1.3}
\begin{center}
    \fontsize{8}{8}
\begin{tabular}{|c|c|c|c||c|c|c|}\hline
$\text{Method}$  & $f^{W^-}_L$ & $f^{W^-}_0$ & $f^{W^-}_R$ & $f^Z_L$ & $f^Z_0$ & $f^Z_R$\\
\hline
$\text{HE LO}$ & $0.216(1)^{+0.1}_{-0.05}$ & $0.555(1)^{+1}_{-1}$ & $0.229(2)^{+1}_{-1}$ & $0.324(1)^{+0.4}_{-0.3}$ & $0.494(0.4)^{+1}_{-1}$ & $0.181(1)^{+0.3}_{-0.4}$\\
\hline
$\text{HE NLOEW}$ & $0.218$ & $0.554$ & $0.228$ & $0.298$ & $0.496$ & $0.206$\\
\hline
$\text{HE NLOQCD}$ & $0.286(2)^{+7}_{-6}$ & $0.515(1)^{+4}_{-5}$ & $0.199(1)^{+2}_{-2}$ & $0.334(1)^{+2}_{-2}$ & $0.475(0.5)^{+2}_{-2}$ & $0.191(1)^{+1}_{-1}$\\
\hline
$\text{HE NLOQCDEW}$ & $0.289$ & $0.513$ & $0.198$ & $0.321$ & $0.475$ & $0.204$\\
\hline\hline
$\text{CS LO}$ & $0.075(2)^{+0.5}_{-1}$ & $0.789(1)^{+1}_{-1}$ & $0.136(2)^{+1}_{-1}$ & $0.243(2)^{+1}_{-0.3}$ & $0.625(1)^{+1}_{-1}$ & $0.132(2)^{+0.3}_{-0.3}$\\
\hline
$\text{CS NLOEW}$ & $0.074$ & $0.790$ & $0.136$ & $0.220$ & $0.632$ & $0.149$\\
\hline
$\text{CS NLOQCD}$ & $0.059(1)^{+1}_{-1}$ & $0.825(1)^{+2}_{-2}$ & $0.115(2)^{+1}_{-1}$ & $0.224(1)^{+1}_{-1}$ & $0.636(1)^{+2}_{-2}$ & $0.140(1)^{+2}_{-1}$\\
\hline
$\text{CS NLOQCDEW}$ & $0.059$ & $0.826$ & $0.115$ & $0.211$ & $0.639$ & $0.150$\\
\hline
\end{tabular}
    \caption{\small $W^-_{}$ and $Z$ fiducial polarization fractions in the
      process $pp \to e^-_{} \nu_e^{}\, \mu^+_{} \mu^-_{} + X$ at LO, NLO EW, NLO QCD and
  NLO QCD+EW at the 13 TeV LHC with the ATLAS fiducial cuts. Results are
  presented for two coordinate systems: the helicity (HE) and
  Collins-Soper (CS) coordinate systems. The PDF uncertainties (in parenthesis) and the scale
  uncertainties are provided for the LO and NLO QCD results, all given on the last digit of the central prediction.}
    \label{tab:coeff_fL0R_WmZ_ATLAS}
 \end{center}
\end{table}
\begin{table}[ht!]
  \renewcommand{\arraystretch}{1.3}
  \begin{center}
    \fontsize{8}{8}
\begin{tabular}{|c|c|c|c||c|c|c|}\hline
$\text{Method}$  & $f^{W^-}_L$ & $f^{W^-}_0$ & $f^{W^-}_R$ & $f^Z_L$ & $f^Z_0$ & $f^Z_R$\\
\hline
$\text{HE LO}$ & $0.248(2)^{+0.02}_{-0}$ & $0.503(1)^{+1}_{-1}$ & $0.249(2)^{+1}_{-1}$ & $0.384(1)^{+1}_{-1}$ & $0.399(1)^{+1}_{-1}$ & $0.217(1)^{+1}_{-1}$\\
\hline
$\text{HE NLOEW}$ & $0.250$ & $0.501$ & $0.249$ & $0.358$ & $0.398$ & $0.245$\\
\hline
$\text{HE NLOQCD}$ & $0.311(2)^{+6}_{-6}$ & $0.474(1)^{+3}_{-4}$ & $0.215(2)^{+2}_{-2}$ & $0.396(1)^{+3}_{-2}$ & $0.369(0.5)^{+3}_{-4}$ & $0.235(1)^{+1}_{-1}$\\
\hline
$\text{HE NLOQCDEW}$ & $0.312$ & $0.473$ & $0.214$ & $0.382$ & $0.368$ & $0.250$\\
\hline\hline
$\text{CS LO}$ & $0.214(2)^{+0.3}_{-0.2}$ & $0.507(1)^{+2}_{-3}$ & $0.279(3)^{+2}_{-2}$ & $0.280(2)^{+0.1}_{-0.1}$ & $0.557(1)^{+0.2}_{-0.02}$ & $0.163(2)^{+0}_{-0.1}$\\
\hline
$\text{CS NLOEW}$ & $0.215$ & $0.504$ & $0.281$ & $0.256$ & $0.564$ & $0.180$\\
\hline
$\text{CS NLOQCD}$ & $0.203(1)^{+1}_{-1}$ & $0.533(1)^{+2}_{-2}$ & $0.264(2)^{+1}_{-1}$ & $0.249(1)^{+2}_{-2}$ & $0.588(1)^{+2}_{-2}$ & $0.163(1)^{+1}_{-1}$\\
\hline
$\text{CS NLOQCDEW}$ & $0.203$ & $0.532$ & $0.265$ & $0.236$ & $0.592$ & $0.171$\\
\hline
\end{tabular}
    \caption{\small Same as \tab{tab:coeff_fL0R_WmZ_ATLAS} but with the CMS fiducial cuts.}
    \label{tab:coeff_fL0R_WmZ_CMS}
  \end{center}
\end{table}

\clearpage
\pagebreak

\subsection{Distributions of fiducial polarization fractions: $W^-Z$ channel}
\label{appen:numres:polar_observables_dist_Wm}

\clearpage
\pagebreak

\section{Off-shell and NLO EW correction effects on fiducial polarization observables}
\label{appen:off_shell_NLOEW_effects}
Here we would like to show the effects of 
various contributions including the $\bar{q}q'$ annihilation channels at NLO EW, 
denoted as $\text{F-EW-}\bar{q}q'$ 
where $\text{F}$ means that the full LO amplitudes are used and the $\bar{q}q'$ corrections 
are calculated as in \eq{eq:qq_corrections}, and similarly the 
quark-photon induced channels denoted as $\text{F-EW-}q\gamma$ 
with the quark-photon induced corrections 
calculated as in \eq{eq:qgamma_corrections}. Results for full LO (denoted $\text{F-LO}$) and for 
full LO plus NLO EW corrections (denoted $\text{F-EW}$) have been provided in \sect{sect:numres:polar_observables_nloqcdew}, but are 
also included in the tables here for easy comparisons. 
It is noted that some results for the $\text{F-LO}$ and the $\text{F-EW}$ shown here
are not identical as those in \sect{sect:numres:polar_observables_nloqcdew}, but agree within the statistical error. 
This is because the full LO results here are obtained using our in-house code while the ones in 
\sect{sect:numres:polar_observables_nloqcdew} are calculated using the {\tt VBFNLO} program. 

Moreover, the DPA LO results, denoted as $\text{D-LO}$, are also shown. This enables one to see the off-shell effects 
at LO by comparing to the $\text{F-LO}$ results. As discussed in \sect{sect:cal:polar_observables}, when we are 
looking at the polarization of a gauge boson, it is interesting 
to separate the EW corrections to the production part from those to the decay part. 
This is possible in our DPA framework because off-shell effects are absent. 
The EW corrections to the $W^\pm$ production part $\bar{q}q' \to W^\pm \mu^+\mu^-$, defined in \eq{eq:Vproduction_corrections}, 
are included in the $\text{D-EW-pV}$ (where $\text{pV}$ denotes production of $V$ boson) results 
presented here. Note that this includes EW corrections to the $\bar{q}q' \to W^\pm Z$ part and also to the 
$Z \to \mu^+\mu^-$ decay. Same things apply to the 
EW corrections to the $Z$ production part $\bar{q}q' \to e \nu_e Z$. To see these effects, one has to compare 
the $\text{D-EW-pV}$ results to the corresponding $\text{D-LO}$ results. 
The EW corrections to the decay $W \to e \nu_e$ or $Z \to \mu^+\mu^-$ are defined in \eq{eq:Vdecay_corrections}. 
They can be seen by comparing the $\text{D-EW-dV}$ (where $\text{dV}$ denotes decay of $V$ boson) 
rows to the corresponding $\text{D-LO}$ rows. 
Finally, the $\text{D-EW}$ entries show the results of the DPA LO plus NLO EW corrections.   

In \tab{tab:coeff_Ai_full_DPA_Wp_Atlas} and 
\tab{tab:coeff_Ai_full_DPA_Z_processWp_Atlas} we show results of the $W$ and $Z$ angular 
coefficients for the $W^+ Z$ channel with the ATLAS fiducial cuts, respectively. 
Similar results with the CMS cuts are provided in \tab{tab:coeff_Ai_full_DPA_Wp_Cms} and 
\tab{tab:coeff_Ai_full_DPA_Z_processWp_Cms}. Both $W^+$ and $Z$ polarization fractions 
with the ATLAS cuts are presented in \tab{tab:coeff_fL0R_full_DPA_Wp_Atlas} 
and with the CMS cuts in \tab{tab:coeff_fL0R_full_DPA_Wp_Cms}. 
Similar results for the $W^- Z$ channel are presented in 
\tab{tab:coeff_Ai_full_DPA_Wm_Atlas} and 
\tab{tab:coeff_Ai_full_DPA_Z_processWm_Atlas} with the ATLAS cuts, 
in \tab{tab:coeff_Ai_full_DPA_Wm_Cms} and 
\tab{tab:coeff_Ai_full_DPA_Z_processWm_Cms} with the CMS cuts, and finally 
in \tab{tab:coeff_fL0R_full_DPA_Wm_Atlas} and \tab{tab:coeff_fL0R_full_DPA_Wm_Cms} for the 
polarization fractions.

We have a few remarks here on the results focusing on the fiducial angular coefficients. Looking at the results for $A_5$, $A_6$, and $A_7$ 
we see that the DPA LO results are all consistent with zero within the statistical error. Taking into the EW corrections to the 
decay part does not change this conclusion. However, the EW corrections to the production part have significant effects and make them non-vanishing, but the results are still very small. If full off-shell effects are included, they become non-vanishing as well, see the 
$\text{F-LO}$ results. In general, the corrections to the decay part have negligible effects compared 
to those to the production part, except for the coefficients $A_3$ and $A_4$. In \tab{tab:coeff_Ai_full_DPA_Z_processWm_Atlas} 
for the $Z$ boson in the $W^+ Z$ channel, we see that the EW corrections to the $Z$ decay have important effects 
and can change the DPA results by $30\%$. This explains why we see significant differences in the $A_3$ and $A_4$ coefficients 
when comparing the $\text{NLOQCD}$ results to the $\text{NLOQCDEW}$ ones for the case of the $Z$ boson in \sect{sect:numres:polar_observables_nloqcdew} and \appen{appen:numres_Wm}. 
However, similar effects are not observed for the $W^\pm$ decays. 

\begin{table}[ht!]
 \renewcommand{\arraystretch}{1.3}
  \begin{center}
\fontsize{7.0}{7.0}
\begin{tabular}{|c|c|c|c|c|c|c|c|c|}\hline
$\text{Method}$  & $A_0$ & $A_1$  & $A_2$ & $A_3$ & $A_4$ & $A_5$ & $A_6$ & $A_7$\\
\hline
$\text{HE F-LO}$ & $1.026$ & $-0.286$ & $-1.315$ & $-0.251$ & $-0.447$ & $-0.0021[3]$ & $-0.0006[4]$ & $-0.0036[3]$\\
$\text{HE F-EW-}\bar{q}q'$ & $1.034$ & $-0.284$ & $-1.321$ & $-0.251$ & $-0.451$ & $-0.0040[4]$ & $-0.0039[5]$ & $0.0030[4]$\\
$\text{HE F-EW-}q\gamma$ & $1.021$ & $-0.286$ & $-1.318$ & $-0.253$ & $-0.434$ & $-0.0021[3]$ & $-0.0006[4]$ & $-0.0035[3]$\\
$\text{HE F-EW}$ & $1.028$ & $-0.284$ & $-1.324$ & $-0.252$ & $-0.438$ & $-0.0039[4]$ & $-0.0038[5]$ & $0.0029[4]$\\
\hline
$\text{HE D-LO}$ & $1.023$ & $-0.326$ & $-1.404$ & $-0.156$ & $-0.445$ & $-0.0001[3]$ & $0.00001[39]$ & $0.0001[3]$\\
$\text{HE D-EW-p}W^+$ & $1.019$ & $-0.325$ & $-1.408$ & $-0.155$ & $-0.437$ & $-0.0019[4]$ & $-0.0033[4]$ & $0.0070[3]$\\
$\text{HE D-EW-d}W^+$ & $1.029$ & $-0.326$ & $-1.415$ & $-0.154$ & $-0.443$ & $-0.00001[35]$ & $0.00003[41]$ & $0.0001[3]$\\
$\text{HE D-EW}$ & $1.025$ & $-0.325$ & $-1.419$ & $-0.152$ & $-0.435$ & $-0.0018[4]$ & $-0.0033[4]$ & $0.0070[4]$\\
\hline\hline
$\text{CS F-LO}$ & $1.397$ & $0.229$ & $-0.945$ & $0.0025[3]$ & $-0.613$ & $-0.0002[4]$ & $0.0021[4]$ & $0.0036[3]$\\
$\text{CS F-EW-}\bar{q}q'$ & $1.399$ & $0.225$ & $-0.958$ & $-0.0004[4]$ & $-0.617$ & $0.0006[4]$ & $0.0059[5]$ & $-0.0030[4]$\\
$\text{CS F-EW-}q\gamma$ & $1.400$ & $0.229$ & $-0.940$ & $0.010$ & $-0.605$ & $-0.0002[4]$ & $0.0021[3]$ & $0.0035[3]$\\
$\text{CS F-EW}$ & $1.402$ & $0.225$ & $-0.952$ & $0.0080[4]$ & $-0.608$ & $0.0006[4]$ & $0.0059[4]$ & $-0.0029[3]$\\
\hline
$\text{CS D-LO}$ & $1.459$ & $0.299$ & $-0.971$ & $-0.073$ & $-0.544$ & $-0.00001[38]$ & $0.00003[38]$ & $-0.0001[3]$\\
$\text{CS D-EW-p}W^+$ & $1.460$ & $0.298$ & $-0.970$ & $-0.069$ & $-0.539$ & $0.0009[6]$ & $0.0038[4]$ & $-0.0070[3]$\\
$\text{CS D-EW-d}W^+$ & $1.465$ & $0.298$ & $-0.981$ & $-0.075$ & $-0.541$ & $-0.00005[33]$ & $0.00001[44]$ & $-0.0001[3]$\\
$\text{CS D-EW}$ & $1.466$ & $0.297$ & $-0.980$ & $-0.071$ & $-0.535$ & $0.0009[5]$ & $0.0038[4]$ & $-0.0070[4]$\\
\hline
\end{tabular}
\caption{\small Fiducial angular coefficients of the $e^+_{}$ distribution for
  the process $pp \to e^+_{} \nu_e^{}\, \mu^+_{} \mu^-_{} + X$ at the
  13 TeV LHC with the ATLAS fiducial cuts. Results are shown for full LO only, 
and also with the $\text{EW-}\bar{q}q'$, $\text{EW-}q\gamma$, and the total 
EW correction included. Similarly, results for DPA LO only, and also with 
the $\text{EW-pV}$, $\text{EW-dV}$, and the total EW correction included are presented. 
The upper rows are for the helicity (HE) coordinate system, while the lower ones 
for the Collins-Soper (CS) coordinate system. The numbers
  in square brackets represent the statistical error, when it is significant.}
\label{tab:coeff_Ai_full_DPA_Wp_Atlas}
\end{center}
\end{table} 

\begin{table}[ht!]
 \renewcommand{\arraystretch}{1.3}
  \begin{center}
    \fontsize{7.0}{7.0}
\begin{tabular}{|c|c|c|c|c|c|c|c|c|}\hline
$\text{Method}$  & $A_0$ & $A_1$  & $A_2$ & $A_3$ & $A_4$ & $A_5$ & $A_6$ & $A_7$\\
\hline
$\text{HE F-LO}$ & $1.035$ & $-0.303$ & $-0.705$ & $0.063$ & $-0.017$ & $-0.0068[4]$ & $-0.0066[4]$ & $0.0032[3]$\\
$\text{HE F-EW-}\bar{q}q'$ & $1.040$ & $-0.305$ & $-0.711$ & $0.051$ & $-0.021$ & $-0.0076[5]$ & $-0.0084[4]$ & $0.0030[3]$\\
$\text{HE F-EW-}q\gamma$ & $1.034$ & $-0.306$ & $-0.710$ & $0.062$ & $-0.016$ & $-0.0067[4]$ & $-0.0065[4]$ & $0.0032[3]$\\
$\text{HE F-EW}$ & $1.039$ & $-0.307$ & $-0.717$ & $0.050$ & $-0.020$ & $-0.0074[5]$ & $-0.0082[5]$ & $0.0029[3]$\\
\hline
$\text{HE D-LO}$ & $0.997$ & $-0.265$ & $-0.720$ & $0.039$ & $0.011$ & $-0.00001[49]$ & $0.00004[41]$ & $0.0001[3]$\\
$\text{HE D-EW-p}Z$ & $0.997$ & $-0.266$ & $-0.727$ & $0.039$ & $0.011$ & $-0.0003[5]$ & $-0.0012[4]$ & $-0.0004[2]$\\
$\text{HE D-EW-d}Z$ & $0.999$ & $-0.267$ & $-0.725$ & $0.024$ & $0.0078[2]$ & $-0.00003[53]$ & $0.00003[45]$ & $0.0001[3]$\\
$\text{HE D-EW}$ & $0.999$ & $-0.268$ & $-0.733$ & $0.025$ & $0.0078[2]$ & $-0.0003[6]$ & $-0.0013[5]$ & $-0.0004[3]$\\
\hline\hline
$\text{CS F-LO}$ & $1.254$ & $0.239$ & $-0.488$ & $-0.061$ & $0.035$ & $-0.0001[5]$ & $0.010$ & $-0.0032[3]$\\
$\text{CS F-EW-}\bar{q}q'$ & $1.260$ & $0.236$ & $-0.494$ & $-0.054$ & $0.023$ & $0.0008[6]$ & $0.012$ & $-0.0030[3]$\\
$\text{CS F-EW-}q\gamma$ & $1.259$ & $0.238$ & $-0.487$ & $-0.059$ & $0.035$ & $-0.0001[5]$ & $0.010$ & $-0.0032[3]$\\
$\text{CS F-EW}$ & $1.266$ & $0.234$ & $-0.493$ & $-0.053$ & $0.023$ & $0.0007[6]$ & $0.012$ & $-0.0029[3]$\\
\hline
$\text{CS D-LO}$ & $1.200$ & $0.305$ & $-0.519$ & $-0.023$ & $0.036$ & $-0.0001[6]$ & $0.00002[30]$ & $-0.0001[3]$\\
$\text{CS D-EW-p}Z$ & $1.205$ & $0.302$ & $-0.521$ & $-0.023$ & $0.036$ & $0.0008[6]$ & $0.0012[3]$ & $0.0004[3]$\\
$\text{CS D-EW-d}Z$ & $1.205$ & $0.307$ & $-0.522$ & $-0.013$ & $0.023$ & $-0.00004[70]$ & $0.00003[25]$ & $-0.0001[2]$\\
$\text{CS D-EW}$ & $1.209$ & $0.303$ & $-0.525$ & $-0.013$ & $0.023$ & $0.0009[7]$ & $0.0012[2]$ & $0.0004[3]$\\
\hline
\end{tabular}
\caption{\small Same as \tab{tab:coeff_Ai_full_DPA_Wp_Atlas} but for
  $\mu^{-}$ distribution.}
\label{tab:coeff_Ai_full_DPA_Z_processWp_Atlas}
\end{center}
\end{table} 

\begin{table}[ht!]
 \renewcommand{\arraystretch}{1.3}
  \begin{center}
\fontsize{7.0}{7.0}
\begin{tabular}{|c|c|c|c|c|c|c|c|c|}\hline
$\text{Method}$  & $A_0$ & $A_1$  & $A_2$ & $A_3$ & $A_4$ & $A_5$ & $A_6$ & $A_7$\\
\hline
$\text{HE F-LO}$ & $0.897$ & $0.088$ & $-0.627$ & $-0.373$ & $-0.488$ & $-0.0019[3]$ & $-0.0008[3]$ & $-0.0031[3]$\\
$\text{HE F-EW-}\bar{q}q'$ & $0.904$ & $0.092$ & $-0.629$ & $-0.375$ & $-0.493$ & $-0.0037[3]$ & $-0.0038[3]$ & $0.0030[3]$\\
$\text{HE F-EW-}q\gamma$ & $0.893$ & $0.089$ & $-0.622$ & $-0.373$ & $-0.475$ & $-0.0018[3]$ & $-0.0008[3]$ & $-0.0030[3]$\\
$\text{HE F-EW}$ & $0.899$ & $0.092$ & $-0.625$ & $-0.374$ & $-0.480$ & $-0.0036[3]$ & $-0.0038[3]$ & $0.0029[3]$\\
\hline
$\text{HE D-LO}$ & $0.900$ & $0.037$ & $-0.740$ & $-0.258$ & $-0.475$ & $-0.00003[41]$ & $0.0001[3]$ & $0.0001[3]$\\
$\text{HE D-EW-p}W^+$ & $0.898$ & $0.039$ & $-0.734$ & $-0.258$ & $-0.468$ & $-0.0018[4]$ & $-0.0030[3]$ & $0.0064[3]$\\
$\text{HE D-EW-d}W^+$ & $0.906$ & $0.037$ & $-0.747$ & $-0.257$ & $-0.474$ & $-0.00003[41]$ & $0.0001[3]$ & $0.0001[3]$\\
$\text{HE D-EW}$ & $0.903$ & $0.040$ & $-0.741$ & $-0.257$ & $-0.467$ & $-0.0018[4]$ & $-0.0030[3]$ & $0.0065[3]$\\
\hline\hline
$\text{CS F-LO}$ & $0.760$ & $0.196$ & $-0.764$ & $0.052$ & $-0.723$ & $-0.00002[28]$ & $0.0021[4]$ & $0.0031[3]$\\
$\text{CS F-EW-}\bar{q}q'$ & $0.758$ & $0.191$ & $-0.775$ & $0.050$ & $-0.728$ & $0.0009[3]$ & $0.0056[5]$ & $-0.0030[3]$\\
$\text{CS F-EW-}q\gamma$ & $0.759$ & $0.196$ & $-0.756$ & $0.059$ & $-0.714$ & $-0.00001[27]$ & $0.0020[4]$ & $0.0030[3]$\\
$\text{CS F-EW}$ & $0.758$ & $0.192$ & $-0.767$ & $0.057$ & $-0.719$ & $0.0008[3]$ & $0.0055[5]$ & $-0.0029[3]$\\
\hline
$\text{CS D-LO}$ & $0.839$ & $0.280$ & $-0.802$ & $-0.027$ & $-0.633$ & $-0.0001[3]$ & $-0.000003[432]$ & $-0.0001[3]$\\
$\text{CS D-EW-p}W^+$ & $0.835$ & $0.278$ & $-0.798$ & $-0.022$ & $-0.629$ & $0.0008[4]$ & $0.0035[5]$ & $-0.0064[3]$\\
$\text{CS D-EW-d}W^+$ & $0.843$ & $0.279$ & $-0.811$ & $-0.028$ & $-0.631$ & $-0.0001[4]$ & $0.00001[46]$ & $-0.0002[3]$\\
$\text{CS D-EW}$ & $0.839$ & $0.277$ & $-0.806$ & $-0.023$ & $-0.627$ & $0.0008[4]$ & $0.0036[5]$ & $-0.0065[3]$\\
\hline
\end{tabular}
\caption{\small Same as \tab{tab:coeff_Ai_full_DPA_Wp_Atlas} but with the CMS fiducial cuts.}
\label{tab:coeff_Ai_full_DPA_Wp_Cms}
\end{center}
\end{table} 

\begin{table}[ht!]
 \renewcommand{\arraystretch}{1.3}
  \begin{center}
    \fontsize{7.0}{7.0}
\begin{tabular}{|c|c|c|c|c|c|c|c|c|}\hline
$\text{Method}$  & $A_0$ & $A_1$  & $A_2$ & $A_3$ & $A_4$ & $A_5$ & $A_6$ & $A_7$\\
\hline
$\text{HE F-LO}$ & $0.858$ & $-0.273$ & $-0.570$ & $0.068$ & $-0.022$ & $-0.0071[4]$ & $-0.0064[3]$ & $0.0028[3]$\\
$\text{HE F-EW-}\bar{q}q'$ & $0.858$ & $-0.278$ & $-0.568$ & $0.056$ & $-0.027$ & $-0.0076[5]$ & $-0.0078[3]$ & $0.0024[3]$\\
$\text{HE F-EW-}q\gamma$ & $0.855$ & $-0.277$ & $-0.576$ & $0.067$ & $-0.022$ & $-0.0070[4]$ & $-0.0063[3]$ & $0.0028[3]$\\
$\text{HE F-EW}$ & $0.855$ & $-0.282$ & $-0.574$ & $0.055$ & $-0.026$ & $-0.0075[5]$ & $-0.0077[3]$ & $0.0024[3]$\\
\hline
$\text{HE D-LO}$ & $0.806$ & $-0.227$ & $-0.592$ & $0.039$ & $0.012$ & $0.00003[39]$ & $0.0001[3]$ & $0.0001[2]$\\
$\text{HE D-EW-p}Z$ & $0.805$ & $-0.229$ & $-0.599$ & $0.040$ & $0.012$ & $-0.0003[4]$ & $-0.0011[3]$ & $-0.0004[3]$\\
$\text{HE D-EW-d}Z$ & $0.803$ & $-0.232$ & $-0.590$ & $0.025$ & $0.0082[3]$ & $0.0001[4]$ & $0.0001[3]$ & $0.00002[18]$\\
$\text{HE D-EW}$ & $0.802$ & $-0.234$ & $-0.596$ & $0.026$ & $0.0082[3]$ & $-0.0002[5]$ & $-0.0011[3]$ & $-0.0004[3]$\\
\hline\hline
$\text{CS F-LO}$ & $1.128$ & $0.296$ & $-0.303$ & $-0.069$ & $0.038$ & $-0.0003[4]$ & $0.010$ & $-0.0028[3]$\\
$\text{CS F-EW-}\bar{q}q'$ & $1.134$ & $0.292$ & $-0.294$ & $-0.063$ & $0.025$ & $0.0005[4]$ & $0.012$ & $-0.0024[3]$\\
$\text{CS F-EW-}q\gamma$ & $1.135$ & $0.294$ & $-0.298$ & $-0.068$ & $0.038$ & $-0.0003[4]$ & $0.010$ & $-0.0028[3]$\\
$\text{CS F-EW}$ & $1.141$ & $0.291$ & $-0.290$ & $-0.062$ & $0.025$ & $0.0005[4]$ & $0.012$ & $-0.0024[3]$\\
\hline
$\text{CS D-LO}$ & $1.065$ & $0.378$ & $-0.337$ & $-0.022$ & $0.039$ & $-0.0001[4]$ & $-0.00002[33]$ & $-0.0001[2]$\\
$\text{CS D-EW-p}Z$ & $1.071$ & $0.374$ & $-0.336$ & $-0.022$ & $0.039$ & $0.0008[4]$ & $0.0011[3]$ & $0.0004[3]$\\
$\text{CS D-EW-d}Z$ & $1.071$ & $0.378$ & $-0.325$ & $-0.013$ & $0.025$ & $-0.0001[4]$ & $-0.0001[3]$ & $-0.00004[18]$\\
$\text{CS D-EW}$ & $1.077$ & $0.374$ & $-0.324$ & $-0.014$ & $0.025$ & $0.0008[3]$ & $0.0011[4]$ & $0.0004[3]$\\
\hline
\end{tabular}
\caption{\small Same as \tab{tab:coeff_Ai_full_DPA_Z_processWp_Atlas} but with the CMS fiducial cuts.}
\label{tab:coeff_Ai_full_DPA_Z_processWp_Cms}
\end{center}
\end{table} 

\begin{table}[ht!]
 \renewcommand{\arraystretch}{1.3}
  \begin{center}
\fontsize{8.5}{8.5}
\begin{tabular}{|c|c|c|c||c|c|c|}\hline
$\text{Method}$  & $f^{W^+}_L$ & $f^{W^+}_0$ & $f^{W^+}_R$ & $f^Z_L$ & $f^Z_0$ & $f^Z_R$\\
\hline
$\text{HE F-LO}$ & $0.355$ & $0.513$ & $0.132$ & $0.222$ & $0.518$ & $0.261$\\
$\text{HE F-EW-}\bar{q}q'$ & $0.354$ & $0.517$ & $0.129$ & $0.215$ & $0.520$ & $0.264$\\
$\text{HE F-EW-}q\gamma$ & $0.353$ & $0.510$ & $0.136$ & $0.223$ & $0.517$ & $0.260$\\
$\text{HE F-EW}$ & $0.352$ & $0.514$ & $0.134$ & $0.216$ & $0.519$ & $0.264$\\
\hline
$\text{HE D-LO}$ & $0.355$ & $0.512$ & $0.133$ & $0.263$ & $0.498$ & $0.239$\\
$\text{HE D-EW-pV}$ & $0.354$ & $0.510$ & $0.136$ & $0.263$ & $0.498$ & $0.239$\\
$\text{HE D-EW-dV}$ & $0.353$ & $0.515$ & $0.132$ & $0.259$ & $0.499$ & $0.241$\\
$\text{HE D-EW}$ & $0.352$ & $0.513$ & $0.135$ & $0.259$ & $0.499$ & $0.241$\\
\hline\hline
$\text{CS F-LO}$ & $0.304$ & $0.699$ & $-0.0025[2]$ & $0.228$ & $0.627$ & $0.145$\\
$\text{CS F-EW-}\bar{q}q'$ & $0.304$ & $0.699$ & $-0.0038[2]$ & $0.212$ & $0.630$ & $0.158$\\
$\text{CS F-EW-}q\gamma$ & $0.301$ & $0.700$ & $-0.0013[2]$ & $0.226$ & $0.630$ & $0.144$\\
$\text{CS F-EW}$ & $0.302$ & $0.701$ & $-0.0025[2]$ & $0.210$ & $0.633$ & $0.157$\\
\hline
$\text{CS D-LO}$ & $0.271$ & $0.729$ & $-0.0005[3]$ & $0.242$ & $0.600$ & $0.158$\\
$\text{CS D-EW-pV}$ & $0.270$ & $0.730$ & $0.0004[3]$ & $0.241$ & $0.603$ & $0.157$\\
$\text{CS D-EW-dV}$ & $0.269$ & $0.732$ & $-0.0014[3]$ & $0.226$ & $0.602$ & $0.172$\\
$\text{CS D-EW}$ & $0.267$ & $0.733$ & $-0.0004[3]$ & $0.225$ & $0.605$ & $0.171$\\
\hline
\end{tabular}
\caption{\small Fiducial polarization fractions of $W^+_{}$ and $Z$ bosons in
  the process $pp \to e^+_{} \nu_e^{}\, \mu^+_{} \mu^-_{} + X$ at the
  13 TeV LHC with the ATLAS fiducial cuts. Results are shown for full LO only, 
and also with the $\text{EW-}\bar{q}q'$, $\text{EW-}q\gamma$, and the total 
EW correction included. Similarly, results for DPA LO only, and also with 
the $\text{EW-pV}$, $\text{EW-dV}$, and the total EW correction included are presented. 
The upper rows are for the helicity (HE) coordinate system, while the lower ones 
for the Collins-Soper (CS) coordinate system. The numbers
  in square brackets represent the statistical error, when it is significant.}
\label{tab:coeff_fL0R_full_DPA_Wp_Atlas}
\end{center}
\end{table} 
\begin{table}[ht!]
 \renewcommand{\arraystretch}{1.3}
  \begin{center}
\fontsize{8.5}{8.5}
\begin{tabular}{|c|c|c|c||c|c|c|}\hline
$\text{Method}$  & $f^{W^+}_L$ & $f^{W^+}_0$ & $f^{W^+}_R$ & $f^Z_L$ & $f^Z_0$ & $f^Z_R$\\
\hline
$\text{HE F-LO}$ & $0.398$ & $0.448$ & $0.154$ & $0.260$ & $0.429$ & $0.312$\\
$\text{HE F-EW-}\bar{q}q'$ & $0.397$ & $0.452$ & $0.151$ & $0.254$ & $0.429$ & $0.317$\\
$\text{HE F-EW-}q\gamma$ & $0.396$ & $0.446$ & $0.158$ & $0.261$ & $0.427$ & $0.312$\\
$\text{HE F-EW}$ & $0.395$ & $0.450$ & $0.155$ & $0.256$ & $0.427$ & $0.317$\\
\hline
$\text{HE D-LO}$ & $0.394$ & $0.450$ & $0.156$ & $0.312$ & $0.403$ & $0.285$\\
$\text{HE D-EW-pV}$ & $0.393$ & $0.449$ & $0.159$ & $0.312$ & $0.403$ & $0.285$\\
$\text{HE D-EW-dV}$ & $0.392$ & $0.453$ & $0.155$ & $0.309$ & $0.402$ & $0.290$\\
$\text{HE D-EW}$ & $0.391$ & $0.452$ & $0.158$ & $0.309$ & $0.401$ & $0.290$\\
\hline\hline
$\text{CS F-LO}$ & $0.491$ & $0.380$ & $0.129$ & $0.262$ & $0.564$ & $0.174$\\
$\text{CS F-EW-}\bar{q}q'$ & $0.492$ & $0.379$ & $0.128$ & $0.246$ & $0.567$ & $0.187$\\
$\text{CS F-EW-}q\gamma$ & $0.489$ & $0.380$ & $0.132$ & $0.260$ & $0.567$ & $0.172$\\
$\text{CS F-EW}$ & $0.490$ & $0.379$ & $0.131$ & $0.244$ & $0.571$ & $0.185$\\
\hline
$\text{CS D-LO}$ & $0.448$ & $0.420$ & $0.132$ & $0.279$ & $0.532$ & $0.189$\\
$\text{CS D-EW-pV}$ & $0.449$ & $0.418$ & $0.134$ & $0.277$ & $0.535$ & $0.187$\\
$\text{CS D-EW-dV}$ & $0.447$ & $0.421$ & $0.132$ & $0.262$ & $0.536$ & $0.203$\\
$\text{CS D-EW}$ & $0.447$ & $0.419$ & $0.134$ & $0.260$ & $0.539$ & $0.201$\\
\hline
\end{tabular}
\caption{\small Same as \tab{tab:coeff_fL0R_full_DPA_Wp_Atlas} but with the CMS fiducial cuts.}
\label{tab:coeff_fL0R_full_DPA_Wp_Cms}
\end{center}
\end{table} 


\begin{table}[ht!]
 \renewcommand{\arraystretch}{1.3}
  \begin{center}
\fontsize{7.0}{7.0}
\begin{tabular}{|c|c|c|c|c|c|c|c|c|}\hline
$\text{Method}$  & $A_0$ & $A_1$  & $A_2$ & $A_3$ & $A_4$ & $A_5$ & $A_6$ & $A_7$\\
\hline
$\text{HE F-LO}$ & $1.110$ & $-0.389$ & $-1.359$ & $-0.127$ & $-0.025$ & $0.0002[4]$ & $0.0035[3]$ & $0.011$\\
$\text{HE F-EW-}\bar{q}q'$ & $1.116$ & $-0.386$ & $-1.371$ & $-0.120$ & $-0.030$ & $0.0016[6]$ & $0.0062[4]$ & $0.010$\\
$\text{HE F-EW-}q\gamma$ & $1.102$ & $-0.387$ & $-1.361$ & $-0.132$ & $-0.016$ & $0.0002[4]$ & $0.0034[3]$ & $0.011$\\
$\text{HE F-EW}$ & $1.107$ & $-0.385$ & $-1.372$ & $-0.126$ & $-0.021$ & $0.0016[6]$ & $0.0061[4]$ & $0.010$\\
\hline
$\text{HE D-LO}$ & $1.087$ & $-0.394$ & $-1.347$ & $-0.162$ & $0.061$ & $0.00001[50]$ & $0.0001[2]$ & $0.0001[4]$\\
$\text{HE D-EW-p}W^-$ & $1.078$ & $-0.390$ & $-1.350$ & $-0.164$ & $0.068$ & $0.0013[6]$ & $0.0025[3]$ & $-0.0016[3]$\\
$\text{HE D-EW-d}W^-$ & $1.091$ & $-0.395$ & $-1.357$ & $-0.161$ & $0.061$ & $0.0001[5]$ & $0.0001[3]$ & $0.0001[4]$\\
$\text{HE D-EW}$ & $1.083$ & $-0.391$ & $-1.359$ & $-0.163$ & $0.068$ & $0.0014[7]$ & $0.0025[4]$ & $-0.0016[4]$\\
\hline\hline
$\text{CS F-LO}$ & $1.578$ & $0.214$ & $-0.893$ & $0.139$ & $-0.123$ & $-0.0028[4]$ & $-0.0027[4]$ & $-0.011$\\
$\text{CS F-EW-}\bar{q}q'$ & $1.579$ & $0.213$ & $-0.909$ & $0.131$ & $-0.123$ & $-0.0040[7]$ & $-0.0056[4]$ & $-0.010$\\
$\text{CS F-EW-}q\gamma$ & $1.578$ & $0.214$ & $-0.887$ & $0.149$ & $-0.124$ & $-0.0028[4]$ & $-0.0026[3]$ & $-0.011$\\
$\text{CS F-EW}$ & $1.580$ & $0.213$ & $-0.902$ & $0.141$ & $-0.124$ & $-0.0039[7]$ & $-0.0055[4]$ & $-0.010$\\
\hline
$\text{CS D-LO}$ & $1.585$ & $0.247$ & $-0.850$ & $0.230$ & $-0.102$ & $-0.0001[5]$ & $-0.00003[30]$ & $-0.0001[4]$\\
$\text{CS D-EW-p}W^-$ & $1.582$ & $0.247$ & $-0.847$ & $0.237$ & $-0.103$ & $-0.0011[6]$ & $-0.0027[3]$ & $0.0016[4]$\\
$\text{CS D-EW-d}W^-$ & $1.590$ & $0.246$ & $-0.860$ & $0.228$ & $-0.101$ & $0.00003[59]$ & $0.000003[348]$ & $-0.0001[4]$\\
$\text{CS D-EW}$ & $1.587$ & $0.247$ & $-0.857$ & $0.235$ & $-0.102$ & $-0.0010[8]$ & $-0.0027[3]$ & $0.0016[4]$\\
\hline
\end{tabular}
\caption{\small Fiducial angular coefficients of the $e^-_{}$ distribution for
  the process $pp \to e^-_{} \bar{\nu}_e^{}\, \mu^+_{} \mu^-_{} + X$ at the
  13 TeV LHC with the ATLAS fiducial cuts. Results are shown for full LO only, 
and also with the $\text{EW-}\bar{q}q'$, $\text{EW-}q\gamma$, and the total 
EW correction included. Similarly, results for DPA LO only, and also with 
the $\text{EW-pV}$, $\text{EW-dV}$, and the total EW correction included are presented. 
The upper rows are for the helicity (HE) coordinate system, while the lower ones 
for the Collins-Soper (CS) coordinate system. The numbers
  in square brackets represent the statistical error, when it is significant.}
\label{tab:coeff_Ai_full_DPA_Wm_Atlas}
\end{center}
\end{table} 

\begin{table}[ht!]
 \renewcommand{\arraystretch}{1.3}
  \begin{center}
    \fontsize{7.0}{7.0}
\begin{tabular}{|c|c|c|c|c|c|c|c|c|}\hline
$\text{Method}$  & $A_0$ & $A_1$  & $A_2$ & $A_3$ & $A_4$ & $A_5$ & $A_6$ & $A_7$\\
\hline
$\text{HE F-LO}$ & $0.989$ & $-0.326$ & $-0.736$ & $-0.014$ & $0.062$ & $0.0030[3]$ & $0.0093[2]$ & $0.0063[4]$\\
$\text{HE F-EW-}\bar{q}q'$ & $0.993$ & $-0.328$ & $-0.742$ & $-0.0082[2]$ & $0.039$ & $0.0037[4]$ & $0.011$ & $0.0065[5]$\\
$\text{HE F-EW-}q\gamma$ & $0.988$ & $-0.327$ & $-0.741$ & $-0.014$ & $0.061$ & $0.0029[3]$ & $0.0092[2]$ & $0.0062[4]$\\
$\text{HE F-EW}$ & $0.992$ & $-0.330$ & $-0.748$ & $-0.0084[2]$ & $0.039$ & $0.0036[4]$ & $0.011$ & $0.0064[5]$\\
\hline
$\text{HE D-LO}$ & $0.965$ & $-0.322$ & $-0.743$ & $-0.019$ & $0.068$ & $-0.0001[2]$ & $-0.00005[17]$ & $0.0001[4]$\\
$\text{HE D-EW-p}Z$ & $0.965$ & $-0.324$ & $-0.750$ & $-0.020$ & $0.070$ & $0.0005[2]$ & $0.0011[2]$ & $-0.0001[5]$\\
$\text{HE D-EW-d}Z$ & $0.967$ & $-0.325$ & $-0.748$ & $-0.012$ & $0.044$ & $-0.0002[3]$ & $0.00002[15]$ & $0.0002[4]$\\
$\text{HE D-EW}$ & $0.967$ & $-0.327$ & $-0.755$ & $-0.014$ & $0.046$ & $0.0004[3]$ & $0.0012[2]$ & $-0.00003[51]$\\
\hline\hline
$\text{CS F-LO}$ & $1.251$ & $0.313$ & $-0.477$ & $0.050$ & $0.048$ & $-0.0044[3]$ & $-0.010$ & $-0.0063[4]$\\
$\text{CS F-EW-}\bar{q}q'$ & $1.258$ & $0.311$ & $-0.481$ & $0.031$ & $0.031$ & $-0.0055[4]$ & $-0.011$ & $-0.0065[5]$\\
$\text{CS F-EW-}q\gamma$ & $1.256$ & $0.310$ & $-0.476$ & $0.050$ & $0.047$ & $-0.0043[3]$ & $-0.0094[2]$ & $-0.0062[4]$\\
$\text{CS F-EW}$ & $1.263$ & $0.308$ & $-0.480$ & $0.032$ & $0.030$ & $-0.0054[4]$ & $-0.011$ & $-0.0064[5]$\\
\hline
$\text{CS D-LO}$ & $1.239$ & $0.361$ & $-0.472$ & $0.058$ & $0.049$ & $-0.0001[3]$ & $0.0001[1]$ & $-0.0001[4]$\\
$\text{CS D-EW-p}Z$ & $1.245$ & $0.357$ & $-0.473$ & $0.061$ & $0.049$ & $-0.0007[3]$ & $-0.0011[2]$ & $0.0001[5]$\\
$\text{CS D-EW-d}Z$ & $1.245$ & $0.362$ & $-0.474$ & $0.038$ & $0.032$ & $-0.0002[4]$ & $0.0002[2]$ & $-0.0002[4]$\\
$\text{CS D-EW}$ & $1.251$ & $0.358$ & $-0.475$ & $0.041$ & $0.031$ & $-0.0009[4]$ & $-0.0011[3]$ & $0.00005[48]$\\
\hline
\end{tabular}
\caption{\small Same as \tab{tab:coeff_Ai_full_DPA_Wm_Atlas} but for 
the $\mu^-$ distribution.}
\label{tab:coeff_Ai_full_DPA_Z_processWm_Atlas}
\end{center}
\end{table} 


\begin{table}[ht!]
 \renewcommand{\arraystretch}{1.3}
  \begin{center}
\fontsize{7.0}{7.0}
\begin{tabular}{|c|c|c|c|c|c|c|c|c|}\hline
$\text{Method}$  & $A_0$ & $A_1$  & $A_2$ & $A_3$ & $A_4$ & $A_5$ & $A_6$ & $A_7$\\
\hline
$\text{HE F-LO}$ & $1.006$ & $-0.079$ & $-0.743$ & $-0.156$ & $-0.0020[3]$ & $0.0004[4]$ & $0.0033[3]$ & $0.010$\\
$\text{HE F-EW-}\bar{q}q'$ & $1.011$ & $-0.075$ & $-0.750$ & $-0.151$ & $-0.0066[3]$ & $0.0018[5]$ & $0.0057[4]$ & $0.0092[3]$\\
$\text{HE F-EW-}q\gamma$ & $0.998$ & $-0.076$ & $-0.735$ & $-0.161$ & $0.0053[3]$ & $0.0004[4]$ & $0.0032[3]$ & $0.010$\\
$\text{HE F-EW}$ & $1.003$ & $-0.072$ & $-0.742$ & $-0.156$ & $0.0010[3]$ & $0.0018[5]$ & $0.0055[4]$ & $0.0091[3]$\\
\hline
$\text{HE D-LO}$ & $0.983$ & $-0.075$ & $-0.718$ & $-0.176$ & $0.083$ & $0.00004[50]$ & $0.00004[24]$ & $0.0001[3]$\\
$\text{HE D-EW-p}W^-$ & $0.976$ & $-0.068$ & $-0.711$ & $-0.178$ & $0.087$ & $0.0014[6]$ & $0.0022[3]$ & $-0.0014[3]$\\
$\text{HE D-EW-d}W^-$ & $0.987$ & $-0.074$ & $-0.724$ & $-0.175$ & $0.084$ & $0.0001[5]$ & $0.00003[26]$ & $0.0001[3]$\\
$\text{HE D-EW}$ & $0.980$ & $-0.068$ & $-0.717$ & $-0.177$ & $0.088$ & $0.0015[6]$ & $0.0022[3]$ & $-0.0014[3]$\\
\hline\hline
$\text{CS F-LO}$ & $1.014$ & $0.215$ & $-0.735$ & $0.182$ & $-0.131$ & $-0.0026[4]$ & $-0.0027[4]$ & $-0.010$\\
$\text{CS F-EW-}\bar{q}q'$ & $1.012$ & $0.214$ & $-0.749$ & $0.175$ & $-0.131$ & $-0.0035[4]$ & $-0.0054[5]$ & $-0.0092[3]$\\
$\text{CS F-EW-}q\gamma$ & $1.009$ & $0.216$ & $-0.725$ & $0.191$ & $-0.133$ & $-0.0025[3]$ & $-0.0027[4]$ & $-0.010$\\
$\text{CS F-EW}$ & $1.007$ & $0.214$ & $-0.739$ & $0.184$ & $-0.133$ & $-0.0034[4]$ & $-0.0053[5]$ & $-0.0091[3]$\\
\hline
$\text{CS D-LO}$ & $1.005$ & $0.257$ & $-0.696$ & $0.263$ & $-0.101$ & $0.00001[45]$ & $-0.00003[33]$ & $-0.0001[3]$\\
$\text{CS D-EW-p}W^-$ & $0.997$ & $0.257$ & $-0.691$ & $0.267$ & $-0.103$ & $-0.0008[5]$ & $-0.0026[4]$ & $0.0014[3]$\\
$\text{CS D-EW-d}W^-$ & $1.008$ & $0.257$ & $-0.704$ & $0.262$ & $-0.100$ & $0.00004[43]$ & $-0.0001[4]$ & $-0.0001[3]$\\
$\text{CS D-EW}$ & $0.999$ & $0.257$ & $-0.699$ & $0.266$ & $-0.102$ & $-0.0008[5]$ & $-0.0026[5]$ & $0.0014[2]$\\
\hline
\end{tabular}
\caption{\small Same as \tab{tab:coeff_Ai_full_DPA_Wm_Atlas} but with the CMS fiducial cuts.}
\label{tab:coeff_Ai_full_DPA_Wm_Cms}
\end{center}
\end{table} 

\begin{table}[ht!]
 \renewcommand{\arraystretch}{1.3}
  \begin{center}
    \fontsize{7.0}{7.0}
\begin{tabular}{|c|c|c|c|c|c|c|c|c|}\hline
$\text{Method}$  & $A_0$ & $A_1$  & $A_2$ & $A_3$ & $A_4$ & $A_5$ & $A_6$ & $A_7$\\
\hline
$\text{HE F-LO}$ & $0.798$ & $-0.288$ & $-0.603$ & $-0.017$ & $0.072$ & $0.0032[3]$ & $0.0094[2]$ & $0.0064[3]$\\
$\text{HE F-EW-}\bar{q}q'$ & $0.797$ & $-0.294$ & $-0.601$ & $-0.012$ & $0.048$ & $0.0039[3]$ & $0.011$ & $0.0065[4]$\\
$\text{HE F-EW-}q\gamma$ & $0.796$ & $-0.291$ & $-0.608$ & $-0.017$ & $0.071$ & $0.0031[2]$ & $0.0093[2]$ & $0.0063[3]$\\
$\text{HE F-EW}$ & $0.795$ & $-0.297$ & $-0.606$ & $-0.012$ & $0.048$ & $0.0038[3]$ & $0.011$ & $0.0063[4]$\\
\hline
$\text{HE D-LO}$ & $0.762$ & $-0.286$ & $-0.616$ & $-0.018$ & $0.073$ & $-0.0001[2]$ & $-0.0001[1]$ & $0.0001[3]$\\
$\text{HE D-EW-p}Z$ & $0.761$ & $-0.289$ & $-0.621$ & $-0.020$ & $0.075$ & $0.0005[2]$ & $0.0012[1]$ & $-0.0001[4]$\\
$\text{HE D-EW-d}Z$ & $0.759$ & $-0.292$ & $-0.613$ & $-0.012$ & $0.048$ & $-0.0001[3]$ & $-0.0001[2]$ & $0.0001[4]$\\
$\text{HE D-EW}$ & $0.758$ & $-0.295$ & $-0.619$ & $-0.013$ & $0.050$ & $0.0005[3]$ & $0.0012[3]$ & $-0.0001[4]$\\
\hline\hline
$\text{CS F-LO}$ & $1.113$ & $0.382$ & $-0.291$ & $0.061$ & $0.050$ & $-0.0046[2]$ & $-0.010$ & $-0.0064[3]$\\
$\text{CS F-EW-}\bar{q}q'$ & $1.120$ & $0.380$ & $-0.281$ & $0.042$ & $0.033$ & $-0.0055[3]$ & $-0.011$ & $-0.0065[4]$\\
$\text{CS F-EW-}q\gamma$ & $1.121$ & $0.379$ & $-0.286$ & $0.061$ & $0.049$ & $-0.0045[2]$ & $-0.0094[2]$ & $-0.0063[3]$\\
$\text{CS F-EW}$ & $1.128$ & $0.377$ & $-0.276$ & $0.042$ & $0.032$ & $-0.0054[3]$ & $-0.011$ & $-0.0063[4]$\\
\hline
$\text{CS D-LO}$ & $1.104$ & $0.442$ & $-0.276$ & $0.063$ & $0.051$ & $-0.00004[25]$ & $0.0001[2]$ & $-0.0001[3]$\\
$\text{CS D-EW-p}Z$ & $1.112$ & $0.438$ & $-0.274$ & $0.065$ & $0.050$ & $-0.0008[3]$ & $-0.0013[2]$ & $0.0001[4]$\\
$\text{CS D-EW-d}Z$ & $1.111$ & $0.442$ & $-0.264$ & $0.041$ & $0.033$ & $-0.0001[4]$ & $0.0001[2]$ & $-0.0001[4]$\\
$\text{CS D-EW}$ & $1.119$ & $0.438$ & $-0.261$ & $0.043$ & $0.033$ & $-0.0008[4]$ & $-0.0012[3]$ & $0.0001[5]$\\
\hline
\end{tabular}
\caption{\small Same as \tab{tab:coeff_Ai_full_DPA_Z_processWm_Atlas} but 
with the CMS fiducial cuts.}
\label{tab:coeff_Ai_full_DPA_Z_processWm_Cms}
\end{center}
\end{table} 

\begin{table}[ht!]
 \renewcommand{\arraystretch}{1.3}
  \begin{center}
\fontsize{8.5}{8.5}
\begin{tabular}{|c|c|c|c||c|c|c|}\hline
$\text{Method}$  & $f^{W^-}_L$ & $f^{W^-}_0$ & $f^{W^-}_R$ & $f^Z_L$ & $f^Z_0$ & $f^Z_R$\\
\hline
$\text{HE F-LO}$ & $0.216$ & $0.555$ & $0.229$ & $0.325$ & $0.494$ & $0.181$\\
$\text{HE F-EW-}\bar{q}q'$ & $0.214$ & $0.558$ & $0.228$ & $0.298$ & $0.497$ & $0.206$\\
$\text{HE F-EW-}q\gamma$ & $0.220$ & $0.551$ & $0.229$ & $0.324$ & $0.494$ & $0.182$\\
$\text{HE F-EW}$ & $0.218$ & $0.554$ & $0.228$ & $0.298$ & $0.496$ & $0.206$\\
\hline
$\text{HE D-LO}$ & $0.244$ & $0.543$ & $0.213$ & $0.338$ & $0.482$ & $0.180$\\
$\text{HE D-EW-pV}$ & $0.247$ & $0.539$ & $0.213$ & $0.340$ & $0.482$ & $0.178$\\
$\text{HE D-EW-dV}$ & $0.242$ & $0.546$ & $0.212$ & $0.310$ & $0.484$ & $0.206$\\
$\text{HE D-EW}$ & $0.246$ & $0.541$ & $0.212$ & $0.312$ & $0.484$ & $0.205$\\
\hline\hline
$\text{CS F-LO}$ & $0.075$ & $0.789$ & $0.136$ & $0.243$ & $0.625$ & $0.132$\\
$\text{CS F-EW-}\bar{q}q'$ & $0.074$ & $0.790$ & $0.136$ & $0.221$ & $0.629$ & $0.150$\\
$\text{CS F-EW-}q\gamma$ & $0.074$ & $0.789$ & $0.137$ & $0.241$ & $0.628$ & $0.131$\\
$\text{CS F-EW}$ & $0.074$ & $0.790$ & $0.136$ & $0.220$ & $0.632$ & $0.149$\\
\hline
$\text{CS D-LO}$ & $0.078$ & $0.793$ & $0.129$ & $0.247$ & $0.620$ & $0.134$\\
$\text{CS D-EW-pV}$ & $0.079$ & $0.791$ & $0.130$ & $0.245$ & $0.623$ & $0.132$\\
$\text{CS D-EW-dV}$ & $0.077$ & $0.795$ & $0.128$ & $0.226$ & $0.622$ & $0.152$\\
$\text{CS D-EW}$ & $0.078$ & $0.794$ & $0.129$ & $0.224$ & $0.626$ & $0.151$\\
\hline
\end{tabular}
\caption{\small Fiducial polarization fractions of $W^-_{}$ and $Z$ bosons in
  the process $pp \to e^-_{} \bar{\nu}_e^{}\, \mu^+_{} \mu^-_{} + X$ at the
  13 TeV LHC with the ATLAS fiducial cuts. Results are shown for full LO only, 
and also with the $\text{EW-}\bar{q}q'$, $\text{EW-}q\gamma$, and the total 
EW correction included. Similarly, results for DPA LO only, and also with 
the $\text{EW-pV}$, $\text{EW-dV}$, and the total EW correction included are presented. 
The upper rows are for the helicity (HE) coordinate system, while the lower ones 
for the Collins-Soper (CS) coordinate system.}
\label{tab:coeff_fL0R_full_DPA_Wm_Atlas}
\end{center}
\end{table} 
\begin{table}[ht!]
 \renewcommand{\arraystretch}{1.3}
  \begin{center}
\fontsize{8.5}{8.5}
\begin{tabular}{|c|c|c|c||c|c|c|}\hline
$\text{Method}$  & $f^{W^-}_L$ & $f^{W^-}_0$ & $f^{W^-}_R$ & $f^Z_L$ & $f^Z_0$ & $f^Z_R$\\
\hline
$\text{HE F-LO}$ & $0.248$ & $0.503$ & $0.249$ & $0.384$ & $0.399$ & $0.217$\\
$\text{HE F-EW-}\bar{q}q'$ & $0.246$ & $0.505$ & $0.249$ & $0.357$ & $0.399$ & $0.244$\\
$\text{HE F-EW-}q\gamma$ & $0.252$ & $0.499$ & $0.249$ & $0.384$ & $0.398$ & $0.218$\\
$\text{HE F-EW}$ & $0.250$ & $0.501$ & $0.249$ & $0.358$ & $0.398$ & $0.245$\\
\hline
$\text{HE D-LO}$ & $0.275$ & $0.492$ & $0.233$ & $0.395$ & $0.381$ & $0.224$\\
$\text{HE D-EW-pV}$ & $0.278$ & $0.488$ & $0.234$ & $0.397$ & $0.380$ & $0.223$\\
$\text{HE D-EW-dV}$ & $0.274$ & $0.494$ & $0.232$ & $0.366$ & $0.380$ & $0.254$\\
$\text{HE D-EW}$ & $0.277$ & $0.490$ & $0.233$ & $0.368$ & $0.379$ & $0.252$\\
\hline\hline
$\text{CS F-LO}$ & $0.214$ & $0.507$ & $0.279$ & $0.280$ & $0.557$ & $0.163$\\
$\text{CS F-EW-}\bar{q}q'$ & $0.214$ & $0.506$ & $0.280$ & $0.258$ & $0.560$ & $0.182$\\
$\text{CS F-EW-}q\gamma$ & $0.215$ & $0.504$ & $0.281$ & $0.277$ & $0.560$ & $0.162$\\
$\text{CS F-EW}$ & $0.215$ & $0.504$ & $0.281$ & $0.256$ & $0.564$ & $0.180$\\
\hline
$\text{CS D-LO}$ & $0.223$ & $0.503$ & $0.274$ & $0.283$ & $0.552$ & $0.165$\\
$\text{CS D-EW-pV}$ & $0.225$ & $0.498$ & $0.277$ & $0.281$ & $0.556$ & $0.163$\\
$\text{CS D-EW-dV}$ & $0.223$ & $0.504$ & $0.273$ & $0.261$ & $0.556$ & $0.184$\\
$\text{CS D-EW}$ & $0.225$ & $0.499$ & $0.276$ & $0.259$ & $0.559$ & $0.182$\\
\hline
\end{tabular}
\caption{\small Same as \tab{tab:coeff_fL0R_full_DPA_Wm_Atlas} but with 
the CMS fiducial cuts.}
\label{tab:coeff_fL0R_full_DPA_Wm_Cms}
\end{center}
\end{table} 

\clearpage
\pagebreak

\acknowledgments

We would like to thank the anonymous referee for their very 
helpful comments and suggestions leading to a substantial 
improvement of the paper over the first version. 
We thank Peter Uwer for an interesting discussion. 
L.D.N. thanks Nguyen Quoc Viet, Vuong Pham Ngoc Hoa, 
and Dieter Zeppenfeld for 
fruitful discussions on polarization observables.   
J.B. acknowledges the support from the
Carl-Zeiss foundation. He also acknowledges the German Academic
Exchange Service (DAAD) for travel support for his stay at ICISE in
2018 as part of a seminar invitation. 
The work of L.D.N. has been partly
supported by the German Ministry of Education and Research (BMBF) under contract no.
05H15KHCAA during the period 2015-2016 when he worked 
in Berlin. This research is funded by the Vietnam
National Foundation for Science and Technology Development (NAFOSTED)
under grant number 103.01-2017.78. L.D.N. also acknowledges the support from
DAAD to perform the final stage of this work at the University of
T\"ubingen under a scholarship. He thanks the members of the 
Institut f\"{u}r Theoretische Physik for their hospitality.  
Parts of this work were performed thanks to the support of the State of Baden-W\"urttemberg
through bwHPC and the German Research Foundation through the grant no. INST 39/963-1 FUGG.


\bibliographystyle{JHEP}
\bibliography{wz_polarization}

\end{document}